\documentclass[reprint,8pt]{elsarticle}
\usepackage[utf8]{inputenc}
\usepackage{hyperref}
\usepackage{mathtools}
\usepackage{footnote}
\makesavenoteenv{tabular}

\usepackage{graphicx,amsmath,amssymb}
\usepackage[margin=34mm]{geometry}
\usepackage{longtable}
\usepackage{booktabs}
\usepackage{soul}
\usepackage{tabularx}
\usepackage{ltablex}
\usepackage{xcolor}
\usepackage{braket}
\usepackage{footnote}
\usepackage{xr}
\usepackage{prettyref}
\usepackage{url}
\usepackage{mathtools}

\DeclarePairedDelimiter\floor{\lfloor}{\rfloor}

\newcommand{\vek}[1]{\boldsymbol{#1}}

\newcommand{\mat}[1]{\mathsf{#1}}

\usepackage{lineno}

\biboptions{sort&compress}

\begin{document}

\begin{frontmatter}


 \title{Nestedness in complex networks: observation, emergence, and implications}


\author[label1,label2]{Manuel Sebastian Mariani}
\ead{manuel.mariani@business.uzh.ch}

\author[label3]{Zhuo-Ming Ren}

\author[label4]{Jordi Bascompte}

\author[label2]{Claudio Juan Tessone}
\ead{claudio.tessone@uzh.ch}

\address[label1]{Institute of Fundamental and Frontier Sciences, University of Electronic Science and Technology of China, Chengdu 610054, PR China}
\address[label2]{URPP Social Networks, University of Zurich, CH-8050 Zurich, Switzerland}
\address[label3]{Alibaba Research Center for Complexity Sciences, Alibaba Business School, Hangzhou Normal University, Hangzhou 311121, China}
\address[label4]{Department of Evolutionary Biology and Environmental Studies, University of Zurich, Winterthurerstrasse 190, CH-8057 Zurich, Switzerland}


\begin{abstract}
The observed architecture of ecological and socio-economic networks differs significantly from that of random networks. From a network science standpoint, non-random structural patterns observed in real networks call for an explanation of their emergence and an understanding of their potential systemic consequences. This article focuses on one of these patterns: nestedness. Given a network of interacting nodes, nestedness can be described as the tendency for nodes to interact with subsets of the interaction partners of better-connected nodes. 
Known since more than $80$ years in biogeography, nestedness has been found in systems as diverse as ecological mutualistic organizations, world trade, inter-organizational relations, among many others. This review article focuses on three main pillars: the existing methodologies to \emph{observe} nestedness in networks; the main theoretical mechanisms conceived to explain the \emph{emergence} of nestedness in ecological and socio-economic networks; the \emph{implications} of a nested topology of interactions for the stability and feasibility of a given interacting system. We survey results from variegated disciplines, including statistical physics, graph theory, ecology, and theoretical economics. Nestedness was found to emerge both in bipartite networks and, more recently, in unipartite ones; this review is the first comprehensive attempt to unify both streams of studies, usually disconnected from each other. We believe that the truly interdisciplinary endeavor -- while rooted in a complex systems perspective -- may inspire new models and algorithms whose realm of application will undoubtedly transcend disciplinary boundaries.
\end{abstract}


\end{frontmatter}



\clearpage
\tableofcontents
\newpage

\clearpage

\section{Introduction}

Network science constitutes one of the pillars of the modern science of complexity. Broadly speaking, network science can be defined as the ``study of the collection, management, analysis, interpretation, and presentation of relational data"~\cite{brandes2013network}.
Given a complex system with many interacting components, a suitable network representation simplifies its analysis by describing it -- in the simplest portrayal -- as a collection of nodes and links connecting them.
The working hypothesis is that while the obtained network represents a simplification of the complex system it was derived from, it carries enough information to allow us to understand the functioning of the system and the emergence of collective phenomena from the interactions of its constituents. 
Leveraging its inherently interdisciplinary nature, network science borrows ideas and tools from a variety of research fields, especially those where networks (also. especially in mathematics and computer science, termed \textit{graphs}) have been long studied: graph theory~\cite{chung1997spectral}, social science~\cite{wasserman1994social}, economics~\cite{jackson2010social}, scientometrics~\cite{price1965networks,small1973co}, computer science~\cite{chartrand1993applied}, and ecology~\cite{jordano1987patterns,bascompte2013mutualistic}, among others.

Perhaps one of the most intriguing features of real networks is the existence of common structural and dynamical patterns that are found in a large number of systems from various domains of science, nature, and technology. The pervasiveness of structural patterns in systems from various fields makes it possible to analyze them with a common set of tools~\cite{barabasi2016network}. A popular example of such widespread structural patterns is the heavy-tailed degree distribution~\cite{newman2010networks} (i.e., the distribution of the number of connections per node). Power-law degree distributions (often termed as ``scale-free") have been reported in a wide variety of different systems, ranging from social and information networks to protein-protein interaction networks (see~\cite{caldarelli2007scale} for a dedicated book).
The ubiquity of heavy-tailed degree distributions has motivated a large number of studies aimed at unveiling plausible mechanisms that explain their emergence ~\cite{barabasi1999emergence,barabasi2016network}, understanding their implications for spreading ~\cite{pastor2001epidemic,pastor2015epidemic}, network robustness~\cite{albert2000error}, synchronization phenomena~\cite{moreno2004synchronization,arenas2008synchronization}, etc.

This review article focuses on one of these widely-observed (and, as a result, widely studied and debated) network structural patterns: \emph{nestedness}.
Informally speaking, nestedness refers to a hierarchical organization
where the set of neighbors of a node is a subset (superset) of the neighbors of lower (larger) degree.
Originally conceived~\cite{hulten1937outline,darlington1957zoogeography} and discovered~\cite{patterson1986nested} in biogeography where it was found in the spatial distribution patterns of species~\cite{patterson1986nested,atmar1993measure}, nestedness has been found in a wide variety of systems, including ecological interaction networks~\cite{bascompte2003nested}, trade networks~\cite{tacchella2012new,konig2014nestedness,saracco2016detecting}, inter-organizational networks~\cite{saavedra2009simple,saavedra2011strong}, firm spatial networks~\cite{bustos2012dynamics,garas2018economic}, interbank payment networks~\cite{soramaki2007topology,konig2014nestedness}, social-media information networks~\cite{borge2017emergence}, among others. 
Nestedness has been observed in two distinct families of networks. 
On the one hand, \textit{bipartite networks} (in which nodes of two categories are connected in some way), like mutualistic networks in ecology.
On the other hand, more recently, nestedness has been found also in unipartite networks, i.e.,~those where all the nodes are of the same nature, like interacting agents in economics. The two streams of research have run mostly in parallel, paying little or no attention to each other. To the best of our knowledge, for the first time, this review provides an in-depth discussion of both research areas. 

The widespread occurrence of a given pattern in systems of different nature naturally leads to a number of questions related to the fundamental mechanisms behind its emergence and its implication for the systems' functioning and preservation.
Based on the massive amount of works on nestedness, this review will address several questions: Which classes of networks exhibit a nested architecture? What are the possible mechanisms behind the emergence of nestedness in ecological and socio-economic networks? Is the emergence of nestedness the result of an optimization process where constituents tend to maximize their fitness~\cite{suweis2013emergence}, or is it a consequence of the assemblage rules of the system~\cite{valverde2018architecture,maynard2018network}? Which mechanisms are responsible for the emergence of nestedness in social systems and trade networks?~\cite{konig2011network,konig2014nestedness}
Is nestedness beneficial or harmful for the stability~\cite{allesina2012stability} and the persistence~\cite{rohr2014structural} of a given ecological community? How is nestedness related to other widely observed network properties, such as modularity~\cite{fortuna2010nestedness,sole2018revealing} and core-periphery structure~\cite{lee2016network}? Given a network, is nestedness a global property, or a property that is only respected by specific subsets of the network?~\cite{strona2013protocol,grimm2017analysing,sole2018revealing,kojaku2018core}
How can we exploit nestedness for predicting future properties of economic systems, such as the gross domestic product of countries~\cite{tacchella2018dynamical} and the appearance of a new firm at a given geographic location~\cite{bustos2012dynamics}? 

Admittedly, some of these questions have been controversial in the literature, and sometimes, contradictory results have been supported by different studies. A complete coverage of all existing theses on each topic would be impossible to attain. However, when necessary, we will try our best to present the most relevant findings that have been used to support competing claims.

This review article aims at providing the reader with a cross-disciplinary perspective on nestedness, the subtleties associated with its observation, the possible mechanisms behind its emergence, and its implications.
Driven by these questions, our review is an interdisciplinary effort where we collected results from network science, statistical physics, ecology,  economics, social sciences, among others, which makes our review fundamentally different from recent ones~\cite{bascompte2007plant,ings2009ecological,ulrich2009consumer,dormann2009indices} and books~\cite{bascompte2013mutualistic} which have focused exclusively on the structure and dynamics of ecological networks. 
We hope that this effort will allow researchers who have been interested in nestedness to learn the methodologies and findings of scholars from other fields, which might accelerate the progress of research on this area and, at the same time, reveal similar organizing principles across disciplines~\cite{saavedra2009simple}. 

A substantial part of the methodologies detailed in this article come from statistical physics and dynamical systems: population dynamics~\cite{bastolla2009architecture,rohr2014structural}, network formation models~\cite{konig2011network,saracco2015innovation}, dynamical system stability analysis~\cite{allesina2012stability}, fitness optimization techniques~\cite{suweis2013emergence,kojaku2018core,sole2018revealing}, prediction based on dynamical systems~\cite{tacchella2018dynamical}. 
Other methodologies and results come from theoretical ecology~\cite{rohr2014structural,valverde2018architecture,maynard2018network}, theoretical economics~\cite{hausmann2011network,konig2014nestedness}, and graph theory~\cite{masuda2004analysis,hagberg2006designing}.
We will always try to provide sufficient background information for all the methodologies and problems presented throughout the review.

\subsection{What are nested networks?} 
\label{sec:nested}

Before detailing the historical developments of nestedness, we provide the definition of perfectly nested networks together with the basic ideas behind nestedness analysis.
In abstract terms, we refer here to systems that are composed of many constituents. 
Said elements are not isolated but they interact with (or are related to) others. 
If these interactions can be described as dyadic, the system admits a representation in terms of a \emph{network}~\cite{jackson2010social,newman2010networks,barabasi2016network}: its constituents are represented as \textit{nodes} (or \textit{vertexes}) and their mutual interactions (or relations) as \textit{links} (or \textit{edges}, or \emph{bonds})\footnote{This is not the only way to represent relational data. For example, temporal~\cite{holme2012temporal,holme2015modern} and multilayer~\cite{boccaletti2014structure,kivela2014multilayer} networks provide  with more sophisticated representations that allow us to take into account temporal effects and multiple types of interactions, respectively.}. 
The set of nodes connected to a certain node constitute its \textit{neighborhood}, and the number of such connections is referred to as its \textit{degree}.

For example, if the network under consideration is a friendship network, two individuals are linked 
when they have amity between them. In a trade network, whose constituents are countries, an edge is present if there exists a trade relationship between two given countries. In the examples above the elements of the system (individuals or countries), while heterogeneous, are of the same nature; it is then said that the network is \textit{unipartite}. 
There are other networks whose constituents can be divided into two distinct classes. 
For example, in a habitat, a set of pollinator species and a set of flowering plants are mutually related if the former helps in the pollination (and feeds from) the latter. 
In this case, relations connect agents of a distinct nature. 
This construct is called a \textit{bipartite} network. 

\begin{figure}[t]
\centering
 \includegraphics[scale=0.25]{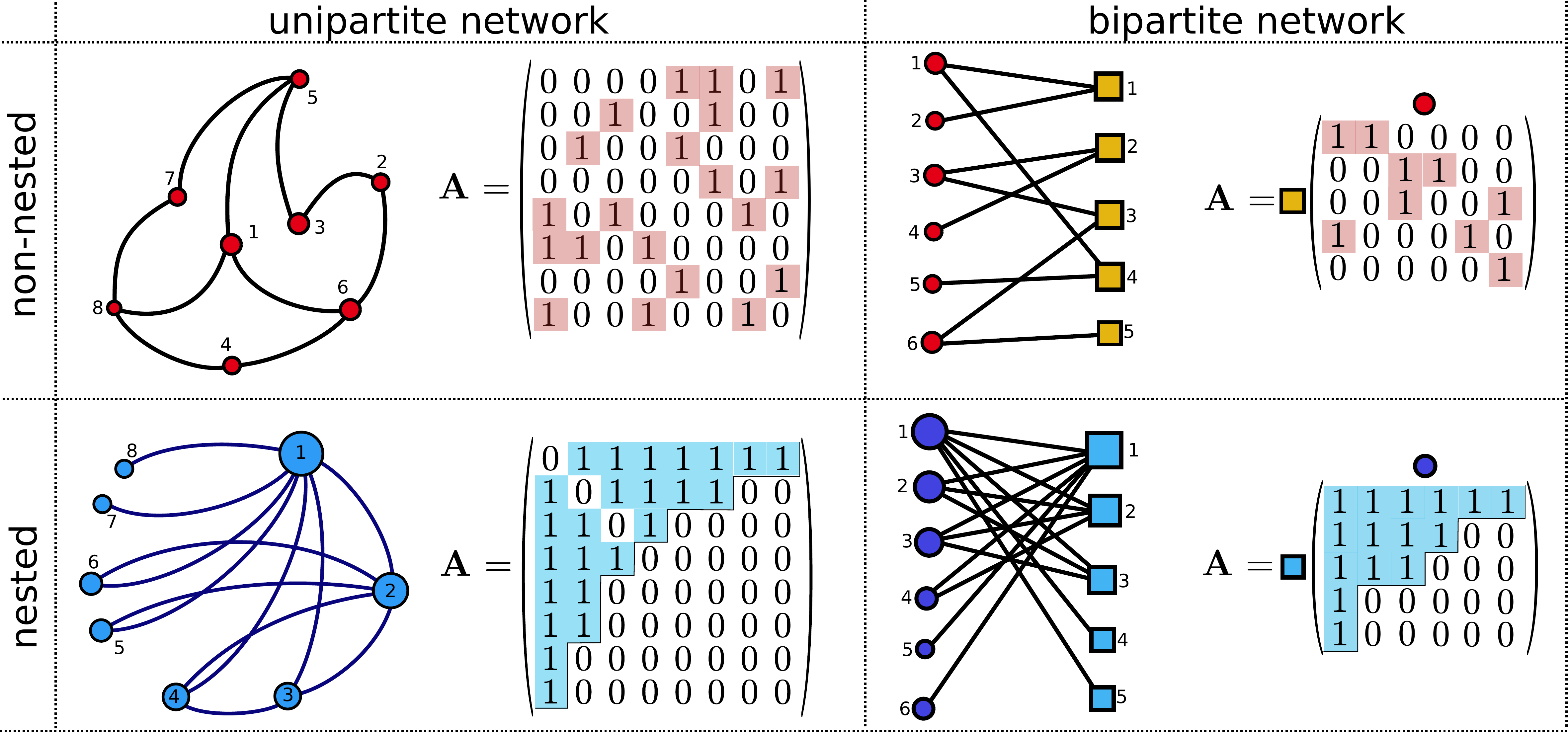}
 \caption{\label{fig:nested-nets} Simple examples of nested and non-nested networks in both cases: unipartite and bipartite networks. The adjacency matrices of the perfectly nested networks (bottom panels) exhibit a clear ``triangular'' shape: a separatrix partitions the matrices into unambiguously-defined filled regions (i.e., regions that only contain ones) and empty regions (i.e., regions that only contain zeros). In this review, we refer to this kinds of matrices as perfectly nested matrices.  }
\end{figure}

A \emph{perfectly nested network} is formally defined in the following terms: 
consider any two nodes $i$ and $j$; if the degree of $i$ is smaller than the degree of $j$, then the neighborhood of $i$ is contained in the neighborhood of $j$.
This definition is generally valid for unipartite and bipartite networks with a single proviso: for the second group, the nodes to be compared must belong to the same class. 
Any network here considered accepts a representation in terms of a binary adjacency matrix $\mat{A}$ whose elements $A_{ij}$ are equal to $1$ ($0$) if the two vertices are connected (not connected).
For the particular case of nested networks, the adjacency matrix has a very peculiar shape (see lower row in Fig.~\ref{fig:nested-nets}): after re-ordering its rows and columns by degree, there exists a monotonic separatrix above which all the elements are equal to one, and zero below\footnote{Strictly speaking, in bipartite network analysis, the matrix that connects nodes of different type is typically referred to as biadjacency matrix~\cite{straka2018ecology} or incidence matrix~\cite{newman2010networks}, whereas the adjacency matrix still refers to the matrix whose elements represent all possible pairs of nodes -- the elements that connect nodes of the same type are, by definition, all equal to zero. To simplify the discussion, when we will present methods for bipartite networks that apply equally well to unipartite networks, we will refer to the biadjacency matrix as the adjacency matrix. This abuse of terminology has no harmful consequences as the elements of the adjacency matrix that connect nodes of the same type are all equal to zero.}. 

The first observation is that this definition can accommodate topologies as different as a star network (a highly centralized topology where one hub is connected to all the other nodes) and a fully-connected network (where each node is connected with everyone else).
It is also clear, however, that this is a very stringent network topology which is seldom (if ever) observed in real-world networks. 
Therefore, \textit{a priori}, this kind of network structure may seem a kind of curiosity. However, research has shown that networks showing this property abound in a wide variety of research fields, and its definition respected to a large extent: 
the vast majority of links are located in nested neighborhoods as defined above. Asking ``how much of a network structure respects the property of nestedness'' gave rise to several measures that aim at quantifying the \textit{degree of nestedness} of a given network, and several null models to quantify the statistical significance of the observed levels of nestedness.
In this review, when we say that a network is ``nested", we mean that it exhibits a level of nestedness (according to a given metric for nestedness) that cannot be explained by a reasonable null model.

From this definition, it is evident that metrics for nestedness and null models used to produce random networks that respect given constraints are of vital importance~\cite{gotelli2000null,ulrich2007null,gotelli2012statistical}.
Different metrics and null models may be responsible for the validation or rejection of the hypothesis of nestedness significance, and different combinations of metrics and null models can lead to different conclusions~\cite{strona2014methods}.
Therefore, a proper definition of null models is crucial, and far from trivial: null models purely based on preserving simple topological properties can be either too restrictive -- when the full degree sequences must be exactly preserved in bipartite networks, for example -- offering little freedom from the observed network~\cite{ulrich2007null,strona2014methods,strona2018bi}; or too lax -- like requiring only that the network density is preserved -- in which only rarely the null hypothesis that the empirical degree of nestedness in a network is significant can be rejected~\cite{connor1979assembly,jonsson2001null,ulrich2009consumer,joppa2010nestedness}. These issues will be discussed in Section~\ref{sec:null} and~\ref{sec:degree_distribution}.

The notion of nestedness introduced above is based on degree.
Unless otherwise stated, when we refer to nestedness throughout this review, we refer to the nestedness by degree defined above. One can also take a different perspective and ask whether a network exhibit a significant level of nestedness when pairwise node orderings build on a node-level scalar property different than the degree -- e.g., species abundance~\cite{krishna2008neutral} and islands' area~\cite{kadmon1995nested}.
In other words, one investigates whether the network's adjacency matrix exhibit a significant degree of nestedness once its rows and columns are sorted by a given property $s$.
This consideration leads to two additional notions of nestedness that have been studied:
\begin{itemize}
\item \emph{Nestedness by other scalar properties.} Consider a node-level scalar property $s$. One can re-define nestedness based on property $s$: the neighborhoods of two given  nodes $i$ and $j$ such that $s_i<s_j$ are nested by property $s$ if and only if the neighborhood of $i$ is a subset of the neighborhood of $j$. The network is highly nested by property $s$ if many of its pairs of nodes' neighborhoods are nested by $s$. In other words, a network is highly-nested by $s$ if, after ordering the adjacency matrix's rows and columns by $s$, the resulting matrix exhibits a clear separation between its filled and its empty region. 
\item \emph{Maximal nestedness.} One seeks to find the ordering of rows and columns that maximizes the level of nestedness of the network. Scholars have especially aimed to find maximal-nestedness structures in the context of nestedness temperature analysis of ecological networks~\cite{atmar1993measure,rodriguez2006new} (see Section~\ref{sec:distance}).
\end{itemize}
Based on the definitions provided above, one can compute, for instance, the level of nestedness of an ecological mutualistic system based on species abundance, and compare the resulting nestedness with the maximal nestedness that can be achieved by the system~\cite{krishna2008neutral}. In biogeography, one can study the nestedness of a species-island network by both degree and island-level environmental variables, including area and distance from the mainland -- see~\cite{kadmon1995nested} for a comparative analysis.

Besides, nestedness-maximizing ranking algorithms have found important applications for the prediction of the future development of countries~\cite{tacchella2012new,tacchella2018dynamical} and the identification of the most vulnerable constituents in ecological networks~\cite{dominguez2015ranking} and world trade networks~\cite{mariani2015measuring}.
When defining existing metrics for nestedness (Section~\ref{sec:metrics}), we will specify which aspect of nestedness they capture. It is worth noticing that metrics for nestedness that take as input a pairwise ordering of the nodes can be applied to measure both nestedness by degree and nestedness by other scalar properties. On the other hand, algorithms that seek to maximize nestedness (like those used to minimize nestedness temperature~\cite{atmar1993measure,rodriguez2006new,lin2018nestedness}) cannot be used to evaluate the level of nestedness by a specific node-level property; however, one can compare the resulting maximal nestedness with the degree of nestedness by a property of interest~\cite{krishna2008neutral}.

\subsection{Historical background}
\label{sec:histo}


While the concept of nestedness was precisely formulated by Patterson and Atmar~\cite{patterson1986nested}, the preceding literature offers studies~\cite{hulten1937outline,darlington1957zoogeography} where the notion of nestedness was already implicitly discussed.
In particular, in his book \emph{Zoogeography}~\cite{darlington1957zoogeography}, Darlington described the potential effect of spatial distance on the patterns of distribution of species on islands. He considered two possible mechanisms of animal over-water dispersal when moving from the continent to islands. He noticed that other things being equal, individuals of a given species might be expected to populate first the nearest island, and then move to the following nearest islands by crossing water gaps of minimal length. This occupation process, referred to as \emph{immigrant pattern} by Darlington~\cite{darlington1957zoogeography} and \emph{selective immigration} by subsequent studies~\cite{cook1995influence,ulrich2009consumer}, naturally results in a nested structure where the species found on the farthest islands (i.e., the species with the higher dispersal rate~\cite{cook1995influence}) are also found in the nearest islands (see Fig.~57 in~\cite{darlington1957zoogeography}).

Besides, between the 70s and the 80s, ecologists became increasingly interested in the spatial distribution patterns of species and statistical null models to assess their statistical significance\footnote{In this respect, it is surprising that a network null model that preserves the individual nodes' degrees, typically referred to as \emph{configuration model} in network science~\cite{newman2010networks}, was already designed in 1979 by Connor and Simberloff~\cite{connor1979assembly}.}~\cite{sale1974overlap,inger1977organization,connor1979assembly,gilpin1982factors}. It was not until 1986 that the notion of \emph{nestedness} was introduced together with the first metric to measure it. This was done by Patterson and Atmar~\cite{patterson1986nested} who found that several species spatial distribution patterns exhibit a highly nested structure that cannot be explained by null models. In a subsequent study, the same authors introduced the popular nestedness temperature~\cite{atmar1993measure}, a metric for nestedness which has been applied in a large number of papers in ecology~\cite{ulrich2009consumer}.

In parallel, graph theory had started investigating \textit{threshold graphs} and \textit{nested split graphs}, which constitute classes of graphs that are equivalent to perfectly nested graphs (see Section~\ref{sec:nested_graph}). Remarkably, the term ``nested bipartite graph'' together with the first mathematical definition of a perfectly nested bipartite graph already appeared in a paper~\cite{hering1971nested} published by the mathematician Franz Hering in 1971. Rigorous results for threshold graphs and nested split graphs are reviewed in the 1995 book by Mahadev and Peled~\cite{mahadev1995threshold}.

In 2003, Bascompte \textit{et al.}~\cite{bascompte2003nested} brought the concept of nestedness to ecological interaction networks by analyzing $25$ plant-pollinator networks and $27$ plant-frugivore networks, finding that most of them exhibit a degree of nestedness that cannot be explained by degree-preserving null models.
Motivated by this finding, scholars have aimed at uncovering the basic mechanisms that are potentially responsible for the emergence of nestedness (see Section~\ref{sec:formation_ecol}). Besides, a large number of works have attempted to understand how nestedness impacts both the structural and the dynamical stability of the system. Scholars have investigated this relation by means of simulating co-extinction cascades~\cite{dominguez2015ranking}, numerical simulations and analytic results on models of mutualistic and competitive interactions~\cite{bastolla2009architecture,rohr2014structural}, random-matrix theory~\cite{allesina2012stability}. We refer to Section~\ref{sec:implications} for the details.

Stimulated by works on ecological networks,
interest in nestedness analysis has arisen in socio-economic networks as well. In 2007, Soramäki \textit{et al.}~\cite{soramaki2007topology} found that the interbank payment network is organized in a topology with a densely connected core and a periphery that is only connected with the core: a structure that is compatible with nestedness.
Saavedra \textit{et al.}~\cite{saavedra2009simple} found that manufacturer-contractor networks exhibit a nested structure, and proposed a parsimonious  model to explain its emergence. K\"onig \textit{et al.}~\cite{konig2014nestedness} found that four economic networks are significantly nested: the Fedwire network of settlements, Austrian inter-bank network, the world trade network and the worldwide arm trade network. Besides, they proposed a network formation model to explain its emergence. 

As for bipartite trade networks, Saracco \textit{et al.}~\cite{saracco2016detecting} found that when analyzing the bipartite country-product trade network by means of standard nestedness analysis tools, a decrease of nestedness took place before the 2007-2008 financial crisis. In parallel, Tacchella \textit{et al.}~\cite{tacchella2012new} have designed a non-linear ranking algorithm to quantify the competitiveness of countries and the sophistication of products. In fact, the algorithm sorts rows and columns of the network's adjacency matrix in such a way that the nested architecture of the system is revealed~\cite{tacchella2012new,dominguez2015ranking}, and it often outperforms genetic algorithms from the ecological literature in maximizing nestedness~\cite{lin2018nestedness}.

\begin{table}[t]
\begin{center}
\resizebox{\textwidth}{!}{%
\begin{tabular}{ |p{3.5cm}|p{3.75cm}|p{3.75cm}|p{3.75cm}| } 
 \hline
\textbf{Symbol}&\textbf{Network science} & \textbf{Graph theory} & \textbf{Ecology} \\ 
 \hline
 $\mathcal{G}$ & Network & Graph & Community \\ 
 \hline
$i\in\{1,\dots,N\}$ & Node & Vertex & Species \\
 \hline
$N$ & Network size & Graph order & Community size \\
 \hline
$(i,j)$ & Link, edge, bond & Edge, arc & Interaction \\
 \hline
 $\mat{A}=\{A_{ij}; A_{ij}=1 \text{ iff } (i,j) \text{ is observed}\}$ & Adjacency matrix & Adjacency matrix & Interaction matrix  \\
 \hline
 $E=\sum_{ij}A_{ij}$ & Number of links/edges & Graph size & Number of pairwise interactions  \\
 \hline
 $\rho=2\,E/(N\,(N-1))$ & Connectance or Density & Graph density & Fill  \\
 \hline
$\mathcal{N}_i=\{j|A_{ij}=1\}$ & Neighborhood & Neighborhood &  Set of interactors \\
 \hline
 $k_i=\sum_{j}A_{ij}$ & Degree & Degree, valency &  Marginal total \\
 \hline
 $\braket{k}=N^{-1}\sum_{j}k_{j}=2\,E/N$ & Average degree &  Average degree & Average number of interactions per species \\
 \hline
Node $i$ with $k_i\gg \braket{k}$ & High-degree node, hub & High-degree node & Generalist \\
 \hline
Node $i$ with $k_i\sim 1$ & Low-degree node & Low-degree node & Specialist  \\
 \hline
$O_{ij}=\sum_{k}A_{ik}\,A_{jk}$ & Common neighbors & Common neighbors & Overlap  \\
\hline
$\Xi_i$ & $i$'s Community & $i$'s Cluster & $i$'s Compartment \\ 
 \hline
\end{tabular}}
\end{center}
\caption{Notation for binary unipartite networks. Mathematically equivalent concepts are expressed through different terms in different disciplines. In this review, we will mostly use the network science nomenclature, yet we will sometimes switch to the ecology language when presenting results in ecology.}
\label{tab:unipartite}
\end{table}

\begin{table}[t]
\begin{center}
\resizebox{\textwidth}{!}{%
\begin{tabular}{ |p{4cm}|p{3cm}|p{3cm}|p{3cm}|p{3cm}| } 
 \hline
\textbf{Symbol}&\textbf{Network science} & \textbf{Graph theory} & \textbf{Ecology} & \textbf{Biogeography} \\ 
 \hline
 $\mathcal{G}$ & Network & Graph & Community & Region, archipelago \\ 
 \hline
 $\mathcal{R,C}$ & Two disjoint sets of nodes & Two disjoint sets of vertices & Set of active and passive species & Set of patches and species \\
  \hline
$i\in\{1,\dots,N\}$ & Class-$\mathcal{R}$'s node, Row-nodes & Class-$\mathcal{R}$'s vertex & Active species & Species \\
 \hline
 $\alpha\in\{1,\dots,M\}$ & Class-$\mathcal{C}$'s node, Column-node & Class-$\mathcal{C}$'s vertex & Passive species & Patches, islands, sites \\
  \hline
 $S=N+M$ &  Network size & Graph order & Community size & \\
 \hline
$(i,\alpha)$ & Link, edge, bond & Edge, arc & Interaction & Presence \\
 \hline
 $\mat{A}=\{A_{i\alpha}; A_{i\alpha}=1 \text{ iff } (i,\alpha) \text{ is observed}\}$ & (Bi-)Adjacency or Incidence matrix & (Bi-)Adjacency or Incidence matrix & Interaction matrix & Presence-absence matrix \\
 \hline
 $E=\sum_{i,\alpha}A_{i\alpha}$ & Number of links/edges & Graph size & Number of pairwise interactions &  \\
 \hline
 $\rho=E/(N\,M)$ & Connectance, Density & Graph density & Fill & Fill \\
 \hline
$\mathcal{N}_i=\{\alpha|A_{i\alpha}=1\}$ & $i$'s Neighborhood & $i$'s Neighborhood & Set of $i$'s partners & Set of sites where $i$ is present \\
\hline
$\mathcal{N}_\alpha=\{i|A_{i\alpha}=1\}$ & $\alpha$'s Neighborhood & $\alpha$'s Neighborhood & Set of $\alpha$'s partners & Set of $\alpha$'s present species  \\
\hline
 $k_i=\sum_{\alpha}A_{i\alpha}$ & Degree & Degree, valency & Row's marginal total or Active species' degree & Species frequency \\
 \hline
   $k_{\alpha}=\sum_{i}A_{i\alpha}$ & Degree & Degree & Column's marginal total or Passive species' degree & Site richness \\
 \hline
 $\braket{k^R}=N^{-1}\sum_{j}k_j$ & Average degree of class-$\mathcal{R}$ nodes & Average degree of class-$\mathcal{R}$ nodes & Average rows' marginal total & Average species frequency \\
 \hline
 $\braket{k^C}=M^{-1}\sum_{\alpha}k_\alpha$ & Average degree of class-$\mathcal{C}$ nodes & Average degree of class-$\mathcal{C}$ nodes & Average columns' marginal total & Average site richness \\
 \hline
Node $i$ with $k_i\gg \braket{k^R}$ & Hub & Hub & Generalist & Ubiquitous species \\
 \hline
 Node $i$ with $k_i\sim 1$ & Low-degree node & Low-degree node & Specialist & Rare species  \\
 \hline
$O_{ij}=\sum_{\alpha}A_{i\alpha}\,A_{j\alpha}$ & Common neighbors & Common neighbors & Rows' Overlap & Rows' Overlap  \\
 \hline
$O_{\alpha\beta}=\sum_{i}A_{i\alpha}\,A_{i\beta}$ & Common neighbors & Common neighbors & Columns' Overlap & Columns' Overlap \\
\hline
$\Xi_i$ & $i$'s Community & $i$'s Cluster & $i$'s Compartment  & \\ 
 \hline
\end{tabular}}
\end{center}
\caption{Notation for binary bipartite networks. Mathematically equivalent concepts are expressed through different terms in different disciplines. In this review, we will mostly use the network science nomenclature, yet we will sometimes switch to the ecology language when presenting results in ecology.}
\label{tab:bipartite}
\end{table}

\subsection{How to read this article}

The main goal of this article is to review three main aspects of nestedness: its observation, the modeling of its emergence, and its implications.
We detail below which Sections specifically deal with each of these three aspects:
\begin{itemize}
\item \textbf{Observing nestedness.} Section~\ref{sec:systems} provides an overview of the classes of real networks of interest for this review. Sections~\ref{sec:metrics} addresses the questions: How to measure the level of nestedness of a given system? How to choose a suitable null model to assess the statistical significance of the observed level of nestedness? How to maximize the degree of nestedness of a given network by properly ranking its nodes?
Section~\ref{sec:mesoscopic} surveys the methods that aim to quantify nestedness at a mesoscopic scale, which shifts the question from assessing whether a given system is nested to finding sub-components of the system such that the nodes that belong to the found components exhibit a nested interaction topology.
\item \textbf{Modeling the emergence of nestedness.} Section~\ref{sec:emergence} deals with possible mechanisms to explain the emergence of nestedness that have been identified in graph theory, ecology, and  economics. Proposed mechanisms range from optimization principles that postulate that nodes aim to maximize their fitness~\cite{suweis2013emergence} or their centrality 
\cite{medan2007analysis,konig2011network,bardoscia2013social,konig2014nestedness} when choosing their interactors, to stochastic processes where nestedness is the outcome of duplication and link randomization processes~\cite{valverde2018architecture}. No agreement on the main mechanisms behind the emergence of nestedness has been reached yet.
\item \textbf{Implications of nestedness.} Section~\ref{sec:implications} deals with both the implications of nestedness for network robustness against targeted attacks, for the stability and feasibility of equilibrium points of mutualistic dynamical processes on interaction networks. In this context, \emph{stability} refers to the ability of the system to return to its original equilibrium state after a small perturbation (\emph{local} stability) or after a perturbation of any magnitude (\emph{global} stability); \emph{feasibility} refers to the existence of equilibrium points such that all species are represented by at least one individual.
The latter problem is critical to the assessment of the impact of nestedness on the co-existence of species in ecological systems. The variety of methods adopted in the literature to tackle these problems has led to variegated conclusions.
\end{itemize}
The Outlook section will point to the current challenges in research in nestedness as well as the most promising unexplored directions in the topic. 

Due to its interdisciplinary nature, our review encompasses methodologies and findings
from diverse fields. 
For readers interested in different aspects of nestedness, we provide the following road-map:
\begin{itemize}
\item A reader interested in the \textbf{ecological} aspects of nestedness analysis may consider reading the following sections: Sections~\ref{sec:systems_ecol} for a short overview of the area; \ref{sec:metrics} to learn about the technical aspects of measuring nestedness; \ref{sec:abundance} to learn about the relation between nestedness and other properties; \ref{sec:formation_ecol} and \ref{sec:bottom1} for the main mechanisms that have been proposed in the ecological literature to explain the observed nestedness of interaction networks; \ref{sec:implications} for the implications of nestedness for systemic robustness, stability, and feasibility; \ref{sec:outlook} for an overview of current challenges and open directions in research on nestedness.
\item A reader interested in the \textbf{socio-economic} aspects of nestedness analysis may consider reading the following sections: Section~\ref{sec:systems_econ} for a brief overview of socio-economic networks of interest; \ref{sec:metrics} to learn about the technical aspects of measuring nestedness; \ref{sec:economic_properties} to learn about the applications of nestedness to predictive problems in socio-economic systems; \ref{sec:formation_socioec} to learn about possible social and economic mechanisms that can explain the emergence of nestedness in socio-economic systems; \ref{sec:outlook} for an overview of current challenges and open directions in research on nestedness. 
\item A reader interested in the \textbf{methodological} aspects of nestedness analysis may consider reading the following Sections: \ref{sec:nested_graph} to learn about the equivalence between perfectly nested networks, nested split graphs, and threshold graphs; \ref{sec:metrics} to learn about the technical aspects of measuring nestedness; \ref{sec:threshold} to learn how to generate perfectly nested networks with different degree distributions; \ref{sec:mesoscopic} to learn the technical aspects and main implications of detecting nestedness at the mesoscopic scale.
\item A reader who is already familiar with the literature of nestedness in ecological systems and would like to catch up with the \textbf{most recent developments} in the topic may consider reading the following Sections: Section \ref{sec:emergence} to discover recent mechanisms that have been proposed to explain the emergence of nestedness in interaction networks; \ref{sec:implications} to discover the recent developments on the implications of nestedness for systemic robustness, feasibility, and stability; \ref{sec:mesoscopic} to learn the technical aspects and main implications of detecting nestedness at the mesoscopic scale; \ref{sec:outlook} for an overview of current challenges and open directions in research on nestedness.
\end{itemize}

The notation of the review for unipartite and bipartite networks is shown in Table~\ref{tab:unipartite} and~\ref{tab:bipartite}, respectively.

\clearpage

\section{Classes of (potentially) nested networks}
\label{sec:systems}

In this Section, we present classes of graph-theoretical, ecological, economic and social systems where a nested architecture has been found. For each of these research fields, we provide a classification of the types of networks involved, and we mention some of the most representative papers that pointed out their nested architecture. The main goal of this Section is to provide readers with a brief introduction to the specific language of the systems studied in this review, without aiming to be exhaustive. For each of the systems presented below, wherever possible, we point to Web repositories where datasets can be found (see Appendix~\ref{appendix:material}).

\begin{table}[t]
\begin{center}
\begin{tabular}{ |p{2cm}|p{3cm}|p{3cm}|p{2cm}|p{3cm}| } 
 \hline
\textbf{Research field} & \textbf{System type} & \textbf{Row nodes} & \textbf{Column nodes} & \textbf{Links} \\
\hline
Ecology & Spatial network, region & Species & Island/site & Species occupy islands/sites  \\
\hline
Ecology & Pollination network & Pollinator animal & Plant & Pollinator pollinates plant  \\
\hline
Ecology & Seed-dispersal network & Frugivore animal & Plant & Frugivore disperses plant's seed \\
\hline
Economics & Interbank payment & Bank & Bank & Bank pays bank  \\
\hline
Economics & Manufacturer-contractor & Manufacturer & Contractor & Manufacturer outsources activity to contractor  \\
\hline
Economics & Trade & Country & Country & Country trades with country \\
\hline
Economics & International trade & Country & Product & Country exports product  \\
\hline
\end{tabular}
\end{center}
\caption{Main networks of interest in this review, together with the interpretation of their nodes and links. Networks where the row- and column-nodes are the same are unipartite.}
\label{tab:systems}
\end{table}

\subsection{Nested networks in graph theory}
\label{sec:nested_graph}

This article will focus on the ``physical'' aspects of nestedness, including the mechanisms that lead to its emergence and its implications for dynamical processes on networks. While rigorous results from graph theory will not constitute a central topic, it is instructive to acknowledge the deep connection between nested networks and classes of graphs that have been widely studied: threshold and split graphs~\cite{mahadev1995threshold}. The fact that many rigorous results have been obtained for these two classes of graphs~\cite{mahadev1995threshold} might inspire their application to the analysis of nested ecological and economic networks. 
Rigorous theorems on graph theory~\cite{bell2008graphs} have been used, for example, by Staniczenko \textit{et al.}~\cite{staniczenko2013ghost} to motivate the spectral radius of the adjacency matrix as a possible metric for nestedness (see Section~\ref{sec:eigenvalue}).
For the sake of completeness, we also mention that triangulated planar graphs are also nested networks~\cite{song2011nested} and can be leveraged for hierarchical clustering of empirical data~\cite{song2012hierarchical}. 
In the following, we narrow down our focus to unweighted unipartite networks (graphs), and define nested split and threshold graphs. 

\subsubsection{Nested split graphs}

\emph{Split graphs} have been originally introduced and studied by F{\"o}ldes and Hammer~\cite{foldes1976split} and, independently, by Chernyak and Chernyak~\cite{chernyak1986recognizing}. A graph $\mathcal{G}$ can be referred to as a split graph if and only if its nodes can be partitioned into a \emph{clique} (or \emph{dominating set}) $\mathcal{K}$ and a \emph{stable set} (or \emph{independent set}) $\mathcal{S}$ -- in this case, one denotes the split graph as $\mathcal{G}(\mathcal{K},\mathcal{S})$. In the graph-theoretical language, a \emph{clique}~\cite{luce1949method} is a subset of nodes such that every pair of nodes that belong to the clique is connected through a link; a \emph{stable set}~\cite{berge1957two} is a subset of nodes such that no two nodes are connected through a link. 

A split graph $\mathcal{G}(\mathcal{K},\mathcal{S})$ is referred to as \emph{nested split graph}~\cite{cvetkovic1990largest} if and only if the neighborhoods of the nodes that belong to the stable set $\mathcal{S}$ are nested: given two nodes $i$ and $j$ that belong to $\mathcal{S}$, if $k_i<k_j$, then the neighborhood of $i$ is included in the neighborhood of $j$\footnote{In graph theory, one says that vertex $j$ dominates vertex $i$~\cite{mahadev1995threshold}.}. By definition, a nested split graph is also a perfectly nested network. Such equivalence will be exploited in Section~\ref{sec:threshold} to introduce a generative algorithm for perfectly nested networks. The fingerprint of a nested split graph is a \emph{stepwise adjacency matrix}~\cite{brualdi1985spectral}: a symmetric binary matrix $\mat{A}$ whose elements $A_{ij}$ satisfy the property that if $i<j$ and $A_{ij}=1$, then $A_{hk}=1$ if $h<k<j$ and $h\leq i$. A stepwise matrix exhibits a peculiar ``triangular'' shape (see Fig.~\ref{fig:nested-nets}).

\subsubsection{Threshold graphs}

Importantly, nested split graphs are equivalent to an important class of graphs: the threshold graphs~\cite{mahadev1995threshold}.
\emph{Threshold graphs} were introduced in the 70s by Chv{\'a}tal and Hammer~\cite{chvatal1977aggregation}.
A graph is referred to as a threshold graph if and only if there exist non-negative real numbers $\{w_i\}, \theta$ such that a subset $\mathcal{U}$ of the nodes is stable if and only if its total weight $w(\mathcal{U}):=\sum_{i\in\mathcal{U}}w_i$ does not exceed the threshold $\theta$: $w(\mathcal{U})\leq\theta$.
One can prove (Theorem 1.2.4 in the book~\cite{mahadev1995threshold}) that given a graph $\mathcal{G}$, the property that $\mathcal{G}$ is a threshold graph is equivalent to the property that $\mathcal{G}$ is a nested split graph. Therefore, nested split graphs, threshold graphs, and perfectly nested networks are equivalent.

\subsection{Ecological networks}
\label{sec:systems_ecol}

As ecologists are typically interested in studying how species interact with one another and with their surroundings, networks naturally emerge as a powerful tool to study ecological systems.
We introduce here the basic concepts of three broad classes of ecological networks~\cite{bascompte2013mutualistic}: mutualistic networks, antagonistic networks, and spatial networks.

\subsubsection{Mutualistic networks} 

Mutualistic systems are typically represented as bipartite networks composed of two kinds of nodes, corresponding to animal and plant species. In such networks, \emph{mutualism} is the key element: the nodes of one class benefit from the interactions with the nodes of the other class. 
Different types of mutualism exist in nature. In line with~\cite{bascompte2013mutualistic}, in this review, we focus on two main types of mutualistic interactions: pollination and seed-dispersal. Both mechanisms involve the dissemination of an agent (pollen or seeds) performed by animals, in exchange for nutrients provided by the plant~\cite{wheelwright1982seed}. Yet, there are fundamental differences between the two types of interactions: for example, pollen has to be transferred to a very specific location (the stigma of a conspecific flower), whereas a similar spatial target is not necessary for seeds. We refer to~\cite{wheelwright1982seed} for an insightful discussion of analogies and differences between the two mechanisms. For our purposes, we consider mutualistic networks as bipartite networks where animals ("active species") interact with plants ("passive species").

The study of mutualistic systems has a long and fascinating history in ecology\footnote{We refer to Chapter One of the book~\cite{bascompte2013mutualistic} for a historical overview of studies of mutualism in ecology.}. Nestedness in both pollination and seed-dispersal mutualistic networks has been found in 2003 by Bascompte \textit{et al.}~\cite{bascompte2003nested}.
That finding has spurred a large number of works that have aimed at understanding both the possible mechanisms behind the emergence of nestedness (see Sections~\ref{sec:formation_ecol}), and the potential implications of nestedness for a given mutualistic system (see Sections~\ref{sec:implications}). Some of the conclusions on the impact of nestedness on systemic stability are based on population dynamics models that describe how the abundances of plant and animal species are affected by the abundances of their interactors, given a topology of interactions. These models of mutualism are described in Section~\ref{sec:mutualistic}.

\subsubsection{Antagonistic networks}

In contrast with mutualistic interactions, antagonistic pairwise interactions benefit one species to the detriment of the other.
We focus here on two classes of antagonistic networks: food webs and host-parasite networks.

\paragraph{Food webs}

While early studies of food webs date back to the first half of the XX century~\cite{bascompte2013mutualistic}, analysis of the structure and stability of food webs intensified in the 70s, also fueled by May's fundamental theoretical work on the stability-complexity relation~\cite{may1972will} (see Section~\ref{sec:random_mat}).
The simplest type of food web is a unipartite directed network where a directed link from species $i$ to species $j$ means that species $j$ preys on species $i$~\cite{newman2010networks}.
If a set of species of interest preys on and is predated by the same sets of species, mostly, one can coarse-grain the network by representing the set of species of interest as a single node, referred to as trophic species~\cite{newman2010networks}. This unipartite directed network is typically referred to as \emph{community food web}. Besides, ecologists have considered subsets of a complete food web by selecting, for example, a specific type of ``resource'' species (e.g., plants), and all the species (e.g., herbivores) that consume the resource species. The resulting bipartite networks are typically referred to as resource-consumer~\cite{bascompte2003nested} or source/sink networks~\cite{ings2009ecological,newman2010networks}.

The ecological literature has reported mixed findings on the nestedness of resource-consumer networks. Bascompte \textit{et al.}~\cite{bascompte2003nested} found that their level of nestedness is significantly lower than that of mutualistic networks. This finding has not been confirmed by a subsequent study by Kondoh \textit{et al.}~\cite{kondoh2010food}, who found that the degree of nestedness of trophic is comparable to that of mutualistic networks. As the nestedness of mutualistic networks has been attracted substantially more attention than that of food webs, we mainly focus on mutualistic networks in this article.

\paragraph{Host-parasitoid networks} 
%
Parasitoids represent a special type of predators: they lay their eggs inside, on the surface of or near their hosts, and the feeding larvae use their host as food~\cite{bascompte2013mutualistic}.
This kind of interaction has been widely studied in ecology. 
As being based on antagonistic interactions, one would expect host-parasitoid networks to exhibit substantially different structural patterns compared to mutualistic networks~\cite{lewinsohn2006structure,bascompte2013mutualistic}. Yet, some studies~\cite{graham2009nestedness,pilosof2014host} found that host-parasite networks can exhibit a nested structure, and scholars have investigated possible dynamical mechanisms that could lead to its emergence~\cite{mcquaid2013host}.

\subsubsection{Spatial networks} From a historical standpoint, following the seminal work by Patterson and Atmar~\cite{patterson1986nested},
spatial networks constitute the first class of systems where nestedness has been extensively studied. The particular class of spatial networks which has been studied by means of nestedness analysis is species distribution bipartite networks: such networks are composed of two classes of nodes, one representing species, and the other one representing habitat patches (such as islands). In this review, spatial networks will be often discussed in the Sections concerning distance-based metrics for nestedness (Section~\ref{sec:distance}) and null models (Section~\ref{sec:null}), yet surveying all the mechanisms that have been proposed to explain the nested structure of spatial networks falls outside the scope of this article.

\subsection{Socio-economic networks}
\label{sec:systems_econ}


\begin{figure*}[t]
	\centering
   \includegraphics[scale=0.8]{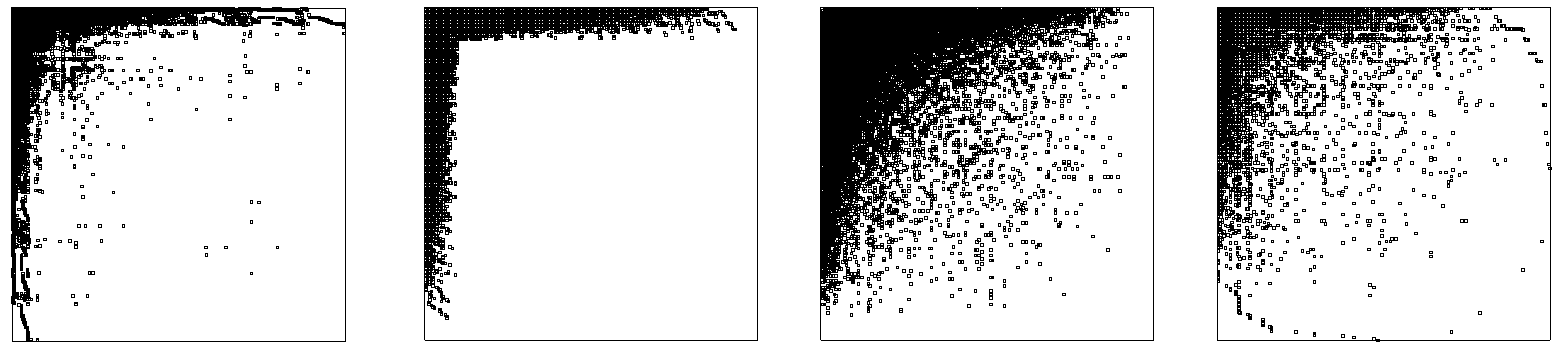}
	\caption{The adjacency matrices of four economic unipartite networks; their rows and columns have been sorted by degree. From left to right,
	the Austrian banking network; the global network of banks obtained from the Bank of International Settlements (BIS) locational statistics, the world trade network and the arm trade network.	All these networks exhibit a significant level of nestedness. Reprinted from~\cite{konig2014nestedness}.}
    \label{fig:konig2}
\end{figure*}

\begin{figure*}[t]
	\centering
   \includegraphics[scale=1]{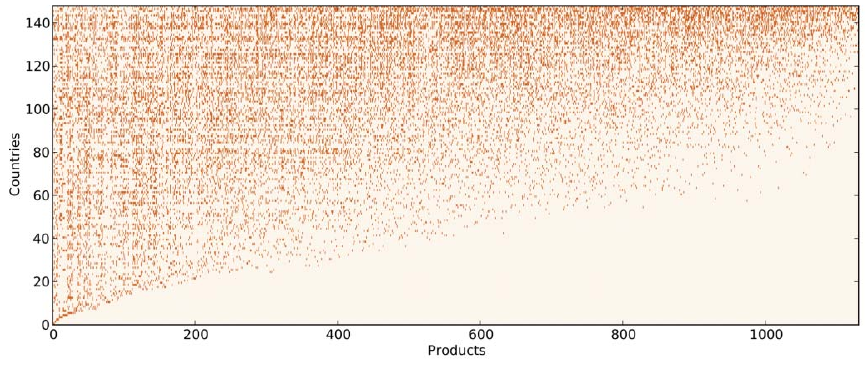}
	\caption{The (bi-)adjacency matrix of the bipartite country-product network (BACI dataset, year 2010), 
	where the rows (countries) and columns (products)
	have been sorted according to the fitness-complexity algorithm (see Section~\ref{sec:nonlinear}).
	The existence of highly-diversified countries (top rows) contradicts the standard view that 
	the wealthiest countries should focus on trading few products with a high degree of specialization~\cite{gaulier2010baci}; 
	instead, the data show that the most developed countries tend 
	This finding is the basis for capability-based models that aim to explain the topology of the 
	country-product network (see Section~\ref{sec:hidden}).
	Reprinted from~\cite{tacchella2012new}.}
    \label{fig:tacchella}
\end{figure*}

Nestedness has been long-studied in ecology, at the point that it is widely regarded as a central structural trait for ecological network analysis. At the same time, it has recently attracted interest in the analysis of social and economic systems~\cite{tacchella2012new,bustos2012dynamics,konig2014nestedness,borge2017emergence}.  

Both unipartite and bipartite socio-economic networks can exhibit significant levels of nestedness. Significant levels of nesteness have been found in various classes of socio-economic unipartite networks (see Fig.~\ref{fig:konig2} for an illustration), including: interbank networks~\cite{konig2014nestedness}, the country-country world trade network~\cite{konig2014nestedness} and the worldwide arm trade network~\cite{konig2014nestedness}. Bipartite networks that exhibit significant nestedness include the country-product export network~\cite{tacchella2012new,bustos2012dynamics,saracco2016detecting} (see Fig.~\ref{fig:tacchella} for an illustration), manufacturer-contractor networks~\cite{saavedra2009simple}, firm-location spatial networks~\cite{bustos2012dynamics}, firm-user-meme networks in social media platforms~\cite{borge2017emergence}.

This is particularly intriguing because it suggests that socio-economic and ecological systems can exhibit qualitatively comparable structures, which can be useful both to understand their common robustness properties, and to design effective strategies to prevent catastrophic events~\cite{may2008complex}.

\subsubsection{Country-level trade networks}

World trade datasets usually feature the volumes of exports of products between countries over multiple years.
From the raw data, one can then build a country-country trade network or a country-product bipartite network, as detailed below.
We refer to~\ref{appendix:material} for the links to publicly available world trade datasets.

\paragraph{Unipartite trade networks}

The structure of the International Trade Network has been widely studied, both from the country-country and from the country-product perspective.
Given publicly available data on the export of products from one country to another, it
is possible to construct unipartite country-country networks where nodes represent
countries, and links between countries represent the existence of a trade exchange of a product between the two (binary network representation)~\cite{squartini2011randomizing}, or the total volume of export between the two countries (weighted network representation)~\cite{squartini2011randomizing2}. Scholars have investigated several structural properties of unipartite trade networks, including their community structure~\cite{zhu2014rise}, their centralization~\cite{de2011world}, their clustering coefficient~\cite{squartini2011randomizing}. 

Importantly, De Benedictis and Tajoli~\cite{de2011world} studied the temporal evolution of the trade network, finding a strong and increasing heterogeneity of the countries' number of partners.
Squartini \textit{et al.}~\cite{squartini2011randomizing} found that most of the structural properties of the binary country-country network can be traced back to its degree sequence, whereas the same does not hold for the weighted representation where higher-order structural patterns cannot be explained by the countries' total export volumes~\cite{squartini2011randomizing2}.
K{\"o}nig \textit{et al.}~\cite{konig2014nestedness} found that the binary country-country network exhibits a significantly nested structure, and proposed a network formation model to explain its topology~\cite{konig2011network,konig2014nestedness} (see Section~\ref{sec:social_climbing2}).
Beyond the widely-studied World Trade network, other unipartite trade networks have been considered in the literature. For instance, some scholars~\cite{akerman2014global,konig2014nestedness} have investigated the topology of the global arm trade network, finding that it exhibits a significantly nested structure.

Besides aggregate trade networks, there is a growing interest in the topology of the trade networks for specific products~\cite{ren2018bridging} and industries~\cite{alves2019nested}. Recent studies suggest that the trade networks of more sophisticated products are typically more centralized~\cite{piccardi2018complexity} and nested~\cite{ren2018bridging} than those of less complex products.
While positive, the correlation between product complexity (as determined by node importance metrics, see Section~\ref{sec:nonlinear}) and trade-network centralization is far from being perfect~\cite{piccardi2018complexity,ren2018bridging}, which suggests that complexity and nestedness may provide complementary information on a product.

\paragraph{Bipartite trade networks}

Another way to look at World Trade data is to study the bipartite country-product network where countries are connected with the products they export.
While the input data for this analysis are typically weighted, scholars have introduced procedures to ``binarize'' the network: for example, one can add a binary link from country $i$ to product $\alpha$ if and only if the relative share of the product in the export basket of country $i$ is larger than the World's average share~\cite{hidalgo2009building}.
Scholars have investigated the nested structure of the country-product 
network~\cite{hausmann2011network,bustos2012dynamics,ermann2013ecological,saracco2015randomizing} (see Section~\ref{sec:degree_distribution}), proposed generative mechanisms to explain 
its emergence~\cite{hausmann2011network,saracco2015innovation} (see Sections~\ref{sec:hidden}-\ref{sec:innovation_novelty}), and investigated its dependence on data regularization procedures and product aggregation schemes~\cite{angelini2018complexity}. Besides, a recent stream of works have introduced and studied 
ranking algorithms for countries and products that enhance the
nestedness of the country-product matrix~\cite{tacchella2012new,pugliese2016convergence,wu2016mathematics,stojkoski2016impact,servedio2018new} (see 
Sections~\ref{sec:nonlinear}, \ref{sec:structural_nodes}).

The nested structure of the country-product network has deep implications for economic theories.
Back in the XIX century, the comparative advantage theory by Ricardo~\cite{ricardo1817principles} predicted that countries benefit from specializing on the products on which they have a comparative advantage.
Besides, based on the found U-shape relationship between metrics for country diversification and income, 
Imbs and Wacziarg~\cite{imbs2003stages} claimed that ``poor countries tend to diversify, and it is not
until they have grown to relatively high levels of per capita income that incentives to specialize take over as the dominant economic force.''
On the other hand, the empirical organization of the country-product network reported by recent literature~\cite{tacchella2012new,bustos2012dynamics} (see Fig.~\ref{fig:tacchella}) indicates that, in fact,
the most developed countries are highly diversified: they exhibit diverse export baskets by having a revealed comparative advantage~\cite{balassa1965trade} over a large number of products. By contrast, developing countries are only competitive in the export of products that are also exported by highly-diversified countries.
Recent predictive schemes~\cite{cristelli2015heterogeneous,cristelli2017predictability,tacchella2018dynamical}
based on network-based ranking algorithms suggest that country diversification is strongly correlated with the ``economic fitness'' of a country, and a country whose economic fitness is larger than that of countries with a similar level of development tend to economically grow in the future (see Section~\ref{sec:fit_gdp}).

\subsubsection{Contractor networks}
\label{sec:contractor}

In many economic systems, production is outsourced, which means that a firm ``outsources'' a part of its
internal activity to an external company~\cite{mccarthy2004impact}. Scholars have identified both benefits and negative effects of outsourcing; we refer the interested
reader to~\cite{mccarthy2004impact} for a review. Importantly for the present article, one can build a
network that connects the firms (``contractors'' or ``designers") to the manufacturing firms (``manufacturers")
that perform parts of the contractors' internal activities~\cite{uzzi1996sources,saavedra2009simple}.

Uzzi~\cite{uzzi1996sources} analyzed the network of resource exchange between New York apparel
firms over almost two years (1990-1991). He was interested in determining whether the structure
of the transaction network can predict firms' failure. He found that (1) firms whose
transaction volume is more heterogeneously distributed across the other 
firms (i.e., more generalist firms) are less likely to fail than firms that tend 
to only interact with few other firms (i.e., more specialized firms); (2) The topology 
of the partners' neighborhood heavily affects the probability that a firm fails:
firms that are less likely to fail have neither too specialist nor too generalist partners.

By analyzing a $15$-year designer-contractor dataset in the New York 
garment industry, a later study~\cite{saavedra2011strong} revealed that the probability
of firm failure is strongly affected by the firm's contribution to the overall nestedness of the manufacturer-contractor network (see Section~\ref{sec:survival}).
Besides, Saavedra \textit{et al.}~\cite{saavedra2009simple} found that manufacturer-contractor 
networks exhibit a significant nested structure, and proposed a parsimonious model with bipartite cooperation to
explain the emergence of these structures (see Section~\ref{sec:two_steps}).

Beyond designer-manufacturer networks, one can study other systems of 
contractors that involve a buyer and a seller. For example, Hern{\'a}ndez
\textit{et al.}~\cite{hernandez2018trust} an 18-month dataset from the 
Boulogne-sur-Mer fish market. This fish market has the
unique feature that the fish is sold through an auction directly
from the buyers to the sellers, and transactions are daily recorded.
Intriguingly, the aggregate buyer-seller network exhibits a significantly 
nested architecture, which excludes the emergence of blocks that would correspond to  ``niche'' markets~\cite{hernandez2018trust}.

\begin{figure*}[t]
	\centering
   \includegraphics[scale=0.4]{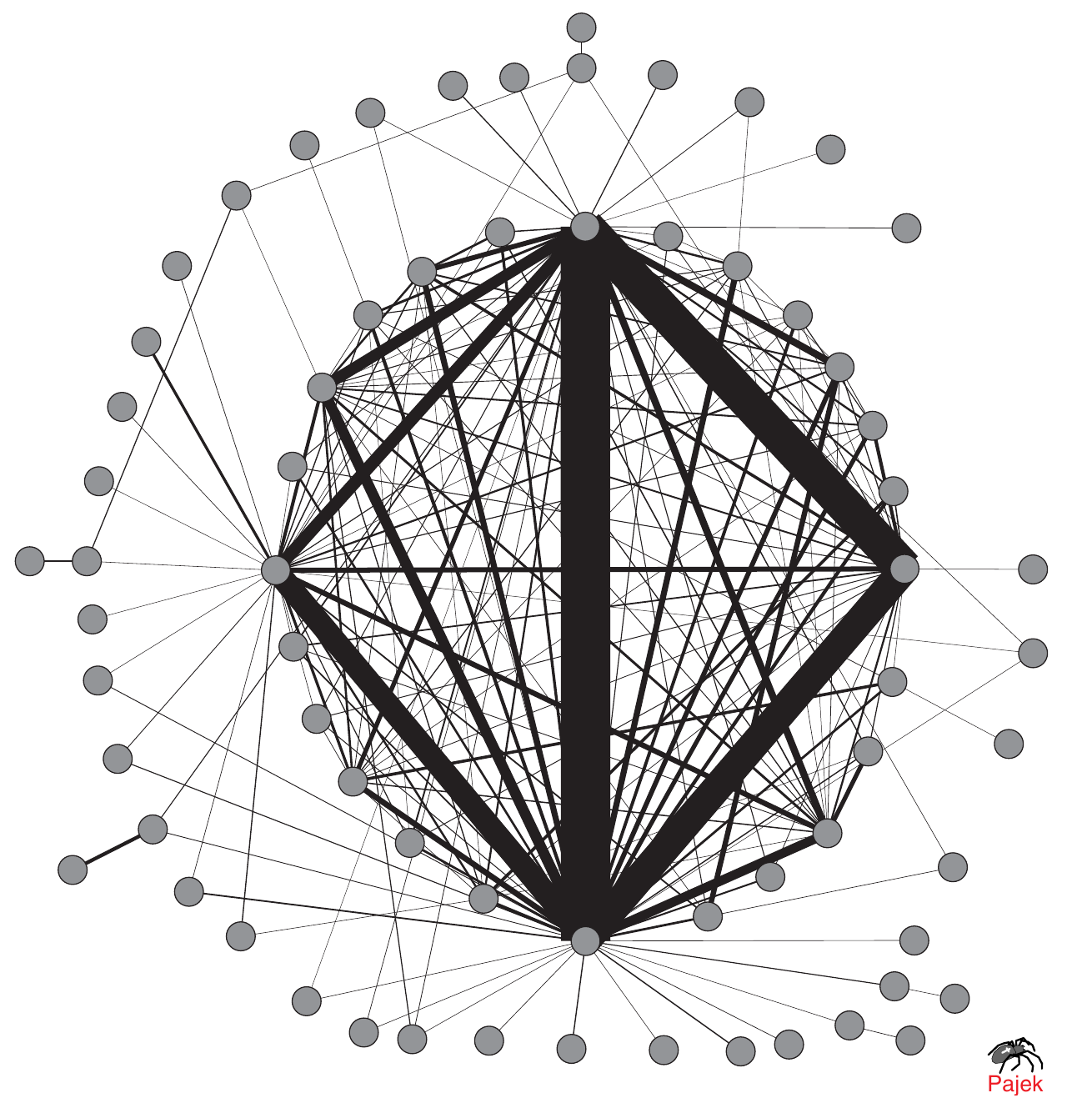}
	\caption{The topology of the US interbank payment network (first quarter of 2004). The figure only shows the undirected links that represent the 75\% of the total transaction flows. The network exhibit a clear core-periphery structure, with a core of densely interconnected nodes, and a periphery of nodes that are only connected with the core nodes. This topology can be considered as a special case of a nested topology (see Section~\ref{sec:core_periphery}). Reprinted from~\cite{soramaki2007topology}.}
    \label{fig:interbank}
\end{figure*}

\subsubsection{Interbank networks}

Understanding the topology of interbank networks is critical to assess their robustness, resilience, and effectiveness.
Soramäki \textit{et al.}~\cite{soramaki2007topology} built an interbank transaction network based on the Fedwire Funds Service,
a real-time gross settlement service. 
The topology of the network is shown in Fig.~\ref{fig:interbank}: it exhibits a clear distinction between a core of densely interconnected
nodes, and a periphery of low-degree nodes that are only connected with nodes that belong to the core. 
Besides, the network is disassortative, meaning that high-degree nodes tend to be connected with lower-degree nodes,
on average, than low-degree nodes (see Section~\ref{sec:disassortativity}).
The network, therefore, exhibits two properties -- core-periphery structure and disassortativity -- that are strongly 
correlated with nestedness (see Sections~\ref{sec:disassortativity}-\ref{sec:core_periphery}); similar structural properties
have been found in diverse interbank transaction and trade networks
by subsequent studies (see~\cite{kyriakopoulos2009network,craig2014interbank,fricke2015core,barucca2016disentangling}, for instance).

Importantly, the core-periphery structure of interbank networks
reveals important insights about financial crises: for example, Fricke and Lux~\cite{fricke2015core} showed that the decline of interbank lendings
during the 2008 financial crisis was mostly due to the core banks reducing their number of active outgoing links;
Kojaku \textit{et al.}~\cite{kojaku2018structural} found that in correspondence of the 2008 crisis, the Italian interbank network transitioned from a structure with multiple core-periphery structures (see Section~\ref{sec:multiple_cp})
to a bipartite structure~\cite{barucca2016disentangling} where transactions are mostly established between the core and periphery banks.
The core-periphery structure can be considered as a particular case of nestedness (see Section~\ref{sec:core_periphery}), which suggests that
interbank networks exhibit a significant level of nestedness as well.
This hypothesis was recently corroborated by K{\"o}nig \textit{et al.}~\cite{konig2014nestedness} who found that both an Austrian interbank network
and the global banking network are significantly nested.

\subsubsection{Spatial networks}

In a similar way as spatial ecological networks connect spatial regions with the species that inhabit them, spatial economic networks connect spatial regions to the  economic activities that are developed in them~\cite{bustos2012dynamics,garas2018economic}. Bustos \textit{et al.}~\cite{bustos2012dynamics} built a Chilean municipality-industry network by considering an industry $i$ as present in a given municipality $\alpha$ if at least one firm classified in industrial classification $i$ declared municipality $\alpha$ as its tax residence.
In this way, approximately $700$ different industries
were connected with Chile’s $347$ municipalities~\cite{bustos2012dynamics}.
They found that the resulting bipartite network exhibits a significantly nested structure, and exploited this property for link prediction, i.e., to predict the appearance and disappearance of industries in municipalities (see Section~\ref{sec:link_prediction}).

In a similar spirit,
Garas \textit{et al.}~\cite{garas2018economic} built a city-firm network where 
$21$ economic activities were connected with $1,169$ cities in the world; also this network turned out to be significantly nested.
From a node-level perspective,
Gao and Zhou~\cite{gao2018quantifying} analyzed a bipartite network connecting industries with the 31 Chinese provinces, finding that 
ranking algorithms that seek to maximize the adjacency matrix nestedness (like the fitness-complexity algorithm~\cite{tacchella2012new}) provide a node score that is highly correlated with province-level macroeconomic indicators.

\subsubsection{Communication networks in organizations}

In social science, the idea that different topologies of communication networks lead to different performances of groups of individuals is not recent.
Already in 1941, Leavitt~\cite{leavitt1951some} tested the performance of groups of individuals arranged in different communication topologies in solving puzzles. 
He found that a network with a star-like structure ("wheel structure'' in~\cite{leavitt1951some}) performed better than more decentralized networks: the presence of a ``generalist'' who could be quickly reached by the other nodes seemed to speed up the problem-solving process. As reminded by Borgatti \textit{et al.}~\cite{borgatti2009network}, later studies have shown that as the complexity of the puzzle increases, decentralized networks tend to perform better.

This idea has motivated researchers to investigate how the structure of organizations impacts their performance and ultimate success.
Importantly, organizations are made by the individuals who coordinate and constantly enhance internal knowledge. Communication networks (with organization members being nodes and communication events, as a proxy for information exchange, being links) play an important role in the structure of organizations.
This fact has originated in the implicit knowledge that resides in the organization, which becomes uncovered by the patterns of communication~\cite{kogut1992knowledge}.  

Kogut and Zander~\cite{kogut1992knowledge} suggested that economic firms are ``social communities in which individual and social expertise is transformed into economically useful products and services by the application of a set of higher-order organizing principles''. 
Thus, firms exist because of a social community structured ``by organizing principles that are not reducible to individuals''~\cite{kogut1992knowledge}.
Firms are better in sharing and transferring knowledge among individuals than markets due to a fundamental dilemma: while higher specialization increases productivity due to the division of labor, it also increases the costs of communication and coordination among the agents \citep{kogut1996firms}. Extant research has shown that the formal and informal networks of communications vastly differ \citep{krackhardt1988informal,krachardt1993informal}. 
The topology of this network, therefore, plays a role in determining the global properties of the organization.

Empirical work has found that intra-firm informal communication networks exhibit properties compatible with nestedness~\cite{rank2008formal,grimm2017analysing}. This is compatible with theoretical arguments mimicking how these networks evolve \cite{konig2011network}. From a managerial perspective, an important question for future research is to assess whether nested communication flows are beneficial or harmful to the organization's efficiency.

\subsubsection{Online communication networks and software development}
\label{sec:online}

Online social networks and social media platforms offer us an unprecedented amount of data on human activity. Often, these data come with fine-grained temporal information, which allows us to study how network topology varies over time.
Valverde and Solé~\cite{valverde2007self} have analyzed the network structure of Open-Source software development communities. They found that the system exhibits a strongly hierarchical structure where ``an elite of highly connected and mutually
communicating programmers control the flow of
information generated by the OS community"~\cite{valverde2007self}.
Besides, the network is disassortative, which means that the average degree of the low-degree nodes' neighbors tends to be larger than that of the high-degree nodes' neighbors.
Both properties are compatible with nestedness; these results suggest that communication networks between online community members might be a future field of application for nestedness analysis.

A different relevant network for analyzing software development is the network of dependencies and conflicts between software packages in a given operating system. The temporal evolution of the modularity of this network has been also investigated by Fortuna \textit{et al.}~\cite{fortuna2011evolution} who found that for the Debian GNU/Linux operating system, both modularity and the number of modules increased over time.

In the context of social-media analysis, Borge-Holthoefer \textit{et al.}~\cite{borge2017emergence} studied the temporal evolution of nestedness and modularity for a bipartite user-hashtag network collected from Twitter. They created the bipartite network by collecting the memes related to the 2011 civil protests in Spain within a two-month time window (see~\cite{borge2017emergence} for the details). They found that a sharp transition from a modular to a nested topology occurs in the vicinity of a critical event -- see Section~\ref{sec:modularity} for details.

More recently, the temporal evolution of topological properties of Twitter communication networks has been investigated by Bastos \textit{et al.}~\cite{bastos2018core}. They focused on the centralization of the user-user communication network (relative to a specific topic, agriculture) in relation to the different level of specialization of shared information (as determined by unsupervised text classification techniques). Intriguingly, they found that a more centralized topology emerges when discussions become more technical, whereas generic discussions unfold over decentralized topologies. Their results confirm the premises of classical diffusion theories (see Chapter 9 in~\cite{rogers2010diffusion}) that posit that more centralized topologies are more appropriate for the dissemination of innovations that involve a high level of technical expertise.

\subsection{Other classes of networks}

Beyond ecological and socio-economic systems, other classes of networks have been found to exhibit nestedness. 
Kamilar and Atkinson~\cite{kamilar2014patterns} recently analyzed various datasets on the distribution patterns of cultural traits across populations of humans, chimpanzees, and orangutans. They found that for humans and chimpanzees, these patterns are significantly nested, meaning that ``cultures with a small
repertoire of traits tend to comprise a proper subset of those traits
present in more complex cultures"~\cite{kamilar2014patterns}. The same does not hold for orangutans, and we refer to the Discussion section in~\cite{kamilar2014patterns} for a detailed discussion of the possible reasons behind this discrepancy.

Cantor \textit{et al.}~\cite{cantor2017nestedness} recently brought nestedness analysis to unipartite biological networks from six different levels of biological organization representing gene and protein interactions, complex phenotypes, animal societies, metapopulations, food webs and vertebrate metacommunities~\cite{cantor2017nestedness}. They found that nestedness emerges at various biological scales, and emphasized the importance of understanding the basic mechanisms behind its emergence.

The core-periphery structure has received much attention in brain networks~\cite{bassett2013task} where it has been found that the separation between temporal core and periphery changes can be used to predict individual differences in learning success. As we shall see in Section~\ref{sec:core_periphery}, a core-periphery topology is a special case of a nested topology.

Johnson \textit{et al.}~\cite{johnson2013factors} analyzed various unipartite and bipartite networks of different types, including food webs, brain networks, metabolic networks, transportation networks and some popular network datasets (like the Zachary's Karate Club network~\cite{zachary1977information} and a collaboration network of Jazz musicians~~\cite{gleiser2003community}). They found that a number of them exhibit significant nestedness -- their detailed results can be found in the appendix S5 of their article~\cite{jonhson2013factors}.

In principle, any unipartite or bipartite network can be analyzed by means of nestedness metrics and null models. While, so far, the emergence and implications of nestedness have been mostly investigated in ecological and, more recently, socio-economic networks, we envision that they may gain additional interest from other research areas.

\clearpage

\section{Observing nestedness: metrics and null models}
\label{sec:metrics}

While perfectly nested networks can be unambiguously defined, they are rarely found in nature.
Yet, many real networks exhibit highly-ordered structures where many pairs of nodes respect the definition of nestedness. 
This means that in such networks, given two nodes $i$ and $j$ such that $k_i<k_j$, most of node $i$'s neighbors are also neighbors of node $j$.
In these highly but imperfectly nested structures, the way we measure the level of nestedness can heavily affect our conclusions
on the significance of the pattern and its relevance. This leads to several questions:
how to best measure the degree of nestedness in real networks that are not 
perfectly nested? How to compare the level of nestedness across datasets of different size,
density, and degree distribution? What is the impact of more basic network properties 
(e.g., degree distribution) on the observed levels of nestedness? 

This Section tackles these questions by providing an overview of the metrics (Section~\ref{sec:metrics1}) and null models (Section~\ref{sec:null}) that have been introduced with the explicit goal to quantify the level of nestedness of a given system. Starting with the unexpected absences metric by Patterson and Atmar~\cite{patterson1986nested}, we will survey various nestedness metrics including the popular nestedness temperature~\cite{atmar1993measure,rodriguez2006new}, NODF~\cite{almeida2008consistent}, and spectral radius~\cite{staniczenko2013ghost}. Most of these metrics were originally introduced in the ecological literature, yet their realm of application has extended to economic and social networks.

In relation to the null models, there has been an intense debate in the ecological community over 
which null model should be preferred to infer the significance of the observed values of
nestedness metrics~\cite{ulrich2007null,ulrich2009consumer,gotelli2012statistical}. We will 
cover the main strength and weaknesses of the available null models. The discussion will
be further deepened in Section~\ref{sec:degree_distribution}, where we will discuss 
the statistical relation between the degree distribution and nestedness.

Besides, as we have pointed out in the Introduction, one can also re-arrange the 
rows and columns of a given networks in such a way to minimize the number of
violations of the nestedness condition. Such reshuffling of rows and columns can be 
interpreted as a node ranking algorithm, and it has important implications for
the identification of vulnerable species in ecological systems~\cite{dominguez2015ranking}, of competitive 
countries in international trade~\cite{tacchella2012new}, and for the effectiveness of targeted attacks
on the network\footnote{The role of ranking algorithm for the identification
of structurally important and vulnerable nodes is discussed in 
Section~\ref{sec:structural_nodes}.}~\cite{dominguez2015ranking,mariani2015measuring}. 
The results of a quantitative comparison of existing ranking algorithms for bipartite networks~\cite{lin2018nestedness} 
are presented in Section~\ref{sec:packing}.

\subsection{Metrics to quantify nestedness}
\label{sec:metrics1}

In line with Ulrich~\cite{ulrich2009consumer}, we classify metrics 
for nestedness into four main categories: gap-counting
metrics (Section~\ref{sec:gap}), overlap
metrics (Section~\ref{sec:overlap}), distance metrics (Section~\ref{sec:distance}), and eigenvalue-based metrics (Section~\ref{sec:eigenvalue}).
In principle, all these metrics apply to any binary matrix. For the sake of generality,
unless otherwise stated, the definitions provided below refer to bipartite binary 
networks. The corresponding definitions for unipartite binary networks can be readily obtained 
by identifying the row-nodes with the column-nodes. Not all the 
metrics below have been generalized to weighted networks, yet we will provide some instances of 
metrics adopted in weighted networks\footnote{W-NODF~\cite{almeida2011straightforward} and spectral 
radius~\cite{staniczenko2013ghost}: see Sections~\ref{sec:overlap} and~\ref{sec:eigenvalue}, respectively.}.

Before defining the metrics, an important caveat that applies essentially to all of them is 
that each of them generally depends on basic network properties such as network size, density, and degree distribution; in order to compare nestedness metrics across networks of different size and density, performing a statistical analysis with a null model is a necessary step -- this fundamental aspect is deepened in Section~\ref{sec:null}. To prevent redundancy, we will avoid repeating this caveat for each of the metrics defined below, yet the conscious reader needs to always keep it in mind.

\subsubsection{Gap-counting metrics}
\label{sec:gap}

Gap-counting metrics build on the observation that the empty and filled regions of a
perfectly nested matrix are perfectly separated, as it is evident from the illustration in Fig.~\ref{fig:nested-nets}. This implies that there are no
"absences'' (i.e., empty spots) in the filled region, and no
"presences'' (i.e., filled spots) in the empty 
region. One can, therefore, evaluate the degree of nestedness of a given network by counting the number of violations of this property.

\begin{figure*}[t]
	\centering
	\scalebox{0.4}[0.3]{\rotatebox{270}{\includegraphics{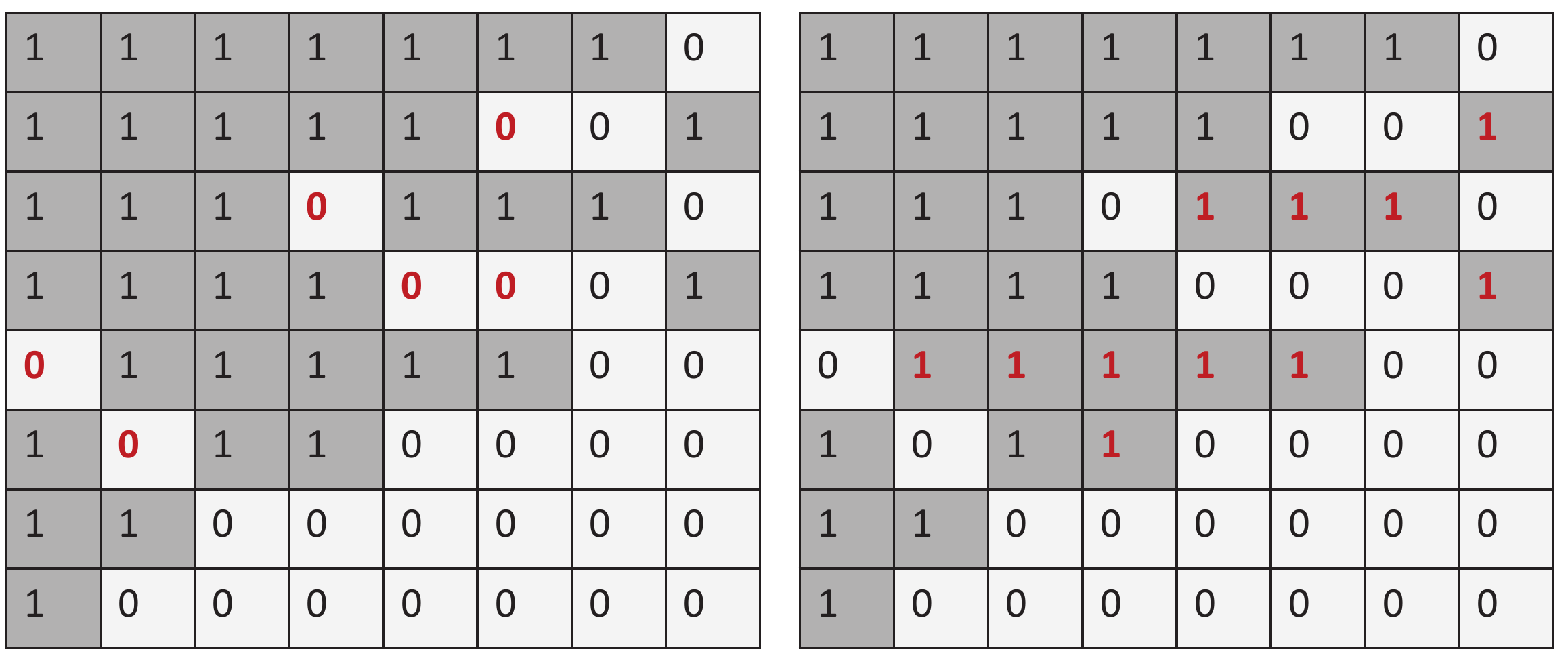}}}
	\caption{An example of calculation of $\mathcal{N}_0$ (left) and $\mathcal{N}_1$ (right panel). Rows and columns are ordered by degree. \emph{Left, $\mathcal{N}_0$.} We scan each row from right to left. For each row $i$, we consider the filled spot (1) that corresponds to the column $\beta_{min}(i)$ with the smallest degree; each zero at the left of such column, if it corresponds to a column $\alpha$ with degree $k_\alpha>k_{\beta_{min}(i)}$, contributes to $\mathcal{N}_0$. \emph{Right, $\mathcal{N}_1$.} We scan each row from left to right. For each row $i$, we consider the empty spot (0) that corresponds to the column $\beta_{max}(i)$ with the largest degree; each one at the right of such column, if it corresponds to a column $\alpha$ with degree $k_{\beta_{max}(i)}>k_\alpha$, contributes to $\mathcal{N}_1$.
  }\label{Fig_N}
\end{figure*}

\paragraph{Unexpected absences and presences} The number of 
unexpected absences $\mathcal{N}_0$ introduced by Patterson
and Atmar~\cite{patterson1986nested} counts how many
times a node $i$ is not a neighbor of a node $\alpha$ that has larger degree than its lowest-degree neighbor $\beta_{min}(i)$ (i.e., $k_{\beta_{min}(i)}:=\min_{\alpha:A_{i\alpha}=1}\{k_{\alpha}\}$).
In formulae,
\begin{equation}
\mathcal{N}_0=\sum_{i} \sum_{\alpha} (1-A_{i\alpha})\,\Theta(k_\alpha-k_{\beta_{min}(i)}).
\end{equation}
In this formula, the sum over $\alpha$ is restricted to the pairs such that $k_\alpha>k_{\beta_{min}(i)}$ through the Heaviside function
$\Theta(k_\alpha-k_{\beta_{min}(i)})$, where $\Theta(x)=1$ if $x>0$, $\Theta(x)=0$ if $x\leq 0$. For each contribution to the sum, the $(1-A_{i\alpha})$ factor excludes node $i$'s neighbors. Importantly, Patterson and Atmar notice that their nestedness metrics is a sum over the contributions from all the individual nodes, which implies that one can compare the contributions to nestedness from different individual nodes. A graphical illustration of the $\mathcal{N}_0$ metric can be found in Fig.~\ref{Fig_N}.

A complementary perspective is offered by the number $\mathcal{N}_1$ of unexpected presences~\cite{cutler1991nested} metric which counts how many times a node $i$ is neighbor of a node $\alpha$ that has smaller degree than its largest-degree non-neighbor $\beta_{max}(i)$ (i.e., $k_{\beta_{max}(i)}:=\max_{\alpha:A_{i\alpha}=1}\{k_{\alpha}\}$).
In formulae,
\begin{equation}
\mathcal{N}_1=\sum_i \sum_{\alpha} A_{i\alpha}\,\Theta(k_{\beta_{max}(i)}-k_\alpha).
\end{equation}
A graphical illustration of the $\mathcal{N}_1$ metric can be found in Fig.~\ref{Fig_N}.

One can also combine unexpected absences and presences into a single nestedness metric.
This was done by Cutler~\cite{cutler1991nested}, who considered one single node $i$ of intermediate degree. In line with the definition of nestedness, absences of node $i$'s neighbors from the neighborhood of nodes with larger degree are counted as unexpected absences ($U^A_i$), whereas occurrences of node $i$'s non-neighbors in the neighborhoods of smaller-degree nodes are counted as unexpected presences ($U^P_i$). One chooses the node $i^*$ that minimizes the sum of unexpected absences and presences $U^T_i=U^A_i+U^P_i$; accordingly, the degree of nestedness is defined as $U:=\min_{i}U^T_i=U^T_{i^*}$. The $U$ metric can be interpreted as the minimum number of steps to convert (either by filling empty spots or by deleting filled spots) the matrix in hand into a perfectly nested matrix.


\paragraph*{Number of departures} 

For an ordered matrix, the number of departures, $D$, is defined as the number of times the absence
of a node is followed by its presence in the neighborhood of the next lower-degree node~\cite{lomolino1996investigating}. Simply,
the number of departures is the number of times the $i$th row does not interact with column $\alpha$ but the $(i+1)$th 
row ($k_{i+1}<k_i$) does. Lomolino~\cite{lomolino1996investigating} estimated
the statistical significance of nestedness for each matrix by 
comparing the measured $D$ values with those obtained for randomly-ordered matrices. In the 
randomization, species distributions are unaltered, but islands are randomly ordered with respect to isolation and area. He calculated the normalized number of departures of each ordered matrix as:
\begin{equation}
	\mathcal{N}=100\,\frac{(R-D)}{R},
\end{equation}	
where $D$ is the number of departures in the ordered matrix and $R$ is the mean number of departures for randomized networks.

\begin{figure*}[t]
	\centering
	\scalebox{0.4}[0.3]{\rotatebox{270}{\includegraphics{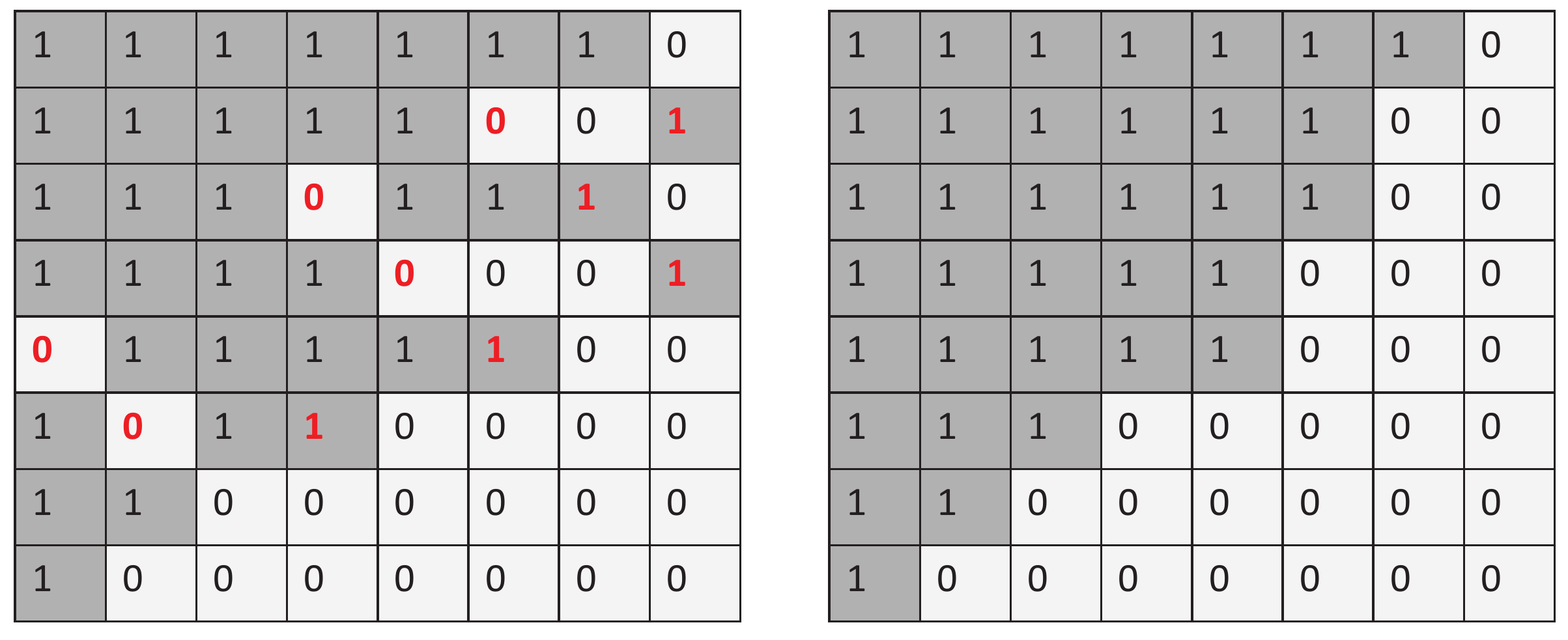}}}
	\caption{An example of calculation of the number of discrepancies, $\mathcal{D}$~\cite{brualdi1999nested}. From the original matrix (left panel), one constructs a corresponding perfectly nested matrix (right panel) by shifting, for each row, all its filled elements to the right. The elements of the observed matrix that do not match the corresponding ones for the perfectly nested matrix are interpreted as ``discrepancies'' (marked in red in the left panel).}\label{fig:discrepancy}
\end{figure*}

\paragraph{Deviations from a perfectly nested matrix}
A perfectly nested matrix is one where the filled elements in each row 
are found as far to the left as possible, and the filled elements in each column
are found as far to the top as possible. Motived by such perfectly nested structure,
Brualdi and Sanderson~\cite{brualdi1999nested} construct, for a given 
adjacency matrix $\mat{A}$, a perfectly nested matrix $\mat{P}$ by shifting the filled 
elements of $\mat{A}$ as far to the left as possible, while keeping their row fixed. Such 
shifting procedure generates matrices where each row-node $i$ has the same degree $k_i$ as the 
original matrix $\mat{P}$, but the first $k_i$ elements of row $i$ from the left are filled.
The discrepancy of a given matrix $\mat{A}$ can then be computed by summing over the 
rows the unexpected absences in each row $i$, i.e., the number of zeros observed in the first $k_i$ elements from the left.
In formulas,
\begin{equation}
\mathcal{D}=\sum_i \sum_{\alpha=1}^{k_{i}}(1-A_{i\alpha}).
\end{equation}
An illustration of the metric is provided in Fig.~\ref{fig:discrepancy}. Brualdi
and Sanderson~\cite{brualdi1999nested} assessed the statistical significance of 
the metric by comparing its observed value with its mean value in randomized networks generated with a null model 
that preserves exactly the row-nodes' and column-nodes' degree (Fixed-Fixed model, see Section~\ref{sec:nine}).

\subsubsection{Distance-based metrics}
\label{sec:distance}

Distance-based metrics are based on a three-step computation. First, one 
determines the ideal line that separates the empty and filled regions of the matrix
in a perfectly nested network. Second, one uses a suitable node-level ranking algorithm to rearrange the rows and columns of the
adjacency matrix in such a way that its nestedness is maximized. Finally, for the 
rearranged matrix, one computes the distance 
of the unexpected elements (empty elements in the filled region, 
and filled elements in the empty region) from the ideal line.
The nestedness temperature of the network is given by the sum of the contributions from all the unexpected elements
of the rearranged matrix: highly nested networks are characterized by low
temperature (see~Eq.~\eqref{temperature} below).
Note that these metrics penalize more heavily unexpected absences and presences that are far away from the ideal line. 
Such assumption was originally motivated by arguments based on theories of island biogeography~\cite{atmar1993measure}, as discussed below.

\paragraph*{Interpretation in terms of fragmented habitats formation}

Distance-based metrics are motivated by considerations based on species distribution
patterns within naturally fragmented habitats~\cite{atmar1993measure}.
Consider an original biota that subsequently fragments into a collection
of islands due to natural causes.
On each resulting island, there will be some species at larger
risk of becoming extinct. Some species will indeed become extinct, and larger islands will be left, typically, with more species than smaller islands. 
The resulting system can be represented as a species-island bipartite network
where each species is connected with the islands where its individuals are found.

If the species extinction order is exactly the same in each island, the resulting 
species-island network would be perfectly nested: each smaller island would
only include a subset of the species that are found in larger islands. In 
this sense, nestedness is a property that signals the existence of \emph{order} in the extinction patterns,
whereas deviations from a perfectly nested structure can be interpreted as ``statistical noise''~\cite{atmar1993measure}.
Therefore, Atmar and Patterson made an explicit connection to statistical physics~\cite{boltzmann1872weitere}
and information theory~\cite{shannon1948mathematical}: similarly as a high-energy configuration of a system with many particles,
the unexpected presence or absence of a species from an island is a ``surprising event''.

\paragraph{Nestedness temperature and Nestedness Temperature Calculator (NTC)}

\begin{figure}[t]
\centering
\includegraphics[scale=0.9]{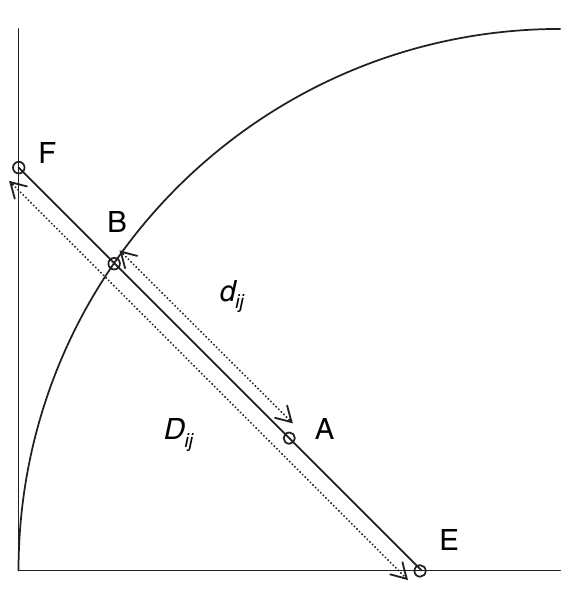}
\caption{A graphical illustration of the terms $d$ and $D$ needed in the temperature calculation~\cite{atmar1993measure,rodriguez2006new}. The rows' and columns' labels are rescaled in such a way that the labels range from zero to one. From an unexpected presence or absence (in the figure, presence A below the line of perfect nestedness), one draws a straight line of slope $-1$. The distance $d_{ij}$ of the unexpected element A is then normalized by $D_{ij}$; the nestedness temperature defined by Eq.~\eqref{temperature} is given by all the unexpected elements' contributions $(d_{ij}/D_{ij})^2$. Reprinted from \cite{rodriguez2006new}.}
\label{fig:distance}
\end{figure}

According to Ulrich~\cite{ulrich2009consumer}, the matrix nestedness
temperature $T$~\cite{atmar1993measure,rodriguez2006new} is ``by far the most 
popular metric for quantifying nestedness". Matrix temperature depends on both the 
determination of the line of perfect nestedness and the matrix packing
algorithm (details below), and different algorithms have been proposed for the temperature computation. The first algorithm proposed is the Nestedness Temperature Calculator (NTC)~\cite{atmar1993measure} which involves the following three steps:
\begin{enumerate}
\item \textbf{Determination of the line of perfect nestedness.}
For a matrix with connectance $\rho=0.5$, it is straightforward
to identify the diagonal that bisects the matrix as the line of perfect
nestedness (see Fig. 2 in~\cite{atmar1993measure}). For a matrix of
connectance different from $0.5$, the NTC software constructs the line of perfect nestedness geometrically (see Fig. 4 in~\cite{atmar1993measure}), whereas the BINMATNEST algorithm uses a function that determines the shape of the perfectly nested matrix (see Eq.~\eqref{shape} below).

\item \textbf{Packing the matrix.} Atmar and 
Patterson~\cite{atmar1993measure} argued
that ``for a matrix's temperature to be calculated, a matrix 
must first be packed to a state of minimum unexpectedness", 
and ``row and column totals cannot be used as a reliable 
guide for packing". By arranging the elements of a given
matrix before computing the nestedness metric, we are not computing 
the nestedness associated with the degree or a specific property, but the 
maximum possible degree of nestedness in the system. We 
discuss the implications of this aspect and possible methods for packing the matrix in Section~\ref{sec:packing}.
\item \textbf{Computing the temperature.} All the unexpected presences
and absences contribute to the matrix temperature. The 
unexpectedness $u_{i\alpha}$ of a single (present or absent) 
link $(i,\alpha)$ is given by $u_{i\alpha}=(d_{i\alpha}/D_{i\alpha})^2$,
where $d_{i\alpha}$ denotes the distance of $(i,\alpha)$ from the
line of perfect nestedness (see Fig.~\ref{fig:distance}), and $D_{i\alpha}$ is geometrically defined as
explained in Fig.~\ref{fig:distance}. 
The total unexpectedness $U$ of the matrix is defined as 
\begin{equation} 
U=\frac{1}{N\,M}\sum_{(i,\alpha)\in\mathcal{U}}\Biggl(\frac{d_{i\alpha}}{D_{i\alpha}}\Biggr)^2,
\label{temperature}
\end{equation} 
where the sum is restricted to the set $\mathcal{U}$ of unexpected element (i.e., empty elements above and filled elements below the line of perfect nestedness).
The matrix temperature $T$ is defined as $T=100\,U/U_{max}$, where $U_{max}=0.04145$. A perfectly nested matrix has zero
temperature, whereas a maximally non-nested matrix has a temperature
equal to $100$.
\end{enumerate}
Variants of the NTC differ from the NTC in one or more of the three elements above: line of perfect nestedness determination, packing algorithm, and temperature computation. Besides,
scholars have been also interested in developing software for a fast 
computation of the nestedness temperature over a large 
number of matrices~\cite{guimaraes2006improving}. Below, we briefly introduce two variants of the NTC: BINMATNEST and the $\tau$-temperature.

\paragraph{BINMATNEST} The BINMATNEST algorithm~\cite{rodriguez2006new} differ from the NTC 
in both the isocline determination procedure and in the packing algorithm.
In particular, Rodr{\'\i}guez-Giron{\'e}s and Santamar{\'\i}a~\cite{rodriguez2006new} 
pointed out the following limitations of the NTC: the line of perfect nestedness is not uniquely defined, and the packing of the matrix is not optimal.
In particular, as for the line definition problem, they pointed out that the definition implemented 
by the NTC does not identify a single line, but a set of curves~\cite{rodriguez2006new}.
To overcome this limitation, they defined the following function:
\begin{equation}
f(x;p)=\frac{0.5}{N}+\frac{N-1}{N}\Biggl(1- \Bigl( 1-\frac{M\,x-0.5}{M-1}\Bigr)\Biggr).
\label{shape}
\end{equation}
The function depends on a parameter, $p$, that can be tuned to match 
the desired level of connectance $\rho$; the resulting line of perfect nestedness obtained in real networks is
essentially indistinguishable from those obtained by the NTC (see Fig. 2 in~\cite{rodriguez2006new}).
The packing algorithm implemented by the BINMATNEST algorithm 
and its relation with that implemented by the NTC is discussed in Section~\ref{sec:packing}.

\paragraph{$\tau$-temperature}
The $\tau$-temperature~\cite{corso2012nestedness} differs from the NTC and the BINMATNEST algorithm in how the temperature is computed. Once the adjacency matrix is packed by a given algorithm, the $\tau$-temperature of the matrix is proportional to the Manhattan distance \cite{corso2012nestedness}
\begin{eqnarray}
D=\sum_{i,\alpha}A_{i\alpha}\,(i+\alpha).
\end{eqnarray}
The $\tau$-temperature is given by the ratio between $D$ and the Manhattan distance observed in randomized networks that preserve the link density of the original network.
Similarly as the Atmar and Patterson's temperature, the smaller the $\tau$-temperature, the higher the matrix's degree of nestedness. The $\tau$-temperature turns out to be positively correlated with the nestedness temperature $T$, yet such correlation is far from being perfect~\cite{corso2012nestedness}.

\subsubsection{Overlap metrics}
\label{sec:overlap}

Overlap metrics are motivated by the definition of nestedness in terms of nodes neighborhoods -- more specifically, by the property that in perfectly nested networks, the neighborhoods of nodes with lower degree is included in the neighborhoods of nodes with larger degree. Based on this consideration, one can then attempt to measure nestedness by evaluating how often neighbors of lower-degree nodes are also neighbors of larger-degree nodes. 

\paragraph{Wright and Reeves' overlap-based metrics} In the context of species-island spatial networks, Wright and Reeves~\cite{wright1992meaning} build metrics for nestedness on the intuition that in a perfectly-nested network, if a species is found on an island, then it should be also found on richer islands.
One can therefore quantify nestedness by aggregating the overlaps between the neighborhoods of pairs of islands. By introducing the overlap $O_{\alpha\beta}=\sum_i A_{i\alpha}\,A_{i\beta}$, they defined the $N_c$ metric as~\cite{wright1992meaning}
\begin{equation}
\mathcal{N}_c=\sum_{(\alpha,\beta)}O_{\alpha\beta}=\frac{1}{2}\sum_{i}k_i\,(k_i-1)=\frac{N}{2}(\braket{k_i^2}-\braket{k_i}).
\label{Nc}
\end{equation}
where $O_{\alpha\beta}=\sum_i A_{i\alpha}\,A_{i\beta}$.
Wright and Reeves~\cite{wright1992meaning} introduce an additional nestedness metric $C$ by normalizing $N_c$ as follows:
\begin{equation}
C=\frac{\mathcal{N}_c-\mathbb{E}[\mathcal{N}_c]}{\mathcal{N}_c^{(max)}-\mathbb{E}[N_c]},
\label{wr}
\end{equation}
where $\mathbb{E}[\mathcal{N}_c]$ is the expected value of $\mathcal{N}_c$ in randomized networks and $\mathcal{N}_c^{(max)}$ is 
the maximum value that can be attained by $\mathcal{N}_c$.
Wright and Reeves~\cite{wright1992meaning} compute $\mathbb{E}[\mathcal{N}_c]$ based on a Equiprobable-Proportional 
null model (see Section~\ref{sec:nine}) where the probability that a given species is found on a given 
island is proportional to the number of species in that island, obtaining $\mathbb{E}[O_{\alpha\beta}]=k_{\alpha}\,k_{\beta}/N$.
Hence, one gets
\begin{equation}
\mathbb{E}[\mathcal{N}_c]=\frac{1}{N}\sum_{(\alpha,\beta)}k_{\alpha}\,k_{\beta}=\frac{1}{2\,N}\sum_{\alpha,\beta}k_{\alpha}k_{\beta}=\frac{\braket{k}^2}{2\,N}.
\end{equation}
It is also possible to compute the maximum value $\mathcal{N}_c^{(max)}$ which is achieved when $O_{\alpha\beta}=\min\{k_{\alpha},k_\beta\}$. One thus obtains~\cite{wright1992meaning}
\begin{equation}
\mathcal{N}_c^{(max)}=\sum_{\alpha,\beta}k_{\beta}\,\Theta(k_{\alpha}-k_{\beta})=\sum_{\beta}k_{\beta}\,(\beta-1).
\end{equation}
Wright and Reeves~\cite{wright1992meaning} observed that the $C$ index
"is free of strong dependence on the size of the
presence-absence matrix, so that it is possible to compare
the relative nestedness of different datasets". On the other hand, Brualdi and Sanderson~\cite{brualdi1999nested} notice that as the $N_c$ metric is entirely determined by the row-column's degrees (see Eq.~\eqref{Nc}), it is of no use if we wish to compare its observed value with its value in randomized networks that preserve exactly the network degree sequence (see also Section~\ref{sec:degree_distribution}).

\paragraph{Nested Overlap and Decreasing Fill (NODF)} Given a pair of row nodes $(i,j)$ of the
same class such that $k_i>k_j$, we expect their number of
common neighbors $O_{ij}=\sum_{\alpha}A_{i\alpha}\,A_{j\alpha}$ to be equal to $k_j$ for 
a perfectly nested network, and smaller than $k_j$ a network that is not perfectly nested.
One can then define the row-NODF $\mathcal{N}^{R}$ as~\cite{almeida2008consistent}
\begin{equation}
\mathcal{N}^{R}=\sum_{(i,j)}\frac{O_{ij}}{k_j}\Theta(k_i-k_j),
\end{equation}
where $\Theta$ denotes the Heaviside function: $\Theta(x)=1$ if $x>0$, $\Theta(x)=0$ if $x\leq 0$.
This metric is maximal when $O_{ij}=k_j$ for all pairs $(i,j)$ such that $k_j<k_i$.
In the same way, one can the define the column-NODF $\mathcal{N}^{C}$ as~\cite{almeida2008consistent}
\begin{equation}
\mathcal{N}^{C}=\sum_{(\alpha,\beta)}\frac{O_{\alpha\beta}}{k_\beta}\Theta(k_\alpha-k_\beta).
\end{equation}
The degree of nestedness is quantified by the total NODF:
\begin{equation}
\eta=\frac{\mathcal{N}^{R}+\mathcal{N}^{C}}{\frac{N\,(N-1)}{2}+\frac{M\,(M-1)}{2}}.
\label{nodf}
\end{equation}

\paragraph{S-NODF} We present here the metric introduced by Bastolla \textit{et al.}~\cite{bastolla2009architecture} to facilitate 
the analytic computation of the impact of network structure on the co-existence of species in mutualistic systems. We refer to this metric as stable NODF (S-NODF) because it is a simple variant of NODF, yet it is more stable with respect to small perturbations of the data, as we shall discuss below.

The definition of rows' nestedness $\eta ^{R}$ reads~\cite{bastolla2009architecture}
\begin{equation}
{\eta }^{R}= \sum_{(i,j)}\frac{ O_{ij}}{\min{\{k_i,k_j\}} }.
\end{equation}
This definition\footnote{Note that in the original paper~\cite{bastolla2009architecture}, there was a typo such that the summation symbol appeared both in the numerator and in the denominator.} differs from the one provided by NODF because it also includes the contribution of pairs of nodes with the same values of degree. In other words,
\begin{equation}
    \eta ^{R}=\mathcal{N}^R+\sum_{(i,j)}\frac{O_{ij}}{k_j}\,\delta(k_i,k_j),
\end{equation}
where $\delta(k_i,k_j)$ denotes the usual Kronecker delta: $\delta(k_i,k_j)=1$ if $k_i=k_j$, $\delta(k_i,k_j)=0$ otherwise.
Compared to the NODF contribution $\mathcal{N}^R$, $\eta^{R}$ avoids penalizing pairs of rows with the same degree. The soundness of including pairs of rows with the same degree can be understood through the following example. Suppose that, in a mutualistic network, two pollinators $i$ and $j$ have a degree of $24$ and $25$, respectively, and suppose that they have $18$ common neighbors (plants). This pair $(i,j)$ contributes to $\mathcal{N}^{R}$ with a contribution equal to $\mathcal{N}_{ij}=O_{ij}/k_i=18/24=0.75$. However, suppose that a new interaction is observed between $i$ and a plant that does not interact with $j$. In the updated network, both $i$ and $j$ have a degree equal to $25$, and their contribution to NODF drops from $0.75$ to zero: a small perturbation of the dataset can lead to a large variation in the pairwise contributions to NODF. 

By contrast, in the example above, the contribution to $\eta^R$ is only marginally affected by the pollinator $i$'s new interaction: when $i$'s degree increases from $24$ to $25$, the contribution $\eta_{ij}=O_{ij}/\min \{k_i,k_j\}$ to $\eta^R$ decreases from $18/24=0.75$ to $18/25=0.72$, but it remains substantially larger than zero. For pairs of nodes of similar degree, the pairwise contributions to S-NODF are more stable with respect to small perturbations of node degree as compared to the pairwise contributions to the original NODF. 

One can compute analogously the columns nestedness; Bastolla \textit{et al.}~\cite{bastolla2009architecture} defined the overall degree of nestedness $\eta$ as:
\begin{equation}
\eta =\frac{{{\eta}^{R}}+{{\eta}^{C}}}{2}.
\label{bastollanodf1}
\end{equation}
 This nestedness index ranges from zero to one, where one corresponds to a perfectly nested network; it is highly correlated with NODF~\cite{bastolla2009architecture}, yet it has the advantage of being more stable with respect to small structural perturbations.
 Differently from Eq.~\eqref{bastollanodf1}, one can normalize $\eta^R+\eta^C$ by the total number of pairs, as it was done for the original NODF (see Eq.~\eqref{nodf}). This leads to the alternative definition:
 \begin{equation}
     \eta =\frac{{{\eta}^{R}}+{{\eta}^{C}}}{\frac{N\,(N-1)}{2}+ \frac{M\,(M-1)}{2}}.
     \label{bastollanodf2}
 \end{equation}
 We argue that the definition of $\eta$ provided by Eq.~\eqref{bastollanodf2} should be preferred to that by Eq.~\eqref{bastollanodf1}, especially when the number of rows is substantially different than the number of columns.

\paragraph{JDM-NODF} Johnson \textit{et al.}~\cite{johnson2013factors} considered a different normalization of nodes' pairwise overlap as compared to NODF and S-NODF. For a unipartite network, they defined the total nestedness as
\begin{equation}
\eta=\frac{1}{N^2}\sum_{i,j}\frac{O_{ij}}{k_i\,k_j}.
\label{eta}
\end{equation}
In a similar way as in~\cite{beckett2014falcon}, we refer to this metric as JDM-NODF after its inventors.
This expression corrects a misprint of the metric introduced by Bastolla et al.~\cite{bastolla2009architecture} (see footnote 8 above), and it allows us to estimate the contribution $\eta_i$ of each node to the global level of nestedness: 
\begin{equation}
    \eta_i=\frac{1}{N}\sum_j \frac{O_{ij}}{k_i\,k_j}.
    \label{individual}
\end{equation}

\paragraph{W-NODF} The weighted NODF (W-NODF)~\cite{almeida2011straightforward} metric is specifically tailored to weighted networks.
To assess the degree of nestedness of the columns of a weighted bipartite network, one compares the weight $w_{i\alpha}$ of each link $(i,\alpha)$ of a given 
column-node $\alpha$ with the weights $w_{i\beta}$ of the links between column-nodes of higher degree ($\beta$ such that $k_\beta>k_\alpha$) and the same partner $i$.
The columns' W-NODF $\mathcal{N}^{C,W}$ is larger if the weights $w_{i\alpha}$ tend to be smaller than the corresponding weights $w_{i\beta}$ for higher-degree column-nodes $k_\beta>k_\alpha$
The definition is the following~\cite{almeida2011straightforward}
\begin{equation}
\mathcal{N}^{C,W}=\sum_{(\alpha,\beta)} \frac{K_{\alpha\beta}}{k_\alpha}\,\Theta(k_{\beta}-k_{\alpha}),
\end{equation}
where $K_{\alpha\beta}$ denotes the number of $\alpha$'s weighted links with lower weights than the links between higher-degree 
column-nodes and the same row-node:
\begin{equation}
K_{\alpha\beta}=\Theta(k_{\beta}-k_{\alpha})\,\sum_{i}A_{i\alpha}\,A_{i\beta}\,\Theta(w_{i\beta}-w_{i\alpha}).
\end{equation}
One can define an analogous weighted NODF $\mathcal{N}^{R,W}$ for the rows, and then define the total weighted NODF $\eta^{W}$
\begin{equation}
\eta^{W}=\frac{\mathcal{N}^{R,W}+\mathcal{N}^{C,W}}{\frac{N\,(N-1)}{2}+\frac{M\,(M-1)}{2}}.
\label{wnodf}
\end{equation}
This equation is analogous to Eq.~\eqref{nodf}, with the difference that the rows' and columns' contributions depends not only on which pairs of nodes interact, but also on the weights of such interactions.

\paragraph{Normalizing overlap metrics based on null models} While one can in principle compare the observed
values of any nestedness metrics with the values observed in randomized networks
obtained with a given null model through standard 
statistical methods (like the $z$-score and $p$-value, see 
Section \ref{sec:null}), scholars have also incorporated
null-models effects directly into the overlap-based nestedness metrics. This
has been done by including a (multiplicative or additive) normalization
term which is typically determined by the expected values of the metrics under a suitable null model. 
The $C$ metric by Wright and Reeves described above (see Eq.~\eqref{wr}) is an example of a normalized metric; 
we describe here analogous normalizations for overlap metrics. 

Based on the JDM-NODF metric defined by Eq.~\eqref{eta}, Johnson \textit{et al.}~\cite{johnson2013factors} calculated the 
expected nestedness $\eta_{CM}$ of a random unipartite network with expected degree sequence equal
to the original network's degree sequence $\{k_1,\dots,k_N\}$, by posing $A_{ij}=k_i\,k_j/(N\braket{k_i})$ in Eq.~\eqref{eta}. We readily obtain $\eta_{CM}=\braket{k^2}/(N\braket{k}^2)$; if one is interested in capturing the nestedness that cannot be explained by the degree sequence, it is therefore useful to consider the normalized JDM-NODF
\begin{equation}
\eta=\frac{\tilde{\eta}}{\eta_{CM}}=\frac{\braket{k}^2}{N\,\braket{k^2}}\sum_{i,j}\frac{O_{ij}}{k_i\,k_j}.
\label{johnson}
\end{equation}
Such metric is used to reveal the relation between nestedness and degree-degree correlations in synthetic and real networks
(see Section~\ref{sec:disassortativity}).

In a similar spirit but with a different methodology, Solé-Ribalta \textit{et al.}~\cite{sole2018revealing} considered a variant of
NODF that compares the observed level of nestedness (as determined by a variant of the NODF function defined by Eq.~\eqref{nodf}) with the expected nestedness under a suitable null model. In formulas, their normalized nestedness metric is given by 
\begin{equation}
\widetilde{\mathcal{N}}=\frac{2}{N+M}\Biggl\{\sum_{i,j}\frac{O_{ij}-\braket{O_{ij}}}{(N-1)\,k_j}\,\Theta(k_i-k_j)+ \sum_{\alpha,\beta}\frac{O_{\alpha\beta}-\braket{O_{\alpha\beta}}}{(M-1)\,k_\beta}\,\Theta(k_\alpha-k_\beta)\Biggr\}.
\label{ibn0}
\end{equation}
Solé-Ribalta \textit{et al.}~\cite{sole2018revealing} evaluated
the expected row-row overlap $\braket{O_{ij}}$ and column-column 
overlap $O_{\alpha\beta}$ as $\braket{O_{ij}}=\braket{O_{ji}}=k_i\,k_j/M$ and
$O_{\alpha\beta}=k_{\alpha}\,k_\beta/N$, respectively\footnote{The two 
expected values $\braket{O_{ij}}=\braket{O_{ji}}=k_i\,k_j/M$ is obtained
by and $O_{\alpha\beta}=k_{\alpha}\,k_\beta/N$ are obtained with the PE and EP null model, respectively. In the PE (EP) model,
the probability that a row-column pair of nodes interact is proportional to the degree 
of the row-node (column-node) -- see Section~\ref{sec:nine} for details.}.
The behavior of this function deviates from that of the original NODF for dense matrices where the expected pairwise overlaps 
with the largest-degree nodes tend to be large and, as a result, the respective contributions to
$\widetilde{\mathcal{N}}$ tend to be small~\cite{sole2018revealing}.

\subsubsection{Eigenvalue-based metrics}
\label{sec:eigenvalue}

What can the eigenvalues of the network's adjacency matrix tell us about nestedness?
The last metric for nestedness considered here is the \emph{spectral radius}~\cite{staniczenko2013ghost}, which builds upon graph-theoretical results on the spectral properties of the adjacency matrix of bipartite networks~\cite{bell2008graphs,bhattacharya2008first}.
The spectral radius $\rho(\mat{A})$ of a network is simply defined as the dominant eigenvalue of the network's adjacency matrix $\mat{A}$~\cite{chung1997spectral}.
Staniczenko \textit{et al.}~\cite{staniczenko2013ghost} considered $\rho(\mat{A})$ as a metric to quantify nestedness.
This interpretation is motivated by the two following theorems: 
\begin{itemize}
\item Consider all the connected bipartite
networks composed of $S$ nodes and $E$ edges. The network with the largest spectral radius is a perfectly nested network. This theorem has been proved by Bell \textit{et al.}~\cite{bell2008graphs}.
\item Consider all the connected bipartite
networks composed of $N$ row-nodes, $M$ column-nodes, and $E$ edges. The network with the largest spectral radius is a perfectly nested network. This theorem has been proved by Bhattacharya \textit{et al.}~\cite{bhattacharya2008first}.
\end{itemize}
While the number of row-nodes and column-nodes is allowed to vary in the first theorem (provided that their sum is equal to $S$), the second theorem considers all networks with a fixed number of both row- and column-nodes.
Therefore, if perfectly nested structures lead to the largest spectral radius for bipartite networks, the spectral radius $\rho(\mat{A})$ can be used to quantify the degree of 
nestedness of a given network: the larger $\rho(\mat{A})$, the more nested the network. The advantage of $\rho(\mat{A})$ is that it can be computed in short time through well-established techniques such as the power-method~\cite{stewart1994introduction,newman2010networks}. Besides, one can readily apply it to weighted networks -- in that case, one simply needs to measure the spectral radius of the weighted adjacency matrix.

While maximal $\rho(\mat{A})$ is associated with a perfectly nested structure, it remains to be determined
whether deviations from the maximal $\rho(\mat{A})$ should be deemed as statistically significant or not. To address this question, Staniczenko \textit{et al.}~\cite{staniczenko2013ghost} defined the $p$-value of a network under a given null model as the probability that 
a network randomized under that null model exhibits a larger $\rho(\mat{A})$. Such definition depends critically on the null model that is chosen for the randomization. For unweighted bipartite networks, Staniczenko \textit{et al.}~\cite{staniczenko2013ghost} considered three null models of different degree of conservativeness (both a model that preserves the degree sequence and two models that do not, see Section~\ref{sec:nine}). For weighted bipartite networks, Staniczenko \textit{et al.}~\cite{staniczenko2013ghost} consider an additional null model that shuffles the weight values in the adjacency matrix $\mat{A}$ but not their position.

\subsection{Including a null model}
\label{sec:null}

Given a network and a structural pattern of interest, one is typically interested in assessing its
statistical significance: is the observed structural pattern compatible with that observed in 
randomized networks which preserve (a small number of) interesting macroscopic properties of the original network?
To address this question for nestedness, scholars have introduced various null models 
and statistical tests, and applied them to the nestedness metrics introduced in Section~\ref{sec:metrics1}.

How to choose the null model has been widely debated in the ecology literature. Back in the 70s, among the first scholars who attempted to infer the significance of structural patterns in ecological networks, Sale~\cite{sale1974overlap} analyzed networks where species are connected with the resources they use, and he generated synthetic random networks by keeping fixed the number of resources used by each species, and allowing the number of species using a resource to vary (Fixed-Equiprobable model, see Section~\ref{sec:nine}).
Connor and Simberloff~\cite{connor1979assembly} used a null model that preserves exactly species' and 
islands' degrees in spatial networks, in the same spirit as the popular configuration model~\cite{molloy1995critical,newman2001random,maslov2002specificity}. 
In 1982, Gilpin and Diamond~\cite{gilpin1982factors} used a null model which preserves, on average, the species' and islands' degree; their model might be considered as a precursor of the popular Chung-Lu model~\cite{chung2002connected}.

A variety of additional randomization procedures have been introduced in the 80s and 90s in the ecological literature~\cite{gotelli2000null,ulrich2009consumer}; the basics of such models are provided in Section~\ref{sec:nine}.
More recently, Squartini and Garlaschelli~\cite{squartini2011analytical} contributed to the significance assessment problem by introducing a framework to compute analytically the expected properties of networks with a given expected degree sequence.
In the same vein as the computation of thermodynamic properties in statistical mechanics~\cite{parisi1988statistical}, their calculation is based on constructing a maximum-entropy ensemble of networks with expected degree sequence equal to the original network's degree sequence (see Section~\ref{sec:maxent}).
This statistical-physics approach has been exploited recently to assess the significance of nestedness in economic~\cite{saracco2015randomizing}, and ecological~\cite{borras2017breaking} networks.

\subsubsection{Nine basic classes of randomization procedures}
\label{sec:nine}

When randomizing a given bipartite adjacency matrix, 
one has essentially two possible choices to make -- one for the row-nodes and one for the column-nodes -- about the preservation
of the individual nodes' degree. In line with Gotelli~\cite{gotelli2000null}, for both row-nodes and column-nodes, we consider three options: (1) preserving 
exactly the individual nodes' degree (the degree is \emph{fixed}); (2) preserving on average the nodes' degree 
(the interaction probability is \emph{proportional} to degree); (3) not preserving
the nodes' degree, and assuming that the all the nodes have the same probability to interact with the nodes of the 
other guild (the pairwise interactions are \emph{equiprobable}). As the choice among such three options 
can be made independently for rows and columns, we have $3^2=9$ resulting classes of null models (see Table~\ref{tab:null}). 

We emphasize that as soon as one has chosen the properties to keep fixed while randomizing the rest, there are multiple alternative possible implementations for the resulting model. Surveying all the possible implementations falls out of the scope of this review, but we shall provide the main ideas of some of them. It is also worth noticing that scholars have started investigating more nuanced scenarios where one can ``tune'' the level of discrepancy of row-nodes' and column-nodes' degree (with respect to the observed degrees) in a continuous manner~\cite{strona2018bi}. These recent developments are not included in this review; nevertheless, they demonstrate that research on null models is still active in ecology, and consensus on the answers to fundamental questions (e.g., which null model is more suitable for a given network and a given structural pattern?) has yet to be achieved.

\begin{table}
\begin{center}
\begin{tabular}{ |c|c|c|c| } 
\hline
 				& Column Equiprobable & Column Fixed  & Column Proportional\\ 
\hline
Row Equiprobable & EE & EF & EP \\ 
\hline
Row Fixed & FE & FF & FP \\
\hline
Row Proportional & PE & PF & PP \\
\hline
\end{tabular}
\end{center}
\caption{Nine basic classes of randomization procedures, corresponding to nine different null models. Each randomization procedure depends on two independent choices. The first choice is whether to preserve the row-nodes' degree. The three possible options are to not preserve it (Row Equiprobable), to preserve it exactly (Row Fixed), or to preserve it on average (Row Proportional). The second choice is whether to preserve the column-nodes' degree. Again, the three possible options are to not preserve it (Column Equiprobable), to preserve it exactly (Column Fixed), or to preserve it on average (Column Proportional). Each pair of choices leads to a specific randomization procedure (see the main text for a detailed explanation of each of them). This Table is inspired by Table 2 in~\cite{gotelli2000null}.}
\label{tab:null}
\end{table}

\begin{figure*}[t]
	\centering
	\scalebox{0.8}[0.8]{\rotatebox{270}{\includegraphics{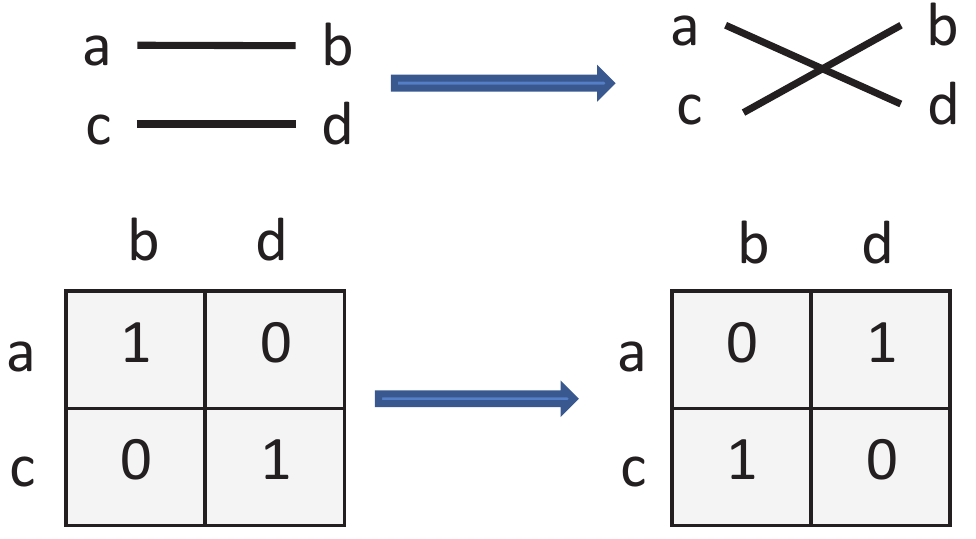}}}
	\caption{An illustration of a popular implementation~\cite{gotelli2001swap,maslov2002specificity} of the Fixed-Fixed null model which randomizes the network by preserving exactly the nodes' degree. One randomly selects two links of the network $(a,b)$ and $(c,d)$; the two selected links are replaced by two new links, $(a,d)$ and $(b,c)$, if these two links did not exist before. In terms of the adjacency matrix, such swap corresponds to a swap of ``checkerboard units"~\cite{stone1990checkerboard}, where the initial and final configurations of the depicted 2x2 matrix are both checkerboard units.}\label{fig:ff}
\end{figure*}

 Below, we provide details for the nine basic null models of Table~\ref{tab:null}, we provide some of the alternative names that have been used in the literature, and we point to some of the papers that have used them. 

\begin{itemize}
\item \textit{Row Equiprobable -- Column Equiprobable (EE) models.} All interactions are equiprobable, 
independently of the row-nodes' and column-nodes' degree. In a possible randomization procedure, the $E$ original links are re-assigned to randomly selected pairs $(i,\alpha)$ of nodes; the probability that pair $(i,\alpha)$ is filled with a link is given by $1/(N\,M)$~\cite{gotelli2000null}.
Alternative names for this model are: SIM1, R00~\cite{gotelli2000null,ulrich2009consumer}, Type I model~\cite{bascompte2003nested}. In ecology, scholars who have used this null model include: Atmar and Patterson~\cite{atmar1993measure}, Gotelli~\cite{gotelli2000null}. In the graph-theoretical language, the EE model generates Erdős–Rényi graphs~\cite{newman2010networks,barabasi2016network}.

\item \textit{Row Equiprobable -- Column Fixed (EF) models.} The column-nodes' degree sequence $\{k_{\alpha}\}$ is kept fixed, whereas their partners are chosen at random. 
In a possible randomization procedure, for each column $\alpha$, the $k_\alpha$ original links are re-assigned to randomly selected row-nodes; the probability that node $i$ is chosen as a partner is given by $1/N$~\cite{gotelli2000null}. An alternative name for this model is SIM3~\cite{gotelli2000null}.  In ecology, scholars who have used this null model include Gotelli~\cite{gotelli2000null}.

\item \textit{Row Equiprobable -- Column Proportional (EP) models.} The probability 
that a pair of nodes interact is proportional to the column-node degree, whereas all row-nodes are equiprobable. 
In a possible randomization procedure, the $E$ original links are re-assigned to randomly selected pairs $(i,\alpha)$ of nodes; the probability that pair $(i,\alpha)$ is filled with a link is given by $k_\alpha/(N\,E)$~\cite{gotelli2000null}.
An alternative name for this model is SIM6~\cite{gotelli2000null}. In ecology, scholars who have used this null model include Gotelli~\cite{gotelli2000null}. 

\item \textit{Row Fixed -- Column Equiprobable (FE) models.} The row-nodes' degree sequence $\{k_{i}\}$ is kept fixed, whereas their partners are chosen at random. In the randomization procedure, for each row $i$, the $k_i$ original links are re-assigned to randomly selected column-nodes; the probability that node $\alpha$ is chosen as a partner is given by $1/M$~\cite{gotelli2000null}. Alternative names for this model are: SIM2, R0, Random0~\cite{gotelli2000null}. In ecology, scholars who have used this null model include: Sale~\cite{sale1974overlap},  Inger and Colwell~\cite{inger1977organization}, Winemiller and Pianka~\cite{winemiller1990organization}, Patterson and Atmar~\cite{patterson1986nested}, Gotelli~\cite{gotelli2000null}. 

\item \textit{Row Fixed -- Column Fixed (FF) models.} The row-nodes' and 
column-nodes' degree are exactly preserved. Alternative names for this model are: SIM9~\cite{gotelli2000null}. In ecology, scholars who have used this null model include: Connor and Simberloff~\cite{connor1979assembly}, Diamond and Gilpin~\cite{diamond1982examination}, Brualdi and Sanderson~\cite{brualdi1999nested}, Gotelli~\cite{gotelli2000null}.
A popular randomization procedure\footnote{Alternative randomization procedures have been introduced in the literature. We refer the interested reader to~\cite{sanderson1998null,gotelli2001swap,miklos2004randomization,strona2014fast,carstens2015proof,fosdick2018configuring} for various randomization procedures and their validation; to Appendix F of the book~\cite{bascompte2013mutualistic} for a general overview in ecology.} to keep a network's degree sequence fixed is based on pairwise link swapping (see~\cite{sanderson1998null,maslov2002specificity} and Fig.~\ref{fig:ff}). In the network science language, the FF model is equivalent to the popular \emph{configuration model}~\cite{newman2010networks}.

\item \textit{Row Fixed -- Column Proportional (FP) models.} The row-nodes' degree sequence $\{k_{i}\}$ is kept 
fixed, whereas their partner column-nodes are chosen with probability proportional to their degree. In a possible randomization procedure, for each row $i$, the $k_i$ original links are re-assigned column-nodes randomly selected with probability $k_\alpha/E$~\cite{gotelli2000null}. Alternative names for this model are: SIM4~\cite{gotelli2000null}. In ecology, scholars who have used this null model include: Coleman \textit{et al.}~\cite{coleman1982randomness}, Graves
and Gotelli \cite{graves1993assembly}, Gotelli~\cite{gotelli2000null}.

\item \textit{Row Proportional -- Column Equiprobable (PE) models.} The probability that a pair of nodes interact 
is proportional to the row-node degree, whereas all column-nodes are equiprobable.  In a possible randomization procedure, 
the $E$ original links are re-assigned to randomly selected pairs $(i,\alpha)$ of nodes; the probability that pair $(i,\alpha)$ is filled with a link is given by $k_i/(M\,E)$~\cite{gotelli2000null}.
An alternative names for this model is SIM7~\cite{gotelli2000null}. In ecology, scholars who have used this null model include Gotelli~\cite{gotelli2000null}. 

\item \textit{Row Proportional -- Column Fixed (PF) models.} The column-nodes' degree sequence $\{k_{\alpha}\}$ is kept fixed, 
whereas their partner row-nodes are chosen with probability proportional to their degree. In a possible randomization procedure, for each column $\alpha$, the $k_\alpha$ original links are re-assigned to row-nodes randomly selected with probability given by $k_i/E$~\cite{gotelli2000null}. Alternative names for this model are: SIM5, R1, Random1~\cite{gotelli2000null}. In ecology, scholars who have used this null model include: Abele and Patton~\cite{abele1976size}, Connor and Simberloff~\cite{connor1979assembly},
Patterson and Atmar~\cite{patterson1986nested},
Gotelli~\cite{gotelli2000null}.

\item \textit{Row Proportional -- Column Proportional (PP) models.} The probability that a pair of nodes interact is 
proportional to both nodes' degrees. In a possible randomization procedure, the $E$ original links are re-assigned to randomly selected pairs $(i,\alpha)$ of nodes; the probability that pair $(i,\alpha)$ is filled with a link is given by $k_i\,k_{\alpha}/E^2$~\cite{gotelli2000null}. An alternative name for this class of models is SIM8~\cite{gotelli2000null}. In Bascompte \textit{et al.}'s implementation~\cite{bascompte2003nested},
the probability that a matrix element $A_{i\alpha}$ is filled is given by $(k_i+k_{\alpha})/2$ -- the resulting model is referred to as Type~II model~\cite{bascompte2003nested}. In ecology, scholars who have used this 
class of null models include:  Gilpin and Diamond~\cite{gilpin1982factors}, Gotelli~\cite{gotelli2000null}, 
Bascompte \textit{et al.}~\cite{bascompte2003nested}, among many others. Beyond ecology, randomization procedures that preserve, 
on average, the network's degree sequence are popular in the network science literature.
Among the null models that belong to the PP family, we find the popular Chung-Lu model~\cite{chung2002connected}, the maximum-entropy models that are detailed in Section~\ref{sec:maxent}, generalized hypergeometric ensembles~\cite{casiraghi2016generalized}. 
\end{itemize}

The careful reader may have noticed that interchanging the roles of row-nodes and column-nodes is equivalent to swapping pairs of classes of null models that fall out of Table~\ref{tab:null}'s diagonal. In other words, if we transpose the original adjacency matrix of the network, the effect of EP models acting on $\mathsf{A}^T$ is equivalent to the effects of the corresponding PE models acting on $\mathsf{A}$. For this reason, the PE and EP models can be seen as equivalent; nevertheless, in line with previous works~\cite{gotelli2000null,strona2018bi}, we preferred to present separately pairs of models that are equivalent upon transposition of the adjacency matrix, thereby implicitly assuming that the assignment of the nodes to $\mathsf{A}$'s rows and columns is fixed\footnote{For an ecological bipartite network where the assignment of the nodes to the $\mathsf{A}$'s rows and columns is fixed (e.g., a mutualistic network where $\mathsf{A}$'s rows reprent insects and columns represent plants), there are ecological reasons to
randomize differently the two groups of nodes. For example, in a plant-pollinator network, plants might be subject to many more phenotypical constraints than insect pollinators, thereby restraining their degree of specialization. Instead, insect pollinators might exhibit a higher flexibility to change their
diet. In a similar fashion, for spatial networks, varying the richness of a species or modifying its spatial distributions correspond to different ecological assumptions.}.

\paragraph*{Which null model to choose?}

When analyzing a collection of empirical networks, the choice of a null model critically affects the fraction of networks that result as ``significantly nested".
Scholars have pointed out that while applying the EE model is the simplest way to randomize a given network, assuming equiprobable interactions might lead to highly unrealistic random networks.
Ulrich~\cite{ulrich2009consumer} summarizes this discontent by pointing out that ``there is growing acceptance that null models that do not consider species-specific differences and variability among sites should not be used in
biogeographic studies~\cite{jonsson2001null,ulrich2007null,ulrich2007disentangling,moore2007toward} and even in
analyses of interaction matrices~\cite{vazquez2006community,vazquez2007parsimony,stang2007both}". 
Based on their results, Joppa \textit{et al.}~\cite{joppa2010nestedness} argue that ``loose constraints invariably lead to the conclusion of significant nestedness", and that such loosely-constrained null models ``create consumers that differ in how generalized their diets are and this
confounds the conclusion".
In other words, there is agreement that the EE model is sensitive to
Type I errors: it can detect a significant nestedness in networks when there is none~\cite{gotelli2012statistical, beckett2014falcon}.

On the other hand, the FF model is the most conservative model, as it preserves exactly both the row-nodes' and 
the column-nodes' degree. Both the FF and the PP model set out to address the following natural question: are networks
still significantly nested when one constrains the degree sequence to be the same (exactly or on average) as that of the original network? 
The use of a more conservative null model substantially mitigates the fraction of networks that are found as 
"significantly nested". Yet, the FF model is more heavily affected by Type II errors~\cite{ulrich2007null};
we will come back to this point in Section~\ref{sec:degree_distribution}.

Another potential issue with the randomization procedures described above (except for the FF model) 
is that they might generate ``degenerate matrices'' where some nodes have a degree equal to zero. To overcome this 
potential shortcoming, one can reject the degenerate networks (as done by the Swappable-Swappable
model~\cite{staniczenko2013ghost,beckett2014falcon}), or 
perform the swapping between two matrix elements only if it does not cause one of the nodes' degree to 
drop to zero (as done by the Cored-Cored model~\cite{beckett2014falcon}).

Finally, using randomized networks to generate random networks has two additional limitations: (1)
it might be computationally slow when one attempts to preserve exactly the nodes' degree; (2) it relies on an arbitrary choice on the number of independent realizations of the randomization procedure.
We would not face those two problems if we were able to compute analytically the expected values of the structural properties of interest. In Section~\ref{sec:maxent}, we will see that this is made possible by a compelling analogy with the problem of computing the average value of macroscopic observables in statistical mechanics.




\subsubsection{Computing the expected nestedness: maximum-entropy approach}
\label{sec:maxent}

Null models based on network randomization have the disadvantage that the results based on them depend on the number of 
performed independent randomizations. Increasing the number of randomizations might make the results more robust, but substantially
increase the computational time, especially for large networks.
To overcome these limitations, a recent stream of literature~\cite{squartini2011analytical,casiraghi2016generalized}
has provided theoretical frameworks to analytically compute the expected properties of random
networks with a set of fixed macroscopic properties, for network properties that can be expressed in terms of the adjacency matrix $\mat{A}$. 
In particular, Squartini and Garlaschelli~\cite{squartini2011analytical} introduced a maximum-entropy framework to analytically calculate the expected value of network structural properties that can be expressed in terms of the network's adjacency matrix $\mat{A}$. 

In the maximum-entropy 
framework, one calculates the expected value of the network properties over a maximum-entropy 
ensemble of networks which preserves, on average, some network properties -- the degree sequences, in all 
the cases considered here. This relieves us of running numerical simulations, which makes the null-model
expectations faster to be computed and independent of the choice on the number of independent realizations of the adopted randomization procedure.
Importantly, Squartini and Garlaschelli's framework has been also generalized to 
weighted~\cite{mastrandrea2014enhanced,squartini2017network} and bipartite networks~\cite{saracco2015randomizing}.
The maximum-entropy method for unipartite networks has been applied to the country-country export network~\cite{squartini2011randomizing} 
and to the interbank network~\cite{squartini2013early}.
Their generalization to bipartite networks has been applied to assess the significance of nestedness in
bipartite World Trade networks~\cite{saracco2015randomizing} and mutualistic networks~\cite{borras2017breaking} (see Section~\ref{sec:degree_distribution}).
We provide here the basic ideas behind the method, and we refer the
interested reader to~\cite{squartini2011analytical,saracco2015randomizing,borras2017breaking} for all the details.

\paragraph{Maximum-entropy approach in unipartite networks}

In the maximum-entropy approach~\cite{squartini2011analytical}, we are interested in computing 
the expected value of a given observable property $\Omega$ of the network over a maximum-entropy ensemble of graphs
with a given set of constrained properties.
Let us denote the value of observable $\Omega$ in the observed network $\mathcal{G}^*$ by $\Omega^*$, 
the value of $\Omega$ in a generic network $\mathcal{G}$ as $\Omega(\mathcal{G})$.
Let us denote by $\vek{\mathcal{C}}$ the vector of constrained properties of the maximum-entropy ensemble that we will construct.
In the following, the constrained property is the complete network degree sequence; in this case, the vector $\vek{\mathcal{C}}$
has $N$ elements, and $\mathcal{C}_i=k_i^*$.

In Squartini and Garlaschelli's framework~\cite{squartini2011analytical}, one seeks to find a probability 
distribution $P(\mathcal{G})$ over the ensemble $\{\mathcal{G}\}$ of possible random graphs that maximizes the entropy
\begin{equation}
S[P(\mathcal{G})]=-\sum_{\mathcal{G}}P(\mathcal{G})\log{(P(\mathcal{G}))}
\end{equation}
under the constraints
\begin{equation}
\begin{split}
&\sum_{\mathcal{G}}P(\mathcal{G})=1,\\
&\sum_{\mathcal{G}}P(\mathcal{G})\,k_i(\mathcal{G})=k_i^* \,\,\,\,\text{for all}\,\,\, i\in\{1,\dots,N\}.
\end{split}
\end{equation}
Basically, we aim to maximize the entropy of the network probability distribution while keeping fixed the average nodes' degree.

The problem is analogous to the entropy maximization in statistical mechanics~\cite{parisi1988statistical} where, given a system with a large number of microscopic constituents, one seeks to maximize the entropy of the probability distribution over the allowed microscopic configurations while keeping fixed the average total energy of the system. Therefore, we know from statistical mechanics that the solution of the problem is given by the canonical distribution
\begin{equation}
P(\mathcal{G})=\frac{\exp{(-H(\mathcal{G},\vek{\theta}))}}{Z(\vek{\theta})},
\end{equation}
where $\vek{\theta}=\{\theta_1,\dots,\theta_N\}$ is a vector of $N$ Lagrange multipliers, and
\begin{equation}
H(\mathcal{G},\vek{\theta})=\vek{\theta}\cdot \vek{k}(\mathcal{G})=\sum_i \theta_i\,k_i(\mathcal{G}).
\end{equation}
is the Hamiltonian of the system, and $Z(\vek{\theta}):=\sum_{\mathcal{G}'}\exp{(-H(\mathcal{G}',\vek{\theta}))}$
is the partition function.
The optimal Lagrange multipliers $\vek{\hat{\theta}}$ can be found by first noticing that the expected value of $A_{ij}$ in the maximum-entropy ensemble is given by 
\begin{equation}
\braket{A_{ij}}=\frac{e^{-\theta_i-\theta_j}}{1+e^{-\theta_i-\theta_j}};
\end{equation}
by enforcing the constraint $\braket{\vek{k}(\mathcal{G'})}=\vek{k}$, we obtain the equation:
\begin{equation}
\sum_{j\neq i}\frac{\exp{(-\hat{\theta}_i-\hat{\theta}_j)}}{1+\exp{(-\hat{\theta}_i-\hat{\theta}_j)}}=k_i^* \,\,\,\,\text{for all}\,\,\, i\in\{1,\dots,N\}.
\label{lagrange}
\end{equation}
Eq.~\eqref{lagrange} corresponds to a system of $N$ equations that has a unique solution $\vek{\hat{\theta}}$. We refer the interested reader to to~\cite{squartini2011analytical} for the full derivation of these equations.

Topological observables can be considered as functions $\Omega(\mat{A})$ of the network's adjacency matrix. By denoting as $\mat{A}^{(exp)}$ the 
matrix whose element $A^{(exp)}_{ij}=\braket{A_{ij}}(\hat{\theta})$, we obtain the expected value of observable $\Omega$ in 
the maximum-entropy ensemble as $\Omega(\mat{A}^{(exp)})$. One can also compute the standard deviation of observable $\Omega$ in the maximum-entropy ensemble as
\begin{equation}
\sigma[\Omega]\simeq \sqrt{\sum_{i,j}\Biggl( \sigma^*[A^{(exp)}_{ij}] \,\frac{\partial \Omega[\mat{A}]}{\partial  A_{ij}}\Bigg|_{A_{ij}=A^{(exp)}_{ij}}\Biggr)^2},
\label{stddev}
\end{equation}
where $\sigma^*[A^{(exp)}_{ij}]=\sqrt{A_{ij}^{(exp)}\,(1-A_{ij}^{(exp)})}$ (see Eq. B31 in~\cite{squartini2011analytical} and its derivation).

\paragraph{Maximum-entropy approach in bipartite networks}

The framework for bipartite networks~\cite{saracco2015randomizing,borras2017breaking} is conceptually analogous. Differently from unipartite networks, one needs two sets $\theta^R$ and $\theta^C$ of Lagrange multipliers for row-nodes and column-nodes, respectively. The canonical distribution is given by
\begin{equation}
P(\mathcal{G})=\frac{\exp{(-H(\mathcal{G},\vek{\theta}^R, \vek{\theta}^C))}}{Z(\vek{\theta}^R, \vek{\theta}^C)},
\end{equation}
where $\vek{\theta}^R=\{\theta^R_1,\dots,\theta^R_N\}$ is the vector of Lagrange multipliers for the $N$ row-nodes, 
$\vek{\theta}^C=\{\theta^C_1,\dots,\theta^C_M\}$ is the vector of Lagrange multipliers for the $M$ column-nodes, 
$Z$ is the normalization factor, and the Hamiltonian function is given by
\begin{equation}
H(\mathcal{G},\vek{\theta}^R, \vek{\theta}^C)=\sum_i \theta^R_i\,k_i(\mathcal{G'})+\sum_\alpha \theta^C_\alpha\,k_\alpha(\mathcal{G'}),
\end{equation}
Again, one obtains the expected values of the adjacency matrix elements as
\begin{equation}
A_{i\alpha}^{(exp)}=\frac{\exp{(-\hat{\theta}_\beta^{C}-\hat{\theta}^R_i)}}{1+\exp{(-\hat{\theta}^C_\beta-\hat{\theta}_i^C)}},
\label{exp_A}
\end{equation}
In the same way as for unipartite networks, the
values of the Lagrange multipliers are determined by fixing the average degree of row-nodes and column-nodes, which leads to the following set of equations~\cite{borras2017breaking}
\begin{equation}
\begin{split}
\sum_{\beta}\frac{\exp{(-\hat{\theta}_\beta^{C}-\hat{\theta}^R_i)}}{1+\exp{(-\hat{\theta}^C_\beta-\hat{\theta}_i^C)}}
&=k_i^* \,\,\,\,\text{for all}\,\,\, i\in\{1,\dots,N\}, \\
\sum_{j}\frac{\exp{(-\hat{\theta}^C_\alpha-\hat{\theta}^R_j)}}{1+\exp{(-\hat{\theta}^C_\alpha-
\hat{\theta}^R_j)}}&=k_\alpha^*\,\,\,\,\text{for all}\,\,\, \alpha\in\{1,\dots,M\}.
\end{split}
\label{lagrange_bi}
\end{equation}
and, consequently, the expected values of an observable $\Omega(\mat{A})$ as $\braket{\Omega(\mat{A})}=\Omega(\mat{A}^{(exp)})$. The standard deviation of $O(\mat{A})$ can be obtained with an equation analogous to Eq.~\eqref{stddev} (see~\cite{borras2017breaking} for details).

\paragraph{Using the maximum-entropy framework to compute the expected nestedness of a network}

The maximum-entropy analytic framework can be used to compute the expected
value and standard deviation of any structural observable $\Omega(\mat{A})$ that explicitly depends on the network's adjacency
matrix $\mat{A}$. Some of the nestedness metrics introduced in Section~\ref{sec:metrics} are explicit
functions of $\mat{A}$. A prominent example is NODF: the analytic calculation of the expected NODF for 
a set of mutualistic networks was carried out by Borràs \textit{et al.}~\cite{borras2017breaking}. 
To perform the calculation, essentially, one replaces $\mat{A}$ with $\mat{A}^{(exp)}$ as determined by Eq.~\eqref{exp_A} in the NODF definition (Eq.~\eqref{nodf}). Importantly, Borràs \textit{et al.}~\cite{borras2017breaking} were also able to compute the standard deviation of NODF, which allowed them to compute the $z$-score (see Section~\ref{sec:tests}) of the empirically observed NODF values.
We refer to Section~\ref{sec:degree_distribution} for a discussion of their results.
Importantly, one can also use the maximum-entropy framework to compute "algorithmic" nestedness metrics (such as the nestedness temperature introduced in Section~\ref{sec:distance}) by sampling from the analytic probability of interaction between nodes (encoded in the $\mat{A}^{(exp)}$ matrix), and then computing the moments of the algorithmic metrics over the set of generated sampled networks.


\subsubsection{Statistical tests}
\label{sec:tests}

The randomization procedures presented in Section~\ref{sec:nine} can be used to generate multiple independent randomized adjacency matrices, and to accumulate thereby statistics on the metrics for nestedness. Alternatively, one can compute analytically the expected degree of nestedness based on a maximum-entropy ensemble of networks with fixed average degree distribution (Section~\ref{sec:maxent}).
How to use this information to quantify the deviation of the degree of nestedness observed in a given real network from that expected in the corresponding random networks? Statistical tests allow us to address this question; we present here the most common statistical tests which are typically used in nestedness analysis.

In the following, we suppose that we are interested in one particular nestedness metric $\mathcal{N}$ and a given observed 
network $\mathcal{G}^*$. We denote by $\mathcal{N}^*$ the value observed in the real network of interest, and by $\mu(\mathcal{N})$ and $\sigma(\mathcal{N})$ the mean and the standard deviation, respectively, of the nestedness metric in the ensemble of random networks that are generated with the chosen null model.
All the statistical tests below are based on the comparison between the observed level of nestedness $\mathcal{N}^*$ and the statistical properties of $\mathcal{N}$ over the ensemble of random networks.

\paragraph{$z$-score}
The $z$-score~\cite{ulrich2009consumer,beckett2014falcon} is calculated as
\begin{equation*}
z(\mathcal{N})=\frac{\mathcal{N}^*-\mu(\mathcal{N})}{\sigma(\mathcal{N})}.
\end{equation*}
It tells us the distance from the mean of a given observed 
value of nestedness $\mathcal{N}^*$, in units of standard deviations. 
By randomizing empirical ecological networks with the FF model, Almeida-Neto \textit{et al.}~\cite{almeida2008consistent} found that $z(\mathcal{N})$ is only weakly correlated with network size and connectance for the overlap metric by Wright and Reeves (see Eq.~\eqref{wr}), the discrepancy index by Brualdi and Sanderson (see Section~\ref{sec:gap}), the NODF metric, 
and matrix temperature (as determined by the NTC algorithm).
The $z$-scores can be also used to compare the information 
provided by different metrics for nestedness (see Fig.~\ref{fig:z_score} below and the related discussion in Section~\ref{sec:degree_distribution}).

\paragraph{$p$-value}
The $p$-value~\cite{beckett2014falcon} of an observed value of
nestedness $\mathcal{N}^*$ is the probability that a randomized matrix exhibits a degree of nestedness that is larger than $\mathcal{N}^*$. Low values of $p$ ($p\to0$) indicate that
the input matrix is highly nested relative to the null distribution; usually, a threshold value $\lambda$ (e.g., $\lambda=0.05$ or $0.01$) is used to
denote a statistically significant level of nestedness ($p<\lambda$). For matrices
where no randomized network is more nested than the input matrix, one can conservatively assign $p<R^{-1}$ where $R$ is the number of 
independently generated random matrices~\cite{beckett2014falcon}.

\subsection{Nestedness maximization: Packing the adjacency matrix}
\label{sec:packing}

In the previous Section, we have introduced metrics to quantify the level of nestedness of a given network.
The discussed metrics are mostly based on the definition of nestedness by degree: a pair of nodes respects the nestedness condition if the neighborhood of the node \emph{with lower degree} is included in the neighborhood of the node \emph{with larger degree}. This definition crucially depends on the ranking of the nodes by their degree.

On the other hand, we have seen that metrics based on temperature (Section~\ref{sec:distance}) require a ``packing'' of
the matrix that re-arranges the adjacency matrix's rows and columns in such a way to maximize nestedness (i.e., minimize the nestedness temperature). 
Such metrics shift the question from whether a given matrix is nested when its rows and columns are ordered by degree to whether a given matrix is nested when its rows and columns are ordered in the way that maximizes nestedness. Answering the latter question requires accurate 
methodologies to produce the optimal ordering of rows and columns; presenting them is the main goal of this Section.

Historically, the first algorithm introduced with the goal of packing the matrix to maximize nestedness is the
Nestedness Temperature Calculator (NTC) by Atmar and Patterson~\cite{atmar1993measure}.
Despite some criticism~\cite{fischer2002treating,rodriguez2006new}, the algorithm has been widely used in the ecological literature
in order to quantify the nestedness of spatial and interaction networks~\cite{ulrich2009consumer}.
On the other hand, it suffers from shortcomings as its determination of the line of perfect nestedness (see Section~\ref{sec:distance}) 
is not unique, and its iterative packing algorithm fails to identify the optimal packing in many empirical networks~\cite{rodriguez2006new}.
The genetic algorithm by Rodr{\'\i}guez-Giron{\'e}s and Santamar{\'\i}a~\cite{rodriguez2006new}, called BINMATNEST, overcomes these limitations 
(Section~\ref{sec:genetic}) and it might be considered as the state-of-the-art approach to maximize nestedness in the ecological literature.

From a different angle, scholars have been interested in quantifying the competitiveness of nations and the sophistication of products in
World Trade through network-based metrics~\cite{hidalgo2009building,tacchella2012new}. 
These metrics take as input the binary adjacency matrix of the network that connects countries with the products they export.
While this ranking problem is seemingly unrelated to the nestedness maximization problem in ecological networks, 
it turns out that one of the most studied country-product ranking algorithm, the fitness-complexity algorithm~\cite{tacchella2012new}, produces 
rankings of countries and products
that reveal a ''triangular shape``~\cite{tacchella2012new} of the adjacency matrix of the country-product network. In other words, the fitness-complexity algorithm
enhances the nestedness of the matrix.
Section~\ref{sec:nonlinear} reviews the algorithm and its variants~\cite{pugliese2016convergence,wu2016mathematics,stojkoski2016impact}.
A natural question emerges: is nestedness temperature better minimized by the genetic algorithm or by the fitness-complexity algorithm?
Section~\ref{sec:best_packing} provides a comparison of the two approaches.

\subsubsection{Genetic algorithms}
\label{sec:genetic}

The problem of sorting the rows and columns of the adjacency matrix in such a way to maximize nestedness is combinatorially difficult:
there are indeed $N!\,M!$ possible permutation of rows and columns, which makes it essentially unfeasible to 
explore all the possible rearrangements and select the one with the lowest temperature.
In similar situations, scholars have often resorted to genetic algorithms inspired by evolution~\cite{whitley1994genetic}.
In this class of computational models, one starts with a set of candidate solutions (''chromosomes``) of the problem of interest,
and recombines them according to pre-defined rules.
Candidate solutions that provide a better solution to the problem -- often, they exhibit larger values
for a target function of interest -- are given larger chance to ``reproduce''.
The BINMATNEST algorithm by Rodr{\'\i}guez-Giron{\'e}s and Santamar{\'\i}a~\cite{rodriguez2006new} is precisely a genetic algorithm
that rearranges the rows and columns of the adjacency matrix with the goal to minimize the nestedness temperature. 

The BINMATNEST genetic algorithm starts from a set of candidate solutions $\{\vek{r}^R,\vek{r}^C\}$ which includes
the ordering by degree, and several other orderings based on routines similar to the NTC algorithm.
The algorithm proceeds by producing an ``offspring'' from a well-performing solution $\vek{w}$. To this end, the 
algorithm randomly selects a ``partner'' solution $\vek{p}$ from the population of candidate solutions. 
Each element $o_i$ of the offspring solution $\vek{o}$ is given by $w_i$ with probability $0.5$, and
it is otherwise determined by a combination of information from both $\vek{w}$ and $\vek{p}$. This 
choice is performed independently for each solution, with the constraint that at least one offspring 
solution must be based on combination. The combination mechanism assigns $o_i=w_i$ for $i={1,\dots,k}$,
where $k$ is a random number uniformly extracted from $\{1,2,\dots, N-1 \}$. For $i\in\{k+1,\dots, N\}$, $o_i=p_i$ if 
that position had not been already assigned -- i.e., if $p_i \not\in \{w_1,\dots w_k\}$; if $p_i \in \{w_1,\dots w_k\}$, the 
element $o_i$ is randomly extracted from all the positions that have not yet been assigned. Finally, each offspring 
solution suffers from a mutation with probability $0.1$: a portion of the vector $\vek{o}$ is randomly selected 
and its elements undergo a cyclic permutation.

The BINMATNEST algorithm produces matrices that exhibit significantly lower 
temperature than those produced by the NTC (see Figs. 4-5 in~\cite{rodriguez2006new}). 
This has been verified in both synthetic and real networks; in real networks, the gap between the temperature by the NTC and the temperature by BINMATNEST tends to be larger for pollination networks than for spatial networks.
Besides, while the temperature by BINMATNEST depends on network density (in particular, the lowest temperature values are achieved by very sparse or very dense networks), the probability that the observed temperature is attained by chance does not, which makes it legitimate to use the $p$-values of nestedness temperature to compare the degree of nestedness of systems of different size and density~\cite{rodriguez2006new}.

\subsubsection{Non-linear iterative algorithms}
\label{sec:nonlinear}

Non-linear iterative algorithms (like the \emph{fitness-complexity algorithm}) were not explicitly 
introduced with the goal to enhance the nestedness of a given matrix. 
The algorithms were originally aimed to rank countries and products in bipartite country-product 
networks where the countries are connected with the products they export \cite{tacchella2012new}. Yet,
they turn out to be effective in enhancing nestedness~\cite{dominguez2015ranking,lin2018nestedness}, which justifies
their place here as methods to pack the adjacency matrix of a given network.

\paragraph{Fitness-Complexity metric and its interpretation}

\begin{figure}[t]
\centering
\includegraphics[scale=1]{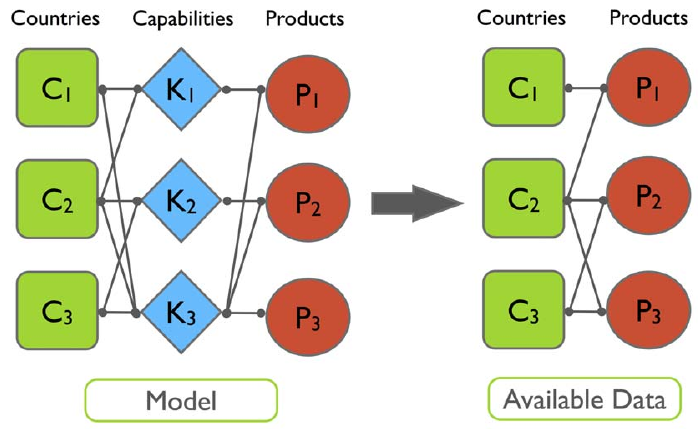}
\caption{The economic complexity interpretation of the bipartite country-product export network. In order to be produced and exported, each product requires a given set of capabilities. Only the countries that possess all the capabilities required to produce a given product can export it. The resulting observed bipartite network (right panel) can be therefore interpreted as a projection of a tripartite network where the capabilities layer cannot be observed. A model based on this simple scheme fits the structural properties of the country-product network (see Section~\ref{sec:hidden}). Reprinted from \cite{cristelli2013measuring}.}
\label{fig:capabilities}
\end{figure}

The algorithm is motivated by simple economic considerations. The economic complexity\footnote{We refer the interested reader to [\url{https://www.mdpi.com/journal/entropy/special_issues/Economic_Complexity}] for a recent special issue of the journal \emph{Entropy} on economic complexity.} approach to macroeconomics views the export of a product as the result of a production process that requires from a country all the necessary capabilities in order to fabricate that product (see Fig. \ref{fig:capabilities} for an illustration). In this view, countries that possess more capabilities are more competitive in the world trade, as they have the possibility to produce and export more products than countries with few capabilities. 

Based on these lines of reasoning, Tacchella \textit{et al.}~\cite{tacchella2012new} defined the \emph{fitness}, $F_i$, of a given country $i$ as the sum of the scores of the products exported by that country:
\begin{equation}
F_i=\sum_{\alpha}A_{i\alpha}Q_{\alpha}.
\label{fit1}
\end{equation}
Naturally, the products' scores $\{Q_{\alpha}\}$ depend on the exporting countries' scores. However, this dependence cannot be linear because a product that is exported by many countries is likely to require few capabilities to be produced, which means that it cannot be a sophisticated product.

Tacchella \textit{et al.}~\cite{tacchella2012new} therefore observe that a product that is exported by specialist countries should be penalized.
One of the simplest mathematical ways to enforce this idea is to define a given product's score as the harmonic mean of the fitness scores of its exporting countries:
\begin{equation}
Q_{\alpha}=\frac{1}{\sum_{i}A_{i\alpha}/F_{i}}.
\label{compl1}
\end{equation}
According to this definition, if a product  $\alpha$ is exported by a low-fitness country $i$, this country's small score $F_i$ gives a large contribution $1/F_i$ to the sum in the denominator, which results in a small product score $Q_{\alpha}$.

Eqs. \eqref{fit1}-\eqref{compl1} are not yet the final equations of the algorithm, because the countries' (products') scores depend on the products' (countries') scores, and we do not know either of them, a priori.
In line with the widely-used power method to compute Google's PageRank~\cite{berkhin2005survey}, one sets a uniform initial condition~\cite{tacchella2012new}
\begin{equation}
\begin{split}
F^{(0)}_i&=1, \\
Q^{(0)}_{\alpha}&=1,
\end{split}
\label{fit_comp_init}
\end{equation}
and then seeks to solve iteratively Eqs. \eqref{fit1}-\eqref{compl1}:
\begin{equation}
\begin{split}
\tilde{F}_i^{(n)}&=\sum_{\alpha}A_{i\alpha}Q_{\alpha}^{(n-1)}\\
\tilde{Q}_{\alpha}^{(n)}&=\frac{1}{\sum_{i}A_{i\alpha}/F_{i}^{(n-1)}}.
\end{split}
\label{fit_comp_norm}
\end{equation}
At each step, the scores are further normalized by their mean:
\begin{equation}
\begin{split}
F_i^{(n)}&=\tilde{F}_i^{(n)}/\braket{\tilde{F}_i^{(n)}}\\
Q_{\alpha}^{(n)}&=\tilde{Q}_{\alpha}^{(n)}/\braket{\tilde{Q}_{\alpha}^{(n)}}.
\end{split}
\end{equation}
Ideally, one would like to define the vector of country and product scores as the stationary point of these iterative equations. This is not always possible: one can show both numerically and analytically that for some shapes of a nested adjacency matrix, the scores of multiple countries and products converge to zero~\cite{pugliese2016convergence,wu2016mathematics}.
Conditional on the level of nestedness and the density of the adjacency matrix, such convergence can be as slow as a power-law of the number of iterations~\cite{pugliese2016convergence,wu2016mathematics}.
Of course, if many countries and products have zero scores, the resulting ranking cannot discriminate their relative importance.
To bypass this issue, scholars have proposed various solutions.
The simplest one is to halt the algorithm after a finite number of iterations, and to check a posteriori that the country and product scores are all larger than zero~\cite{mariani2015measuring}.
Other scholars~\cite{pugliese2016convergence,lin2018nestedness} suggested to use a convergence criterion that relies on the convergence of the \emph{ranking} of the nodes, and not on the convergence of their score. 

Before introducing variants of the fitness-complexity algorithm, we stress the reason why the algorithm produces highly-nested adjacency matrices.
The algorithm indeed not only rewards generalist countries and specialist products, but also ranks the products in such a way that a product's score is mostly determined by the score of the least-fit exporting countries.
Therefore, given a pair of countries, the fittest one (i.e., in the ordered matrix, the one whose corresponding adjacency matrix's row lies above) is typically able to export additional products that are not exported by the least-fit country; these additional products tend to have a higher complexity score (i.e., their corresponding adjacency matrix's columns tend to lie more on the right) than those exported by the least-fit country, which is in agreement with the definition of nestedness.

\begin{table}
\begin{center}
\begin{tabular}{ |p{2.5cm}|p{2.5cm}|p{2.5cm}|p{2.5cm}|p{2.5cm}| } 
\hline
Type of network & System 			& Fitness score 	& Complexity score & Refs. \\ 
\hline
Economic	 & Country-product export 	& Country competitiveness & Product specialization & \cite{tacchella2012new} \\ 
\hline
Economic 	& Country-food production	& Country competitiveness & Food specialization	& \cite{tu2016data}  \\ 
\hline 
Knowledge production & Country-research field 	& Country competitiveness & Field complexity	& \cite{cimini2014scientific}\\
\hline 
Social & User-page engagement in Facebook	& User engagement & Page impact & \cite{zaccaria2019poprank}\\
\hline
Ecological 	& Plant-pollinator 		& Pollinator importance	& Plant vulnerability	& \cite{dominguez2015ranking} \\
\hline
\end{tabular}
\end{center}
\caption{Applications of the fitness-complexity algorithm and its variants to diverse systems. 
For each system, we provide a brief interpretation of both Fitness and Complexity score,
and we refer to the mentioned references for all the details.}
\label{tab:fcm}
\end{table}

We conclude this introduction to the fitness-complexity algorithm by emphasizing that the algorithm can be applied to any bipartite network (see Table~\ref{tab:fcm}).
For example, Dom{\'\i}nguez-Garc{\'\i}a and Mu{\~n}oz \cite{dominguez2015ranking} have applied the algorithm\footnote{To stress the different interpretation of the scores by the algorithm, the fitness-complexity algorithm is dubbed as ``MusRank'' by Dom{\'\i}nguez-Garc{\'\i}a and Mu{\~n}oz~\cite{dominguez2015ranking}.} to mutualistic networks, with the goal to study its ability to identify structurally important nodes -- more details are provided in Section~\ref{sec:structural_nodes}. In mutualistic networks, countries and products are naturally replaced by active and passive species, respectively; the fitness and complexity scores represent the active species' importance and the passive species' vulnerability, respectively~\cite{dominguez2015ranking}.

\paragraph{Generalized fitness-complexity algorithm}

According to the fitness-complexity algorithm, the score of a product is largely determined by the scores of the least-fit countries.
This dependence can be sharpened by introducing an exponent in the complexity definition \cite{pugliese2016convergence,mariani2015measuring}.
By keeping the initial condition of the original algorithm (Eq. \eqref{fit_comp_init}) and the score normalization after each iteration (Eq. \eqref{fit_comp_norm}), the equations of the resulting \emph{generalized fitness-complexity algorithm} are~\cite{pugliese2016convergence,mariani2015measuring}
\begin{equation}
\begin{split}
F_i^{(n)}(\gamma)&=\sum_{\alpha}A_{i\alpha}Q_{\alpha}^{(n-1)}\\
Q_{\alpha}^{(n)}(\gamma)&=\frac{1}{\sum_{i}A_{i\alpha}/(F_{i}^{(n-1)})^{\gamma}},
\end{split}
\label{fit_comp_gen}
\end{equation}
where $\gamma$ is a parameter of the method.
The original fitness-complexity algorithm is obtained for $\gamma=1$.
Given the interpretation of the algorithm in terms of economic capabilities, we are mostly interested in the $\gamma>0$ range.
Increasing $\gamma$ increases the dependence of product score on the fitness of the least-fit exporting country.
This results in more nested adjacency matrices (see Fig. 3 in~\cite{mariani2015measuring}), which proves 
to be beneficial for the identification of structurally important nodes, as we shall analyze in Section \ref{sec:structural_nodes}. 
At the same time, the rankings obtained with $\gamma>1$ tend to be more sensitive to structural perturbations of the network's structure \cite{mariani2015measuring}, which is a drawback especially for systems where a non-negligible fraction of the links might be unreliable, such as the World Trade~\cite{battiston2014metrics,tacchella2018dynamical}. A different generalization of the fitness-algorithm was introduced by Zaccaria~\textit{et al.}~\cite{zaccaria2019poprank} to rank users' engagement and pages' impact in Facebook; they found that the resulting algorithm (which they called \textit{PopRank}) can reliably predict the future activity of a Facebook page.

\paragraph{Minimal extremal metric}

While the generalized fitness-complexity algorithm introduced in the previous paragraph allows us to fine-tune the dependence of
a product score on the score of the least-fit exporting countries, it is instructive to consider the limit case
where the products' score is \emph{entirely} dependent of the least-fit exporting country's score.
By keeping the initial condition of the original algorithm (Eq. \eqref{fit_comp_init}) and the score normalization after each iteration (Eq. \eqref{fit_comp_norm}), the equations 
of the resulting \emph{minimal extremal metric} are~\cite{wu2016mathematics}
\begin{equation}
\begin{split}
F_i^{(n)}&=\sum_{\alpha}A_{i\alpha}Q_{\alpha}^{(n-1)}\\
Q_{\alpha}^{(n)}&=\min_{i:A_{i\alpha=1}}{\{F_i^{(n)}\}}.
\end{split}
\label{mem}
\end{equation}
This metric corresponds to the limit $\gamma\to\infty$ of the generalized fitness-complexity metric, and it is interesting for two main reasons: (1) for empirical networks, with respect to the FCM, it improves the nested packing of the adjacency matrix
(see Fig. 4 in~\cite{wu2016mathematics}); (2) for perfectly nested networks, the vector of node scores is related to the network's degree sequence through a simple, exact mathematical relation (see Eq.~11 in~\cite{wu2016mathematics}), which allows us to build a simple
intuition on the convergence properties of non-linear ranking algorithms.

\paragraph{A variant with improved convergence properties}

To improve the convergence properties of the fitness-complexity algorithm, Stojkoski, 
Utkovski and Kocarev \cite{stojkoski2016impact} defined a modified fitness-complexity algorithm.
By keeping the initial condition of the original algorithm (Eq. \eqref{fit_comp_init}) and the score normalization after each iteration (Eq. \eqref{fit_comp_init}), the equations of their proposed algorithms (hereafter referred to as SUK Fitness-Complexity algorithm, after its authors) read~\cite{stojkoski2016impact}
\begin{equation}
\begin{split}
F_i^{(n)}&=\sum_{\alpha}A_{i\alpha}Q_{\alpha}^{(n-1)}\\
Q_{\alpha}^{(n)}&=\frac{1}{\sum_{i}A_{i\alpha}(N-F_{i}^{(n-1)})}.
\end{split}
\label{suk}
\end{equation}
The score product by the metric can be viewed, approximately, as a second-order expansion of the original fitness-complexity algorithm\footnote{This can be seen by expanding the inverse $1/F$ of country score around $F=f/N$, where $f=N\,F>0$ is the real number such that $f/N$ equates the country score (for example, if a country has fitness equal to one, $f=N$). We obtain that $1/F\simeq
2\,f\,(N-f\,F/2)/N^2$. Therefore, each contribution to product score can be expressed, approximately, in terms of the difference between the total number of countries $N$ and country fitness $F$. In the special case $f=2$ ($F=2/N$)~\cite{stojkoski2016impact}, each contribution becomes precisely equal to $N-F$, i.e., the term used in Eq.~\eqref{suk}.}.
Importantly, product and country scores cannot converge to zero in the SUK algorithm.
This is a simple consequence of the fact that
\begin{equation}
\sum_{i}A_{i\alpha}(N-F_{i}^{(n-1)})=N\,k_{\alpha}-\sum_i A_{i\alpha}\,F_{i}^{(n-1)},
\end{equation}
which implies
\begin{equation}
Q_{\alpha}^{(n)}=\frac{1}{\sum_{i}A_{i\alpha}(N-F_{i}^{(n-1)})}\geq \frac{1}{N\,k_\alpha}>0.
\end{equation}
The algorithm was used by Stojkoski \textit{et al.}~\cite{stojkoski2016impact} to assess the impact of services on the rankings by economic complexity metrics -- a similar goal was also pursued through the fitness-complexity algorithm by a recent paper~\cite{zaccaria2018integrating} co-authored by World Bank members. Stojkoski \textit{et al.}~\cite{stojkoski2016impact} found that services tend to be ranked higher than goods in product rankings, and high-fitness countries tend to have more developed service sectors. By using the fitness-complexity algorithm, Zaccaria \textit{et al.}~\cite{zaccaria2018integrating} found that including services trade in the economic fitness approach can substantially alter the rankings of countries and products.

\begin{figure*}[t]
	\centering
{\includegraphics[scale=0.95]{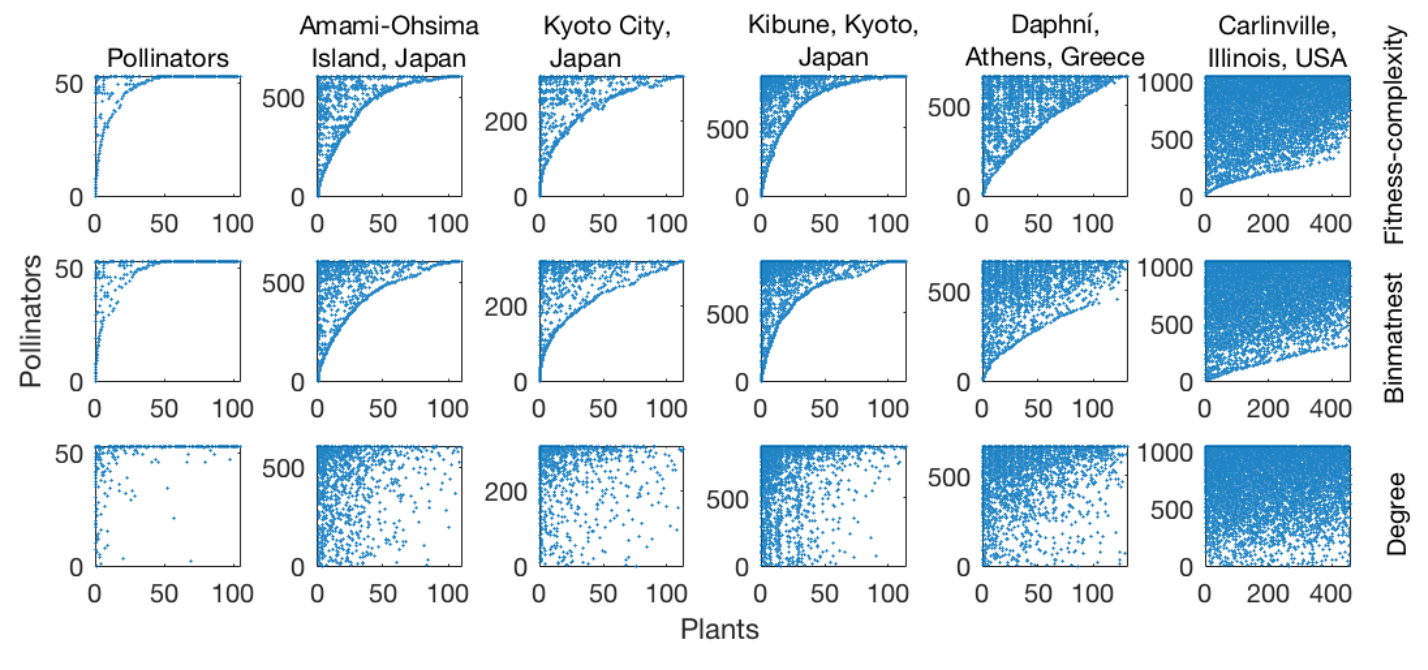}}
	\caption{Interaction matrices of six mutualistic bipartite networks when rows and columns are ranked by the fitness-complexity algorithm (upper panels), by the BINMATNEST algorithm (intermediate panels), and by degree (lower panels). 
    The matrices ranked by the fitness-complexity and BINMATNEST algorithms exhibit a higher degree of nestedness than those packed by degree. Reprinted from~\cite{lin2018nestedness}.}
\label{fig:packing}
\end{figure*}

\subsubsection{Nestedness temperature minimization: BINMATNEST or fitness-complexity?}
\label{sec:best_packing}

So far, we have independently introduced two approaches -- genetic algorithms 
and iterative non-linear ranking algorithms -- to rank the nodes of a given matrix. 
We now ask ourselves: how do the two methods perform in minimizing nestedness temperature or, equivalently,
in maximizing the adjacency matrix's degree of nestedness?
Preliminary results by Dom{\'\i}nguez-Garc{\'\i}a and Mu{\~n}oz~\cite{dominguez2015ranking} suggested that the fitness-complexity algorithm ``packs the matrices'' substantially better than the widely-used NTC (see Fig.~5 in~\cite{dominguez2015ranking}). They concluded that the fitness-complexity algorithm ''should be used (rather than existing ones) to measure nestedness in bipartite matrices"~\cite{dominguez2015ranking}.
On the other hand, shortcomings of the NTC were already pointed out by Rodr{\'\i}guez-Giron{\'e}s and Santamar{\'\i}a~\cite{rodriguez2006new}, which motivated them
to introduce the genetic algorithm described in Section~\ref{sec:genetic}. 

How does the fitness-complexity algorithm perform when compared with the BINMATNEST algorithm?
A detailed comparison of the matrices packed by BINMATNEST with those packed by the fitness-complexity algorithm has been performed recently~\cite{lin2018nestedness}.
Lin \textit{et al.}~\cite{lin2018nestedness} found that the matrices as ranked by fitness-complexity and BINMATNEST are substantially better ``packed'' than those ranked by degree (see Fig.~\ref{fig:packing}).
More surprisingly, they found that the temperature of the matrices ranked by the fitness-complexity algorithm is lower than that of the matrices ranked by BINMATNEST for the majority of the mutualistic bipartite networks that they analyzed. The only networks where BINMATNEST turned out to substantially outperform the fitness-complexity algorithm were characterized by small size and high density.

Lin \textit{et al.}~\cite{lin2018nestedness} concluded that beyond its application in trade networks, the fitness-complexity algorithm has the potential to become a standard tool in nestedness analysis.
It remains open to assess whether variants of the fitness-complexity algorithm can further reduce nestedness temperature, and the impact of improved algorithms for nestedness minimization on the acceptance or rejection, based on suitable null models, of the hypothesis that a network is nested.

\subsection{Bottom-line: How to measure nestedness?}
\label{sec:bottom_measure}

Scholars have introduced several distinct metrics to measure nestedness. 
Some of them (like NODF~\cite{almeida2008consistent} and spectral radius~\cite{staniczenko2013ghost}) measure the level of nestedness by degree, whereas others (like the nestedness temperature by the NTC~\cite{atmar1993measure}, by BINMATNEST~\cite{rodriguez2006new}, and by the fitness-complexity algorithm~\cite{lin2018nestedness}) measure the maximal level of nestedness in the system, and they require a reordering of the rows and columns of the (bi)adjacency matrix.

Generally, these metrics depend on basic systemic properties such as network size, density, and degree distribution. This makes it necessary to assess the statistical significance of their observed values based on a null model. How to choose the null model is controversial: scholars have warned against the risk of both an excessive number of false positives, if the null model is too loose (e.g., Equiprobable-Equiprobable model), or an excessive number of false negatives, if the null model is too conservative (e.g., Fixed-Fixed model).

There is no universal answer on which combination of nestedness metric and null model should be adopted to investigate the presence and implications of nestedness~\cite{bascompte2013mutualistic}. A valid operational strategy is to always make sure that obtained results on nestedness hold for different metrics and similar null models. 
Information about publicly available software for the implementation of nestedness metrics together with null models for statistical significance tests is provided in~\ref{appendix:material}.

\clearpage

\section{Relation between nestedness and other systemic properties}

Nestedness is a network structural property, which means that it depends on the adjacency matrix $\mat{A}$ of the network of interest.
Two questions emerge: how is this property related 
to other known network properties (such as degree distribution, assortativity, modularity, etc.)?
How is the property related to important properties of real systems that are not directly related to network structure?
Section~\ref{sec:network_properties} addresses the first question, revealing a tight bound between nestedness and
the network's degree distribution~\cite{borras2017breaking} and disassortativity~\cite{johnson2013factors}. 
Besides, nestedness turns out to be a generalization of the core-periphery 
structure~\cite{lee2016network}, and it has a multifaceted relation with modularity~\cite{olesen2007modularity,borge2017emergence}.

Section~\ref{sec:ecological_properties} discusses the relation between nestedness and ecological properties
that are not captured by the topology of interactions. These properties include macroecological properties, 
species relative abundances, and forbidden links~\cite{krishna2008neutral,bascompte2013mutualistic}.
Section~\ref{sec:economic_properties} focuses on economic systems, and it describes the relation between nestedness and 
various economic properties. We mostly focus on 
the predictive power of nestedness for the success or failure of firms~\cite{saavedra2011strong}, the appearance and disappearance of links in spatial and trade networks~\cite{bustos2012dynamics},
the future economic development of countries~\cite{cristelli2017predictability}, and on the relation between nestedness and hidden capabilities in world trade~\cite{hausmann2011network,bustos2012dynamics}.

\subsection{Nestedness and other network properties}
\label{sec:network_properties}

This Section studies the relation between nestedness and other network properties: degree 
distribution (Section~\ref{sec:degree_distribution}), disassortativity (Section~\ref{sec:disassortativity}), core-periphery 
structure (Section~\ref{sec:core_periphery}), modularity (Section~\ref{sec:modularity}). For each of these 
network properties, we will describe its relevance in network analysis, introduce the main metrics to
measure it, and discuss its relation with nestedness.

\subsubsection{Nestedness and degree distribution}
\label{sec:degree_distribution}

\begin{figure*}[t]
	\centering
\includegraphics[scale=0.5]{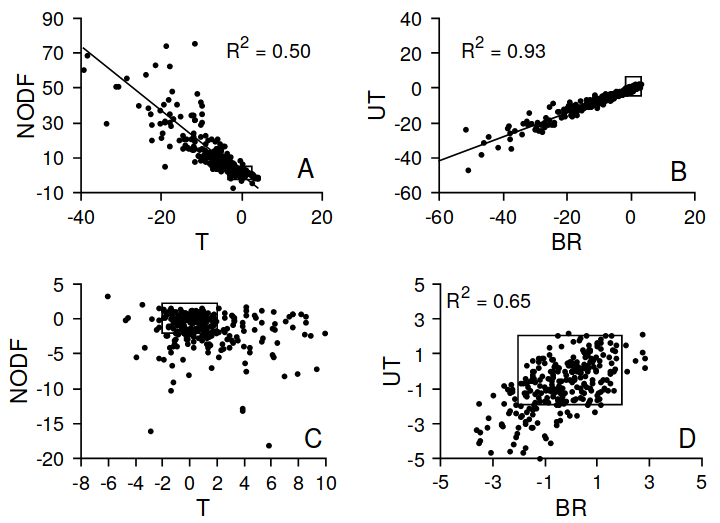}
	\caption{Pairwise comparisons of the $z$-scores by four different nestedness metrics (NODF, nestedness temperature $T$, number of discrepancies $BR$, number of unexpected transformations $U_T$ -- see Section~\ref{sec:metrics} for all the definitions) for the spatial datasets collected by Atmar and Patterson (see Appendix~\ref{appendix:material}). The $z$-scores are based on the EE model which only preserves network size and density (panels A-B),
	and the FF model which preserves exactly the degree sequence (panels C-D).
	The squares delimit the regions where nestedness is not significative according to any of the two metrics.
	Reprinted from~\cite{ulrich2009consumer}.}\label{fig:z_score}
\end{figure*}

A natural question which invariably arises for any network structural pattern is whether the property can be simply explained by the network's degree distribution~\cite{newman2010networks} or by other local, higher-order network properties~\cite{orsini2015quantifying}. 
Almost as soon as the first instances of nested structures were reported~\cite{patterson1986nested}, ecologists have started investigating whether the observed nested patterns could be explained by the degree distribution alone. The question has been addressed by means of both the degree-preserving randomization procedures introduced in Section~\ref{sec:nine} (e.g., PP and FF model), and by the maximum-entropy techniques described in Section~\ref{sec:maxent}.

\paragraph{Biogeographic networks}

As nestedness was first introduced in the biogeography literature~\cite{patterson1986nested}, the first attempts to study the relation between network nestedness and degree sequence were carried out for bipartite species-island networks.
For spatial networks, 
Brualdi and Sanderson~\cite{brualdi1999nested} found that the observed number of 
discrepancies (a metric for nestedness introduced in Section~\ref{sec:gap}) in $33$ empirical
networks is compatible with the number of discrepancies observed in randomized networks generated with the FF model.
They argued that the main reasons why nested patterns were widely found in previous studies~\cite{simberloff1991nestedness,cook1995influence} was that ``violation of the row or column sums (or both) did not sufficiently constrain the sample space containing the nested species subsets".

Ulrich \textit{et al.}~\cite{ulrich2009consumer} performed an extensive analysis on the $286$ empirical
networks previously collected by Patterson and Atmar (see~\ref{appendix:material}). They focused on the FF model which preserves the degree 
sequence of both kinds of nodes in bipartite networks (see Section~\ref{sec:nine}). They both measured the 
correlation between the $z$-scores of different nestedness metrics (NODF, number of discrepancies, 
temperature, unexpected number of transformations $U_T$) and assessed the number of empirical networks that exhibit
a statistically significant nestedness (see 
Fig.~\ref{fig:z_score}). Out of the $286$ empirical networks, they found $113$ of them to 
be significantly nested according to at least one of the considered nestedness metrics. 
Intriguingly, they found only $11$ networks to be significantly nested according to all 
the four considered nestedness metrics. Ulrich \textit{et al.}~\cite{ulrich2009consumer} also found
that the EE null model (i.e., a null model that only preserves network size and density,
see Section~\ref{sec:nine}) produces high values of $z$-score for most of the networks, 
regardless of the adopted nestedness metric: the choice of the null model has a strong impact 
on our conclusions on the significance of a nested pattern. As expected, more constrained models lead to fewer significantly-nested networks.

\paragraph{Ecological interaction networks}

Already the first work by Bascompte \textit{et al.}~\cite{bascompte2003nested} found that for 20\% of the analyzed pollination networks,
nestedness can be explained by a PP null model that preserves, on average, animals' and plants' degree; they also found that
this percentage drops to zero when restricting the analysis to sufficiently large networks.
Joppa \textit{et al.}~\cite{joppa2010nestedness} also found that most mutualistic networks are more nested than one would expect by degree sequence alone.
They pointed out that the networks that are more nested than expected by degree sequence are typically the large ones. From a different perspective, Medan \textit{et al.}~\cite{medan2007analysis} computed analytically the degree sequence associated with a particular line of perfect nestedness.

More recently, Borràs \textit{et al.}~\cite{borras2017breaking} used the maximum-entropy approach for bipartite networks 
(see Section~\ref{sec:maxent}) to assess the relation between nestedness and degree 
sequence in $167$ mutualistic networks. Their conclusions are different from those 
by previous studies~\cite{bascompte2003nested,joppa2010nestedness}.
Based on NODF and spectral radius, they found that among
all the analyzed networks, only a tiny fraction of them exhibit $z$-score 
larger than two. They concluded 
that in mutualistic networks, the ``observed nested structure of the ecological
communities studied is, in fact, a mere consequence of the degree sequences of the
two guilds". They interpreted this result as a consequence of the
fact that for networks with highly-heterogeneous degree distribution, 
disassortative structures tend to have larger entropy than non-disassortative ones~\cite{johnson2010entropic}. This 
happens because a specialist has many more possibilities to be connected with a generalist than with 
another specialist. As nestedness and disassortativity are strongly related (see Section \ref{sec:disassortativity}), it
follows that for a highly-heterogeneous degree distribution, maximizing entropy can lead to nested structures~\cite{borras2017breaking}.

\begin{figure*}[t]
	\centering
\includegraphics[scale=0.35]{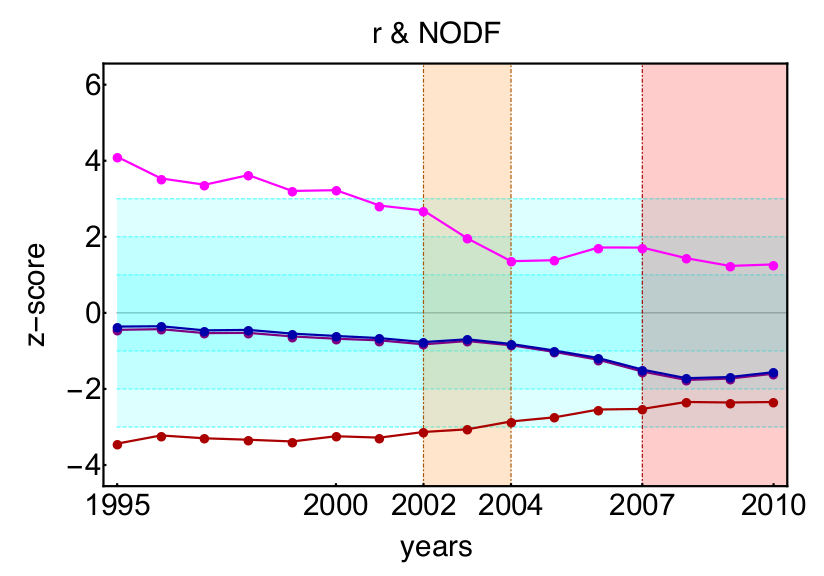}
	\caption{Temporal evolution in the World Trade for the $z$-score of three quantities: NODF (blue dots),
	row-contribution to NODF (pink dots), assortativity (brown dots). 
	The NODF and its contributions are based on the bipartite country-product export network; the 
	$z$-scores are based on the maximum entropy model described in Section~\ref{sec:maxent} which preserves, on average, the nodes'
	degree. The row-contribution to NODF (i.e., the country-contribution) shows a non-trivial trend: it is significant in the first years
	but it gradually declines, falling below the $z=2$ line after $2003$.
	Assortativity shows a monotonous trend as well, yet it remains significatively negative ($z<2$)
	over the whole observation time span. Adapted from~\cite{saracco2016detecting}.}
	\label{fig:saracco}
\end{figure*}

\paragraph{World Trade networks}
Results on the significance of nestedness in World Trade are intriguing. Using the maximum-entropy approach
described in Section~\ref{sec:maxent},
Saracco \textit{et al.}~\cite{saracco2015randomizing} quantified the significance of the 
nestedness (as measured by NODF, see Section~\ref{sec:overlap}) of the bipartite country-product
export network (NBER dataset) over the period 1963-2000.
The authors found that the $z$-scores of the network's NODF (determined numerically through the maximum-entropy framework described in Section~\ref{sec:maxent}) are always
smaller than two. They concluded that the degree sequence allows us to reproduce the nestedness of the World Trade bipartite network.
An analysis of the World Trade over the period 1995–2010 (see Fig.~\ref{fig:saracco}) shows again that network NODF is
explained by the network degree sequence \cite{saracco2016detecting}. Nevertheless, the row-contribution
to NODF (see Section~\ref{sec:overlap}) exhibits statistically significant values ($z$-score above two) from 1995 to 2002.
The $z$-score declines from 1995 to 2002, lying on an almost constant plateau below $z=2$ over
the period 2003-2010. Based on similar findings for other structural properties, Saracco \textit{et al.} \cite{saracco2016detecting} argued that the increasing randomness of the world trade network over the 2003-2007 period might be interpreted as an early sign of the 2007-2008 financial crisis.

\paragraph{Nestedness: significant pattern or a consequence of the degree sequence?}

The question is tightly related to the choice of the null model and its implications. Should one prefer unconstrained models (such as the EE model, see~\ref{sec:nine}) where the degree of the nodes is allowed to vary, or should one prefer a constrained null model (like the FF model, see~\ref{sec:nine}) where the nodes' degree is fixed exactly (like in the FF model, see~\ref{sec:nine} or on average (like in the PP model and in the maximum-entropy approach, see Sections~\ref{sec:nine}-\ref{sec:maxent})?
Ecologists have widely debated this question.
We can identify two main viewpoints in the literature.

Scholars have warned against the possible overestimation of Type-I and Type-II errors for unconstrained and constrained models, respectively.
It has been widely recognized~\cite{ulrich2009consumer} that unconstrained models where all species have the same interaction probability might make it too easy for a network to achieve a statistically significant degree of nestedness. In other words, given the null hypothesis that the null model explains the observed degree of nestedness, one risks rejecting the null hypothesis when it is true (Type-I error~\cite{gotelli2012statistical}).
A long stream of works~\cite{brualdi1999nested,saracco2015randomizing,borras2017breaking,straka2018ecology} have employed constrained null models where the degree of the nodes is fixed, exactly or on average. As a result, the fact that a null model that preserves the degree sequence can generate networks of statistically comparable nestedness as the real ones is known since long time~\cite{brualdi1999nested}.

On the other hand, it has been also recognized that in constrained models (such as the FF model), fixing exactly the degree sequence might make it too hard for a network to achieve a statistically significant degree of nestedness. In other words, given the null hypothesis, one risks accepting the null hypothesis when it is false (Type-II error~\cite{gotelli2012statistical}). 
Some scholars~\cite{presley2010comprehensive} advocated the use of models that do not constrain the degree sequence of both kinds of nodes, based on the following argument, summarized by Gotelli and Ulrich~\cite{gotelli2012statistical}: with constrained models, ``if the biological processes (e.g. competition) affect the constrained elements
(e.g. matrix row totals) then the effect of interest has been smuggled into the test, 
which reduces the sample space and leads to excessive type II errors".
Because of the small size of the sample space, constrained models can lead to paradoxes: for example,
a perfectly nested matrix might be classified as maximally non-nested by the FF model
if it is the only possible matrix with that degree sequence (see the Supplementary Information in~\cite{staniczenko2013ghost}).

This debate points out the importance of considering carefully the properties of the ensemble of random graphs used to assess the significance of observed patterns: classifying
a network as non-significantly nested might be a consequence of the limited number of networks in the ensemble of 
networks with the same (or similar) degree sequence. A detailed analysis of Type I and Type II errors for both 
constrained and unconstrained null models can be found in~\cite{ulrich2007null}: they found that in a set of non-random matrices
that contains nestedness, the FF
model rejected the
null hypothesis only less than 10\% of the times for most of the nestedness metrics. In light of this discussion and recent developments~\cite{strona2018bi}, 
we recommend the detailed analysis of Type I vs Type II errors as a necessary step to be performed before adopting a given null model for the analysis.
If a null model is not able to classify perfectly or almost perfectly 
nested networks as significantly nested, concluding that nestedness is not significant
based on its results might be misleading.
A similar analysis might also explain the reasons behind the contrasting results
in previous literature~\cite{bascompte2003nested,ulrich2009consumer,joppa2010nestedness,borras2017breaking}.

\subsubsection{Nestedness and disassortativity}
\label{sec:disassortativity}

A network exhibits nestedness if the neighborhood of a node is contained in the neighborhoods of the nodes with higher degrees.
A direct consequence is that low-degree nodes tend to only interact with high-degree
nodes, whereas high-degree nodes interact with both other high-degree nodes and with
low-degree nodes. As a result, the average degree of the nodes that interact with high-degree
nodes tends to be lower than that of the nodes that interact with low-degree nodes.
This property is typically referred to as \emph{disassortativity} in network science~\cite{newman2002assortative,newman2010networks}.
Therefore, nestedness is expected to be significantly correlated with plausible metrics for disassortativity; this expectation has been confirmed by 
analysis of synthetic and empirical networks~\cite{abramson2011role,johnson2013factors}.
Below, we provide more details on these results.

\paragraph*{Assortativity and disassortativty}

\emph{Assortativity} in networks generally refers to the tendency of nodes to connect with nodes that are similar to them. By contrast, \emph{disassortativity} refers to the tendency of nodes to connect with nodes that are different from them (see~\cite{newman2010networks}, Section 7.13). Of course, such general definitions strongly depends on which properties we consider when measuring the similarity between two given nodes. There can be two types of assortativity: assortativity by enumerative node properties (e.g., measuring the tendency of students in a given school to establish friendships with students of the same gender) and assortativity by scalar node properties (e.g., measuring the tendency of students in a given school to establish friendships with students of similar age or parental income)~\cite{newman2010networks}.  

From a network perspective, it is instructive to measure the assortativity by degree, which aims to
quantify the tendency of nodes to connect with nodes of similar degree~\cite{newman2010networks,barabasi2016network}.
Assortativity by degree has been found in systems as diverse as social networks~\cite{ugander2011anatomy}, 
scientific co-authorship networks~\cite{newman2002assortative},
paper~\cite{bornmann2010scientific} and patent~\cite{mariani2018early} citation networks, World Wide Web~\cite{foster2010edge},
among many others. 
On the other hand, disassortativity has been found in metabolic networks~\cite{barabasi2016network}, 
the Internet~\cite{pastor2001dynamical}, food webs~\cite{newman2002assortative,foster2010edge}, among many others. 

To measure the degree of assortativity in a given network, a commonly employed metric is 
the Pearson's linear correlation coefficient $r$ between the degrees of pairs of nodes that belong
to a given edge. Assortative (disassortative) networks are characterized by positive (negative) 
values of $r$, meaning that the degree of a link's node is positively (negatively) correlated with the degree of the other node that forms the link.


\begin{figure*}[t]
	\centering
\includegraphics[scale=2]{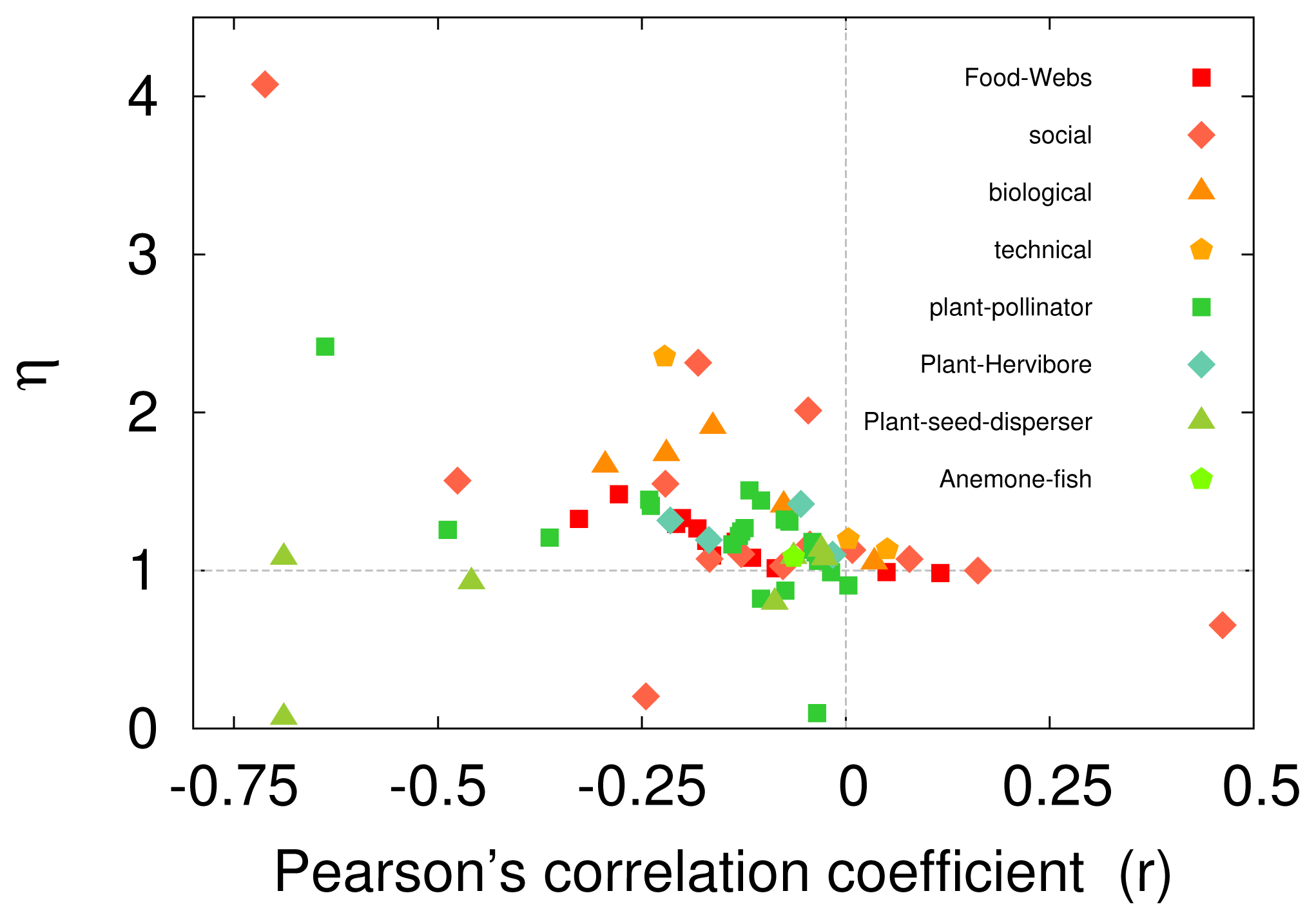}
	\caption{The relation between nestedness 
	(as measured by the JDM-NODF $\eta$ defined by Eq.~\eqref{eta}) and disassortativity
	(as measured by the linear correlation $r$ between the degrees of the two nodes that belong to each link)
	for empirical networks of various kinds.
	The most nested networks (large $\eta$) are all disassortative ($r<0$), whereas the nestedness of assortative networks ($r>0$)
	tends to be comparable with that of random networks ($\eta\simeq 1$).
	Reprinted from~\cite{johnson2013factors}.}\label{fig:disassortativity}
\end{figure*}

\paragraph*{Nestedness and disassortativity}

Abramson \textit{et al.}~\cite{abramson2011role} found that perfectly nested network exhibit negative assortativity coefficient $r$.
The coefficient decreases (i.e., the network becomes more disassortative) as network density increases.
Johnson \textit{et al.}~\cite{johnson2013factors} showed that network disassortativity is significantly correlated with nestedness, as their metric $\eta$ for nestedness (defined by Eq. \eqref{johnson}) is negatively correlated with the Pearson's linear correlation coefficient $r$ between the degrees of pairs of nodes that form the network links. They showed this both in synthetic networks with tunable assortativity coefficient (see Fig.~2 in~\cite{johnson2013factors}) and in $60$ unipartite and bipartite empirical networks of diverse nature (see Fig.~\ref{fig:disassortativity}).
They concluded that ``disassortative networks are typically nested and nested networks are typically 
disassortative".

\subsubsection{Nestedness and core-periphery structure}
\label{sec:core_periphery}

A core-periphery structure is a network structure composed of a ”core” of nodes
that are connected with all the other nodes, and a ”periphery” of nodes that
tend to be only connected with the nodes in the core.
Such structure has been investigated and observed in a wide variety of systems, including international trade~\cite{nemeth1985international,smith1992structure}, social networks~\cite{borgatti2000models,cattani2008core}, human brain network~\cite{betzel2017multi,battiston2017multiplex}, mutualistic networks~\cite{diaz2010changes,ruggera2016linking}, R\&D networks~\cite{tomasello2017rise}, financial networks~\cite{barucca2016disentangling}, among many others. We refer the interested reader to~\cite{csermely2013structure} for a recent review on the structure and dynamics of core-periphery networks.
Some scholars have observed that compared with the community detection problem (see~\cite{fortunato2010community,fortunato2016community} and Section~\ref{sec:modularity}), ``other types of mesoscale structures [\dots] have received much less attention than they deserve"~\cite{rombach2017core}, and that the core-periphery structure is one of those.

In a core-periphery structure, the peripheral nodes' neighborhoods are included by construction in the core nodes' neighborhoods; therefore, it is natural to conjecture a relation between such structure and nestedness. This relation has been explicitly found by Lee~\cite{lee2016network} in both empirical and synthetic networks. It suggests that the methods and implications of nestedness analysis might be highly relevant to all systems where core-periphery structures are typically found.

\paragraph*{Detection of core-periphery structures}

Before investigating the relation between nested and core-periphery structures, we provide a brief introduction to the methods to detect core-periphery structures in networks. We describe the Borgatti-Everett (BE) approach~\cite{borgatti2000models} and its subsequent generalizations~\cite{lee2016network, rombach2017core} that have been used to investigate the relation between core-periphery structures and nestedness~\cite{lee2016network}.
We start by introducing the Borgatti-Everett (BE) approach~\cite{borgatti2000models} to detect a single core-periphery structure in unipartite networks\footnote{While we will focus on the Borgatti-Everett method in this Section, we refer the interested reader to \cite{holme2005core,boyd2010computing,zhang2015identification,kojaku2018core,rombach2017core} for alternative methods for the detection of core-periphery structures.}. 
To identify core and peripheral nodes, Borgatti and Everett introduced, for each node $i$, a binary variable $x_i$ such that $x_i=1$ and $x_i=0$ for core and peripheral nodes, respectively.
They also considered an ideal core-periphery structure where the core nodes are connected with all the nodes, whereas the peripheral nodes are only connected with the core nodes. In terms of the $\vek{x}$ variables, the corresponding adjacency matrix $\mat{A}^{CP}$ can be simply written as $A_{ij}^{CP}=x_i+x_j-x_i\,x_j$ -- i.e., $A_{ij}^{CP}=1$ if $i$ or $j$ belong to the core, whereas periphery-periphery links are strictly forbidden.
A simple metric to quantify the similarity between the network's adjacency matrix and the ideal core-periphery structure is the quality function
\begin{equation}
Q^{CP}=\sum_{(i,j)}A_{ij}\,A^{CP}_{ij}(\vek{x}).
\label{qcp}
\end{equation}
To distinguish among core and peripheral nodes, one seeks to find the vector $\vek{x}$ that maximizes $Q^{CP}$.

Borgatti and Everest~\cite{borgatti2000models} recognized that the assumption of a binary distinction between core and peripheral nodes might be simplistic to accurately describe real systems' nodes. To make the model more flexible, they considered a continuous variant where each node is endowed with a ``coreness'' value $c_i$. The elements of the ideal core-periphery network's adjacency matrix are then defined as
$A_{ij}^{CP}=c_i\,c_j$. Such matrix elements: (1) are large when both $i$ and $j$ have a high coreness value; (2) are small when both $i$ and $j$ have a small coreness value; (3) assume intermediate values when one of the two nodes has a small coreness, the other one has a large coreness value.
The maximization of the resulting quality function~\cite{borgatti2000models}
\begin{equation}
Q^{CP}=\sum_{(i,j)}A_{ij}\,c_i\,c_j
\end{equation}
gives as output the optimal vector of coreness scores. Interestingly, as recognized by Borgatti and Everett~\cite{borgatti2000models}, this optimal vector is equivalent to the Bonacich eigenvector centrality~\cite{bonacich1987power}. 

\begin{figure}[t]
\centering
\includegraphics[scale=0.95]{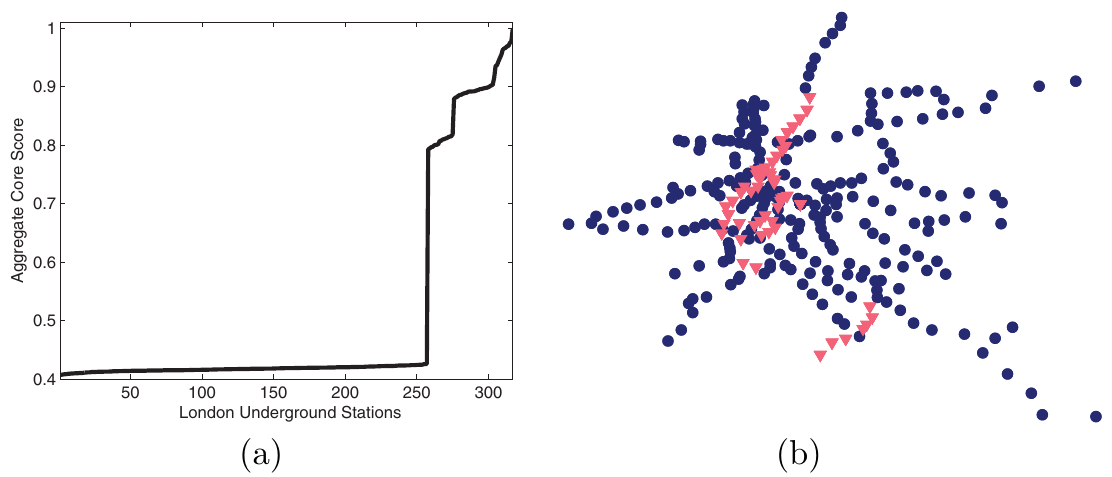}
\caption{Core nodes and core-periphery structure of the London Underground stations.
The node-level aggregate core score $C$ (defined in Eq.~\eqref{node_core}) reveals that there are $60$ important stations in the system (panel (a)).
The structural importance of this stations (pink triangles)
compared to the non-core stations (blue circles) can be appreciated in panel (b), where the stations' position is determined by their
geographical location. The two clusters of cores
correspond to central stations in London (north cluster) and to the stations close to Waterloo station (south cluster).
Reprinted from \cite{rombach2017core}.}
%
\label{fig:cp}
\end{figure}

Rombach \textit{et al.}~\cite{rombach2017core} generalize this approach by introducing a parametrization $c_i(a,b)$ of the core score in terms of two parameters $a$ and $b$. They choose a ``sharp'' function that well separates the core from the peripheral nodes:
\begin{equation}
c_{i}(a,b)=\begin{cases}
\frac{i\,(1-a)}{2\,\floor*{b\,N}} \,\,\,\,&\text{if} \,\,\,\, i\in\{1,\dots,\floor*{b\,N}\}, \\
\frac{(1-\floor*{b\,N})\,(1-a)}{2\,(N-\floor*{b\,N})}+\frac{1+a}{2} \,\,\,\,&\text{if}\,\,\,\,i\in\{\floor*{b\,N}+1,\dots N\}.
\end{cases}
\label{corescore}
\end{equation}
The parameter $b\in[0,1]$ determines the size of the core (all the nodes are in the core when $b=0$, whereas the core is empty when $b=1$). The parameter $a$ determines the score gap between the lowest-score core node and the highest-score peripheral node. When $a=1$, such score gap is maximal, $C_i$ is a discontinuous function, and the nodes are either core nodes ($c_i=1$) or peripheral nodes ($c_i=0$), without intermediate coreness values. In line with Borgatti and Everett~\cite{borgatti2000models}, the assignment of the nodes to the core or the periphery of the network is performed in such a way to maximize the core-periphery function $Q^{CP}(a,b)=\sum_{(i,j)}A_{ij}\,c_i(a,b)\,c_j(a,b)$.
The
aggregate core score of node $i$ is defined as the weighted average of $c_i(a,b)$ over all the possible values for the parameter pair $(a,b)$; parameter pairs $(a,b)$ that lead to larger values of the core-periphery quality function $Q^{CP}$ give larger contribution to $C_i(a,b)$: 
\begin{equation} C_i(a,b)=Z\,\sum_{(a,b)} c_i(a,b)\,Q^{CP}(a,b),
\label{node_core}
\end{equation}
where the normalization factor $Z$ ensures that $\max_i C_i=1$. The aggregate core score of a node is informative about its centrality in the network (see Fig.~\ref{fig:cp}).

Eq.~\eqref{corescore} can be generalized to bipartite networks~\cite{lee2016network}. This is done by introducing two vectors $\{c_i^R(a^R,b^{R})\}$ and $\{c_{a}^{C}(a^C,b^{C})\}$ of coreness scores for row-nodes and for column-nodes, respectively. The model has now four parameters in total: two parameters, $\{a^R,b^{R}\}$ and $\{a^C,b^{C}\}$, parametrize the coreness score of row-nodes and column-nodes, respectively.
In line with~\cite{rombach2017core}, one can again define the aggregate core score of row-node $i$ and column-node $\alpha$ as~\cite{lee2016network}
\begin{equation}
\begin{split} 
C_i^{R}& =Z^{R}\,\sum_{(\vek{a},\vek{b})} c_i(a^{R},b^{R})\,Q^{CP}(\vek{a},\vek{b}), \\
C_i^{C}& =Z^{C}\,\sum_{(\vek{a},\vek{b})} c_i(a^{C},b^{C})\,Q^{CP}(\vek{a},\vek{b}), \\
\end{split}
\end{equation}
where $(\vek{a},\vek{b})=(a^{R},b^{R},a^{C},b^{C})$, and we defined the quality function
\begin{equation}
Q^{CP}(a,b)=\sum_{i,\alpha}A_{i\alpha}\,c_i(a^{R},b^{R})\,c_{\alpha}(a^{C},b^{C}).
\end{equation}
The degree of core-periphery organization $\xi$ of the network can be defined as~\cite{lee2016network}
\begin{equation}
\xi=\frac{\sum_{i,\alpha}A_{i\alpha}\,C_i^R\,C_{\alpha}^C}{E\,\sum_i C_i^R \sum_\alpha C_\alpha^C}.
\label{cp_xi}
\end{equation}
The numerator of this expression has the same form as Eq.~\eqref{node_core}, with the difference that it uses the aggregate core scores which do not depend on the specific $(\vek{a},\vek{b})$ parameters.


\paragraph*{Nestedness and core-periphery structure}

Lee~\cite{lee2016network} compared the NODF nestedness metric (defined by Eq.~\eqref{nodf}) with the core-periphery metric $\xi$ (defined by Eq.~\eqref{cp_xi}) on 89 mutualistic bipartite networks (both plant-pollinator and seed dispersal) and in synthetic networks with tunable nestedness. In mutualistic networks, he found a strong correlation between NODF and $\xi$; however, for those systems, both NODF and $\xi$ turn out to be strongly correlated with both edge density (see Fig.~5 in~\cite{lee2016network}). To factor out the impact of edge density on the NODF-$\xi$ correlation, Lee considers a model where nestedness can be tuned while preserving the network's edge density. Such analysis reveals again a strong correlation between NODF and $\xi$, but both metrics turn out to be strongly correlated with the variance of the degree distribution (see Fig.~6 in~\cite{lee2016network}). To factor out both the impact of edge density and the impact of degree heterogeneity on the NODF-$\xi$ correlation, Lee~\cite{lee2016network} considered a randomization procedure (analogous to the FF model, see Section~\ref{sec:nine}) that preserves exactly the degree sequence. Again, NODF and $\xi$ are strongly correlated, which suggests that while not equivalent, nestedness and core-periphery are two closely related network properties, regardless of the network's degree distribution.

\subsubsection{Nestedness and modularity}
\label{sec:modularity}

Modularity aims to quantify how well a network can be partitioned into different groups of nodes (referred to as \emph{modules}) such 
that nodes that belong to the same group are
more likely to be connected than nodes that belong to different groups. In the network science language, the
identification of modules can be seen as a technique of community detection\footnote{A large number of 
community detection methods have been introduced in the literature, involving techniques as
diverse as random-walk minimum description length~\cite{rosvall2008maps}, statistical inference 
techniques~\cite{holland1983stochastic,karrer2011stochastic}, spectral methods~\cite{krzakala2013spectral}, among many others.
To quote Hric and Fortunato, ``as long as there will be networks, there will be people
looking for communities in them"~\cite{fortunato2016community}. We refer to~\cite{fortunato2010community,fortunato2016community} for comprehensive reviews of community detection techniques.}~\cite{fortunato2016community}.

\begin{figure}[t]
\centering
\includegraphics[scale=0.95]{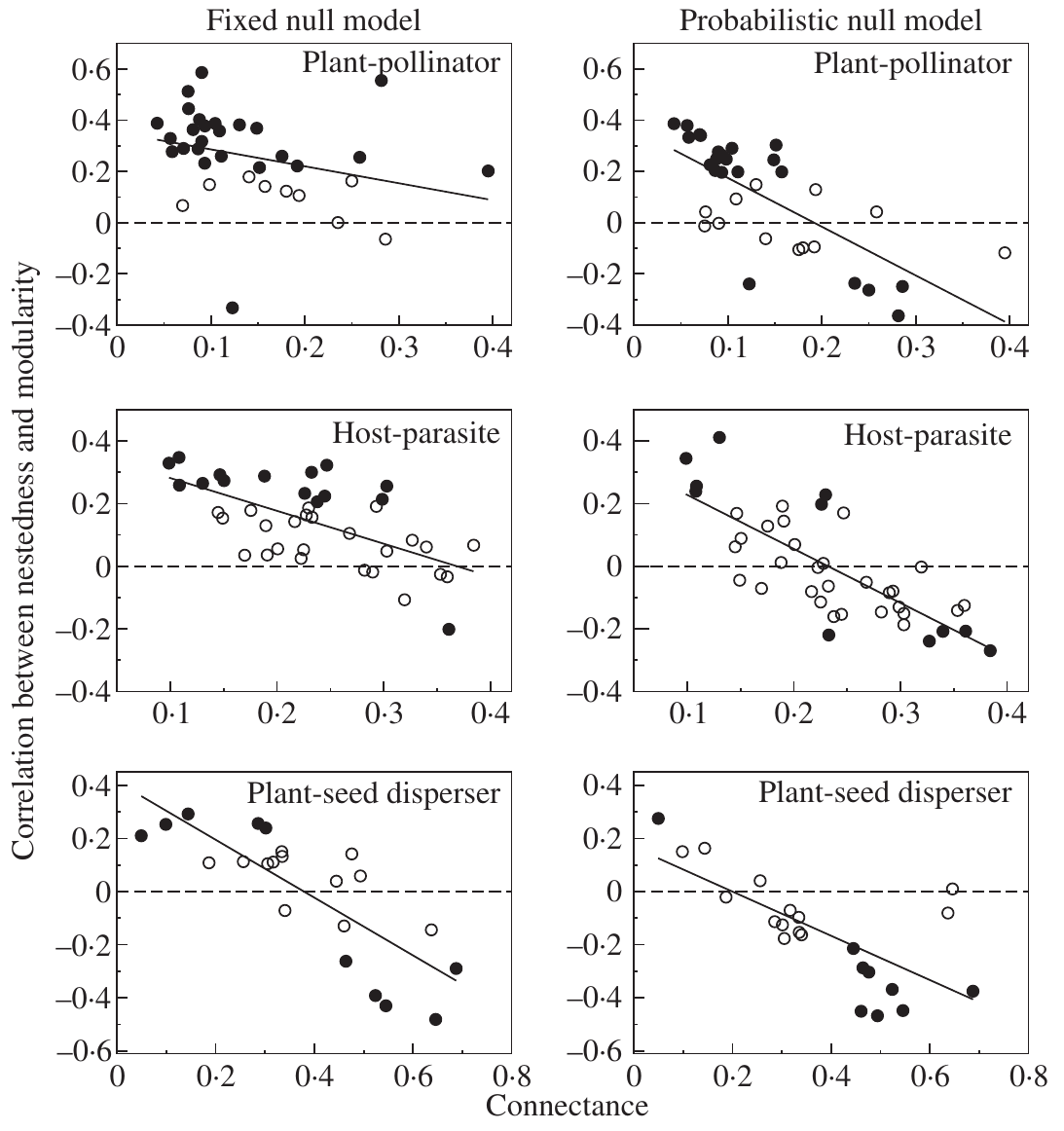}
\caption{Nestedness-modularity correlation as a function of network connectance -- each dot represents a network. The nestedness and modularity values represented
here are averages over $100$ randomizations of the original networks, 
performed with two null models: The \emph{fixed} null model~\cite{gotelli2000null} that preserves exactly the network's degree 
sequence (see the FF model in Section~\ref{sec:nine}), and the \emph{probabilistic} null 
model~\cite{bascompte2003nested} that, on average, preserves the
observed total number of interactions and the network's degree sequence (see the PP model in Section~\ref{sec:nine} and
\cite{fortuna2010nestedness}) for implementation details. Black and white circles represent significant and non-significant
correlations, respectively. Reprinted from \cite{fortuna2010nestedness}.}
\label{fig:nemo_correlation}
\end{figure}

It turns out that the relation between nestedness and modularity 
strongly depends on network connectance: low-density (high-density) networks exhibit a
positive (negative) correlation between nestedness and modularity 
(see Fig.~\ref{fig:nemo_correlation} below and the related discussion). Beyond 
this correlational evidence, it is interesting to study the joint dynamics of these 
two properties. Borge-Holthoefer \textit{et al.}~\cite{borge2017emergence} recently found 
that a social system can transition from a modular to a nested structure as a consequence of
a special event (see Fig.~\ref{fig:modtonest} below). Besides, a module can exhibit an internal 
nested structure, and one can design structural functions to detect subsets of nodes that exhibit 
a nested pattern of interaction; this topic will be addressed in Section~\ref{sec:mesoscopic}.

\paragraph{Modularity}
Modularity \cite{newman2004finding} is one of the most studied properties of complex networks.
The modularity score $Q$ of a given network quantifies how well the network can be partitioned into blocks (or communities, or compartments) that have a high internal density of links and few connections with other blocks.
For a undirected unipartite network, to compute the modularity of a given partition $\vek{\Xi}=\{\Xi_i\}$ of the network into blocks -- $\Xi_i$ denotes here the block to which node $i$ is assigned to -- we compare the number of edges $A_{ij}$ between two nodes $i$ and $j$ that belong to the same community with the expected number of edges $E_{ij}$ according to a suitable null model. 
In formulas, for an undirected network, the modularity $Q$ is defined as
\begin{equation}
Q=\frac{1}{2\,E}\sum_{i,j}(A_{ij}-E_{ij})\,\delta(\Xi_i,\Xi_j),
\label{modularity}
\end{equation}
where $E_{ij}=k_i\,k_j/(2\,E)$ is the expected number of edges between $i$ and $j$ under the configuration model (or, equivalently, the PP null model), and the delta 
function $\delta(\Xi_i,\Xi_j)$ restricts the sum to the pairs of nodes that belong to the same community.

Modularity can be readily generalized to directed~\cite{malliaros2013clustering} and bipartite networks~\cite{barber2007modularity}.
The most popular extension of modularity to bipartite networks is arguably Barber's modularity~\cite{barber2007modularity}.
In Barber's modularity, one uses the fact that in bipartite networks, nodes of one type can only connect with nodes of the other type.
Therefore, only pairs of nodes of dissimilar type contribute to the bipartite modularity $Q$:
\begin{equation}
Q=\frac{1}{E}\sum_{i,\alpha}(A_{i\alpha}-E_{i\alpha})\,\delta(\Xi_i,\Xi_\alpha),
\end{equation}
where $E_{i\alpha}=k_i\,k_\alpha/E$ is the expected number of edges between $i$ and $\alpha$ according to the configuration model (PP model, in the language of Section~\ref{sec:nine}).
Heuristic algorithms developed to maximize the modularity function for unipartite networks can be extended to maximize Barber's modularity as well.

Modularity optimization is one of the most popular community detection techniques. Research
on modularity has focused on evaluating its ability to reconstruct ground-truth communties 
in synthetic networks~\cite{lancichinetti2008benchmark,lancichinetti2009community,yang2016comparative}, designing 
fast and effective heuristics to optimize it~\cite{blondel2008fast,sobolevsky2014general}, unveiling 
its incapability to detect small modules under particular circumstances (''resolution limit``) ~\cite{fortunato2007resolution}, quantifying 
its statistical significance~\cite{lancichinetti2011finding}, generalizing it to multilayer
and temporal networks~\cite{mucha2010community}, among other problems.

\paragraph{Nestedness and modularity in ecological networks}

Two natural questions arise: Do ecological networks exhibit modularity? If yes, what is the relation between nestedness and modularity?
Early studies attempting to uncover the modular (''compartmentalized``) structure of ecological networks 
date back to the 80s~\cite{pimm1980food}, and related studies followed in the 90s and early 2000s~\cite{raffaelli1992compartments,dicks2002compartmentalization}. 
These works adopted simple statistics to quantify the ''compartmentalization``
of the system, and concluded that empirical food webs~\cite{raffaelli1992compartments} and mutualistic networks~\cite{dicks2002compartmentalization} can 
be compartmentalized.
More recently, based on the modularity function defined above, Olesen \textit{et al.} \cite{olesen2007modularity} evaluated the modularity (optimized
through a simulated annealing algorithm) of $51$ pollination networks. They found that all networks of more than $50$ species 
were significantly modular, whereas all networks with less than $50$ species were not significantly modular.
In addition, they found no significant correlation between modularity and nestedness temperature.
They argued that nestedness emerges from the assembly of distinct modules “glued together by interactions among modules”. 

Fortuna \textit{et al.} \cite{fortuna2010nestedness} analyzed  95 ecological communities including plant–animal mutualistic networks and host–parasite networks.
First, they found a significant correlation between nestedness and modularity for plant–pollinator networks ($r=0.363, p=0.035$) but
not for plant–seed disperser ($r=0.151, p=0.503$) and host-parasite ($r=0.066, p=0.689$) networks.
Importantly, by using two statistical null models, they found that the relation
between nestedness and modularity strongly depends on network connectivity: at low (high) connectivity, the two properties
tend to be positively (negatively) correlated -- in other words, the correlation between the nestedness-modularity
correlation and network connectance is negative (Fig.~\ref{fig:nemo_correlation}). 

Beyond the overall correlation between nestedness and modularity, another important question is the
quantification of the level of nestedness within the modules. While one may naively 
compute the level of nestedness within the modules detected by modularity-optimization algorithms, 
recent studies~\cite{sole2018revealing} suggest that this approach can lead to misleading results.
This problem is the main topic of Section~\ref{sec:mesoscopic}.

Besides structural analysis, it is natural to investigate how such a compartmentalized structure affects dynamical processes on the network. 
The question is typically addressed by means of dynamical models and stability analysis~\cite{stouffer2011compartmentalization,grilli2016modularity}. Recently, Gilarranz \textit{et al.}~\cite{gilarranz2017effects} tackled this question with an experimental setup that allowed them to study the secondary effects of a localized perturbation in a networked population of springtail (Folsomia candida) microarthropods. They found that a modular topology can limit the impact of a localized perturbation. More research is needed to assess the impact of different network topologies on the spreading of a localized perturbation.

\begin{figure}[t]
\centering
\includegraphics[scale=0.95]{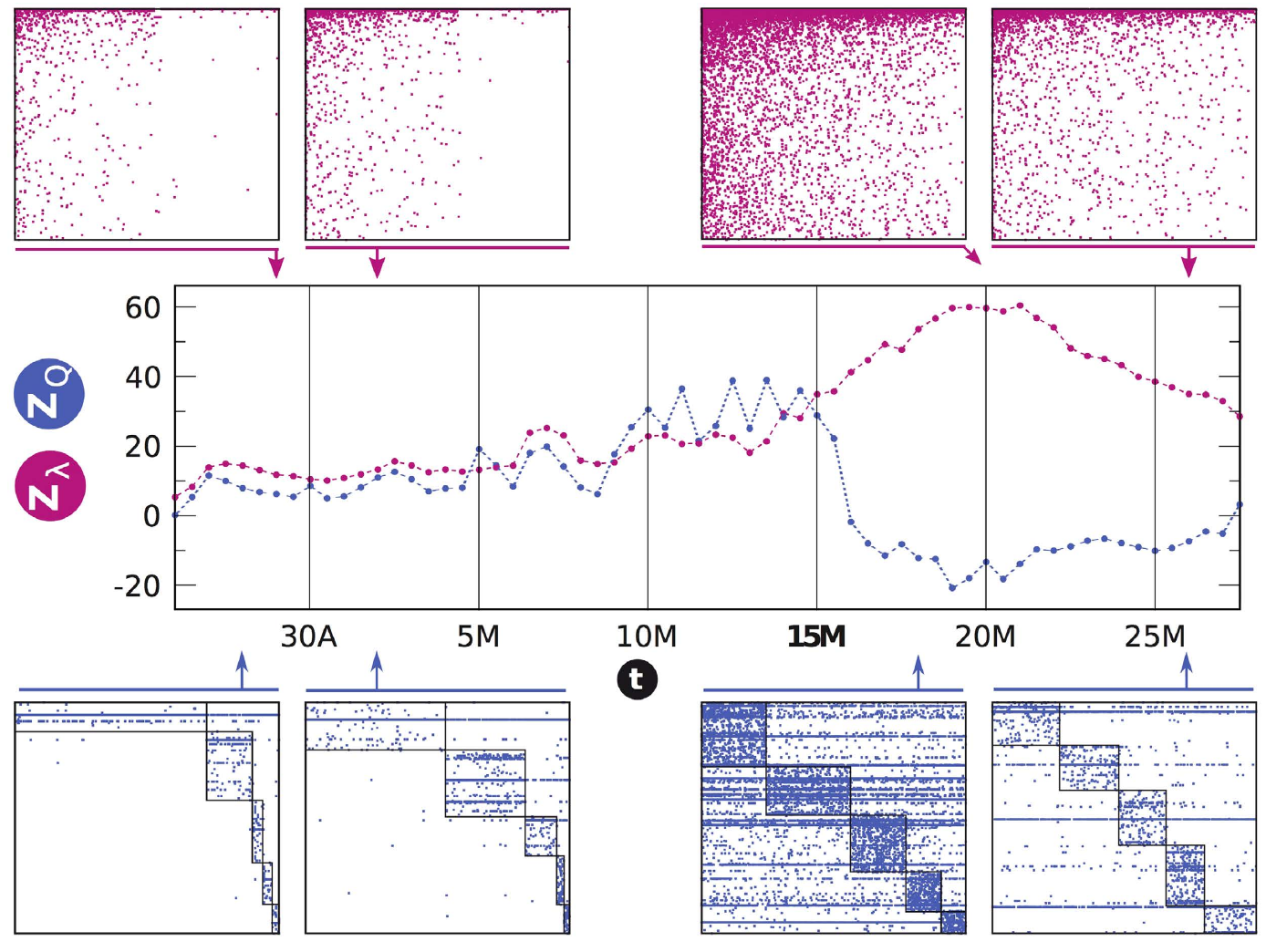}
\caption{Modular-to-nested transition in the bipartite user-meme network that only includes memes related to the 2011 civil protests in Spain. 
The central panel shows the simultaneous temporal evolution of nestedness $z_{\lambda}$ and modularity $z_{Q}$ over a two-month period (April-May 2011) -- both quantities are $z$-scores obtained by comparing the observed level of nestedness and modularity, respectively, with a suitable null model.
The other panels show the network's adjacency matrix when rows and columns are ordered in order of decreasing degree (top panels) and according to the detected modules (bottom panels).
Importantly, around the 15 May 2011, we observe the abrupt modular-to-nested transition described in the main text.
Reprinted from \cite{borge2017emergence}.}
\label{fig:modtonest}
\end{figure}

\paragraph{Modular-to-nested transition in communication dynamics}

While several socio-economic networks exhibit nestedness (see Section~\ref{sec:systems}), how such nested patterns emerge remains elusive. 
Borge-Holthoefer \textit{et al.}~\cite{borge2017emergence} found that in communication networks, the emergence of collective attention on a given topic can manifest itself as a transition from a phase where the system exhibits both significant nestedness and modularity to a phase where the system only exhibits significant nestedness -- such kind of evolution is referred to as \emph{modular-to-nested transition}~\cite{borge2017emergence}.
In other words, when collective attention narrows down to a single topic, the system abruptly transitions from a phase where the attention is dispersed across different sub-topics (modular structure) to a phase where all the attention is concentrated on a given topic (nested structure, no significant modularity).

Borge-Holthoefer \textit{et al.}~\cite{borge2017emergence} uncovered this phenomenon by analyzing a two-month time-resolved user-meme network from Twitter that only includes memes related to the 2011 civil protests in Spain (see~\cite{borge2017emergence} for details on data collection).
They found that in a first phase, both nestedness and modularity exhibit a growing trend. However, around the climax of the civil protests (May 
15–17), they observed a transition point after which nestedness keeps increasing, whereas modularity severely decreases. Such a transition marks the beginning of a phase where network topology is radically different from the initial topology (Fig. \ref{fig:modtonest}).
This result suggests that in communication networks, a nested architecture can emerge as a result of the emergence of a consensus on a given topic. A more recent study~\cite{bastos2018core} also focused on temporal variations of the topology of a Twitter communication network, and found that centralization tends to increase when the discussion becomes more specialized.

\subsection{Nestedness and ecological properties}
\label{sec:ecological_properties}

Given an observed macroscopic pattern, it is important to assess how that pattern is related to other system-specific macroscopic properties.
In ecology, an interaction network is embedded in an external environment; exploring how network nestedness (and modularity) is related to the macroecological environmental properties (such as precipitation, mean annual temperature, and temperature variability) has been the subject of various studies in ecology (see Section~\ref{sec:macroecological}).
The level of nestedness of spatial networks has been also related to their beta diversity~\cite{whittaker1960vegetation}, an important property that quantifies the relative magnitude of local and regional diversity (see Section~\ref{sec:beta}).

Besides, from a complex systems perspective, an essential challenge is to explain how the observed structural pattern of interest arose from the interactions of its constituents~\cite{mitchell2006complex,mitchell2009complexity,newman2011complex}. 
To this aim, one can adopt two (possibly complementary) strategies. 
From a network science perspective~\cite{newman2010networks, barabasi2016network},
one can study dynamic mechanisms of network growth that seek to reproduce the observed structural patterns~\cite{barabasi1999emergence,konig2011network,medo2011temporal,konig2014nestedness}, and statistically validate mechanisms that explain the dynamics of the system~\cite{medo2014statistical}.
This aspect is further developed in Section~\ref{sec:emergence} where we introduce several distinct network formation mechanisms that aim to explain the observed level of nestedness in ecological and socio-economic systems.

From an ecological perspective, one seeks to explain how the observed individual interactions are mediated by species individual traits. Explaining what drives pairwise interactions can, in turn, explain why macroscopic structural patterns of interest have been found. Ecologists have identified species abundance~\cite{krishna2008neutral,santamaria2007linkage}, the existence of forbidden links~\cite{krishna2008neutral,santamaria2007linkage}, and phylogenetic signal~\cite{rezende2007effects} as factors that can, to some extent, explain the observed degrees of nestedness (see Section~\ref
{sec:abundance}).

\subsubsection{Nestedness and macroecological properties}
\label{sec:macroecological}

We summarize here some of the results on the relation between nestedness and macroecological properties, i.e.,
ecological properties that characterize the system as a whole.
Nestedness and modularity display linear relationships with precipitation in plant-pollinator networks \cite{trojelsgaard2013macroecology}.
In a similar spirit, Schleuning \textit{et al.} \cite{schleuning2014ecological} focused on modularity, 
and found that modularity is negatively and positively correlated with mean annual temperature and temperature seasonality, respectively, in
weighted seed-dispersal networks.
Welti and Joern \cite{welti2015structure} examined 
$46$ bipartite mutualistic networks and $22$ bipartite
trophic networks. They found that nestedness decreases with temperature variability 
between years. On the other hand, no significant evidence was found for an
influence of environmental variables on the level of nestedness of trophic networks \cite{welti2015structure}.

Dalsgaard \textit{et al.} \cite{dalsgaard2013historical} studied how climate change affects
nestedness and modularity of pollination networks. They found that the
magnitude of this influence is different among islands and mainland regions. 
On the mainland, they found a positive association between nestedness 
and Quaternary climate change. The same was not found on islands, where Quaternary climate-change had no effects on nestedness.
Sebasti\'{a}n-Gonz\'{a}lez \textit{et al.} \cite{sebastian2015macroecological} found that 
nested structures tend to be located ``in areas with a high degree of human impact, 
high temperature seasonality, low precipitation, and, especially on the mainland, high stability in precipitation". Takemoto
and Kajihara \cite{takemoto2016human} confirmed that nestedness tends to increase with human impact, whereas modularity follows the opposite trend.
Finally, we refer to~\cite{pellissier2018comparing} for a recent review on the variation of structural patterns of ecological networks along environmental gradients.

\subsubsection{Spatial networks: Nestedness and beta diversity}
\label{sec:beta}

Beta diversity is a key concept in biogeography. Broadly introduced by Whittaker~\cite{whittaker1960vegetation} as ``The extent of change in community composition, or degree of community differentiation, in relation to a complex-gradient of environment, or a pattern of environments", beta diversity has been quantified through various metrics and its definition has been revisited several times. We refer to~\cite{tuomisto2010diversity} for an exhaustive review. In this Section, we introduce the original definition of beta diversity provided by Whittaker~\cite{whittaker1960vegetation} and discuss the relation between beta diversity and nestedness metrics. In addition, we will mention recent developments on the topic.

For a given geographical region composed of multiple sites, Whittaker’s $\beta$ is defined as the ratio between the regional and local diversities. In formulas, the regional diversity (usually referred to as ``gamma'' diversity) can be quantified as the total number of species in the geographical region, $\gamma=N$.
The local diversity ("alpha'' diversity) can be quantified as the average number of species per geographic site, $\alpha=M^{-1}\,\sum_{\alpha=1}^{M}k_\alpha=\braket{k^C}$. Whittaker’s $\beta$ is defined as the ratio between regional and local diversity: $\beta:=\gamma/\alpha=N/\braket{k^{C}}=N\,M/E=\rho^{-1}$. Essentially, Whittaker’s $\beta$ can be interpreted as a quantification of ``how many times as rich in effective species the dataset is than one of its constituent compositional units"~\cite{tuomisto2010diversity}. If one adopts Whittaker's $\beta$ to quantify beta diversity, the relation between nestedness and beta diversity is determined by the relation between the adopted nestedness metric and network density. For example, the nestedness metric by Wright and Reeves (Eq.~\eqref{wr}) is negatively correlated with $\beta$~\cite{wright1992meaning}.

Beta diversity can be also quantified in terms of the dissimilarities between the species compositions of different sites. For a two-site region, one can define the dissimilarity between the two sites through the S{\o}rensen index~\cite{sorensen1948method}
\begin{equation}
    \beta_{sor}=\frac{b+c}{2\,a+b+c},
    \label{sorensen}
\end{equation}
where $a$ denotes the number of species found in both sites;
$b$ denotes number of species that found in the first site but not in the second site;
$c$ denotes number of species that found in the first site but not in the second site.
This definition can be readily generalized to multiple-site regions~\cite{baselga2010partitioning}.

Importantly, ecologists have debated the factors determining the level of beta diversity of a given geographical area. Baselga~\cite{baselga2010partitioning} emphasized that beta diversity is modulated by two properties of the region: nestedness and turnover of species between sites. It is indeed intuitive that perfectly nested regions must exhibit some degree of diversity: in line with the definition of nestedness, some of the species that are present in richer sites cannot be found in poorer sites. At the same time, even a low-nestedness region can exhibit beta diversity if different species populate the different sites ("turnover of species") -- see Fig.~1 in~\cite{baselga2010partitioning} for an illustration. 

Stimulated by these considerations, a question arises: how to disentangle the relative contributions of nestedness and species turnover to beta diversity?
The S{\o}rensen index defined by Eq.~\eqref{sorensen} is, in principle, affected by both factors. Following Baselga~\cite{baselga2010partitioning}, one can consider a dissimilarity index that is only affected by species turnover and not by nestedness, like the Simpson index~\cite{simpson1943mammals}. 
For a two-site region, the Simpson index $\beta_{sim}$ is defined as~\cite{baselga2010partitioning}
\begin{equation}
    \beta_{sim}=\frac{\min{\{b,c\}}}{a+\min{\{b,c\}}}.
    \label{simpson}
\end{equation}
If two sites are occupied by the same number of species, nestedness does not contribute to the region's beta diversity. Therefore, the region's beta diversity is entirely determined by species turnover, and $\beta_{sor}=\beta_{sim}$.
Such equality does not hold, in general, for regions composed of two sites that have a different number of species. In that case, one can quantify the contribution of nestedness to beta diversity as $\beta_{nes}=\beta_{sor}-\beta_{sim}$, which implies that the overall diversity $\beta_{sor}$ is modulated by the joint contribution of species turnover ($\beta_{sim}$) and nestedness ($\beta_{nes}$). By performing the subtraction and using Eqs.~\eqref{sorensen}-\eqref{simpson}, one obtains~\cite{baselga2010partitioning}
\begin{equation}
    \beta_{nes}=\beta_{sor}-\beta_{sim}=\frac{\max{\{b,c\}}-\min{\{b,c\}}}{2\,a+\max{\{b,c\}}+\min{\{b,c\}}} (1-\beta_{sim}).
\end{equation}
We refer to~\cite{baselga2010partitioning} for the generalization of these indexes to regions with multiple sites ($M>2$), and to~\cite{baselga2010partitioning,baselga2012relationship} analogous decomposition of the $\beta$ diversity for $\beta$-diversity metrics different than the S{\o}rensen index (e.g., Jaccard dissimilarity~\cite{baselga2012relationship}). Beta-diversity metrics and their contributions from nestedness and species turnover can be found in the R package \url{betapart}~\cite{baselga2012betapart}.

\subsubsection{Interaction networks: Nestedness and relative species abundance distribution}
\label{sec:abundance}

Ecologists have recognized since a long time that species abundance plays an important role in shaping the structure of mutualistic networks~\cite{jordano1987patterns,bascompte2013mutualistic}. 
A natural question arises: can species abundance explain the observed levels of nestedness? 
The question has been investigated in the context of neutral theory~\cite{azaele2016statistical}, a theoretical framework that assumes that all 
species within a particular trophic level\footnote{\emph{Trophic level} refers to the set of
all species that belong to the same level in the food chain~\cite{azaele2016statistical}. Neutral theory is 
limited to communities of species that belong to one specific trophic level.} have
the same chances of reproduction and death regardless of their peculiar
traits~\cite{rosindell2011unified}. In other words, according to neutral theory, all species are functionally
equivalent: if an individual of a given species dies, it is replaced by a new individual, 
regardless of the species it belongs to. This assumption and neutral-theory 
stochastic models\footnote{A detailed description of the models and implications
of neutral theories goes out of the scope of the present review, but it can be 
found in the review article by Azaele \textit{et al.}~\cite{azaele2016statistical}.} lead 
to expected relative species abundance (RSA) distributions~\cite{etienne2005new}. By assuming that the 
species-species interaction probabilities only depend on species abundance, one can 
check whether the abundance distributions predicted by neutral theory can explain the observed degree of nestedness~\cite{krishna2008neutral}, 

While such a parsimonious explanation of nestedness is tempting, it might neglect other important 
factors. It is known indeed that unobserved interactions are, in most cases, the consequence
of biological constraints that make that interaction physically impossible -- see~\cite{olesen2010missing} for 
an evaluation of the number of such ``forbidden links'' in interaction networks\footnote{Among the unobserved links,
forbidden links are fundamentally different from those links (``missing links''~\cite{olesen2010missing}
or ``neutral forbidden links``~\cite{canard2012emergence}) that were not recorded simply because we did not observe the system over a long
enough time span. }. Can we explain the observed 
nestedness through species abundance, without including forbidden links? If we include the forbidden links, 
can we achieve a more accurate explanation of the observed nestedness? 
Another caveat is necessary for the use of the word ``explanation". Even if species abundance alone
was able to explain, in a statistical sense, species degree and network nestedness, it is still debatable
whether generalists are generalists because they are more abundant or, by contrast, more abundant
species are more abundant because they are more generalists~\cite{santamaria2007linkage}.

With these caveats in mind, one can investigate whether the Relative Species Abundance (RSA) distribution predicted by neutral
theory can explain the observed nestedness of mutualistic networks~\cite{krishna2008neutral}. The Relative Species Abundance (RSA) distribution of a given
community is denoted as the sequence $\{n_1,\dots,n_S\}$ of the species' population sizes, where $S$ is the total number of species.
To uncover the impact of the RSA distribution on nestedness, Krishna \textit{et al.}~\cite{krishna2008neutral} proceeded in two steps. First, they generated the plant and animal RSA 
distributions based on the theoretical prediction of neutral theory~\cite{etienne2005new}. We denote the two 
distributions as $\{p_1,\dots,p_M\}$ and $\{a_1,\dots,a_N\}$, respectively, where $M$ ($N$) denotes the total number of plant (animal) species.
Second, they assumed that the probability $P_{i\alpha}$ that animal $i$ interacts with plant $\alpha$ is proportional 
to their relative abundance: $P_{i\alpha}= a_i\,p_{\alpha}/(\sum_j a_j \,\sum_{\beta}p_{\beta})$. They built 
mutualistic networks -- which they refer to as \emph{neutral mutualistic networks} -- by establishing a link
between a plant and animal if a number of interactions above a given threshold was observed.
Finally, they investigated whether the resulting networks exhibit a nested structure.

Besides, they aimed at establishing whether the RSA distribution could explain the nestedness observed in mutualistic networks. To this end, they observed that if a given factor explains nestedness, it should produce a ranking of the nodes that well approximates the ranking by the nestedness temperature calculator, which is aimed at maximizing the degree of nestedness (see Sections~\ref{sec:distance} and \ref{sec:packing}).
Based on this observation, the problem of assessing whether the RSA distribution explains nestedness reduces to comparing the temperature $T_{ab}$ of the matrix where species are ranked by their abundance with the temperature $T$ of the matrix 
that is produced by the nestedness temperature
calculator.

The results of their investigation are two-fold.
First, they found that neutral mutualistic networks are almost perfectly nested.
Probabilistic animal-plant interactions consistent with neutral theory~\cite{etienne2005new} are, therefore, among 
the possible mechanisms that lead to the emergence of nestedness.
Second, they found that in real mutualistic networks, the RSA distribution can explain only part of the observed nestedness,
with values of the fraction of explained nestedness $(100-T_{ab})/(100-T)$ around $0.6-0.7$~\cite{krishna2008neutral}.
Importantly, they found that the agreement between expected and observed nestedness can be substantially improved by accounting for the forbidden links.

Santamaría and Rodríguez-Gironés~\cite{santamaria2007linkage} found 
that a log-normal neutral model of interactions (i.e., a model that produces a 
log-normal RSA distribution) fits well the observed nestedness 
in empirical pollinator-plant networks. However, they pointed out
that the log-normal neutral model should be rejected as a parsimonious explanation of
nestedness because of three factors~\cite{santamaria2007linkage}: (1) the fact that the 
model fits the empirical nestedness does not tell us anything about the causal direction 
of the abundance-degree relationship: it is not excluded that more abundant species 
might be more abundant because they are generalists, and not the other way around;
(2) the RSA distribution might explain nestedness under the assumption that the species-species
interaction probability is proportional to species abundance; however, this assumption is not
supported by empirical studies~\cite{ollerton2003pollination}; (3) As we pointed out above, several studies~\cite{jordano2003invariant,olesen2010missing} 
have shown that forbidden links play a critical role in shaping the topology of interaction networks.
 Hence, while of undeniable relevance, the connection between nestedness, species abundance, and forbidden links alone does not allow us to infer the generative mechanisms that have led to those macroscopic properties.

Beyond relative species abundance distribution, Rezende \textit{et al.}~\cite{rezende2007non} aimed to detect the presence of aphylogenetic signal for plants and animals assemblages for 36 plant-pollinator and 23 plant-frugivore mutualistic networks. To detect phylogenetic signal, they compared the properties of the empirical networks with those numerically generated based on the structure of the species phylogenetic history~\cite{rezende2007effects,rezende2007non}. They found a fundamental difference between species degree (i.e., number of interactors per species) and species strength (i.e., number of recorded interactions): differently from species strength, species degree exhibits a strong phylogenetic signal~\cite{rezende2007non,bascompte2013mutualistic}.
Besides, they found that the distance between two given species in the interaction network is significantly correlated with their distance in the phylogeny: for most communities, phylogenetically-similar species tend to play similar roles in the interaction network. In a different article, Rezende \textit{et al.}~\cite{rezende2007effects} found that phylogenetic effects can significantly contribute to nestedness.
Overall, information on relative abundance and phenology can be sufficient to predict the level of nestedness of a network, yet it cannot predict the detailed network structure~\cite{vazquez2009evaluating}.

\subsection{Nestedness and economic properties}
\label{sec:economic_properties}

Some scholars have argued that the ultimate goal of understanding a complex system is to accurately predict its future behavior~\cite{ren2018structure}. 
Predicting the future evolution of a complex system can be difficult due to various factors, including chaotic behavior, our incomplete knowledge of the fundamental laws that describe the system's behavior, our lack of knowledge of the system's state vector -- we refer to~\cite{baldovin2018role} for a survey. The network science literature has designed a wide spectrum of techniques either to predict the future links that will appear in a network, or to infer the connections that are missing in noisy data (see~\cite{lu2011link,ren2018structure,squartini2018reconstruction} for recent reviews on the topic).
The nested structure of socio-economic systems has motivated scholars to leverage such structure to predict the future behavior of the system, at the level of both individual links and nodes. 
These studies aimed to predict the future success of firms~\cite{saavedra2011strong} (Section~\ref{sec:survival}) and countries~\cite{cristelli2015heterogeneous,cristelli2017predictability} (Section~\ref{sec:fit_gdp}), and to predict future links in economic 
networks~\cite{bustos2012dynamics} (Section~\ref{sec:link_prediction}). 
Besides, in a similar spirit as neutral models in ecology (see Section~\ref{sec:abundance}), scholars have attempted to explain the observed level of nestedness in economic networks based on hidden capabilities that drive the formation of pairwise interactions (see Section~\ref{sec:hidden}).

\subsubsection{Individual nestedness contributions and the survival of firms}
\label{sec:survival}

As nestedness arises as a result of node-level interactions, it is natural
to ask whether some nodes give a larger contribution to the nested 
structure of a given system. 
Should this be the case, one can further ask (1) whether the strongest contributors to nestedness play a more significant role to network persistence (measured as the fraction of initial nodes that did not become extinct by the end of population dynamics simulations~\cite{saavedra2011strong}) than weaker contributors; (2) whether stronger contributors experience a larger or smaller probability of becoming extinct.

Saavedra \textit{et al.}~\cite{saavedra2011strong} quantified the individual contribution to network nestedness from each individual node. Given a node $i$, its contribution $C_i$ to nestedness is determined by comparing the observed nestedness with the nestedness of randomized networks obtained by only randomizing node $i$'s links. More specifically, $C_i=(\mathcal{N}-\mu_i(\mathcal{N})) /\sigma_i(\mathcal{N})$, where $\mu_i(\mathcal{N})$ and $\sigma_i(\mathcal{N})$ represent the mean and the standard deviation, respectively, of nestedness $\mathcal{N}$ over the networks where node $i$'s links have been randomized. 

Saavedra \textit{et al.}~\cite{saavedra2011strong} found that both in ecological mutualistic networks and in economic manufacturer-contractor networks, individual contributions to nestedness are heterogeneously distributed, and some nodes give a significantly larger contribution than the others. Through simulations on mutualistic networks based on a model of mutualistic interactions\footnote{We refer to~\cite{saavedra2011strong} for details. The model is essentially the same that has been applied in other works~\cite{bastolla2009architecture,rohr2014structural} to uncover the impact of nestedness on systemic feasibility and dynamical stability -- see Section~\ref{sec:mutualistic}.}, they found that the strongest contributors to nestedness are also those whose extinction has the largest 
impact on network persistence\footnote{A recent study~\cite{dominguez2015ranking} indicates that
non-linear iterative algorithms perform better than individual nestedness contributions in identifying the 
most important and vulnerable network nodes under targeted attack. See Section~\ref{sec:structural_nodes} for a detailed presentation.}. The results of numerical simulations indicate that, ironically, the strongest contributors are also the most vulnerable nodes to extinction~\cite{saavedra2011strong}.

While the analysis of the persistence and stability of ecological communities almost invariably requires a population dynamics model, datasets on social and economic activity often feature fine-grained temporal information, which can relieve us from the need of numerical simulations. In particular, one can translate the questions above to economic systems: have the firms with the strongest contributions to nestedness the highest risk to fail? Saavedra \textit{et al.}~\cite{saavedra2011strong} addressed this question by analyzing a manufacturer-contractor 
bipartite network (see Section~\ref{sec:contractor}) over a $15$-years time span. In agreement with their model-based findings on mutualistic networks, they found that for both manufacturers and contractors, the strongest contributors to nestedness are also the firms that are less likely to survive.

The work by Saavedra \textit{et al.}~\cite{saavedra2011strong} highlights at least two 
important factors. First, the similarity of the findings in ecological and 
economical networks point to the generality of the role of nestedness individual contributions for node survival. Second, ecologists often lack time-stamped data over a long enough time span. As a result, 
works on the implications of network topology on systemic stability
are typically based on population dynamics and dynamical-system stability analysis (see Section~\ref{sec:implications}). On the other hand, socio-economic data are often time-stamped.
Socio-economic data might, therefore, allow us to validate mechanisms on the emergence of nestedness and its implications, which might give insights that are also relevant to ecologists.

\subsubsection{Nestedness and prediction of economic links}
\label{sec:link_prediction}


The study by Saavedra \textit{et al.}~\cite{saavedra2011strong} focused on the prediction of the survival of firms. At a finer scale, one might attempt to leverage the nested structure to predict the future links of a system.
The argument is the following: suppose that we are confident that a given system exhibits a significantly nested structure. If we found a link in a matrix region where there are only a few other links, we might argue that this link constitutes an outlier in the overall systemic structure and, as a consequence, it is more likely to disappear in the future.
In an analogous way, if a link is missing in a dense region of the adjacency matrix, it might be more likely that that link will appear in the future.


Building on a previous work in ecology~\cite{maron2004can}.
Bustos \textit{et al.}~\cite{bustos2012dynamics} tested this idea in both 
the country-product bipartite network over the 1985-2009 period,
and in a Chilean firm-location bipartite network over the 2005-2008 period. For both networks, 
they found that the level of nestedness was relatively stable over time and statistically 
significant with respect to the Proportional-Proportional model by~\cite{bascompte2003nested} (see Section~\ref{sec:nine}) and a dynamic null model
(see Figs.~1f-g in~\cite{bustos2012dynamics}). Besides, they found that the coefficients of the logistic model~\cite{bustos2012dynamics}
\begin{equation}
M_{i\alpha}(t)=a\,k_i(t)+b\,k_{\alpha}(t)+c\,k_i(t)\,k_\alpha(t)+\epsilon_{cp}(t)
\label{hidalgo}
\end{equation}
are all highly significant: the interaction term $c\,k_i(t)\,k_\alpha(t)$ is essential to 
explain the presence of links. This interaction term is compatible with the nested structure of the network, as it penalizes links between two nodes with low degree.
The observed temporal persistence of nestedness, together with the explanatory power of the model~\eqref{hidalgo}, suggests that nestedness can be used to predict the future links in both systems.

\begin{figure*}[t]
	\centering
\includegraphics[scale=0.25]{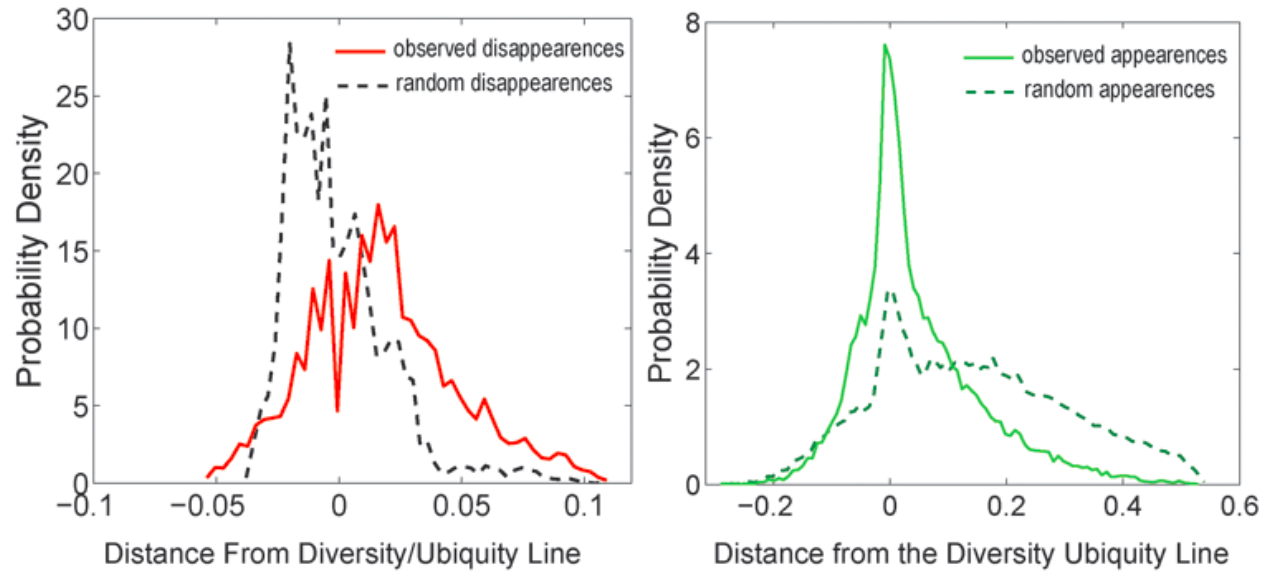}
	\caption{Fraction of appearances and disappearances as a function of the distance from the diversity-ubiquity line for the country-product network. 
	Observed appearances (disappearances) tend to be closer to (farther away from) the diversity-ubiquity line than expected by chance.
	Adapted from~\cite{bustos2012dynamics}.}\label{fig:hidalgo2}
\end{figure*}

To test this hypothesis, Bustos \textit{et al.}~\cite{bustos2012dynamics} compared, for each pair of years, the ability of their model to predict the appearance and disappearance of links\footnote{ Bustos \textit{et al.}~\cite{bustos2012dynamics} defined an appearance of a link $(i,\alpha)$ as an increase in exports per capita from
less than 5\% of the world average, for five consecutive years, to more than 25\%, for at least five years. 
They defined the disappearance of a link $(i,\alpha)$ as the decrease in exports per capita of a country 
from 25\% or more than the world average to 5\% or less, for at least five years.}. The model produces 
an Area Under the Curve (AUC)\footnote{The AUC is one of the most used metrics to benchmark prediction accuracy. We 
refer to~\cite{lu2011link,bustos2012dynamics}
for its definition.} substantially larger than 0.5 for both appearances and disappearances, 
which implies better predictability than random predictions. Interestingly, it predicts disappearances significantly better than 
appearances (average AUC equal to $0.81$ against $0.62$, see Fig. 3c in~\cite{bustos2012dynamics}).
Besides, they studied the structural position of appearances and disappearances in the adjacency matrix. 
In particular, they measured their distance from the diversification-ubiquity line, which denotes the shape of
a perfectly nested matrix with the same number of edges as the matrix in exam. They found that the observed 
appearances tend to lie significantly closer to the diversification-ubiquity line than expected by chance, whereas observed
disappearances tend to lie outside of the perfectly nested shape (see Fig.~\ref{fig:hidalgo2}). 
The nested structure of the network is therefore highly informative on future changes in the system.

More recently, Medo \textit{et al.}~\cite{medo2018link} aimed to leverage the nested structure of World Trade and mutualistic networks to reconstruct missing links in these systems.
More specifically, they
considered the number of violations of the nestedness condition as defined by the following equation~\cite{grimm2017analysing,medo2018link}
\begin{equation}
    V : = (1-\Theta(k_j-k_i)) \sum_{\alpha} A_{j\alpha} (1-A_{i\alpha}).
\end{equation}
Based on this quantity, they defined the link-prediction score of a link $(i,\alpha)$, $s_{i\alpha}$, as the difference between the number of violations of the nestedness condition  \textit{after} $(i,\alpha)$ is added to the network, $V^{(i,\alpha)}$, and the number of violations of the original network, $V$: $s_{i\alpha}=V^{(i,\alpha)}-V$.
They validated this index in both synthetic networks with a tunable degree of nestedness and empirical networks, finding that the resulting link-prediction method can outperform some of the state-of-the-art link prediction techniques for both highly-nested synthetic networks and some of the empirical networks that they analyzed.

Beyond the nestedness-based method by Bustos \textit{et al.}~\cite{bustos2012dynamics} and Medo \textit{et al.}~\cite{medo2018link}, scholars attempted to predict future links in World Trade based on country-level development paths in the product space~\cite{hidalgo2007product,zaccaria2014taxonomy}, and on recommender systems~\cite{vidmer2015prediction}.
We point out that a comprehensive evaluation of different methods for link prediction in World Trade is currently missing. Such comparative analysis would be particularly relevant as World Trade data are typically noisy due to diverse factors~\cite{battiston2014metrics,tacchella2018dynamical}, which can affect the reliability of structural metrics of country and product importance~\cite{battiston2014metrics, mariani2015measuring,wu2016mathematics} and the resulting economic development forecasts~\cite{tacchella2018dynamical}.

\subsubsection{Nestedness maximization and the future development of countries}
\label{sec:fit_gdp}

As we have discussed in Section \ref{sec:nonlinear}, ranking algorithms that produce nested adjacency matrices (such as the fitness-complexity algorithm~\cite{tacchella2012new}) have been extensively applied to world trade networks, aiming to quantify the countries' competitiveness and the products' level of sophistication.
The fitness score of a country can be interpreted as a proxy for its number
of available ``capabilities", to be interpreted as the building blocks needed to produce and export
different types of products~\cite{hidalgo2009building} (see Sections~\ref{sec:nonlinear} and~\ref{sec:hidden}). Besides, the countries' fitness score is positively correlated with their gross domestic product per capita (GDPpc), yet this correlation is far from one~\cite{tacchella2012new,cristelli2015heterogeneous}.

Importantly, the deviations from the linear-regressed trend are informative about the future development of countries~\cite{cristelli2015heterogeneous}. Suppose indeed that a country has a relatively ordinary GDPpc, and a relatively large fitness score. This means that while the level of development of that country is not yet comparable with the leading economies, the country possesses a wide set of capabilities which might allow it to economically grow in the future.
The recent history has seen one such example: China. In 1995, China had a substantially larger fitness than countries with a similar GDPpc and, in the subsequent years, China's GDPpc experienced a drastic increase~\cite{cristelli2013measuring}.

\begin{figure*}[t]
	\centering
\includegraphics[scale=0.95]{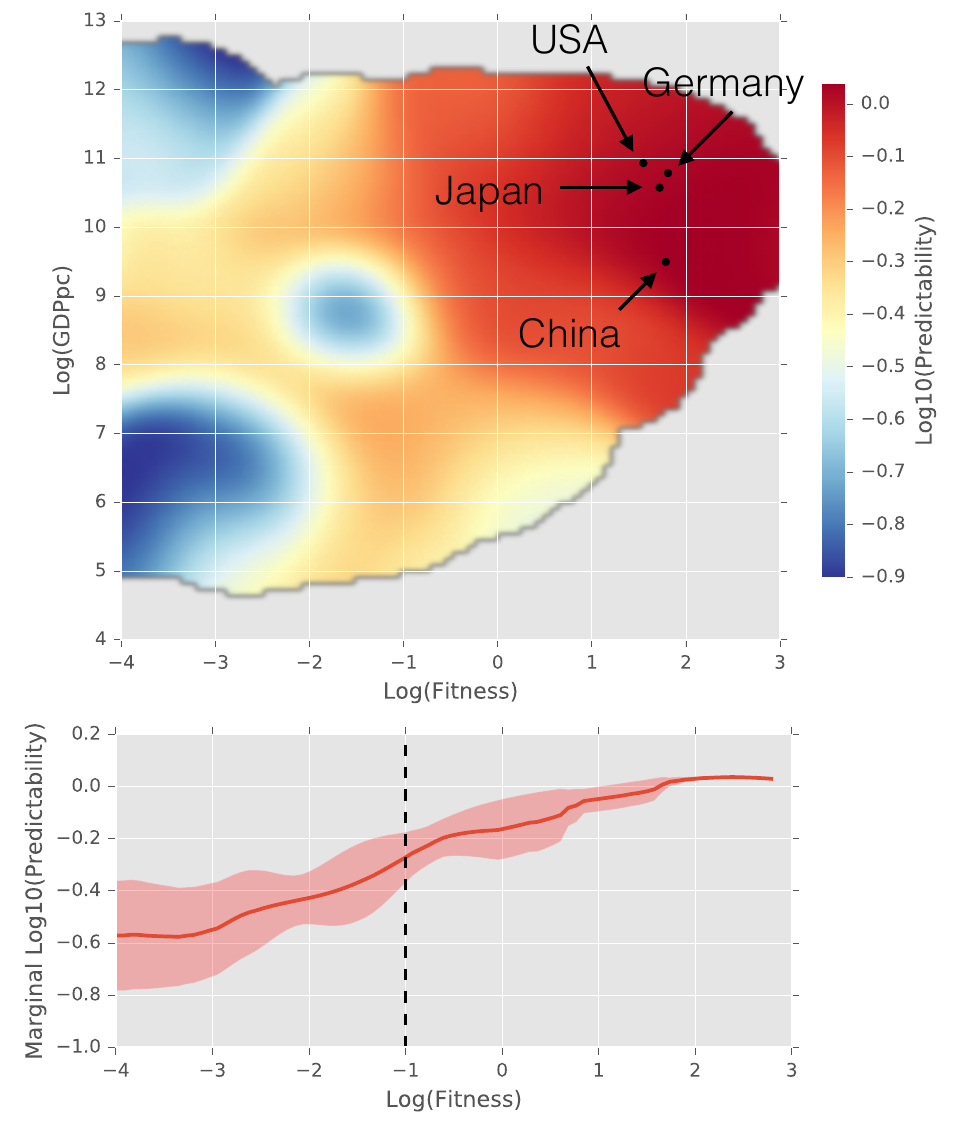}
	\caption{Economic-complexity predictions of the national Gross Domestic Product per capita (GDPpc): Level of predictability in the fitness-GDPpc plane (top panel), and as a function of fitness (bottom panel).
	The GDPpc of high-fitness countries is significantly more predictable than that of low-fitness countries.
	Reprinted from~\cite{cristelli2017predictability}.}\label{fig:fit_gdp}
\end{figure*}

The hypothesis that a country's high fitness score to economic growth in the future has been first tested by Cristelli \textit{et al.}~\cite{cristelli2015heterogeneous}. They analyzed $16$ years of World Trade (1995-2010), and sought to predict the future development of the countries based on their position in the GDPpc-fitness plane (see Fig.~\ref{fig:fit_gdp}). Of course, given the inherent complexity of the world production, it is not obvious that the history from the past is informative about the future of the system. 
If the current fitness of a country is informative about its future development, we would observe highly-predictable regions in the fitness-GDPpc plane. That would signal some degree of ``stationarity'' in those regions, and the possibility to attempt to predict the future~\cite{cristelli2017predictability}. 

Borrowing from the methodologies from the dynamical systems literature~\cite{lorenz1969atmospheric}, Cristelli \textit{et al.}~\cite{cristelli2015heterogeneous} built the Selective Predictability Scheme (SPS) which analyzes the past history of the system to evaluate, for each region of the plane, its level of predictability\footnote{The predictability of a region $\mathcal{R}$ over a $t$-years time window is essentially the heterogeneity of 
the region $\mathcal{R}(t+\Delta)$ where countries that were in region $\mathcal{R}$ at time $t$ are located at time $t+\Delta t$.
In practice, it is convenient to split the plane into rectangular boxes and study the predictability of each of the resulting boxes~\cite{cristelli2015heterogeneous}.}~ and the most likely future position in the plane for countries that are located in that region.
Intriguingly, there turn out to be both highly-predictable and low-predictability regions in the fitness-GDPpc plane~\cite{cristelli2015heterogeneous}. Importantly, the predictability of a region is an increasing function of fitness, and high-fitness countries tend indeed to grow their GDPpc in the future~\cite{cristelli2015heterogeneous,cristelli2017predictability}.
When visualizing the dynamics of countries through animations, one can detect by eye the regular upward movement of the countries in the high-fitness region~\cite{noorden2015physicists}.

A recent work by the World Bank~\cite{cristelli2017predictability} systematically compared the predictions of the SPS based on country fitness with those by the International Monetary Fund (IMF). They found that, when one only considers countries that are in the predictable region of the fitness-GDPpc plane, the SPS predictions significantly outperform the IMF predictions. The performance of the SPS and IMF are comparable when one considers all the countries. With respect to the IMF growth model which builds upon many variables, the SPS method is very parsimonious, as it only involves one parameter: country fitness.
In a subsequent work, Tacchella \textit{et al.}~\cite{tacchella2018dynamical} found that proper data sanitation can substantially increase the volume of the predictable region; on the cleaned data, the predictions by the SPS method over a five-year time horizon outperform those by the IMF even when one considers both high- and low-fitness countries. Besides, the predictions by the SPS can be combined with those by the IMF to substantially improve the GDPpc forecasts.
Myanmar, Cambodia, and India turned out to be the three countries whose GDPpc's forecast benefited most from the combination of SPS and IMF results~\cite{tacchella2018dynamical}.

These works have inspired other scholars~\cite{liao2017ranking,pugliese2017complex,vinci2018economic} to investigate the relation between a country's fitness and its future economic development.
To conclude, we mention two additional aspects of this framework. First, in a similar way, one can study the dynamics of products on a plane formed by the score of the product by the fitness-complexity algorithm and a suitable indicator of product added value~\cite{angelini2017complex}.
Second, one can build score-GDPpc planes for the scores produced by other structural metrics based on the bipartite country-product network. In particular, Liao \textit{et al.}~\cite{liao2017ranking} measured the overall predictability\footnote{The overall predictability of the plane is obtained by aggregating the predictability of its regions, and weighting each region by the number of observed events -- see Section 7.2 in~\cite{liao2017ranking} for details.} of the score-GDPpc plane for the country-level scores by three different network-based ranking metrics: degree, fitness, and the Economic Complexity Index (ECI)~\cite{hausmann2014atlas} -- see Fig.~15 in~\cite{liao2017ranking} for an illustration. They found that country fitness as quantified by the fitness-complexity algorithm leads to the largest overall predictability, whereas no regular dynamics emerges in the ECI-GDPpc plane.
The obtained results suggest a link between nestedness maximization (see Section~\ref{sec:packing}) and economic predictions: rankings of countries that produce highly-nested matrices, such that the one by fitness, can be highly informative for the countries' future development.

\subsubsection{Nestedness and hidden economic capabilities}
\label{sec:hidden}

In the economic complexity approach to World Trade analysis~\cite{hidalgo2009building,tacchella2012new,cristelli2015heterogeneous}, the observed bipartite country-product export network is interpreted as the result of a projection of a tripartite country-capability-product network where countries are connected with the capabilities they possess, and products are connected with the capabilities they require in order to be produced (see Fig.~\ref{fig:capabilities}).
Hausmann and Hidalgo~\cite{hausmann2011network} leveraged this intuition to build a model of the World Trade where a given country exports a given product if and only if the country possesses all the capabilities required to produce that product. In practice, this model can be considered as a ``neutral model'' of development~\cite{hausmann2011network,bustos2012dynamics} in a similar spirit to neutral models in ecology (see Section~\ref{sec:abundance}): the interactions driven by a hidden property -- species abundance and country capabilities in ecological and trade networks, respectively -- might explain the network's structural patterns and, in particular, its nestedness.

Formally, let us denote by $\kappa\in\{1,\dots,K\}$ the available capabilities. We denote by $\mat{A}$ the bipartite country-product adjacency matrix, by $\mat{C}$ the country-capability matrix ($C_{i\kappa}=1$ if country $i$ possesses capability $\kappa$), by $\mat{P}$ the product-capability matrix ($P_{\alpha\kappa}=1$ if product $\alpha$ requires capability $\kappa$ in order to be made). With this notation, $A_{i\alpha}=1$ if and only if $\sum_\kappa C_{i\kappa}\,P_{\alpha\kappa}=\sum_\kappa P_{\alpha\kappa}$, i.e., if country $i$ possesses all the capabilities necessary to produce $\alpha$.
Therefore, the model assumes that the availability of capabilities is the only factor that determines the level of diversification of each country: each country will produce all the products for which that country has the required capabilities. While in contrast with standard economic theories that prescribe country specialization as a determinant for wealth~\cite{ricardo1817principles}, this assumption well matches the empirical nestedness of the country-product matrix~\cite{bustos2012dynamics,tacchella2012new}.

The matrices that involve the capabilities, $\mat{C}$ and $\mat{P}$, are unobservable; nevertheless, one can formulate a probabilistic rule to generate their elements, and verify a posteriori the agreement between the generated country-product matrices and the observed ones.
Hidalgo and Hausmann~\cite{hidalgo2009building} assumed that country $i$ possesses capability $\kappa$ with probability $r$, whereas it does not possess it with probability $1-r$; analogously, product $\alpha$ requires capability $\kappa$ with probability $q$, whereas it does not require it with probability $1-q$. Hence, each country $i$ ends up with a given number of available capabilities $c_i=\sum_{\kappa} C_{i\kappa}$, and each product with a given number of required capabilities $p_{\alpha}=\sum_{\kappa} P_{\alpha\kappa}$. The model has three parameters: $r,q,C$. Nevertheless, to calibrate the model to a real system, only two of these parameters are free, whereas the third one is fixed by the condition that the average degree of the model-generated networks matches that of the original network.

It is natural to investigate how countries' diversification and products' ubiquity are related to their number of available and required capabilities, respectively.
The average number of products exported by country $i$ (or $i$'s average diversification), $\overline{k_i}$, can be written in terms of $i$'s number of available capabilities, $c_i$, as
\begin{equation}
\overline{k_i(c_i)}=\sum_{x=1}^C \mathbb{P}[c_i\to x]\,M(x),
\end{equation}
where $\mathbb{P}[c_i\to x]$ denotes the probability that a country with $c_i$ capabilities exports a product that requires $x$ capabilities, whereas $M(x)$ denotes the number of products that require $x$ capabilities. Given that each capability is extracted independently, the probability $\mathbb{P}[c_i\to x]$ is simply given by $\mathbb{P}[c_i\to x]=(c_i/C)^x$. The fraction of products that require $x$ capabilities is given by the binomial distribution:
\begin{equation}
\frac{M(x)}{M}={C\choose x}\,q^x\,(1-q)^{C-x},
\end{equation}
which is why the model was referred to as \emph{binomial model} by Hausmann and Hidalgo~\cite{hausmann2011network}.
Therefore, we obtain
\begin{equation}
\overline{k_i(c_i)}=M\,\sum_{x=0}^C \Bigl(\frac{c_i}{C}\Bigr)^x\,{C\choose x}\,q^x\,(1-q)^{C-x}.
\end{equation}
By using the binomial series, we can approximate the previous expression as
\begin{equation}
\overline{k_i(c_i)}\simeq M\,\Bigl(q\,\frac{c_i}{C}+1-q \Bigr)^C.
\end{equation}
In a similar way, one can obtain expressions for product ubiquity as a decreasing function of the number of required capabilities, and the expected functional forms of the country diversification and product ubiquity distributions,

While this model produces country-product networks that are significantly more nested than expected by chance, it cannot generate networks with a degree of nestedness comparable to the observed one in World Trade data~\cite{bustos2012dynamics}.
To overcome this limitation, Bustos \textit{et al.}~\cite{bustos2012dynamics}
introduced an alternative capability-based model where the number of capabilities each country has is a random number extracted from the uniform distribution in $[0,R]$, whereas the number of capabilities required by each product is a random number extracted from the uniform distribution in $[0,Q]$. Because of such uniform distributions, the model is referred to as the \emph{uniform model} by Bustos \textit{et al.}~\cite{bustos2012dynamics}. The model has again two free parameters: $R$ and $Q$. The probability that country $i$ has capability $\kappa$ is given by $P_{i\kappa}=\min\{1,R\,i/c_i\}$. The uniform model generates bipartite adjacency matrices that are significantly more nested than those generated by the binomial model, and the level of nestedness of the generated matrices is comparable with that observed in real country-product matrices~\cite{bustos2012dynamics}.

An alternative capability-based model was proposed by Tacchella \textit{et al.}~\cite{tacchella2016build}, who assumed that capabilities have a different ``usefulness", defined as the number of products that require that capability. The possibility of varying the usefulness of individual capabilities makes such model more general than the binomial model which assumes that the number of capability required by each product follows a binomial distribution. A power-law distribution of usefulness can be used to generate networks with levels of nestedness comparable with those observed in real data~\cite{tacchella2016build}, which suggests that the most diversified countries might be those endowed with the most ``useful'' capabilities.

\clearpage 

\section{Emergence of nestedness: rewiring and formation mechanisms}
\label{sec:emergence}

Network structural properties are of interest not only because they deepen our understanding of the organization of complex systems, but also because 
distinct macroscopic patterns result from different microscopic mechanisms of network growth.
Candidate mechanisms to describe the evolution of the system might need to be ruled out if they lead to network topologies
that mismatch those observed in real data.
Following the seminal work by Barab{\'a}si and Albert \cite{barabasi1999emergence}, explaining the
scale-free degree distribution of many real systems has received an enormous amount of 
attention in the network science literature \cite{dorogovtsev2013evolution}. 
Models with a node-level hidden variable (typically referred to as \emph{fitness}~\cite{caldarelli2002scale}) provide an explanation for the scale-free structure of real networks that does not resort to preferential attachment mechanisms~\cite{caldarelli2002scale}. Importantly, a fitness model with a threshold linkage rule~\cite{masuda2004analysis,hagberg2006designing} can generate perfectly nested networks with various degree distributions (see Section~\ref{sec:threshold}).

In parallel, ecologists have devoted a massive effort to understand how nested networks
can emerge in interaction networks. 
Mechanisms that have been proposed include (but are not limited to) competition load minimization~\cite{bastolla2009architecture}, fitness optimization~\cite{zhang2011interaction,suweis2013emergence}, trait-matching~\cite{saavedra2009simple}, speciation and divergence~\cite{valverde2018architecture}, invasion dynamics~\cite{maynard2018network}. Some of these mechanisms
build on optimization problems where species aim to maximize their own fitness~\cite{zhang2011interaction,suweis2013emergence} or the community-level fitness~\cite{suweis2013emergence}, whereas 
other mechanisms~\cite{valverde2018architecture,maynard2018network} do not involve optimization.
Section~\ref{sec:formation_ecol} surveys the mechanisms for the emergence of nestedness in socio-economic networks.

More recently, scholars have started investigating the emergence of nestedness in socio-economic networks as well.
In unipartite socio-economic network, this effort has revealed that nestedness can emerge as a result of a process where individual actors strive to maximize the centrality of their
interaction partners, i.e., their social status~\cite{konig2011network,bardoscia2013social,konig2014nestedness}, through a dynamical process that has been referred to as ``social climbing game''~\cite{bardoscia2013social}. From this standpoint, highly nested structures correspond to highly hierarchical, ``low-temperature'' structures where, from a statistical-physics perspective, social mobility faces high energetic costs~\cite{bardoscia2013social}. In bipartite country-product export networks, capability-based mechanisms~\cite{hausmann2011network,bustos2012dynamics} can generate networks with a level of nestedness compatible with the observed one. Nestedness in World Trade has been also explained in terms of network formation mechanisms that feature both innovation and specialization in the countries' development process~\cite{saracco2015innovation}.
Section~\ref{sec:formation_socioec} surveys the mechanisms for the emergence of nestedness in socio-economic networks.



\subsection{Graph-theoretic mechanisms}
\label{sec:threshold}

Without even considering ecological or socio-economic processes, perfectly nested topologies can simply arise from a fitness model based on a threshold linkage rule~\cite{caldarelli2002scale,masuda2004analysis,hagberg2006designing}. In the following, we first review the threshold-based fitness model (Section~\ref{sec:threshold_model}), and then describe how threshold models can be used to generate nested networks with various degree distributions (Section~\ref{sec:emerging1}).

\subsubsection{Threshold model}
\label{sec:threshold_model}

The threshold model~\cite{caldarelli2002scale} is perhaps one of the simplest generative models for threshold networks, i.e., perfectly nested networks (see Section~\ref{sec:nested_graph}).
Fitness models~\cite{caldarelli2002scale} are heterogeneous random graph models where each node $i$ is endowed with one parameter $\eta_i$, referred to as fitness. The fitness value of each node is randomly drawn from some probability distribution $P(\eta)$; for each pair of nodes $i$ and $j$, the probability $P_{ij}$ that the two nodes are connected by a link is only a function of $\eta_i$ and $\eta_j$: $P_{ij}=P(\eta_i,\eta_j)$.
The threshold model~\cite{caldarelli2002scale} assumes that $P(\eta_i,\eta_j)=\Theta(\eta_i+\eta_j-\zeta)$, i.e., $i$ and $j$ are connected if and only if the sum of their fitness values is larger than a threshold value $\zeta$.

Caldarelli \textit{et al.}~\cite{caldarelli2002scale} studied the threshold model with an exponential fitness distribution $P(\eta)=\exp{(-\eta)}$, with $\eta\in[0,\infty)$. They found that the model generates networks with a power-law degree distribution. Back in the early 2000s, in the rising wave of interest in scale-free networks spurred by the seminal work by Barab{\'a}si and Albert~\cite{barabasi1999emergence}, this result was intriguing: to generate a scale-free network, neither growth nor preferential attachment are necessary elements. By adopting a generating-function approach, Boguñá and Pastor-Satorras~\cite{boguna2003class} performed an extensive analytic investigation of the threshold model with exponential fitness distribution. They found that the degree distribution essentially decays as $k^{-2}$, yet there is an accumulation point at $k=N$ which corresponds to a small fraction of generalist nodes that are connected to all the other nodes in the network. Besides, they found that the average degree of the neighbors of nodes of degree $k$ tends to decay as $k^{-1}$, which means that the network is strongly disassortative. As we have seen in Section~\ref{sec:disassortativity}, disassortativity is one of the fingerprints of nestedness.

Masuda \textit{et al.}~\cite{masuda2004analysis} emphasized that the threshold model generates threshold graphs (i.e., nested networks), and drew analogies with other threshold social phenomena that are typically investigated by means of complex contagion models~\cite{granovetter1978threshold,watts2002simple,kempe2003maximizing}.
They considered various fitness distributions ("weight'' distributions in~\cite{masuda2004analysis}), and they were interested in determining which fitness distributions can explain the heavy-tailed degree distributions often found in real networks. While the power-law degree distribution stemming from an exponential fitness distribution was already documented by Boguñá and Pastor-Satorras~\cite{boguna2003class}, Masuda \textit{et al.}~\cite{masuda2004analysis} found that power-law tails of the type $k^{-2}$ can also arise from a logistic, Gaussian, and a Pareto fitness distribution.
They point out that power-laws with exponent $\gamma=-2$ might be a general feature of thresholding models, as opposed to power-laws with exponent $\gamma=-3$ that emerge from the Barab{\'a}si-Albert preferential attachment model~\cite{barabasi1999emergence}.

\subsubsection{Growing perfectly nested networks with arbitrary degree distribution}
\label{sec:emerging1}

As the threshold model~\cite{caldarelli2002scale} can generate perfectly nested networks for different choices of the fitness distribution~\cite{masuda2004analysis}, one might wonder whether it would be possible to generate perfectly nested networks with arbitrary network topology.
Hagberg \textit{et al.}~\cite{hagberg2006designing} provided a recipe to do that, approximately, for large enough networks.
To this end, they introduced a creation-sequence mechanism for the generation of a threshold network, i.e., a perfectly nested network. The mechanism is simple: before creating the network, one partitions the nodes into the dominant and the independent set (see Section~\ref{sec:nested_graph}). At each time step, a new node is added to the network. If the newcomer is a dominant node, it connects to all the preexisting nodes. If the newcomer is an independent node, it remains isolated. It is evident that this growth mechanism generates a perfectly nested structure.
Therefore, if we define a node-level binary variable $x_i$ which is equal to one and zero for dominant and independent nodes, respectively, the growth of the perfectly nested network can be encoded in a creation sequence $\mathcal{S}=\{x_i\}$. If we consider networks composed of five nodes, a star is represented by the sequence $\mathcal{S}=00001$, whereas the complete graph by the sequence $\mathcal{S}=11111$.

\begin{figure}[t]
\centering
\includegraphics[scale=0.95]{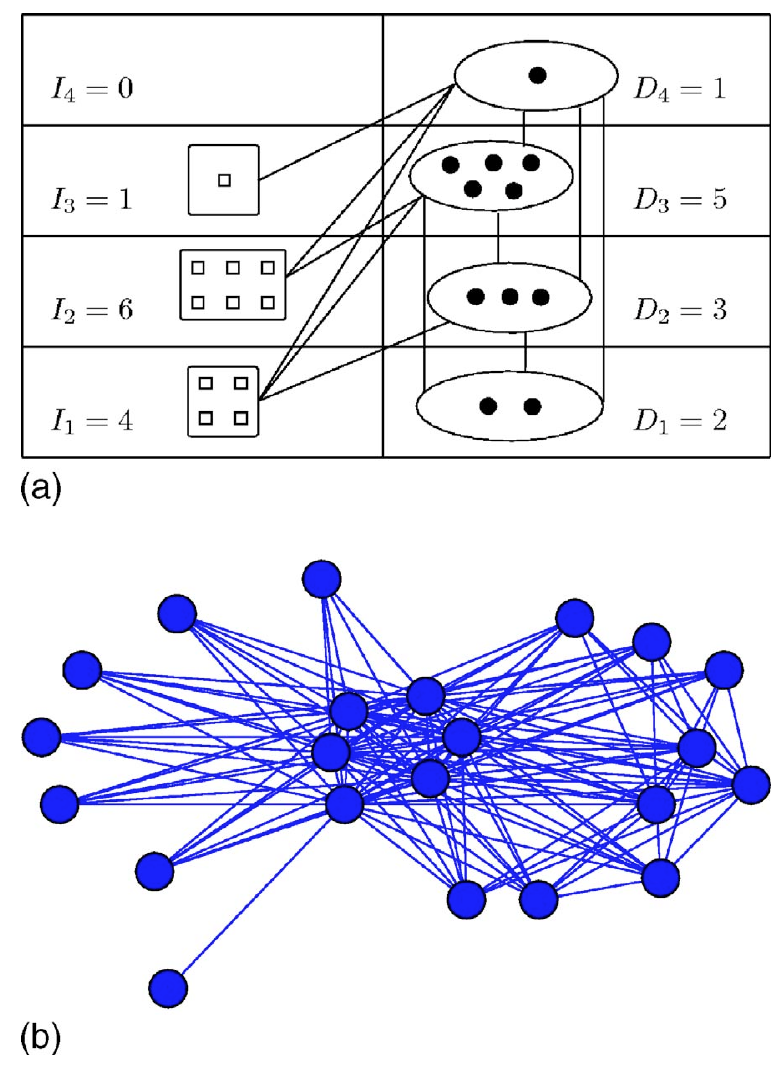}
\caption{Creation-sequence procedure to generate perfectly nested networks. From bottom to top in panel (a), one starts from $D_1=2$ dominant nodes, and subsequently adds $I_1=4$ independent nodes, $D_2$ dominant nodes, and so forth. When introduced into the system, independent nodes stay isolated, whereas dominant nodes connect to all the preexisting nodes. This results in a perfectly nested structure (panel (b)).
Reprinted from \cite{hagberg2006designing}.}
\label{fig:cake}
\end{figure}

An illustration of this growth mechanism is provided in Fig.~\ref{fig:cake}.
Interestingly, one can also show that this procedure is equivalent to the threshold model introduced in Section~\ref{sec:threshold}. This means that, given the set $\{\eta_i\}$ of nodes' fitness values and the threshold value $\zeta$, one can obtain a unique creation sequence that generates the corresponding perfectly nested network.
At the same time, given a creation sequence $\mathcal{S}$, one can find a unique set of fitness values and a threshold value that produce the corresponding perfectly nested network. We refer the interested reader to~\cite{hagberg2006designing} for the proof of the equivalence. 

By exploiting the growth dynamics presented above, we can generate perfectly nested structures with arbitrary degree distribution. To do that, the independent nodes are used to enforce the target degree distribution; indeed, given a creation sequence, the degree of a given independent node $i$ is simply equal to the number of dominant nodes that are introduced after $i$. The dominant nodes turn out to have a large degree and they do not respect the pre-imposed degree sequence; however, their impact on the overall degree sequence is relatively small for sufficiently large networks.

\subsection{Ecological mechanisms}
\label{sec:formation_ecol}

The reasons behind the emergence of nestedness in ecological networks have been widely debated. 
Some works have interpreted the observed levels of nestedness in empirical networks as the potential result of optimization processes where species aim to maximize either their own fitness~\cite{medan2007analysis,zhang2011interaction,suweis2013emergence} or the community's fitness~\cite{suweis2013emergence}.
In this spirit, Medan \textit{et al.}~\cite{medan2007analysis} found that nestedness can emerge when species switch their interaction in the attempt to increase the centrality of their neighbors (see Section~\ref{sec:snm}). 
Saavedra \textit{et al.}~\cite{saavedra2009simple} found that a parsimonious bipartite-network formation model that features simple cooperation rules between the two classes of nodes can generate networks that exhibit levels of nestedness comparable to the empirically observed ones (see Section~\ref{sec:two_steps}). 
Bastolla \textit{et al.}~\cite{bastolla2009architecture} mentioned that if a species entering a community tends to minimize its competition load when establishing links, a nested structure would emerge.
Zhang \textit{et al.}~\cite{zhang2011interaction} found that an interaction switch aimed at replacing the least profitable interactions (in terms of per-capita abundance gain) with more profitable ones can lead to nested structures comparable with the empirical ones.
Suweis \textit{et al.}~\cite{suweis2013emergence} found that nestedness could emerge as a result of an evolutionary process where species aim to maximize their own fitness or the community's fitness (see Section~\ref{sec:fitness}).

Other works~\cite{valverde2018architecture,maynard2018network} have focused on nestedness and other structural properties as consequences of the generative rules of the system, without any optimization mechanism involved. These works pointed out that nestedness can be observed in networks generated with selective mechanisms that do not act to adapt the species to their local environment.
From this viewpoint, nestedness and other structural patterns might be viewed as analogous to spandrels in architecture: elements that might be perceived as functional key components of the system, yet they are mostly the byproduct of other structural components~\cite{valverde2018spandrels}. In this spirit, Valverde \textit{et al.}~\cite{valverde2018architecture} showed that nestedness can emerge as a result of a stochastic dynamics with speciation and divergence (see Section~\ref{sec:speciation}).
Maynard \textit{et al.}~\cite{maynard2018network} pointed out that the presence (or absence) of nestedness might be the consequence of the system's specific assemblage rules, and seemingly minor differences in the way the network is assembled can lead to drastically different topologies (see Section~\ref{sec:immigration}).

In the following, we detail some of the above-mentioned mechanisms. We stress that the property alone that these mechanisms theoretically lead to a nested topology does not guarantee their empirical relevance. For example, even if a nested topology allows species to minimize their competition load~\cite{bastolla2009architecture} and maximize their fitness~\cite{suweis2013emergence} according to analytic results on population-dynamics models, we have no guarantee that the observed structure of interaction networks has emerged for this reason. In other words, from the property that a topology (in our case, a nested topology) performs specific functions (in our case, minimization of competition loads and maximization of species' fitness), we cannot infer that the empirically observed structures were ``selected'' in order to perform those functions. A similar argument holds for mechanisms that feature no optimization of species-level or community-level properties~\cite{valverde2018spandrels,maynard2018network}: the fact alone that a given mechanism can theoretically lead to nestedness does not imply that that mechanism acted to shape the observed empirical networks.

\subsubsection{Self-organizing network model}
\label{sec:snm}

The self-organizing network model (SNM) introduced by Medan \textit{et al.}~\cite{medan2007analysis} is one of the simplest models that lead to the emergence of nestedness. The model aims to explain the nested topology of ecological networks through a rewiring mechanism where the nodes tend to connect to central nodes in the system. We shall see that rewiring models based on a similar idea have been also applied to social systems~\cite{konig2011network,bardoscia2013social} (see Section~\ref{sec:social_climbing}), which points out nodes' effort to connect to central nodes as a possible general mechanism behind the emergence of nestedness in diverse systems.

The SNM takes as input the number of species $S$, the total number of links $E$, and a set of forbidden links. One starts by distributing the $E$ links across randomly-selected pairs of nodes. Starting from such random topology, we sequentially perform two iterative steps: (1) we select a row/column of the adjacency matrix, and we randomly pick a $1$ and a $0$ from the selected row/column. (2) We swap the two elements if and only if the following three conditions are met: the degree of the new partner is larger than the degree of the previous partner; the swap does not cause the previous partner to become extinct; the new link is not a forbidden link.

After a large number of iterations, this process leads to a perfectly nested structure, except when some pathological conditions occur\footnote{For example, when there exist specialized interactions among two nodes with degree one. Such a link will never be rewired (as it would leave one of the two nodes with degree zero). This pathological case may appear either during the algorithmic process, or in the initial
configuration when one starts from an empirical network.}. As real networks are rarely perfectly nested, one can stop the iterations as soon as the network reaches the same level of nestedness as observed in the real network we aim to model. Interestingly, the model leads to truncated power-law degree distributions similar to those observed in real networks, regardless of the fraction of forbidden links~\cite{medan2007analysis}. The SNM was used by Burgos \textit{et al.}~\cite{burgos2007nestedness} to assess the impact of nestedness on systemic robustness against node removal -- see Section~\ref{sec:robustness}. A generalization of the SNM to incorporate phylogenetic data was introduced by Perazzo \textit{et al.}~\cite{perazzo2014study} to uncover the influence of phylogenetic effects on nestedness.

\subsubsection{Bipartite cooperation: specialization and interaction}
\label{sec:two_steps}

One of the common key aspects of both mutualistic ecological networks and manufacturer-contractor systems is that
nodes benefit from interacting with each other.
Motivated by this observation, Saavedra \textit{et al.}~\cite{saavedra2009simple} introduced a parsimonious network formation model based on bipartite cooperation. They used the model to explain the structural properties of both animal-plant mutualistic networks
and of manufacturer-contractor networks; to fix ideas, in the following, we refer to the two classes of nodes as animals and plants, yet the 
model equally applies to bipartite networks of manufacturers and contractors.
The model features two mechanisms:
\begin{itemize}
 \item \emph{Specialization} determines the number of partners, $k_\alpha$, a given plant $\alpha$ interacts with. It is determined by
 \begin{equation}
  k_\alpha=1+\floor*{(L-M)\frac{r_\alpha \,\lambda_\alpha}{\sum_\beta r_\beta\,\lambda_\beta}},
 \end{equation}
where $r_{\alpha}$ is the reward trait associated with interactions with plant $\alpha$ (randomly extracted from the uniform distribution
in $[0,1]$) and $\lambda_\alpha$ is a parameter (randomly extracted from an exponential distribution) that accounts for exogenous factors such as geographic effects and population diversity.
 \item \emph{Interaction} determines which animals interact with each plant.
 Each animal $i$ is endowed with a foraging trait $f_i$ randomly extracted from the uniform distribution in $[0,1]$.
 Plants are sorted in order of increasing reward trait $r$, whereas animals are sorted in order of decreasing foraging trait $f$.
 Besides, each link $l_\alpha$ is endowed with an external factor $\lambda_{l_{\alpha}}$ randomly extracted from the same exponential distribution as $\lambda_\alpha$.
 The dynamics starts by connecting the first plant $\alpha_1$ with $k_{\alpha_1}$ animals.
 For the other plants $\alpha$ and their links $l_\alpha$, if $r_\alpha>\lambda_{l_{\alpha}}$, then the plant connects with the first animal in $\mathcal{A}'$, the set of animals
 that have not yet been linked to by another plant $\beta\neq\alpha$.
 Conversely, if $r_\alpha\leq\lambda_{l_{\alpha}}$, then the plant connects with an animal in $\mathcal{A}''$,
 the set of animals that already gained at least one connection in previous time steps.
 If the plant has already connected to all the animals in $\mathcal{A}'$ or $\mathcal{A}''$, it only connects to animals in the 
 non-saturated set.
\end{itemize}

The model can reproduce more than $70\%$ of the structural metrics (degree distribution, modularity, and nestedness)
of empirical networks~\cite{saavedra2009simple}.
This percentage is much larger than that achieved by previous models based on trait-matching~\cite{santamaria2007linkage}
and preferential attachment~\cite{guimaraes2007build}.
For both pollination networks and manufacturer-contractor networks from the New York garment industry, the model can generate networks
with levels of nestedness compatible with the empirically observed ones (see Fig. 2 in~\cite{saavedra2009simple}).

\subsubsection{Fitness maximization based on population dynamics} 
\label{sec:fitness}

In an adaptive system, it is plausible to consider a mechanism where a node tends to delete its less profitable connection and to replace it with a new, more profitable connection. We have already leveraged this idea when introducing the Self-organizing Network Model (SNM, Section~\ref{sec:snm}), and we shall explore analogous mechanisms in the context of socio-economic systems (see Section~\ref{sec:social_climbing}).
When assuming that nodes strive to maximize their centrality, or fitness, the resulting model critically depends on how we define node centrality or fitness in the first place.
In this Section, we focus on fitness-maximization mechanisms where species strive to maximize their abundance as determined by the stationary state of a suitable population dynamics model~\cite{zhang2011interaction,suweis2013emergence}.
This class of mechanisms is akin to variational principles in physics, where one seeks to find the optimal state for a given system by maximizing (or minimizing) a suitable function of the variables that describe the system's configuration. As an example, the equilibrium state of a thermodynamic system can be found by maximizing the entropy of the probability distribution of its configurations~\cite{parisi1988statistical}.

Zhang \textit{et al.}~\cite{zhang2011interaction} proposed a rewiring mechanism where, starting from a given topology, at each step, a species is randomly selected. The selected species deletes its connection that brings the smallest increase in its per-capita population, as determined by the stationary state of a model of mutualistic population dynamics (see~\cite{zhang2011interaction} for the details). The deleted connection is replaced ("interaction switch"~\cite{zhang2011interaction}) by a randomly chosen connection. Zhang \textit{et al.}~\cite{zhang2011interaction} found that this simple behavioral mechanism leads to the emergence of nested structures that exhibit levels of nestedness relatively close to those observed in real networks.

Suweis \textit{et al.}~\cite{suweis2013emergence} studied a dynamics where, starting from a random species-species interaction matrix, the species rewire their interactions with the goal to maximize their abundance. They showed that such dynamics results in a nested structure of the interaction network. Fitness optimization, therefore, may be a suitable explanation of the nested patterns found in mutualistic networks.
In the following, we provide the essential elements of the work by Suweis \textit{et al.}~\cite{suweis2013emergence}, referring to the original paper for all the details and additional insights.

\paragraph*{Linear and non-linear population dynamics}
To evaluate the impact of an interaction swap on the individual species' abundances and on the community's total population, Suweis \textit{et al.}~\cite{suweis2013emergence} considered both a linear dynamics and a non-linear population dynamics. 
We denote by $N$ and $M$ the number of animal and plant species, respectively; we denote by $\{x_1,\dots,x_N\}$ and $\{x_{N+1},\dots,x_{N+M}\}$ the abundances of the animal and plant species, respectively\footnote{This notation is different from the one that we defined in Table~\ref{tab:bipartite}, where we labeled animals and plants through Latin ($i\in\{1,\dots,N\}$) and Greek letters ($\alpha\in\{1,\dots,M\}$), respectively. However, we adopt it in this Section as it is convenient for a concise description of the model.}. 
One can define an interaction matrix $\mat{M}$ composed of four blocks:
\begin{equation}
\mathsf{M}= 
    \left[\begin{array}{cc}
    \mat{\Omega}^{AA} & \mat{\Gamma}^{AP} 	\\
    \mat{\Gamma}^{PA} & \mat{\Omega}^{PP} 	
    \end{array}\right]
\end{equation}
The blocks $\mat{\Omega}^{PP}$ and $\mat{\Omega}^{AA}$ represent the plant-plant and animal-animal competitive interactions, repsectively; the blocks $\mat{\Gamma}^{PA}$ and $\mat{\Gamma}^{AP}$ represent the plant-animal mutualistic interactions.
The linear dynamics is given by~\cite{suweis2013emergence}
\begin{equation}
\frac{d\vek{x}}{dt}=\vek{x}(\vek{\alpha}-\mat{M}\,\vek{x}),
\end{equation}
where $\vek{\alpha}$ represents the vector of intrinsic growth rates.
The non-linear dynamics is described by the equation~\cite{suweis2013emergence}
\begin{equation}
\frac{d\vek{x}}{dt}=\vek{x}\,\Biggr(\vek{\alpha}-d\,\vek{x}-\frac{\mat{M}\,\vek{x}}{h+\mat{\Theta}\vek{x}}\Biggl),
\end{equation}
where $d$ represents a self-interaction strength, $h$ is a parameter referred to as \emph{handling time}\footnote{The handling time is defined as the amount of time needed for the plant-animal interaction to take place, which limits the total benefits for a given species from the interactions with its mutualistic partners. We refer to Section~\ref{sec:mutualistic} for more details about this mutualistic model and its implications for systemic stability and feasibility.}, and $\mat{\Theta}$ is a matrix whose elements are defined as $\Theta_{ij}=1$ if $i$ and $j$ are neither both plants nor both animals, $\Theta_{ij}=0$ otherwise.
It is worth noticing that this model can be obtained from the model with competition and mutualism studied by~\cite{bastolla2009architecture,rohr2014structural} (see Eq.~\eqref{mutualistic_model} and the related discussion below) by switching off interspecific competitive interactions. One can write down the equations for the stationary point and the local stability conditions for this dynamics -- we refer to the Supplementary Information of \cite{suweis2013emergence} for details.

\paragraph{Community-level and species-level optimization dynamics}

\begin{figure}[t]
\centering
\includegraphics[scale=0.95]{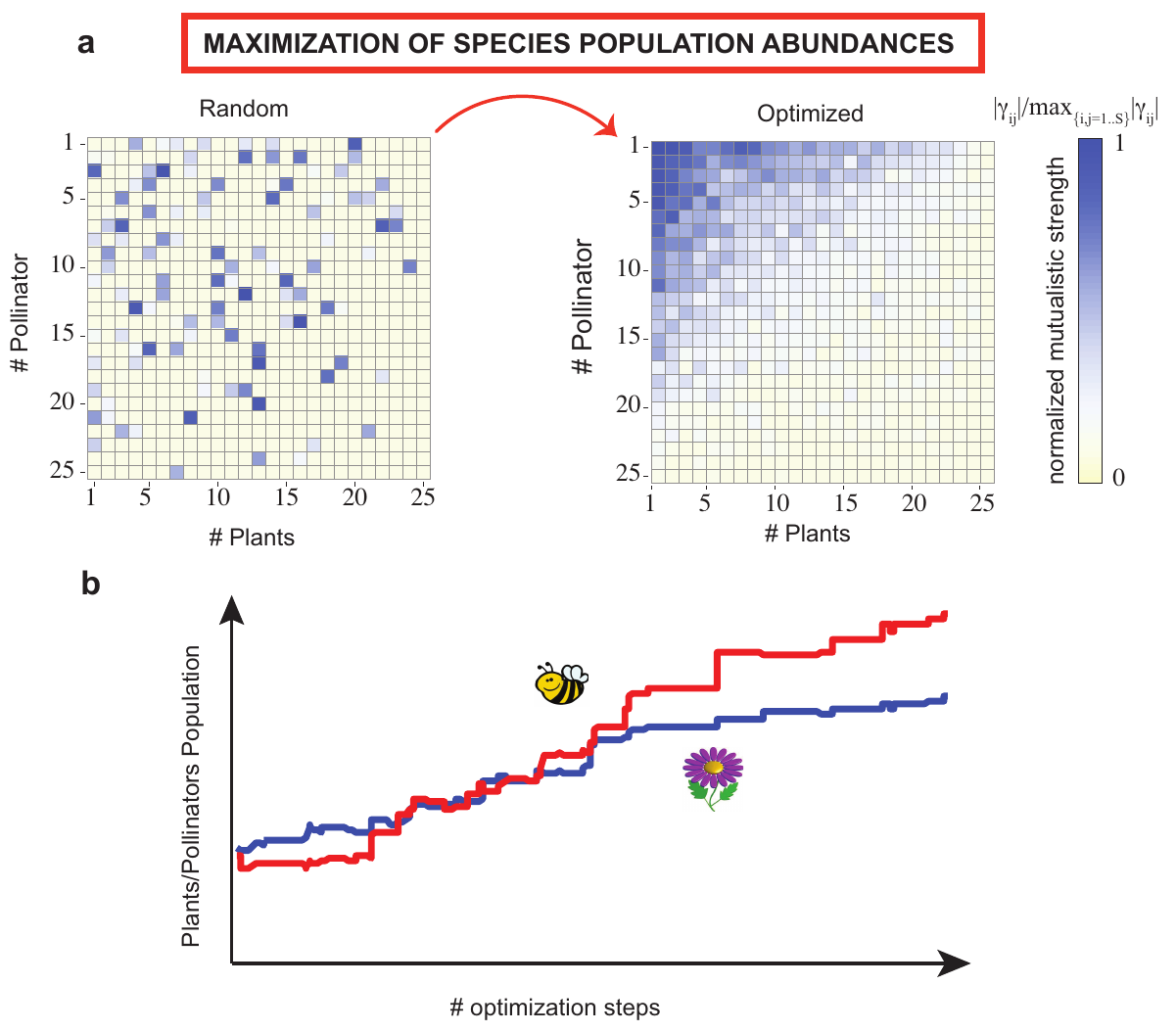}
\caption{Effects of the species-abundance maximization dynamics by Suweis \textit{et al.}~\cite{suweis2013emergence}. As the number of optimization steps increases, \textbf{a} the structure of the interaction matrix gradually approaches a nested structure; \textbf{b} the population of both plants and pollinators increases.
Reprinted from \cite{suweis2013emergence}.}
\label{fig:optimization}
\end{figure}

The \emph{community-level optimization} dynamics starts with $x_i=1$ for each species $i$, and a randomly generated interaction matrix $\mat{M}$.
We assume that the vector $\vek{\alpha}$ of intrinsic growth rates stays constant during the optimization dynamics.
At each time step $n$ of the optimization dynamics, we randomly select a pair $(i,j)$ of interacting species ($\gamma_{ij}\neq 0$) and a pair $(l,m)$ of non-interacting species ($\gamma_{lm}=0, (l,m)\neq (i,j)$). We propose a swap between the two interactions: $\tilde{\gamma}_{lm}^{(n+1)}=\gamma_{ij}^{(n)}$ and $\tilde{\gamma}_{ij}^{(n+1)}=\gamma_{lm}^{(n)}$. To decide whether to accept the swap, we compute the new equilibrium point of the system: in the linear dynamics, $\vek{\tilde{x}^{*}}(n+1)=[\mat{\widetilde{M}}(n+1)]^{-1}\,\vek{\alpha}$, where the two original matrix elements have been swapped in $\mat{\widetilde{M}}$. We accept the swap only if $\sum_{i=1}^{A+P}\tilde{x}^{*}_i(n+1) \geq \sum_{i=1}^{A+P}x^{*}_i(n)$; in this case, $\Gamma(n+1)=\tilde{\Gamma}(n+1)$. If, by contrast, $\sum_{i=1}^{A+P}\tilde{x}^{*}_i(n+1) < \sum_{i=1}^{A+P}x^{*}_i(n)$, we set $\Gamma(n+1)=\Gamma(n)$.

This community-level optimization dynamics corresponds to a rewiring of the interaction matrix $\mat{\Gamma}$ aimed at maximizing the total population $\sum_i x_i$ of the system. It can be interpreted as a group selection process, as opposed to a process where species aim to maximize their own abundance. The latter mechanism can be included by accepting a potential interaction swap not if the system's total population increases, but if the abundance of the species involved in the interaction increases (\emph{species-level optimization}). More specifically, at each step $n$ of the optimization process, one select a species $i$ and two of its links $(i,j)$, $(i,k)$. Again, one proposes an interaction switch $\tilde{\gamma}_{ij}^{(n+1)}=\gamma_{ik}^{(n)}$ and $\tilde{\gamma}_{ik}^{(n+1)}=\gamma_{ij}^{(n)}$. As opposed to the community-level optimization process, the swap is only accepted if it would not result in a decrease of species $i$'s abundance (as determined by the steady state of the population dynamics under consideration). 

Importantly, both optimization dynamics lead the interaction matrix from the initial random topology to a highly nested topology. To analytically understand this phenomenon, one can consider a mean-field scenario where interaction and competition strength parameters are homogeneous across all pairs of species.
Within the mean-field scenario, the fact that the above-described optimization processes lead to nested topologies is a consequence of two properties: (1) both optimization dynamics presented above tend to increase the total population of the system\footnote{While this is true by construction for the community-level optimization, it may look counterintuitive for the species-level optimization process. If species aim at maximizing their own abundance, why does the total population increase? To answer this question, one can use a perturbative argument to show that in the limit $\gamma\sim \omega \ll 1$, at each optimization step, the total population increase is approximately equal to the abundance increase of the species involved in the rewiring. Increasing the abundance of the species involved in the interaction, therefore, results in increasing the total population of the system. We refer the reader interested in this intriguing equivalence to Sections 4.3 and 5 of the Supplementary Information of \cite{suweis2013emergence}.}; (2) the total population of the system is significantly correlated with the nestedness of the interaction matrix. In the mean-field scenario, the total population $x^{tot}$ of the system ($x^{tot}=\sum_i x_i$) depends linearly on nestedness according to a relation $x^{tot}\simeq B_1\,O^{tot}+B_2$, where $O^{tot}$ is the total overlap of the interaction matrix (proportional to the sum of all the node-node pairwise overlaps $O_{ij}=\sum_l A_{il}\,A_{jl}$), and $B_1,B_2$ are two constants that depend on the interaction parameters $\gamma, \omega$, on the network connectance $C$ and on the number of species $S$. As nestedness (as measured by NODF) increases with the total overlap $O^{tot}$ of the interaction matrix, it follows that nestedness increases with the total population of the system~\cite{suweis2013emergence}.

\subsubsection{Nestedness as an evolutionary spandrel: Speciation and divergence}
\label{sec:speciation}

Valverde \textit{et al.}~\cite{valverde2018architecture}
recently introduced a simple generative model for mutualistic networks based on two mechanisms: speciation and divergence.
The model focuses on weighted networks, where the weight of an interaction between animals and plants may represent the number of observed interaction within a given time interval.
In an evolutionary ecology spirit, the \emph{speciation} process is essentially a duplication process where a species splits into two separate species.
In the model, with a given probability $\pi_A$, an animal species $i$ undergoes a \emph{speciation} process where a new (daughter) animal species $j$ is created with the same links of the original (parent) species $i$: if we denote as $\omega_{i\alpha}$ the weight of the interaction between animal $i$ and plant $\alpha$, we have $\omega_{i\alpha}=\omega_{j\alpha}$ for all plant species $\alpha\in\{1,\dots,M\}$. With a given probability $\pi_A$, a plant species $\beta$ undergoes an analogous speciation process. 

Species \emph{divergence} is basically a weight randomization process.
The weights between parent and daughter species are then redistributed. In particular, we generate a random number $\mu\in(0,1)$ and we update the weights as follows: $\omega_{i\alpha}\to(1-\mu)\,\omega_{i\alpha}, \omega_{j\alpha}\to \mu\,\omega_{j\alpha}$. Besides, with a given probability $p$, each link $(i,\alpha)$ can change weight: $\omega_{i\alpha}\to\omega_{i\alpha}+\xi$, where $\xi\in(\beta,\beta)$ is a relatively small random number, and $p,\beta$ are parameters of the model. The model also features a maximum total weight for plants and animals, and a threshold minimal value for links' weights. 

\begin{figure}[t]
\centering
\includegraphics[scale=0.95]{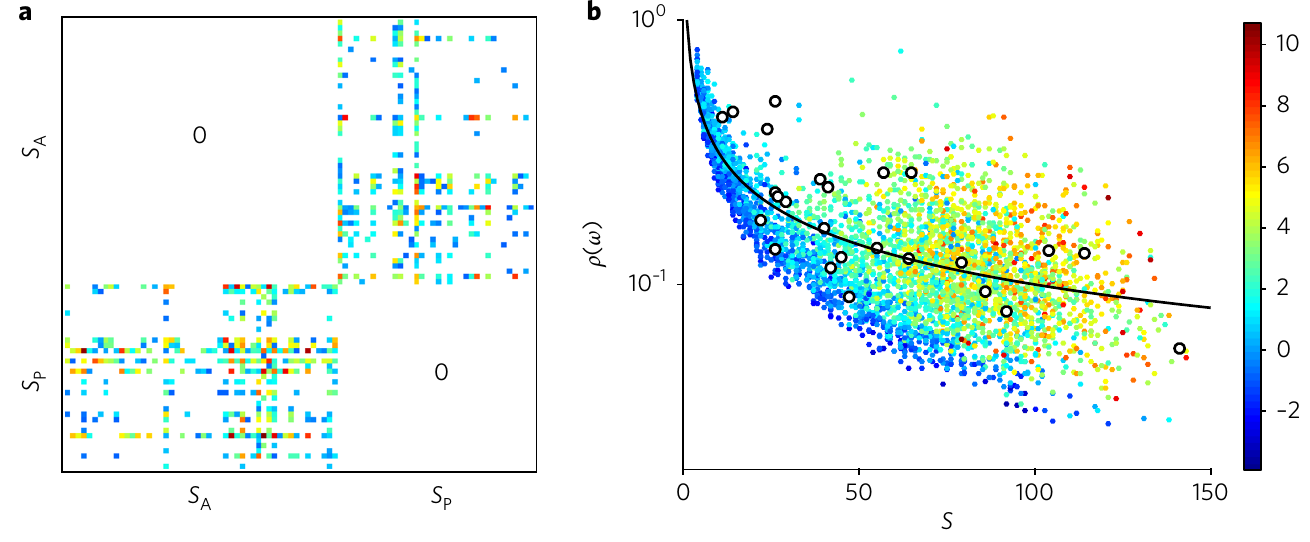}
\caption{Nestedness in the speciation-divergence model. Panel \textbf{a} illustrates a bipartite adjacency matrix obtained with the model. Panel \textbf{b} represents the
spectral radius $\rho(\mat{A})$ (see Section~\ref{sec:eigenvalue}) of 4000 networks generated with the speciation-divergence model (for different values of the input parameters) as a function of the number of species $S$. The circled dots correspond to real datasets, whereas the continuous line represents the analytic expectation for a random network.
Reprinted from \cite{valverde2018architecture}.}
\label{fig:valverde}
\end{figure}

The networks generated according to this rule exhibit various degrees of nestedness, from low to high, conditional to the values of the input parameters (Fig.~\ref{fig:valverde}). Based on this result, Valverde \textit{et al.}~\cite{valverde2018architecture} point 
out that network nestedness and other structural properties (such as the network's heterogeneous degree and
weight distribution, and link asymmetry~\cite{valverde2018architecture}) can be viewed 
as ``evolutionary spandrels", i.e., as structures that are the consequence of the
system's building rules, have well-defined non-random features, and reveal some of 
the underlying rules of construction. Importantly, these findings indicate 
that nestedness can emerge as a result of building rules that do not involve species-level or community-level optimization, 
as opposed to the optimization rules that we investigated in the previous Sections.

Evolutionary mechanisms to explain the emergence of nestedness were also previously considered by Takemoto and Arita~\cite{takemoto2010nested}.
They assumed that starting from a small initial network, new plants can emerge from randomly selected existing ones through mutation.
The traits of the new plants are similar to those of the original ones. Based on the assumption that
plant-animal interactions are driven by 
trait-matching mechanisms, some of the animals that interact with the new plant are inherited from those that interacted with the ancestral plant.
Besides, a plant can acquire new traits and new, randomly selected partners.
This mechanism generates nested networks as well, and the resulting networks were reported~\cite{takemoto2010nested} to match the nestedness of empirical networks more accurately than
those generated by the bipartite cooperation model described in Section~\ref{sec:two_steps}.

\subsubsection{Invasion dynamics with different assemblage rules}
\label{sec:immigration}

The results by Valverde \textit{et al.}~\cite{valverde2018architecture} indicate that network formation mechanisms without optimization can explain the emergence of nestedness in mutualistic networks. In a similar spirit, Maynard \textit{et al.}~\cite{maynard2018network} considered a community of $N$ competing species. They were interested in finding slightly different stochastic mechanisms of network growth that lead to drastically different network topologies. If such similar but different mechanisms can be identified, one might argue that widely different observed topologies might be the result of different assemblage rules, without the need for introducing formation mechanisms based on optimization or selection for stability. 

By denoting as $x_i=x_i(t)$ the abundance of species $i$ at time $t$, they considered a competitive dynamics of the form
\begin{equation}
\frac{dx_i}{dt}=x_i(t)\,\Biggl(r_i+\alpha_i\,\sum_{j}A_{ij}\,x_j \Biggr),
\label{competition}
\end{equation}
where $r_i$ represents the intrinsic growth rate and $\alpha_{i}<0$ represents the competition parameter.
They considered a simple network formation process where one starts with a single species and, at each step, considers a potential ``invader". They introduced two mechanisms, referred to as \emph{immigration} and \emph{radiation}, that differ in the way the invader interacts with the preexisting species (see next paragraph).
If the potential invader satisfies the invasion condition (see Eq.~\eqref{invader} below), it is added to the system. Once it has been added to the system, one runs the dynamics described by Eq.~\eqref{competition} until a stable point is reached. The addition of the invader perturbs the system, and some species might become extinct as a result. Once the stationary point is reached, one repeats this procedure by proposing a new invader.

An invader $i$ is added to the system if and only if 
\begin{equation}
r_i+\alpha_i\,\sum_{j}A_{ij}\,x_j>0.
\label{invader}
\end{equation}
The validity of this condition depends on the coefficients $A_{ij}$. Maynard \textit{et al.}~\cite{maynard2018network} considered two mechanisms:
\begin{itemize}
\item In the \emph{immigration} scenario, the $A_{ij}$ coefficients for the potential invader $i$ are random numbers extracted from the uniform distribution $\mathcal{U}(-1,0)$ for $j\neq i$, whereas $A_{ii}=1$. If the potential invader satisfies the invasion condition (Eq.~\eqref{invader}), then it is added to the system. Otherwise, one replaces randomly-chosen coefficients $A_{ij}$ with new random numbers extracted from $\mathcal{U}(-1,0)$ until the invasion condition is met and the invader can enter the system.
\item The \emph{radiation} scenario differs from the immigration scenario in the way the coefficients are replaced if the invasion condition is not met. To allow the potential invader $i$ to enter the system, in the radiation scenario, one randomly selects preexisting species and slightly perturbs their coefficients, until the invasion condition for species $i$ is met.
\end{itemize}
The two mechanisms lead to drastically different levels of nestedness in the generated interaction networks. While the nestedness of immigration-generated networks is not distinguishable from that of randomized networks, the nestedness of radiation-generated networks tends to be significantly larger than that of their randomized counterparts. This points out that: (1) different assemblage mechanisms can lead to different network topologies, without the need for considering optimization mechanisms; (2) Radiation-like assemblage mechanisms might be responsible for the levels of nestedness observed in empirical networks. 
We will come back to these points in Section~\ref{sec:bottom1}.

\subsection{Social and economic mechanisms}
\label{sec:formation_socioec}

Compared to ecological networks, scholars have started investigating the emergence of nestedness in socio-economic networks much more recently.
In this Section, we focus on network formation and rewiring mechanisms that aim to explain the emergence of nestedness
in unipartite socio-economic networks (Sections~\ref{sec:social_climbing2}-\ref{sec:social_climbing}) and in bipartite country-product networks (Section~\ref{sec:innovation_novelty}).

\subsubsection{Network formation based on centrality maximization}
\label{sec:social_climbing2}

\begin{figure}[t]
\centering
\includegraphics[scale=0.95]{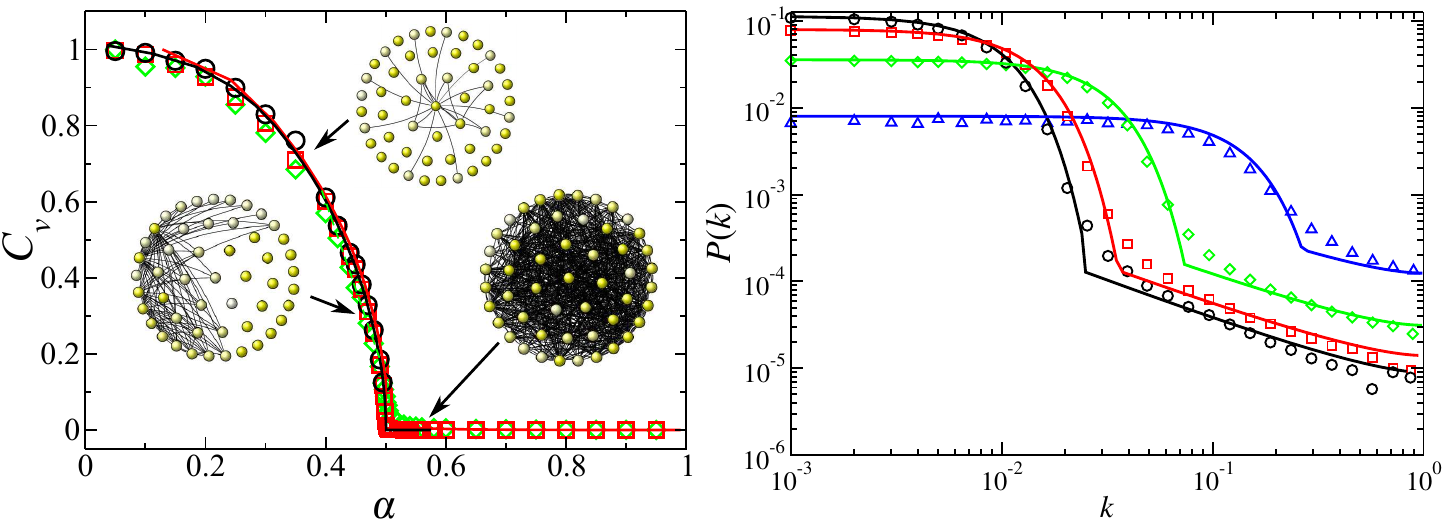}
\caption{Transition from a fully connected topology ($\alpha>1/2$) to a nested, highly centralized topology ($\alpha<1/2$) in the growing network model based on social climbing~\cite{konig2011network}. The transition manifests itself in the discontinuity of the derivative of the eigenvector centralization $C_v$ at the critical point $\alpha_c=1/2$ (panel (a)). The degree distributions of the networks generated by the model (panel (b)) exhibit a two-step relaxation which combines an exponential (low $k$) and a power-law (large $k$); the different curves in panel (b) correspond to different values of $\alpha$.
Reprinted from \cite{konig2011network}.}
\label{fig:social_climbing2}
\end{figure}

In~\cite{konig2011network}, the authors introduced a network formation model where the agents attempt to maximize strategically their centrality in the network in presence of link decay. Therefore, it is a network formation model where the number of edges changes over time until it reaches a stationary value which depends on the model parameter $\alpha$ (see below).
In that dynamics, one starts from an arbitrary topology. At each time step, a randomly selected node has two options. With probability $\alpha$, it creates a link to the most central node that is not already among its neighbors. With probability $1-\alpha$, it deletes its connection with the least-central of its neighbors. The dynamics has one parameter, $\alpha$, and it converges to a stationary, perfectly nested topology, provided that the system started from a perfectly nested topology.

The authors in \cite{konig2011network} noted that this evolutionary dynamics leads to a nested network independently of the exact notion of centrality that is used. Later, Schoch \textit{et al.}\cite{schoch2017correlations} showed formally that in a nested network all notions of centrality are ranked in the same way in a nested network. An insightful consequence is that agents do not become central because of their actions; they do so because they turn into the best prospective connection for other agents who seek to increase their own centrality. 
This simple mechanism reinforces the nestedness in the network; 
concurrently, it promotes central agents to become even more central, and it confines peripheral agents to remain in those positions.
Further, when the network is nested, agents can find the best connection target even with local information.  

In the model, one can write down the master equation~\cite{barabasi1999mean} for the degree distribution $p_k(t)$ for the nodes in the independent sets as\footnote{As in~\cite{konig2011network}, we neglect all the contributions from the nodes that belong to the dominating set. These contributions are nevertheless $\mathcal{O}(1/N)$ small.}~\cite{konig2011network}
\begin{equation}
\frac{\partial p_k(t)}{\partial t}=\mathbb{P}[k+1\to k]\,p_{k+1}(t)+
									\mathbb{P}[k-1\to k]\,p_{k-1}(t)-
                                    (\mathbb{P}[k\to k-1]+\mathbb{P}[k\to k+1])\,p_{k}(t),
\end{equation}
where $\mathbb{P}[k+1\to k]=(1-\alpha)/N$ is the probability that a node deletes one connection in the time unit, whereas $\mathbb{P}[k-1\to k]=\alpha/N$ is the probability that a node acquires one connection. The smallest degree increase is given by $\delta k=1/N$. One can therefore expand the previous Equation in powers of $\delta k$, obtaining (for $k>0$)~\cite{konig2011network}
\begin{equation}
\frac{\partial p_k(t)}{\partial t}=(1-2\,\alpha)\,\frac{\partial p_k(t)}{\partial k}+\delta k\,\frac{\partial^2 p_k(t)}{\partial k^2}+\mathcal{O}(\delta k^2)
\label{master}
\end{equation} 
For $\delta k\ll|1-2\,\alpha|$, the $\mathcal{O}(\delta k)$ diffusion term can be neglected and we are left with a drift equation whose asymptotic solution is either a fully-connected network ($\alpha>1/2$) or an empty network ($\alpha<1/2$). Around $|1-2\,\alpha|\simeq \delta k$, i.e., when $\alpha\simeq 1/2+\mathcal{O}(1/N)$, the diffusion term is not anymore negligible, and the dynamics is described by the Fokker-Planck equation
\begin{equation}
\frac{\partial p_k(t)}{\partial t}=(1-2\,\alpha)\,\frac{\partial p_k(t)}{\partial k}+\frac{\partial^2 p_k(t)}{\partial k^2}.
\end{equation} 
In the regime $\alpha<1/2$, one solves Eq.~\eqref{master} to derive the full degree distribution $p_k$. One finds that $p_k$ has a two-step behavior: the distribution is exponential up to a critical degree value $k^*$ which can be analytically determined, whereas it exhibits a power-law tail $\sim k^{-1}$ for $k>k^*$ (see Fig.~\ref{fig:social_climbing2}b).

It is interesting to look at the centralization of the eigenvector centrality vector as a function of the model parameter $\alpha$ (see Fig.~\ref{fig:social_climbing2}a). Intriguingly, the model exhibits a sharp phase transition at $\alpha=1/2$: the system has an even degree distribution for $\alpha>1/2$, which results in a centralization of the eigenvector centrality vector close to zero. By contrast, the system becomes highly-centralized for $\alpha<1/2$, which results in a significant level of nestedness~\cite{konig2011network}. 
To conclude, it is worth noticing the analogy between social-climbing models and ecological network formation models where species attempt to maximize their fitness (like the SNM, see Section~\ref{sec:snm}), which points out that explaining the same structural pattern across different systems might unveil common formation mechanisms.

\subsubsection{Network formation in a two-stage game}
\label{sec:two_stage}

Later, in \citep{konig2014nestedness}, a dynamic network formation model was developed that can explain the observed nestedness of real-world economic networks: e.g.,~interbank loans, trade in conventional goods and arms trade between countries (cf.~Fig.~\ref{fig:konig2}).  
Within the framework of the random utility model \cite{manski1977structure}, links in the network are formed on the basis of agents' centrality and have an exponentially distributed lifetime. 
In the model, at each period of time, agents play a two-stage game. In the first
stage, as in \cite{ballester2006s}, agents compute their effort level which is proportional to their \emph{Bonacich centrality}~\citep{bonacich1987power}. In the second stage, a randomly chosen agent can update her linking strategy by creating a new link as the best response to the current network. Links do not last forever but have an exponentially distributed life time. The most valuable links (i.e. the ones with the highest Bonacich centrality) decay at a slower rate than those that are less valuable. As a result, this model considers the formation of economic networks as the result of a tension between search for new linking opportunities and volatility that leads to the decay of existing links.

The dynamics runs as follows. The system is composed of $N$ agents. At any time $t$, there is a network $\mathcal{G}(t)$ whith adjacency matrix $\mathsf{A}$ that describes the connections between agents. Then, the two-stages of the game are applied. 
In the first stage, each player selects an effort level $x_i$ to put into the game. It is assumed that the payoff $\pi_i$ she receives is 
\begin{equation}
  \pi_{i} = x_{i}-\frac{1}{2}x_{i}^{2}+\lambda
  \sum_{j=1}^{N}A_{ij}x_{i}x_{j},  
  \label{eq:general-payoff}
\end{equation}
where $\lambda$ is a parameter that weights the relative contribution of the idiosyncratic component and network interdependencies to the total payoff. Ballester~\textit{et al.}~\cite{ballester2006s} found that the equilibrium effort levels that maximize the payoff of participant $i$ is a function of $i$'s Bonacich centrality $b_i(\lambda)$: $\pi_i^*(G,\lambda) = b_i(\lambda) / 2$.  In this stage of the game, agents select their equilibrium effort levels.

In the second stage, the network topology is updated.  First, an agent $i$ has the opportunity to create a link $(i,j)$ -- i.e., the network $\mathcal{G}(t)$ transitions to another one $\mathcal{G}(t) \oplus (i,j)$ -- with a rate
\begin{equation}
\nu_{+}(G(t) \oplus (i,j)) \propto \exp \left( \pi_i^*(G(t) \oplus (i,j),\lambda)  / \zeta \right).
\end{equation}
Second, links have an expected lifetime which is exponentially distributed. In particular, the rate of link decay becomes
\begin{equation}
\nu_{-}(G(t) \ominus (i,j)) \propto \exp \left( \pi_i^*(G(t) \ominus (i,j),\lambda)  / \zeta \right).
\end{equation}
The parameter $\zeta$ controls the level of randomness in the network dynamics -- essentially, it plays the role of an inverse temperature.

Stochastic stability was used to identify the network's stationary state; the authors~\cite{konig2014nestedness} found that the resulting network topology is fully nested when no randomness exists in the agent's decisions. 
The degree of nestedness is preserved to a large extent for small temperature. In particular, in the limit $\zeta \to \infty$, we recover the centrality-maximization dynamics~\cite{konig2011network} described in Section~\ref{sec:social_climbing2}.

\subsubsection{Network rewiring based on social climbing}
\label{sec:social_climbing}

In contrast to the previous model, the social climbing game~\cite{bardoscia2013social} considers a static system composed of a fixed number of nodes and links.
This model is a highly stylized model of a society where individuals seek to maximize their social status by forming new connections and deleting old ones in a strive to maximize the centrality of their interaction partners. The essential role of centrality as a driver for success has been proven in several studies from different angles. For instance, central individuals in the social network tend to score higher in education tests~\cite{calvo2009peer}; individuals' centrality in communication networks is a predictor for employees' future promotions and resignations~\cite{feeley1997predicting,feeley2010erosion,yuan2016promotion}; central individuals tend to have the largest impact on the opinions of their peers according to linear opinion formation models~\cite{friedkin1991theoretical}.
In light of these studies, the assumption that individuals are, to some extent, driven by the maximization of their centrality is a reasonable one~\cite{taylor2012social,bardoscia2013social}.

Consider a system composed of $N$ individuals and a fixed number $E$ of undirected links between them.
Bardoscia \textit{et al.}~\cite{bardoscia2013social} introduced a model where each agent $i$ is characterized by a ``social capital'' utility function, $u_i$, defined as
\begin{equation}
u_i=\sum_{j}A_{ij}\,k_j+\mu\,k_i,
\end{equation}
where $\mu$ is a parameter of the model that rules the relative strength of the two terms.
At each step of the dynamics, one selects randomly a node $i$ together with one of its neighbors (i.e., a node $j$ such that $A_{ij}=1$) and one of the selected neighbor's neighbors (a node $l$ such that $A_{lj}=1$). If $i$ was already connected to $l$, nothing happens. If $i$ was not connected to $l$, the link $(i,j)$ is replaced by $(i,l)$ with probability
\begin{equation}
\mathbb{P}[(i,j)\to(i,l)]=\frac{e^{\beta\,\Delta u_i}}{1+e^{\beta\,\Delta u_i}},
\label{rewiring}
\end{equation}
where $\Delta u_i$ denotes the variation in node $i$'s utility caused by the possible link replacement.

\begin{figure}[t]
\centering
\includegraphics[scale=0.95]{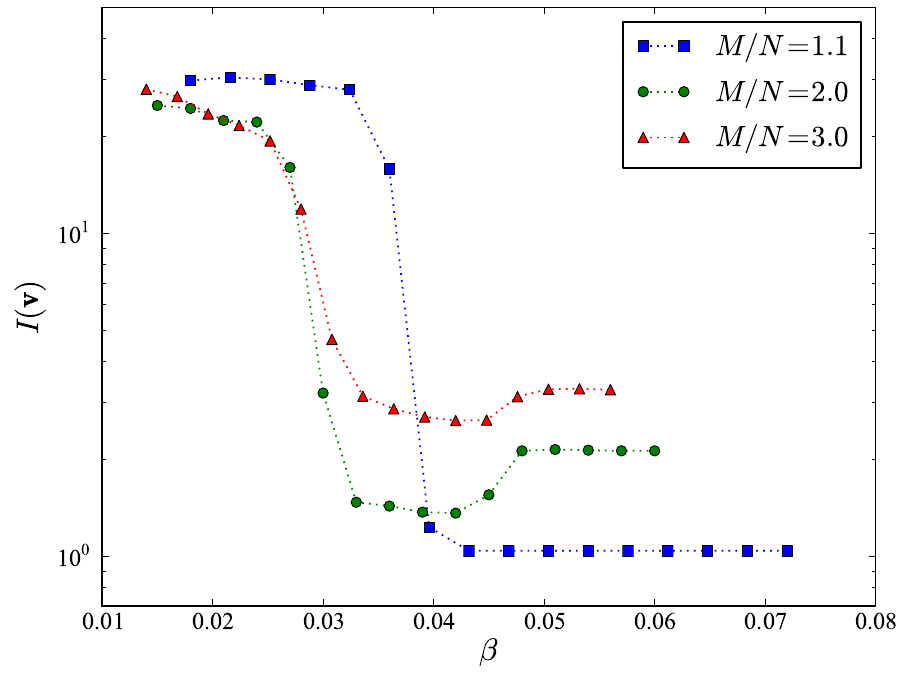}
\caption{Transition from an even, ``disordered'' topology (high-temperature, low $\beta$) to a highly centralized topology (low-temperature, high $\beta$) in the social climbing rewiring dynamics~\cite{bardoscia2013social}. The transition manifests itself in the drop of the inverse participation ratio $I$ of the network degree sequence as a function of the inverse temperature $\beta$; different symbols stand for different values of average degree $\braket{k}=2\,L/N$.
Reprinted from \cite{bardoscia2013social}.}
\label{fig:social_climbing1}
\end{figure}

The rewiring probability defined by Eq.~\eqref{rewiring} encodes the tendency of individuals to connect to individuals who bring them large utility gains. The strength of this tendency is ruled by the intensity parameter $\beta$, which can be interpreted as an inverse temperature, as we shall see below.
In a ``frozen'' society ($\beta\gg 1$), the nodes always rewire their connections when this leads to an increase of their utility. In a ``high-temperature'' society ($\beta\simeq 0$), the rewiring probability approaches $1/2$ regardless of the utility gain $\Delta u$, which makes the rewiring of links essentially independent of the candidate new neighbor's centrality.

The interpretation of the intensity parameter $\beta$ as the temperature of a statistical-mechanics system is a compelling feature of the model. To understand it, one can show that (1) when a rewiring $(i,j)\to(i,l)$ is performed, the variation $\Delta U$ of the total utility $U:=\sum_m u_m$ can be expressed as twice the variation $\Delta u_i$ of node $i$'s local utility: $\Delta U=2\,\Delta u_i$~\cite{bardoscia2013social}; (2) the dynamics described above is ergodic. Hence, the dynamics based on Eq.~\eqref{rewiring} converges to the equilibrium of a system described by the Hamiltonian function~\cite{bardoscia2013social}
\begin{equation}
H=-U=-\sum_i k_i^2-\mu\,\sum_i k_i =-N\,\braket{k^2}-N\,\mu\,\braket{k},
\end{equation}
with fixed average degree $\braket{k}$ and temperature $2/\beta$.
In contrast to the dynamic processes described in Sections~\ref{sec:social_climbing2} and \ref{sec:two_stage}, this one preserves the network size and density throughout the dynamics.

The results of numerical simulations based on this model show that as $\beta$ increases, the system abruptly transitions from a rather even degree distribution to a highly unequal degree distribution where few individuals are connected with the majority of the others, whereas most of the individuals have only few connections. A useful quantity to visualize this transition is the degree-sequence's inverse participation ratio $I(\vek{v})$ -- i.e., a proxy for the effective number of nodes who received connections\footnote{Given a normalized $N$-component vector $\vek{v}$ ($\sum_i v_i^2=1$), its inverse participation ratio $I(\vek{v})$ is given by $I(\vek{v})=(\sum_i v_i^4)^{-1}$. A vector that is completely localized (i.e., a vector that has one component equal to one and all the others equal to zero) has $I(\vek{v})=1$. A fully delocalized vector (i.e., a vector whose components are all equal to $1/\sqrt{N}$) has $I(\vek{v})=N$.} 
-- which displays an abrupt jump from a $\mathcal{O}(N)$ value to a $\mathcal{O}(1)$ value (see Fig.~\ref{fig:social_climbing1}). Besides, high-$\beta$ (low-temperature) systems are also characterized by an increased temporal stability of the ranking of the nodes by degree (see Fig.~6 in~\cite{bardoscia2013social}), which is representative of a low-mobility society where pre-defined hierarchies are likely to be preserved.
We have seen that ``low-temperature'' nested structures can be interpreted as the fingerprints of ``order'' in biogeographic systems (\cite{atmar1993measure}, see Section~\ref{sec:distance}); similarly, highly nested hierarchies in society can be interpreted as the manifestation of a ``low-temperature'' societal dynamics where it is hard for an individual to climb well-established ladders, regardless of his/her individual capabilities.

\subsubsection{Emergence of nestedness in the country-product network: Innovation and novelty}
\label{sec:innovation_novelty}

While capability-based models~\cite{hausmann2011network,bustos2012dynamics} (see Section~\ref{sec:hidden}) can generate nested 
country-product export networks, they do not describe the detailed dynamics that leads the countries to diversify their export baskets.
To describe this process, in a similar spirit to non-linear preferential attachment models in
network science~\cite{krapivsky2001organization}, Saracco \textit{et al.}~\cite{saracco2015innovation} introduced 
a network formation model with three steps:
\begin{itemize}
\item A country is selected with probability $P_i\sim k_i^{a}$, where $a>0$ is a parameter of the
model. Such probability reflects the intuition that more diversified countries are more likely to further diversify their export basket. 
\item The chosen country $i$ selects a product $\alpha$ that already belongs to its export basket
according to the probability $P(\alpha|i)\sim k_\alpha^b$, where $b>0$ is a parameter of the model.
The rationale behind this rule is that the more ubiquitous a product is, the more countries possess the apposite capabilities to make it, the higher the possibility of generating an innovation from that product. 
\item Given the selected country-product pair $(i,\alpha)$, the country has now the last choice to make. It considers two types of potential new products: all the products $\tilde{\alpha}$ that belong to the neighborhood of $\alpha$ in the product space~\cite{hidalgo2007product,zaccaria2014taxonomy}, and that were not already exported by country $i$; a new product $\alpha_{new}$ sprouting out from $\alpha$. Within this set $\{\alpha',\alpha_{new}\}$ of candidate products, country $i$ chooses the new product $\tilde{\alpha}$ to export according to the probability $P(\tilde{\alpha}|i,\alpha)\sim (k_{\tilde{\alpha}}+k_0)^c$, where $c>0$ and $k_0>0$ are parameters of the model. 
\end{itemize}

The model has four parameters: $a,b,c,k_0$. The parameter $k_0$ determines whether countries tend to move in the existing adjacent of already exported products (diversification, small values of $k_0$), or rather prefer to innovate and introduce new products in the system (innovation, large $k_0$).
The model can generate bipartite country-product networks whose values of row-nestedness and column-nestedness (see Section~\ref{sec:overlap}) are compatible with those observed in the real networks (see Fig. 4 in~\cite{saracco2015innovation}).

\subsection{Bottom-line: Why does nestedness emerge in empirical networks?}
\label{sec:bottom1}

As we emphasized in Section~\ref{sec:formation_ecol}, the fact alone that a topology can theoretically arise as a result of a given dynamics does not guarantee that that dynamics shaped the emergence of that topology.
The mechanisms investigated above are all able to generate nested networks, yet it remains unclear  which of them best describe the formation of real-world systems.
In both ecological and socio-economic networks, this uncertainty can be ascribed to the focus of most extant studies on static, time-aggregate datasets of ecological communities and trade networks, and on processes that can theoretically lead to the observed topologies. In ecology, recent works~\cite{saavedra2016nested,robinson2018flower} have started investigating the temporal dynamics of communities.
Crucially, increasing efforts in this direction may allow us to single out those mechanisms that likely led to the emergence of the observed topologies, while ruling out mechanisms that lead to realistic network topologies but do not reproduce the observed network dynamics.

From an interdisciplinary standpoint, we envision that the validation of network formation mechanisms (and, therefore, of potential explanations for nestedness) will likely be performed soon in socio-economic networks, where typically one has access to fine-grained temporal data. Existing studies on socio-economic networks have both introduced candidate mechanisms to explain the emergence of nestedness (Sections~\ref{sec:formation_socioec}), and studied the detailed temporal dynamics of nestedness (see Section~\ref{sec:modularity} and, in particular, Fig.~\ref{fig:modtonest}).
Validating the proposed mechanisms through robust statistical analysis, perhaps in a similar way to procedures adopted for growing information networks~\cite{medo2014statistical}, is a natural step for future research.

\clearpage

\section{Implications of nestedness for systemic stability and feasibility}
\label{sec:implications}

In occasion of the 100th anniversary of the British Ecological Society, Sutherland \textit{et al.}~\cite{sutherland2013identification} have collected a list of 100 fundamental ecological questions whose solution has the highest potential to advance ecological science.
Among them, we find: ``How does the structure of ecological interaction networks affect ecosystem functioning and stability"?~\cite{sutherland2013identification}
The main goal of this Section is to address this question for nested structures: how does nestedness impact on the robustness of the system against extinctions and targeted attacks? Does nestedness facilitate or harm the stable co-existence of species? 

Driven by these questions, we focus here on two aspects of systemic stability: network \emph{robustness against external perturbations} and \emph{dynamical stability}.
Network robustness against structural perturbations (Section~\ref{sec:robustness}) is typically defined in terms of node-removal processes. The core idea behind it is that when we remove a node in the network, other nodes might become extinct because they were only connected with the removed node -- a phenomenon known as ``co-extinction"~\cite{koh2004species}. Sequentially removing all the nodes (either in order of decreasing degree or in order of decreasing score as determined by some ranking algorithm) therefore leads to a process where eventually all the nodes become extinct.

Besides, the topology of an interaction network has important consequences on dynamical processes that act on it. Ecologists are especially interested in revealing whether some widely-observed network structural patterns can increase the dynamical stability of the system, i.e., its capability to return at its equilibrium point after a perturbation of the species' population. They are also interested in whether observed topological patterns facilitate the co-existence of species and, therefore, the community's feasibility. The impact of nestedness on dynamical stability and feasibility has been studied both with a population-dynamics approach~\cite{bastolla2009architecture,thebault2010stability,james2012disentangling,rohr2014structural} and with a random-matrix theory approach~\cite{allesina2012stability,allesina2015stability}.

In the population-dynamics approach, one considers a dynamic model with competition and mutualism (which acts on the underlying species-species interaction network), and seeks to find whether different network topologies facilitate or impede the stable co-existence of species. We will review this approach in Section~\ref{sec:population}.
We will then focus on the results from random-matrix theory on the local stability of random, mutualistic, and nested interaction matrices (Section~\ref{sec:rmt}). We will finally review the obtained results within the various approaches and draw a bottom-line on the impact of nestedness on systemic stability and feasibility, based on the results obtained so far (Section~\ref{sec:bottom2}).

\subsection{Robustness against structural perturbations}
\label{sec:robustness}

Understanding the co-extinction patterns triggered by the loss of a given species is central to the preservation of complex systems. In ecological systems, it can indeed inform decisions on which species to give higher priority for research and management in order to preserve a given ecosystem -- we refer to~\cite{kaiser2015integrating} for a review on the application of ecological networks to applied conservation practices. In economic and financial systems, understanding how the failure of an actor (like a financial institution) can trigger large-scale cascades of failures is essential to prevent the collapse of the global financial system~\cite{battiston2012debtrank,battiston2016complexity}.

Importantly, the topology of a network affects its robustness against structural perturbations, which can be performed by sequentially removing the nodes according to pre-established strategies.
In the following, we investigate two main questions related to topological robustness: (1) When we remove sequentially the nodes in order of degree, do nested arrangement lead to a faster or a slower systemic breakdown? (2) Do more nested arrangements of the adjacency matrix (as produced by non-linear ranking algorithms introduced in Section \ref{sec:nonlinear}) lead to faster or slower systemic breakdown? The two questions will be addressed in Section \ref{sec:structural_robustness} and \ref{sec:structural_nodes}, respectively.

\subsubsection{Topological robustness of nested networks}
\label{sec:structural_robustness}

Consider a bipartite network. We label the two kinds of nodes as ``active'' and ``passive species", yet the lines of reasoning presented in this section equally apply to any bipartite network.
If we delete some active (passive) species from the network, some passive (active) will become extinct as a result. The loss of single species can even trigger the co-extinction of several other species.
Such a process, called \emph{co-extinction cascade}, has been extensively studied in the ecological literature~\cite{memmott2004tolerance,burgos2007nestedness}. In this Section, we focus on the impact of a nested structure on the speed of network disruption under such dynamics.
In particular, scholars have debated whether an increase in the level of nestedness in a given network is associated with an enhanced robustness. We shall see in the following that this property only holds for specific removal strategies of the nodes.

\paragraph{Quantifying systemic robustness: attack-tolerance curve (ATC)}

To quantify the impact of network topology on network robustness under node deletion processes, we introduce the attack-tolerance curve (ATC) \cite{memmott2004tolerance}.
This curve represents the fraction of surviving passive species as a function of the fraction of deleted active species. The systemic robustness $R$ of a given network is defined as the area under the ATC. By definition, $R\in(0,1)$. A ``robust'' system is characterized by $R$  values relatively close to one. This corresponds to a system where, even when a substantial number of active species is deleted, a considerable number of passive species are still alive. By contrast, a ``fragile'' system exhibits relatively small $R$ values: even the removal of few active species is enough to substantially decrease the number of passive species in the system.

It follows from the definition of ATC that the robustness $R$ depends not only on network topology, but also on the order in which the active species are sequentially deleted. Burgos \textit{et al.}~\cite{burgos2007nestedness} considered three different removal strategies:
\begin{itemize}
\item $[+\to-]$ (from generalists to specialists) removal strategy. The active species are removed in order of decreasing degree.
\item $[-\to+]$ (from specialists to generalists) removal strategy. The active species are removed in order of increasing degree.
\item Random removal strategy. The active species are removed in a random order.
\end{itemize}
While all the three cases are important from a theoretical standpoint, experimental evidence strongly suggests that in mutualistic networks, the specialist species have also higher likelihood to extinguish \cite{aizen2012specialization}, which suggests that the $[-\to+]$ strategy is the most relevant one for this class of networks.
Importantly, we shall see in the following paragraph that the $[-\to +]$ strategy is also the one for which the topological robustness benefits from a nested topology to the largest extent\footnote{Motivated by this consideration, Burgos \textit{et al.}~\cite{burgos2009understanding} introduced a nestedness metric based on the difference between the robustness coefficients corresponding to attack strategies $[-\to +]$ and $[+\to -]$.}.

\begin{figure}[t]
\centering
\includegraphics[scale=0.75]{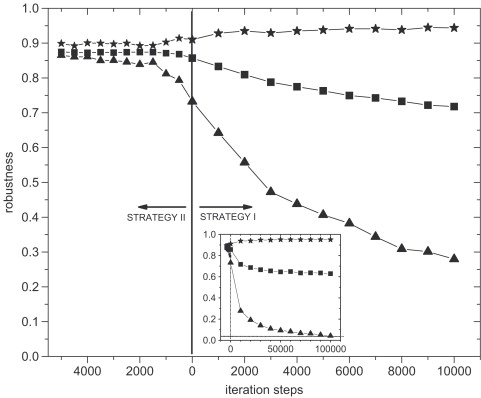}
\caption{Robustness as a function of the number of iterations of two rewiring strategies, for the mutualistic system studied by Clements and Long~\cite{clements1923experimental}. Strategy I corresponds to the Self-organizing Network Model (SNM, see Section~\ref{sec:snm}): the nodes replace their partners with more generalist partners. This rewiring strategy gradually increases the level of nestedness in the system. Strategy II is the opposite: the nodes strive to connect with more specialized nodes. Increasing nestedness increases the robustness against $[-\to +]$ attacks (from specialists to generalists, filled stars), whereas it decreases the robustness against $[+\to-]$ (filled triangles) and random attacks (filled squares). The inset focuses on Strategy I, by displaying results for a larger number of iterations.
Reprinted from \cite{burgos2007nestedness}.}
\label{fig:atc}
\end{figure}

\paragraph{Nestedness and topological robustness}

 Memmott \textit{et al.}\cite{memmott2004tolerance} studied the co-extinction of species in plant-pollinator networks, and they found that the $[-\to+]$ removal strategy (from specialists to generalists) leads to a significantly slower network breakdown with respect to the random removal strategy and the $[+\to-]$ (from generalists to specialists) removal strategy. They concluded that mutualistic networks are robust with respect to preferential node deletion, in particular when specialists have a higher probability to become extinct. 

Burgos \textit{et al.}~\cite{burgos2007nestedness} moved one step forward by using the self-organizing network model (SNM, see Section \ref{sec:snm} and \cite{medan2007analysis}) to gauge the impact of nestedness on systemic robustness. As we have seen in Section~\ref{sec:snm}, the SNM allows us to gradually move from a completely random topology to a perfectly nested topology. 
The authors found that more nested structures are both more robust against the $[-\to+]$ removal strategy, and more fragile against the $[+\to-]$ removal strategy. 
This is intuitive: in a large perfectly nested structure, if we lose the most specialist species $\alpha$, no other species will become extinct as $\alpha$ was only connected with the most generalist species. By contrast, if we lose the most generalist species, the most specialist species are likely to become extinct as they were only connected with the most generalist species.
Burgos \textit{et al.}~\cite{burgos2007nestedness} noticed that if we accept that the only systems that we can observe in nature are those with a higher level of robustness, the omnipresence of nested structure in mutualistic networks might be a manifestation of the more vulnerable position of specialist species.

\paragraph{Incorporating interaction strength and link rewiring}

While the removal strategies studied by Burgos \textit{et al.}~\cite{burgos2007nestedness} revealed the implications of nestedness for the topological robustness of a given system, they are simplified descriptions of real species losses.
Several subsequent studies have aimed at incorporating more realistic effects into the co-extinction cascade process.
Kaiser-Bunbury \textit{et al.}~\cite{kaiser2010robustness} pointed out that analyses like those carried out by Memmott \textit{et al.}~\cite{memmott2004tolerance} and Burgos \textit{et al.}~\cite{burgos2007nestedness} assume that all the removed species contribute equally to the fraction of removed species, regardless of their abundance and interaction frequency.
To overcome this potential limitation, they defined the \emph{normalized interaction strength} $L_{i\alpha}^{(N)}$ between a given animal $i$ and a given plant $\alpha$ as the total number of visits per flower per hour, and the \emph{absolute interaction strength} $L^{(A)}_{i\alpha}$ as $L^{(N)}_{i\alpha}$ times the mean number of flowers per square meters. The total interaction strength of animal $i$ and plant $\alpha$ are defined as $N^{(A)}_i=\sum_{\alpha}L^{(A)}_{i\alpha}$ and $N^{(A)}_\alpha=\sum_{i}L^{(A)}_{i\alpha}$, respectively. The total interaction strength of the network is given by $N=\sum_{i}N^{(A)}_i$; in the co-extinction cascade process, the removal of species $i$ leads to a loss $N^{(A)}_i$ in the total interaction strength.

Building on this setting, Kaiser-Bunbury et al~\cite{kaiser2010robustness} analyzed $12$ two-weeks snapshots of two pollination networks. They found that when the active species are removed progressively according to the strategy $[+\to-]$, the attack tolerance curve of the total interaction strength $N^{(A)}$ as a function of the active-species total interaction strength displays a sigmoidal shape where the removal of few animals caused a sudden drop in the total interaction strength. Even more, 
they studied removal strategies that include the possibility of rewiring -- the links observed across the $12$ snapshot are considered as ``potential links", and these potential links were used as rewiring options for disrupted links during the removal process. 
Networks with potential rewiring turned out to be significantly more stable than non-rewired networks, which suggests that the possibility of rewiring might confer additional robustness to ecological communities.

\subsubsection{Nestedness maximization and the structural importance of individual nodes}
\label{sec:structural_nodes}

\begin{figure}[t]
\centering
\includegraphics[scale=0.95]{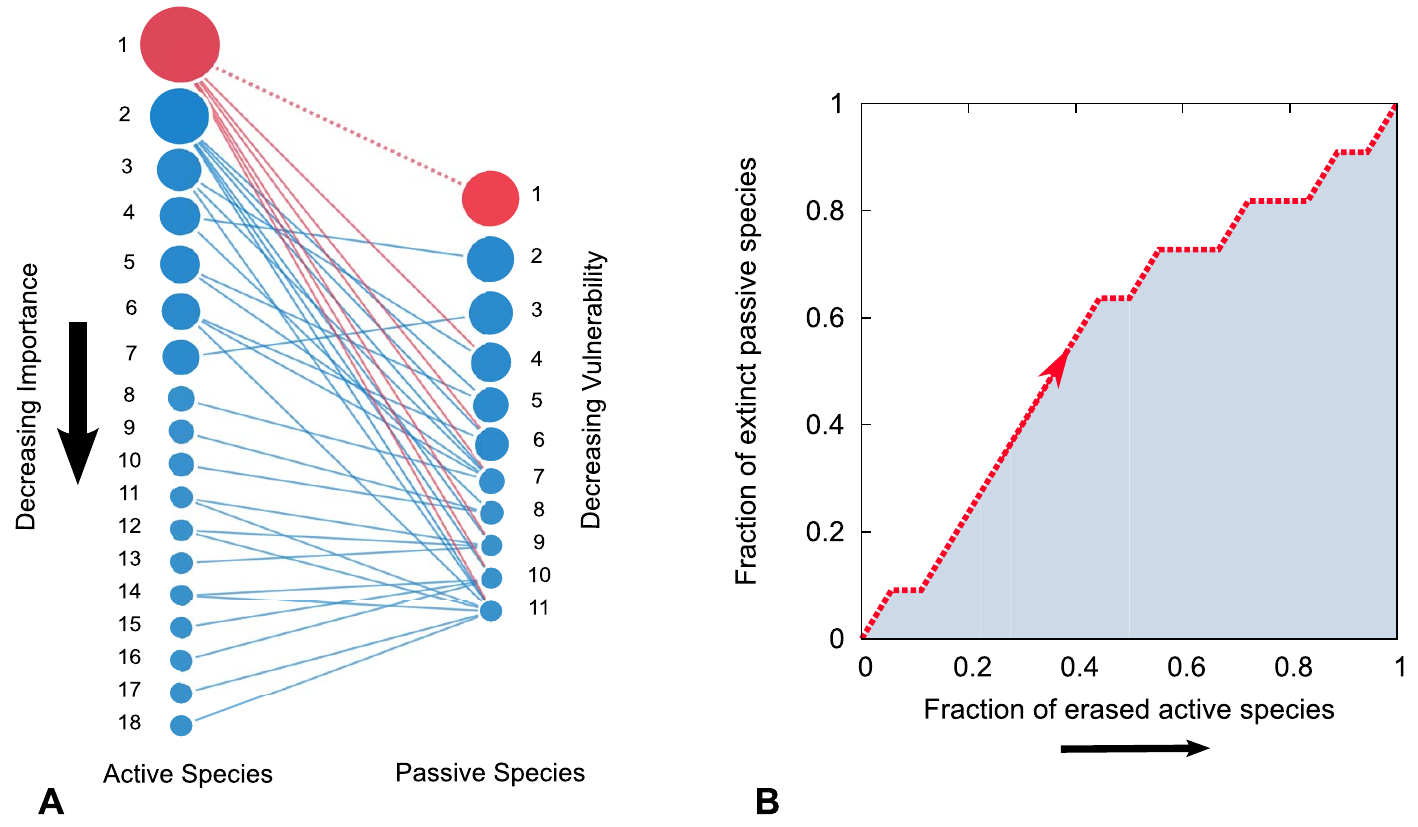}
\caption{Illustration of the node removal process (panel A) used to calculate the extinction area metric (panel B).
From the bipartite network that connects active and passive species, active species are sequentially removed in order of decreasing ``importance'' as determined by a given ranking algorithm (panel A). One records the size of co-extinction cascades as the fraction of direct extinct species increases; the area under this curve is the extinction area of the ranking algorithm. The larger the algorithm's extinction area, the more accurate the algorithm is in identifying the structurally important active species.
One can define a specular extinction area based on a process where passive species are progressively removed in order of increasing vulnerability~\cite{dominguez2015ranking}.
The same procedure can be applied to the country-product bipartite network by interpreting countries and products as active and passive nodes, respectively~\cite{mariani2015measuring}.
Reprinted from \cite{dominguez2015ranking}.}
\label{fig:dominguez1}
\end{figure}

\begin{figure}[t]
\centering
\includegraphics[scale=0.75]{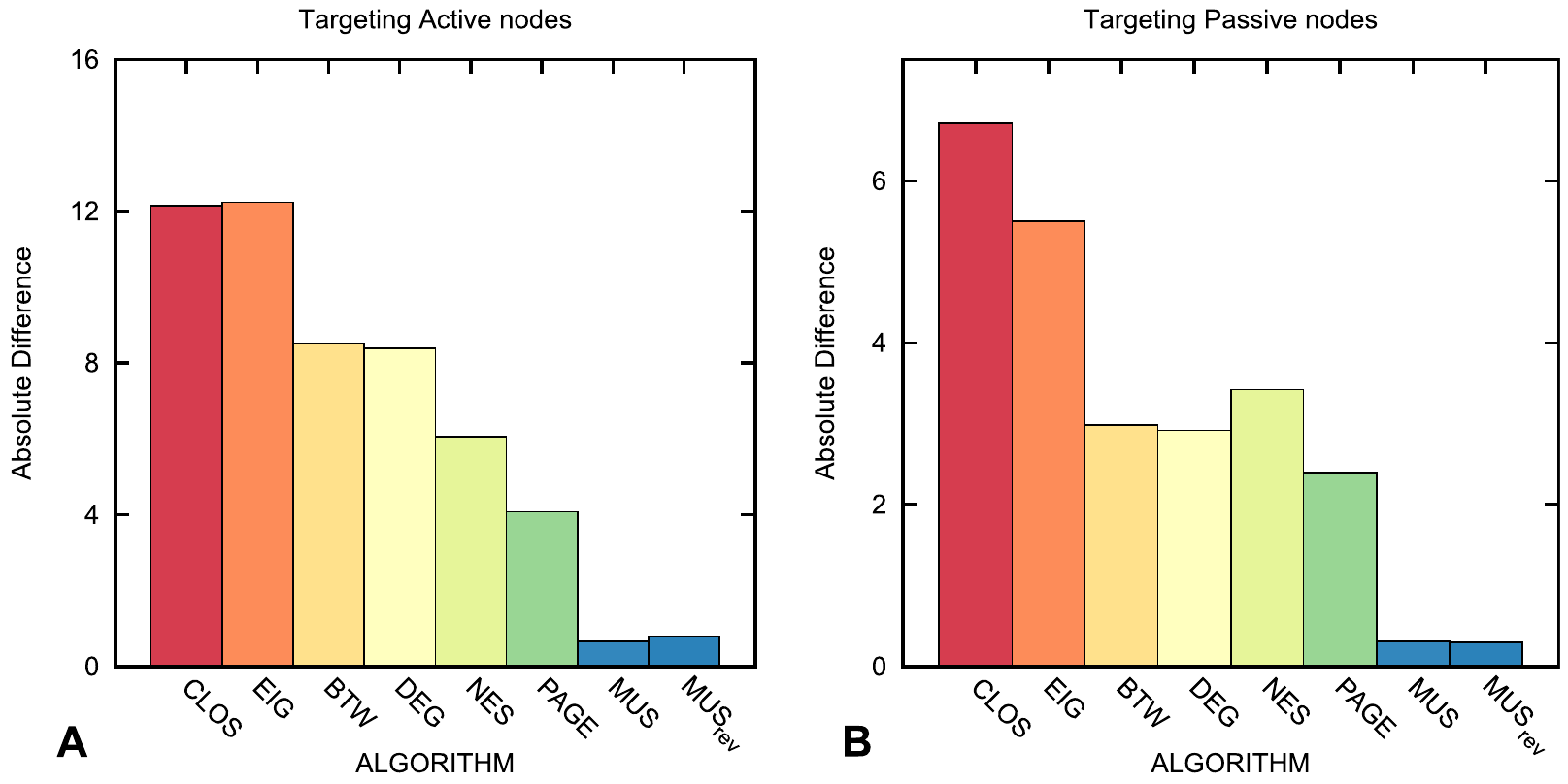}
\caption{The ability of different ranking algorithms to quantify node importance for co-extinction cascades as quantified by the extinction area of the co-extinction cascades that they trigger. The Figure shows the difference between the optimal extinction areas obtained by a genetic algorithm and the extinction areas attained by the algorithms, for attacks targeting active (left panel) and passive species (right panel).
The metrics included in the analysis are: closeness centrality ($CLOS$), eigenvector centrality ($EIG$), betweenness centrality ($BTW$), degree centrality ($DEG$), inverse contributions to nestedness ($NES$), Google's PageRank ($PAGE$), fitness-complexity algorithm ($MUS$), ``reversed'' fitness-complexity algorithm ($MUS_{rev}$). We refer to~\ref{appendix:ranking} for a detailed description of the metrics.
The fitness-complexity algorithm gets substantially closer to the maximal extinction area than all the other metrics. Reprinted from~\cite{dominguez2015ranking}.}
\label{fig:dominguez2}
\end{figure}



Section~\ref{sec:robustness} dealt with the topological robustness of different network structures, for different node removal strategies.
The analysis was restricted to removal strategies based on the ranking of the nodes by their degree.
One can nevertheless design node removal strategies that are based on node centrality metrics other than degree.
Crucially, the sequential deletion of the nodes according to different centrality metrics lead to different co-extinction patterns~\cite{allesina2009googling,dominguez2015ranking,mariani2015measuring}. 

An ideal algorithm would provide us with an accurate ranking of the active (passive) species by their importance (vulnerability) in the system~\cite{dominguez2015ranking}.
In this respect, we would expect an accurate algorithm to rank first (last) the active (passive) species that cause the biggest damage when removed. The rationale behind this assumption is that an important active species interacts with many passive species, from the generalist to the specialist ones; hence, removing it is likely to cause the secondary extinction of some of the specialist passive species.
The main goal of this Section is to present the obtained results on the performance of different ranking algorithms with respect to this benchmark.

\paragraph{Benchmarking ranking algorithms: extinction areas (EAs)}

The \emph{extinction area}~\cite{allesina2009googling} quantifies the ability of a ranking algorithm to produce a ranking of the nodes that leads to a fast network fragmentation when the nodes are deleted in order of ranking. The EA is closely related to the ATC introduced above. When the active (passive) nodes are progressively removed from the system, one defines the extinction curve (EC) as the fraction of extinct passive (active) nodes as a function of the fraction of removed nodes. The EA is defined as the area under the EC, and it lies within the range $(0,1)$. The larger the EA, the faster the network breakdown due to the node removal process -- see Fig. \ref{fig:dominguez1} for an illustration.
As emphasized above, the EA critically depends on the removal strategy; we consider two removal strategies:
\begin{itemize}
\item Targeting active nodes. We remove the active species in order of decreasing score as determined by a given ranking algorithm.
\item Targeting passive nodes. We remove the passive species in order of increasing score as determined by a given ranking algorithm.
\end{itemize}
We expect a good metric for active species' importance and passive species' vulnerability to maximize the respective EA. We stress that while all traditional centralities produce the same ranking in perfectly nested networks~\cite{schoch2017correlations}, the problem becomes non-trivial as soon as imperfect nested structures are considered, such as those observed in empirical datasets.

\paragraph{Results on mutualistic networks}

Following Allesina and Pascual \cite{allesina2009googling}, two recent works \cite{dominguez2015ranking,mariani2015measuring} have compared different ranking algorithms with respect to their extinction areas in bipartite networks.
Inspired by the literature on economic complexity \cite{tacchella2012new}, Dom{\'\i}nguez-Garc{\'\i}a and Mu{\~n}oz~\cite{dominguez2015ranking} noticed that different ranking algorithms produce different shapes for the adjacency matrix when the matrix's rows and columns are ordered by ranking (see also our previous discussion in Section~\ref{sec:packing}).

One can expect ranking algorithms that lead to more nested arrangements of the adjacency matrix to exhibit larger extinction areas than other ranking algorithms.
Indeed, as soon as we remove the first few top-ranked nodes from a highly nested structure, the most specialist nodes are likely to become extinct as a result. The same is not guaranteed for a ranking algorithm that does not lead to a packed adjacency matrix.

In line with this expectation, Dom{\'\i}nguez-Garc{\'\i}a and Mu{\~n}oz \cite{dominguez2015ranking} found that in mutualistic networks, the fitness-complexity metric exhibits systematically larger extinction areas than all the other metrics, including closeness centrality, eigenvector centrality, betweenness centrality, degree centrality inverse contributions to nestedness, Google's PageRank -- see Fig. \ref{fig:dominguez1} for the main result, and Appendix~\ref{appendix:ranking} for the definitions of the existing metrics. The fitness-complexity metric achieves a similar performance as a computationally expensive genetic algorithm specifically designed by the authors in order to find a ranking of the nodes that maximizes the extinction area. Dom{\'\i}nguez-Garc{\'\i}a and Mu{\~n}oz~\cite{dominguez2015ranking} concluded that the fitness-complexity algorithm should be used to construct rankings of species by importance in mutualistic networks. 

\paragraph{Results on country-product export networks}

The ranking evaluation procedure described above with the terminology of mutualistic networks can be applied  to the country-product export network as well.
In this case, the countries (products) play the role of active (passive) species \cite{mariani2015measuring}.
Mariani \textit{et al.} \cite{mariani2015measuring} found that in world trade networks, the fitness-complexity metric exhibit larger extinction areas than the degree centrality and the method of reflections (a linear ranking algorithm introduced by Hidalgo and Hausmann~\cite{hidalgo2009building} with the goal to quantify the economic complexity of countries and products). 
Besides, generalizations of the fitness-complexity metric that lead to more nested matrices exhibit even larger extinction areas.
This confirms that non-linear ranking algorithms are the most effective ones in ranking the nodes by their structural importance for co-extinction cascades in bipartite networks.

By exploiting the analogy with ecological networks, in the same way as vulnerable passive species only interact with important active species, ``vulnerable'' (less complex) products are only exported by ``important'' (diversified) countries. 
The fitness-complexity algorithm and its generalizations naturally incorporate this property through the non-linear dependence of product score on country score: indeed, if a product is exported by a low-fitness country, it receives a high penalization by the algorithm (see Eq.~\eqref{fit_comp_norm} and related discussion). This is desirable inasmuch the product is more likely to be affected by the extinction of high-fitness countries. Non-linear algorithms, therefore, emerge as natural methods to rank the nodes by importance in both mutualistic networks~\cite{dominguez2015ranking} and country-product networks~\cite{mariani2015measuring}.

\subsection{Stability and feasibility of systems with competition and mutualism in the population dynamics approach}
\label{sec:population}

Understanding how network topology and dynamics affect systemic stability is a key problem in network science~\cite{gao2016universal}.
Following the observation that the structure of ecological interaction networks is highly non-random \cite{bascompte2003nested}, scholars have sought to understand how the topology of interaction networks impacts the co-existence of species in the system ("feasibility") and the ability of the system to return to its equilibrium state after a perturbation ("stability"). 
In absence of empirical temporal data, this impact can be evaluated through numerical simulations or analytic insights based on population dynamics models. To this end, it is essential to introduce a model that incorporates the key elements of mutualistic systems: competition, mutualism, finite interaction handling time~\cite{okuyama2008network,bastolla2009architecture}. Such a model is introduced in Section~\ref{sec:mutualistic}.

To properly assess the impact of topology on the stable co-existence of species, \emph{structural stability} is a key notion.
According to the structural-stability paradigm, when analyzing the outcomes of a population dynamics model,
we need to determine not whether nestedness favors biodiversity for a given parameterization of the dynamics, but to determine which topological pattern leads to the stable co-existence of species for the largest region of parameter values~\cite{rohr2014structural}. Considering only specific, arbitrary parameterizations of the dynamics can indeed lead to contradictory conclusions on the impact of nestedness on biodiversity~\cite{james2012disentangling,saavedra2013disentangling,james2013james,rohr2014structural} -- we refer to~\cite{rohr2014structural,bascompte2018structural} for detailed discussions of this point.

In this Section, we start by reviewing the different notions of local and global stability for a given dynamical system, together with the notion of feasibility (Section~\ref{sec:feasibility}). We discuss the global and feasible equilibrium points of interacting systems that feature both competitive and mutualistic interactions, and the impact of nestedness on the range of dynamics parameter that can lead to a stable and feasible equilibrium (Section~\ref{sec:mutualistic}).

\subsubsection{Stability and feasibility of equilibrium points}
\label{sec:feasibility}

The first studies on the stability theory of ordinary differential equations date back to the early XX century, with the seminal works by Poincaré and Lyapunov~\cite{vulpiani2010chaos}.
A fundamental problem in dynamical systems theory is to understand the late-time behavior of a given system. Among all the possible types of asymptotic behavior for a given dynamical system, the simplest one is the convergence toward an equilibrium point. In the following, we review the notion of local and global stability for an equilibrium point.

We narrow our focus to dynamic models described by a set of coupled ODEs in the form
\begin{equation}
 \frac{dN_i}{dt}=N_i\,f_{i}(\vek{N}),
 \label{odes}
\end{equation}
where $\vek{N}=\{N_1,\dots N_S\}$ is the set of population sizes, $S$ is the number of species, 
and $f_{i}(\vek{N})=N_i^{-1}\,(dN_i/dt)$
represents species $i$'s per-capita growth rate.
The functions $f_i$ typically depend on a set of parameters that
determine the interactions between species.
For example, we will sometimes consider a linear per-capita growth rate in the 
form $f_{i}(\vek{N})=\alpha_i-\sum_j\,\beta_{ij}\,N_j$, where $\alpha_i$ represents the species' \emph{intrinsic growth rates}
and $\{\beta_{ij}\}$ represent the interaction coefficients. Such dynamics represents a generalization to $N$ species of the equations originally proposed independently by Lotka~\cite{lotka1910theorie} and Volterra~\cite{volterra1926fluctuations} to study the competitive dynamics of a system composed of two species. Despite its apparent simplicity, this dynamical system can exhibit limit cycle periodic
behavior and chaotic behavior conditional on the parameters and the number of included species\footnote{We refer the interested reader to the books~\cite{murray2002mathematical,vulpiani2010chaos} for details.}.

In the rest of this Section, we introduce three properties of equilibrium points: local stability, global stability, and feasibility.
Importantly, insights from random matrix theory on the persistence of ecologic communities (see Section~\ref{sec:rmt}) are mostly based on the local stability~\cite{allesina2012stability,allesina2015stability} and, recently, feasibility~\cite{grilli2017feasibility,stone2018feasibility}. Recent insights from population dynamics (\cite{rohr2014structural}, see Section~\ref{sec:population}) focused on globally stable and feasible equilibrium points.

\paragraph{Local stability}
A given equilibrium point $\vek{N}^{*}$ is \emph{locally stable} if, after a small external perturbation, the system returns to the equilibrium point~\cite{vulpiani2010chaos}.
To determine whether a given equilibrium point $\vek{N}^{*}$ is locally stable or not, one needs to find the eigenvalues of the so-called ``community matrix"~\cite{levins1968evolution} whose elements are defined as 
\begin{equation}
M_{ij}(\vek{N}^*)=\frac{\partial\,(N_i\,f_i(\vek{N}))}{\partial N_j} \bigr \rvert_{\vek{N}=\vek{N}^{*}}.
\label{jacobian}
\end{equation}
The community matrix corresponds to the system's Jacobian matrix evaluated at the equilibrium point $\vek{N}^*$.
A given equilibrium point is locally stable if and only if the real parts of the eigenvalues of $M_{ij}(\vek{N}^*)$ are all negative.
For a linear per-capita growth rate in the form $f_{i}(\vek{N})=\alpha_i-\sum_j\,\beta_{ij}\,N_j$, we obtain $M_{ij}=-N_i^*\,\beta_{ij}$ for each equilibrium point $\vek{N}^*$~\cite{saavedra2017structural}; such equilibrium point is locally stable if $\beta_{ij}\geq 0$ for all $i,j$.

\paragraph{Feasibility and global stability} 
A given equilibrium point $\vek{N}^{*}$ is \emph{feasible} if $f(\vek{N}^{*})=0$
and $N^{*}_i>0$ for all species $i$~\cite{goh1979stability,rohr2014structural,saavedra2017structural}. A feasible equilibrium point describes, therefore, a scenario where all species can co-exist.
A given equilibrium point $\vek{N}^{*}$ is \emph{globally stable} if $\vek{N}^{*}$ is a global attractor,
and the trajectories of the dynamic system converge to $\vek{N}^{*}$ regardless of the initial condition~\cite{rohr2014structural,saavedra2017structural}.
The fundamental difference between local and global stability is that while local stability studies
whether a system will return to the original equilibrium point after an infinitesimally small perturbation,
global stability studies whether the system will return to the original equilibrium point after perturbations of any 
magnitude~\cite{rohr2014structural}.
Is a given equilibrium point globally stable?
In general, no: we will provide a simple example of an equilibrium point that is not globally stable in Section \ref{sec:two}.
Linear algebra theorems nevertheless provide us with sufficient conditions for an equilibrium point to be stable.
It is convenient to introduce the following definitions: 
\begin{itemize} 
\item A matrix $\mat{M}$ is \emph{positive-defined} if and only if the eigenvalues of $\mat{M}+\mat{M}^T$ are all positive.
 \item A matrix $\mat{M}$ is \emph{Volterra-dissipative}
if and only if there exists a diagonal matrix $\mat{D}$ such that $\mat{D}\,\mat{M}+\mat{M}^T\,\mat{D}$ is positive-defined. 
 \end{itemize}
The following theorems hold~\cite{saavedra2017structural}: 
\begin{itemize}
 \item If a matrix is positive-defined, then it is Volterra-dissipative. 
 \item For a model with linear per-capita growth rate, $f_{i}(\vek{N})=\alpha_i-\sum_j\,\beta_{ij}\,N_j$,
 we can arrange the interaction coefficients in a $S\times S$ matrix, which we denote as $\mat{B}$.
 If $\mat{B}$ is Volterra-dissipative, then (1) Any feasible equilibrium is globally stable;
 (2) There exists one unique globally stable equilibrium point\footnote{Such equilibrium point is not necessarily feasible,
 as some of the population sizes may be equal to zero.}.
\end{itemize}
The previous theorems imply that if the matrix of interaction strengths is positive-defined, then the system has a unique globally stable equilibrium.
In Section~\ref{sec:mutualistic} we shall see that, in models with competition and mutualism, this property can be used to set an upper bound to the level of mutualistic strength up to which
global stability is guaranteed.

\subsubsection{Stability and feasibility in a two-species competition model}
\label{sec:two}

\begin{figure}[t]
\centering
\includegraphics[scale=0.75]{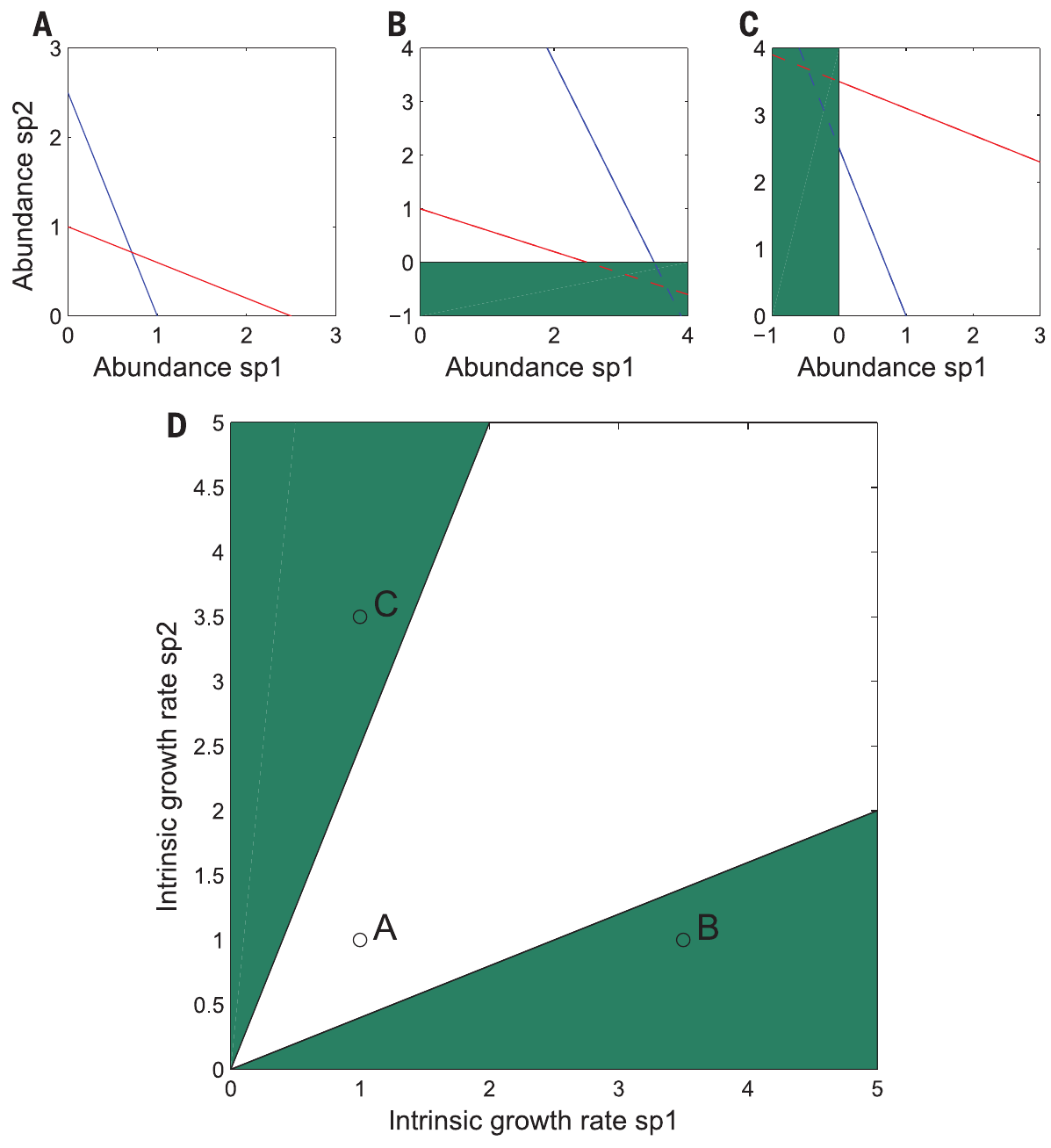}
\caption{Global stability and feasibility of the equilibrium points of the simple mutualistic dynamics described by Eq.~\eqref{lotka}. Panels A to C represent the isoclines $f_1(N_1,N_2)=0$ and $f_2(N_1,N_2)=0$ for three different parametrizations of the intrinsic growth rates $(\alpha_1,\alpha_2)$; their intersection point is the equilibrium point of the dynamics. Among the three scenarios A-C, only A represents a feasible equilibrium point ($N_1>0$ and $N_2>0$). Panel D shows that the intrinsic growth rates space can be partitioned into regions where the equilibrium is feasible (white area), and regions where the equilibrium is not feasible (green areas). For this simple dynamics, the feasibility domain is given by Eq.~\eqref{feasibility}. The three points A, B, C corresponds to the three above panels; their position in the intrinsic growth rates space confirms that only A is feasible. 
Reprinted from \cite{rohr2014structural}}
\label{fig:feasibility}
\end{figure}

To get an intuition about the implications of the different stability definitions and of the feasibility condition, we consider here a simple two-species competition model. The proposed example and the calculations below closely follow those presented in~\cite{rohr2014structural}.
The system evolves according to the generalized Lotka-Volterra equations:
\begin{equation}
\begin{dcases}
 \frac{d N_1}{dt}=N_1\,(\alpha_1-\beta_{11}\,N_1-\beta_{12}\,N_2)	\\
 \frac{d N_2}{dt}=N_2\,(\alpha_1-\beta_{21}\,N_1-\beta_{22}\,N_2)
\end{dcases}
\label{lotka}
\end{equation}
In line with the notation of Eq.~\eqref{odes}, the previous equations can be written in the compact form
\begin{equation}
 \frac{d \vek{N}}{dt}=\vek{N}\,\vek{f}_{\alpha,\mat{B}}(\vek{N}),
\end{equation}
where we defined $\vek{f}(\vek{N})=(f_1(\vek{N}),f_2(\vek{N}))$, $f_1(N_1,N_2)=\alpha_1-\beta_{11}\,N_1-\beta_{12}\,N_2$, $f_2(N_1,N_2)=\alpha_2-\beta_{21}\,N_1-\beta_{22}\,N_2$.
It is also convenient to introduce the matrix $\mat{B}$ whose elements are the interaction coefficients $\beta_{ij}$
The equilibrium point $(N_1^*.N_2^*)$ is found by imposing $f_1(N_1^*,N_2^*)=0$ and $f_2(N_1^*,N_2^*)=0$, which has the solution
\begin{equation}
\begin{dcases}
 N_1^*=\frac{\alpha_1\, \beta_{22}-\alpha_2\,\beta_{12}}{\beta_{11}\,\beta_{22}-\beta_{21}\,\beta_{12}} \\
 N_2^*=\frac{\alpha_2\, \beta_{11}-\alpha_1\,\beta_{21}}{\beta_{11}\,\beta_{22}-\beta_{21}\,\beta_{12}}
 \end{dcases}
 \label{equilibrium}
\end{equation}
According to the theorem presented in Section~\ref{sec:feasibility}, if the matrix $\mat{B}$ is positive-defined, then this
equilibrium point is globally stable. Therefore, we narrow our attention to $\{\beta_{ij}\}$ values such that
$\beta_{11}\,\beta_{22}-\beta_{21}\,\beta_{12}>0$.
Graphically, the equilibrium point $(N_1^*,N_2^*)$ corresponds to the intersection point between the isoclines $f_1(N_1,N_2)=0$ and $f_2(N_1,N_2)=0$
in the $(N_1,N_2)$ plane (see Fig.~\ref{fig:feasibility}A-C). The equilibrium point is feasible if the intersection point lieas in the quadrant with $N_1>0$ and $N_2>0$.
Importantly, different settings of the model parameters $r_1,r_2, \mat{\alpha}$ lead to different isoclines
in the $(N_1,N_2)$ plane; some parameter settings may lead to a feasible equlibrium point (see Fig.~\ref{fig:feasibility}A), whereas
other parameter settings may not (see Fig.~\ref{fig:feasibility}B-C).

The equilibrium point $(N_1^*,N_2^*)$ is feasible if and only if both $N_1^*>0$ and $N_2^*>0$, which leads to the conditions :
\begin{equation}
 \begin{cases}
  \alpha_1\, \beta_{22}-\alpha_2\,\beta_{12}>0\\
  \alpha_2\, \beta_{11}-\alpha_1\,\beta_{21}>0
 \end{cases}
\end{equation}
If all the parameters $\{\beta_{ij}\}$ are positive, we can compress the two previous conditions into a single condition for the ratio $r_1/r_2$
between the intrinsic growth rates:
\begin{equation}
 \frac{\beta_{12}}{\beta_{22}}<\frac{\alpha_1}{\alpha_2}<\frac{\beta_{11}}{\beta_{21}}.
 \label{feasibility}
\end{equation}
Such condition is verified in the region of the $(\alpha_1,\alpha_2)$ plane bounded by the two straight lines $\alpha_2=(\beta_{12}/\beta_{22})\,\alpha_1$ and 
$\alpha_2=(\beta_{11}/\beta_{21})\,\alpha_1$.

The results on the simple two-species competition model described by Eq.~\eqref{lotka} shed light on two fundamental facts~\cite{rohr2014structural}:
\begin{itemize}
 \item The feasibility of a given equilibrium point critically depends on the dynamics parameters.
 Some sets of dynamics parameters may lead to a feasible equilibrium point (such as in Fig. \ref{fig:feasibility}A) or it may not (such as in Figs.~\ref{fig:feasibility}B-C).
 \item One can identify a ''feasibility domain`` in the intrinsic growth rates' plane where the dynamics' equilibrium point is
 feasible (see Fig. \ref{fig:feasibility}D). Crucially, the size of such domain depends on the dynamics parameters.
\end{itemize}
Section~\ref{sec:mutualistic} shows that these properties also hold for a mutualistic model of interaction.
Based on these properties, Rohr \textit{et al.}~\cite{rohr2014structural} put forward the notion of \emph{structural stability}: 
when studying the dynamical stability of a given ecological system, the question is not whether the dynamics' equilibrium point is feasible, given a set of
dynamics parameters. The main question is whether the parameter region composed of the parameters that lead to a feasible equilibrium point is large,
and how the size of such region depends on network topology~\cite{rohr2014structural}. Next Sections will investigate this fundamental question for a model with competition and mutualism, and reveal the strong dependence of its answer on the level of nestedness in the network.

In a similar spirit to the simple two-species described above, Bastolla \textit{et al.}~\cite{bastolla2005biodiversity1,bastolla2005biodiversity2} analyzed more complex competitive systems, and  derived analytical conditions that must be fulfilled by the moments of the growth rate distribution in order for the system to be feasible. The framework introduced by Bastolla \textit{et al.}~\cite{bastolla2005biodiversity1,bastolla2005biodiversity2} has been later extended~\cite{bastolla2009architecture} to systems with both competition and mutualism.
We refer to~\cite{bascompte2018structural} for a detailed presentation of the results in~\cite{bastolla2005biodiversity1,bastolla2005biodiversity2}. 

\subsubsection{Stability and feasibility in models with competition and mutualism}
\label{sec:mutualistic}

The previous Section considered a simple, two-species competition model.
Following Rohr \textit{et al.}~\cite{rohr2014structural}, we extend here those insights to the $N$-species model with (inter- and intra-specific) competition and mutualism studied by Bastolla \textit{et al.}~\cite{bastolla2009architecture}.
The model incorporates network topology, interspecific competition, and mutualistic interactions with finite handling time (see below).
To introduce the main elements of the model, we start from a simpler two-species model~\cite{wright1989simple}, and later describe the full $N$-species model~\cite{bastolla2009architecture}.

\paragraph{A two-species model with intraspecific competition and mutualism} Before describing the results for an ecosystem composed of multiple species,
let us consider a simple two-species world where the two species represent animals and plants, respectively. Our simplified world features both intraspecific competitive interactions and mutualistic interactions between plants and animals.
For an isolated ``animal'' species whose $A$ individuals compete for the same resource, it is natural to consider a logistic population model
\begin{equation}
 \frac{dA}{dt}=A\,(\alpha-\beta\,A),
\end{equation}
where $\alpha$ is the species' intrinsic per-capita growth rate, and $\beta$ is a parameter that quantifies the strength of
intra-species competitive interactions. To simplify the discussion, we refer to the individuals of this species as ''animals``.

Mutualistic interactions can be incorporated by introducing a second species, whose number of individuals is denoted as $P$ -- we refer to the individuals of this species as ''plants``.
The interactions between animals and plants are mutualistic, which results in a positive contribution to $dA/dt$ from the plants.
It is plausible to assume the benefits for animals from mutualistic interactions to increase with the number $P$ of plants.
However, Wright~\cite{wright1989simple} noticed that this contribution cannot be unbounded.
Let us consider the situation where the animals are bees and the plants are the flowers that they pollinate.
A bee needs a certain amount of time to travel to the plant, collect enough pollen from a flower, and travel back to its nest to provision its
larval cells.
Such time -- referred to as \emph{handling time} \cite{wright1989simple} -- limits the benefits that the animal species can obtain from its mutualistic interaction with the plant species.

To incorporate both handling time and mutualistic interactions, Wright~\cite{wright1989simple} assumed a dynamic equation of the form
\begin{equation}
 \frac{dA}{dt}=A\,\Biggl( \alpha-\beta\,A + \frac{c\,\gamma\,P}{1+h\,\gamma\,P} \Biggr).
 \label{wright}
\end{equation}
The form of the mutualistic interaction term $c\,\gamma\,P/(1+h\,\gamma\,P)$ is in line with the empirical observations by Holling on the capability of a consumer to process a given amount of food resource~\cite{holling1959some}.
The non-linear functional response in the form $\gamma\,P/(1+h\,\gamma\,P)$ is also referred to as Type II response~\cite{holling1959some}. 
Importantly, if $h=0$, the contribution from mutualistic interactions is unbounded. By contrast, for any $h>0$,
the mutualistic term is bounded by $h^{-1}$, the value it assumes in a world with a large number $P\gg 1$ of plants or, equivalently,
in a world with strong mutualistic interactions ($\gamma\gg 1$).
In a purely-mutualistic world with $\alpha=0, \beta=0$, and $\gamma \gg 1$, the parameter $h$ determines the rate at which mutualistic interactions
are converted into new animals: $dA=A\,c\,(dt/h)$. 
The model described by Eq.~\eqref{wright} constitutes the basis for the model with competition and mutualism studied by Bastolla \textit{et al.}~\cite{bastolla2009architecture}.

\paragraph{Generalizing the model to an arbitrary number of species}

Different types of $N$-species mutualistic population-dynamics models have been studied in the literature.
A model without handling time for a fully mixed population (without network structure) was considered 
by Bascompte \textit{et al.}~\cite{bascompte2006asymmetric} to analytically derive conditions for a system to be 
feasible and stable.
Commenting on~\cite{bascompte2006asymmetric}, Holland
\textit{et al.}~\cite{holland2006comment} pointed out the necessity of including the saturation 
of the mutual benefit two species can gain from their interaction, an aspect 
that was already stressed by May~\cite{may1981theoretical} and included in Wright's model~\cite{wright1989simple} presented above. 
Okuyama \textit{et al.}~\cite{okuyama2008network} considered a model with mutualistic interactions, intraspecific competitive interactions, network structure, and finite handling time.

Here, we present the version of the model used by~\cite{bastolla2009architecture,rohr2014structural}.
In the model, there are two 
classes of species: plants and animals.
We assume that there are $N$ animal species and $P$ plant species whose populations are denoted as $\vek{A}=\{A_1,\dots,A_N\}$ and $\vek{P}=\{P_1,\dots,P_M\}$, respectively.
The dynamical equations for the evolution of the species' populations are~\cite{bastolla2009architecture,rohr2014structural}
\begin{equation}
\begin{dcases}
  \frac{dP_i}{dt}= P_i\,\Biggl(\alpha_i^{(P)}-\sum_j \beta_{ij}^{(P)}\,P_j + \frac{\sum_j \gamma_{ij}^{(P)}\,A_j}{1+h\,\sum_j \gamma_{ij}^{(P)}\,A_j} \Biggr)\\
  \frac{dA_i}{dt}= A_i\,\Biggl(\alpha_i^{(A)}-\sum_j \beta_{ij}^{(A)}\,A_j + \frac{\sum_j \gamma_{ij}^{(A)}\,P_j}{1+h\,\sum_j \gamma_{ij}^{(A)}\,P_j} \Biggr)\\
\end{dcases}
\label{mutualistic_model}
\end{equation}
where $\{\alpha_i^{(A)},\alpha_i^{(P)}\}$ 
represent the intrinsic growth rates, $\{\beta_{ij}^{(A)},\beta_{ij}^{(P)}\}$ are the parameters
that quantify the strength of competitive interactions between species of the same kind;
$\{\gamma_{ij}^{(A)},\gamma_{ij}^{(P)}\}$ are the parameters that quantify the
strength of beneficial (mutualistic) interactions between animal and plants; $h$ is the
\emph{handling time}~\cite{wright1989simple} which implies an upper bound to the benefits a species can receive through interaction with species of different type\footnote{The handling time here plays the same role as in Eq.~\eqref{wright}.}. Recent works~\cite{saavedra2013estimating,rohr2014structural} further parameterized the mutualistic interaction coefficients $\gamma_{ij}$ as 
\begin{equation}
\gamma_{ij}(\delta)=\gamma_0\,\frac{A_{ij}}{k_i^\delta},
\label{tradeoff}
\end{equation}
where $\gamma_0$ represents the mutualistic strength, and $\delta$ is a parameter called \emph{mutualistic trade-off}~\cite{saavedra2013estimating}. The mutualistic trade-off factor $k_i^{-\delta}$ reduces the benefit from mutualistic interactions with species that have many interactors, which is justified by field observations~\cite{vazquez2007species}.

Crucially, the model defined by Eq.~\eqref{mutualistic_model} features interspecific competitive interactions.
To keep the competitive interactions as simple as possible, some works~\cite{bastolla2009architecture,rohr2014structural} adopted a mean-field description of the interspecific competition by posing $\beta_{ii}^{(P)}=\beta_{ii}^{(A)}=1$ (\emph{intraspecific} competition), and $\beta_{ij}^{(P)}=\beta_{ij}^{(A)}=\lambda<1$ for $i\neq j$ (\emph{interspecific} competition). Essentially, species of a type are assumed to compete with equal strength with all the species of the same type.
The mean-field assumption can be relaxed: for example, Gracia-Lázaro \textit{et al.}~\cite{gracia2017joint} studied a similar metapopulation model, but they assumed that species do compete more intensely with species that are interested in the same resources. More specifically, they assumed that a pairwise competition term between two plant species $i$ and $j$ proportional to the number of common interaction partners of $i$ and $j$. 

Variants of this population dynamics model and alternative models can be also relevant.
Besides competitive and bipartite mutualistic interactions, we notice that recent field experiments~\cite{losapio2017facilitation} indicate that plant-plant facilitation interactions can have a substantial impact on the structure and robustness of pollination networks, which may motivate the inclusion of these interactions in population-dynamics models. 
One can also consider population dynamics models where the foraging effort of a given pollinator on a given plant is a heterogeneous and adaptive variable~\cite{valdovinos2013adaptive}. The presence of adaptive foraging can substantially alter the impact of nestedness on species persistence -- we refer the interested reader to~\cite{valdovinos2016niche} for the details.

\paragraph*{Preliminary results on the model with competition and mutualism}

The model defined by Eq.~\eqref{mutualistic_model} and its variants have been used in several works to assess how specific network structures affect the co-existence of species in mutualistic and trophic networks. More specifically, early works based on the population dynamics approach focused on the question: given a parameterization of the intrinsic growth rates and the interaction parameters, how does a given network structure impact on the properties of the dynamics' equilibrium point?
We mention here some of the most prominent results obtained with the goal of addressing this research question.

Okuyama and Holland~\cite{okuyama2008network} considered the model without interspecific competition (i.e., in the notation of the previous paragraph, $\lambda=0$), and they assumed that the degree distribution $p_k$ of mutualistic communities follows a power-law $p_k=C\,k^{-\gamma}$. They found that the resilience of mutualistic communities only depends on nestedness when the power-law exponent $\gamma$ is large enough; in this case, resilience correlates positively with nestedness. More recently, Morone \textit{et al.}~\cite{morone2018kcore} found that the model without interspecific competition features a sharp transition from a ``feasible'' phase (where the average number of individuals per species is larger than zero) to a ``collapsed'' phase (where the average number of individuals per species is equal to zero). The critical point that separates the two phases such transition can be analytically estimated based on the maximal $k$-core of the network~\cite{morone2018kcore}. 

Bastolla \textit{et al.}~\cite{bastolla2009architecture} used the model with interspecific and competitive interaction to show both analytically and numerically that in real mutualistic networks, nestedness has a positive impact on the system's biodiversity increase due to mutualism. 
More specifically, the analytic framework developed by Bastolla \textit{et al.}~\cite{bastolla2009architecture} clarified that in a competitive system, introducing mutualism alters the ``effective competition'' between species; crucially, the potential mitigation of competitive effects is modulated by network topology.
The resulting increase of biodiversity due to mutualism can be quantified as the derivative of the predicted maximum number of species with respect to the mutualism-to-competition ratio $R$, calculated at $R=0$ -- see Eq. (6) in \cite{bastolla2009architecture}. The mutualism-to-competition ratio $R$ is expressed in terms of the model's parameters, and it grows as the relative strength of mutualistic interaction increases; $R=0$ corresponds to the absence of mutualism. Importantly, nestedness is positively correlated with the so-defined increase of biodiversity, which led Bastolla \textit{et al.}~\cite{bastolla2009architecture} to conclude that ``nestedness reduces effective interspecific competition and enhances the number of coexisting species".

As the analytic theory developed in~\cite{bastolla2009architecture} has been already detailedly reviewed in Appendix G in~\cite{bascompte2013mutualistic} and in~\cite{bascompte2018structural}, we will not present its details here, and focus on the framework developed by Rohr \textit{et al.}~\cite{rohr2014structural} in the following.
It is worth mentioning that the conclusion that nestedness increases the biodiversity was later challenged by James \textit{et al.}~\cite{james2012disentangling} in a paper that has been subsequently debated~\cite{saavedra2013disentangling,james2013james}; however, the conclusions by~\cite{james2012disentangling} are based on a specific choice of the parameters of the dynamics, and they are based on an incorrect formula for the predicted maximum number of species in the community~\cite{bascompte2018structural}. 

Thébault and Fontaine~\cite{thebault2010stability} consider a variant of the dynamics described by Eq.~\eqref{mutualistic_model} that can describe both mutualistic and trophic interaction. The core difference between mutualistic and trophic interactions is the sign of the term that represents animal-plant interactions. They found that network structure has an opposite effect on the stability of mutualistic networks as compared to the stability of trophic networks.
In particular, nestedness has a positive effect on the resilience of mutualistic networks, defined as the speed at which the community returns to equilibrium after a perturbation in the system.
By contrast, it has a strongly negative effect on the persistence of trophic networks, defined as the proportion of existing species after equilibrium is reached.

Rohr \textit{et al.}~\cite{rohr2014structural}  emphasized that when we fix the model parameters and we find a feasible fixed point, (1) it is not guaranteed
that it is globally stable, i.e., that the system's dynamics will converge toward that state 
for any given initial condition; (2) it is not guaranteed that we will find it for other
parameterizations of the intrinsic growth rates. Besides, for a given system and two 
given network topologies (e.g., a random and a nested one), it is possible to find a
set of intrinsic growth rates such that one of the two is persistent and the other 
one is not, and another set such that the opposite holds -- see Fig. 2 in~\cite{rohr2014structural} for an illustration.

\paragraph{Domain of feasibility in the mutualistic model} 

The notion of \emph{structural stability} has been long-known in developmental biology~\cite{alberch1989logic} and engineering~\cite{siljak1978large} -- we refer to~\cite{bascompte2018structural} for a historical overview.
Early precursors applied the notion of structural stability to ecological models of a few interacting species~\cite{vandermeer1975interspecific,sole1992structural,case2000illustrated}. More recently, structural stability has been applied by Bastolla \textit{et al.}~\cite{bastolla2005biodiversity1,bastolla2005biodiversity2} for purely competitive systems, and by Bastolla \textit{et al.}~\cite{bastolla2009architecture} for systems with competition and mutualism. 

The key idea behind structural stability analysis in ecology is that when studying a population dynamics, we should not pre-impose an arbitrary 
parameterization of the intrinsic growth rates, but study a large portion 
of possible growth rate values, and quantify the size of the region of the growth
rate space -- called \emph{feasibility domain} -- where the system reaches a feasible equilibrium point~\cite{rohr2014structural}.
For the two-species competition model described in Section~\ref{sec:two}, the feasibility domain 
can be determined analytically (see Eq.~\eqref{feasibility} and Fig.~\ref{fig:feasibility}). A similar
derivation for the mutualistic model defined by Eq.~\eqref{mutualistic_model} is not possible. 
An exhaustive numerical search is also impossible due to the large size of the parameter
space of intrinsic growth rates.

To determine the size of the feasibility domain, there are at least two approaches, a geometrical and a probabilistic one.
In the geometric approach,  for a linear dynamics of the form $\vek{\dot{N}}=-\vek{N}(\vek{\alpha}+\mat{B}\,\vek{N})$ ruled by an interaction matrix $\mat{B}$, the feasibility domain is defined as~\cite{song2018guideline}
\begin{equation}
 D_F(\mat{B})=\{\vek{\alpha}=-\mat{B}^{-1}\,\vek{N}^*|N_i^*>0 \,\,\,\forall\, i\}
\end{equation}
By using expressions for solid angles in dimension larger than three~\cite{ribando2006measuring}, one can compute the volume $\Omega(\mat{B})$ of $D_F(\mat{B})$
as~\cite{saavedra2016nested,song2018guideline}
\begin{equation}
 \Omega(\mat{B})=\frac{1}{(2\,\pi)^{S/2}\sqrt{|\det{(\mat{B})}|}}\int\dots\int_{\vek{N}^*>0} \exp{\Biggl( -\frac{1}{2}\vek{N}^{*T}\,\mat{B}^T\,\mat{B}\,\vek{N}^*\Biggr)}\,d\vek{N}^*;
\end{equation}
we refer to~\cite{song2018guideline} for the details, and to~\cite{cenci2018structural} for the generalization to non-linear population dynamics models.
A simpler approach to estimate the size of the feasibility domain is to compute the average probability $\omega(\mat{B})$
that a randomly selected species $i$ is feasible, i.e., $N_i^*>0$~\cite{rohr2014structural,song2018guideline}.
In order for an equilibrium point to be feasible, all species must be feasible; therefore,
for sufficiently weak interactions, the feasibility domain size $\Omega(\mat{B})$
can be simply estimated as $\Omega(\mat{B})\simeq (\omega(\mat{B}))^{S}$~\cite{song2018guideline}.
This approach has been adopted by Rohr \textit{et al.}~\cite{rohr2014structural} to assess the impact of nestedness on
the feasibility domain size (see Section~\ref{sec:impact_nestedness}).

\paragraph{Critical value of mutualism for global stability in the mutualistic model}

A feasible equilibrium point is not necessarily globally stable, and a globally stable equilibrium point is not necessarily feasible.
The existence of a unique globally stable equilibrium point 
is only guaranteed when the interaction matrix $\mat{B}$ is positive-defined, i.e., when all its eigenvalues have a positive real 
part (see Section~\ref{sec:feasibility} and~\cite{rohr2014structural}).
Numerically, one finds that in the linearized version of the model with competition and mutualism (i.e., the model defined by Eq.~\eqref{mutualistic_model} with handling times equal to zero), the real part of the largest eigenvalue of $\mat{B}$
tends to decrease with the mean mutualistic interaction strength $\overline{\gamma}$.
Hence, the convergence to a unique globally stable equilibrium is guaranteed up to a critical value $\hat{\gamma}$ of the interaction 
strength; such point is numerically determined as the smallest $\overline{\gamma}$ value such that one of the 
eigenvalues of $\mat{B}$ reaches zero~\cite{rohr2014structural,saavedra2016nested}.
The larger $\hat{\gamma}$, the stronger mutualistic interactions can be tolerated 
by the system without losing global stability, i.e., the more globally stable the system~\cite{saavedra2016nested}.
Therefore, one can study the impact of a structural property on the global stability of the system by studying its relation with
$\hat{\gamma}$: a positive correlation would imply that that structural property favors the global stability of the 
system\footnote{In this discussion, we have referred to the linearized version of the model ($h=0$).
The critical value $\hat{\gamma}$ for the non-linear dynamics is larger than or equal to the critical value for the linearized dynamics,
which implies that the critical value $\hat{\gamma}$ determined with the linearized dynamics serves as an upper bound to the
range of mutualism that allows for global stability. We refer to~\cite{rohr2014structural} for all the details.}.

\subsubsection{The impact of nestedness on feasibility and global stability}
\label{sec:impact_nestedness}

\begin{figure}[t]
\centering
\includegraphics[scale=0.9]{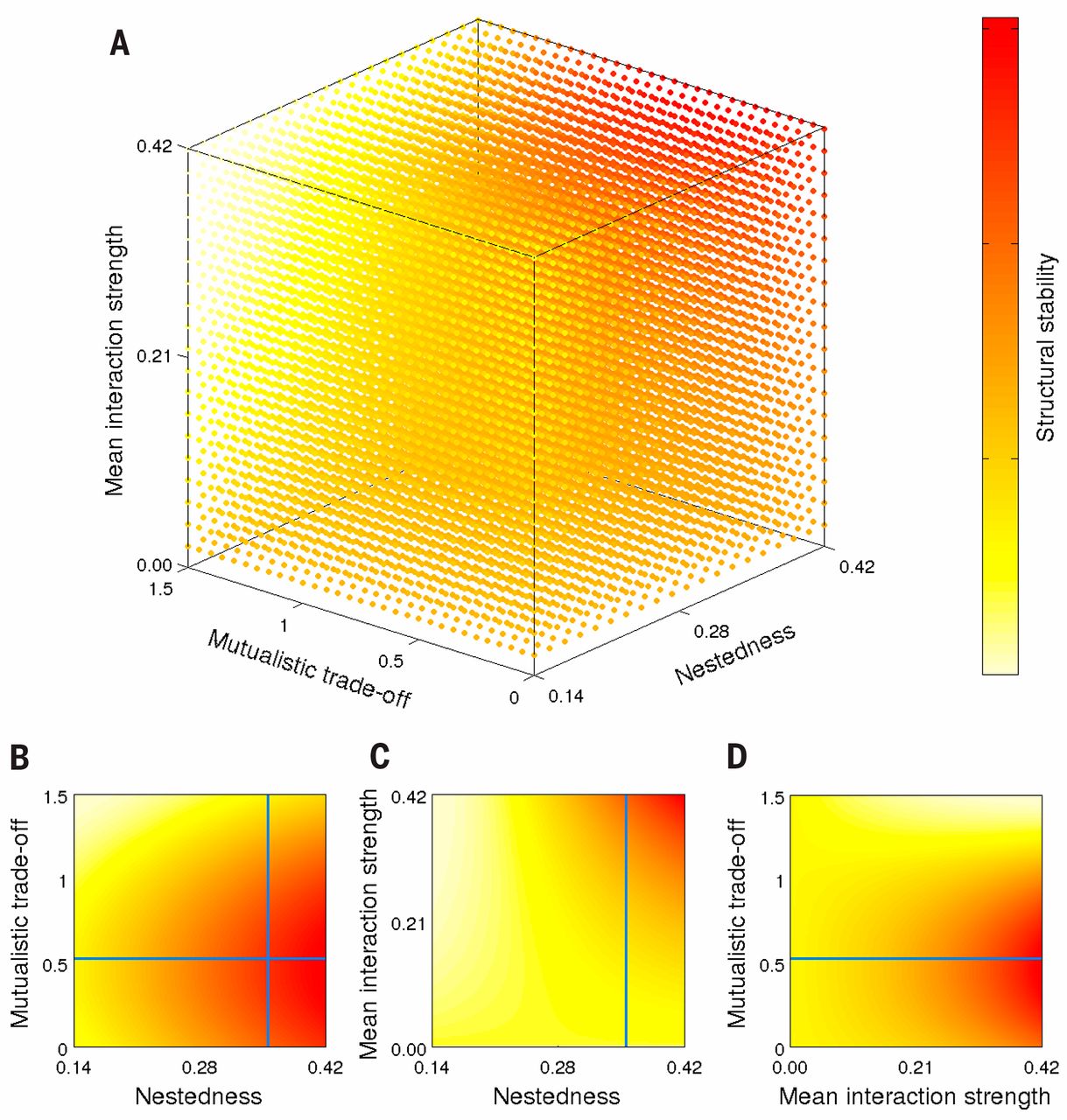}
\caption{The overall contribution of nestedness, mutualistic interaction strength, and mutualistic trade-off to the fraction of surviving species. The larger such contribution (determined by Eq.~\eqref{partial}), the larger the region of the intrinsic growth rate space which is compatible with the stable co-existence of species. The results on this empirical network (mutualistic interactions located in the grassland asclepiads in South Africa -- results for different datasets are qualitatively similar) show that more stable structures tend to be associated with larger nestedness, larger interaction strength, and smaller mutualistic trade-off. The panels below show two-dimensional slices of the three-dimensional feature space above. Reprinted from \cite{rohr2014structural}.}
\label{fig:structural_stability}
\end{figure}

The structural stability framework allows us to address one of our original questions: what is
the impact of nestedness on the feasibility of a given system?
Based on our above considerations, to address this question for a given empirical network, one can generate model networks with different
level of nestedness through a resampling procedure that preserves the total number of species 
and the expected number of interactions, and then study how the probability $\omega$ that a randomly-selected species is
feasible is affected by nestedness~\cite{rohr2014structural}.
This probability $\omega$ tends to be smaller the closer we are to the boundaries of the feasibility domain. To control for this possible confounding factor, first, 
Rohr \textit{et al.}~\cite{rohr2014structural} determined the center $\{\vek{\alpha}_{S}^{(A)},\vek{\alpha}_{S}^{(P)}\}$ of the feasibility domain. 
This vector is the vector that can tolerate the maximal perturbation before leaving the feasibility domain, and it can 
be determined analytically -- see the Supplementary Material of~\cite{rohr2014structural} for details. 
Second, they quantified the deviation of a given set of intrinsic growth rates from the structural
vector of intrinsic growth rates.  Intuitively, this deviation $\eta$ (which is an angle in the intrinsic 
growth-rates space) quantifies how far a set of intrinsic growth rates is from the ``center'' of the 
feasibility domain, and it can be used as a control variable when assessing the impact of 
network topology on the feasibility domain size.

Rohr \textit{et al.}~\cite{rohr2014structural} studied the impact of various factors on 
the fraction of surviving species: deviation from the structural vector of plants $\eta_{P}$ and 
animals $\eta_A$, network nestedness $\mathcal{N}$ (as determined by NODF), mutualistic 
trade-off $\delta$ (see Eq.~\eqref{tradeoff} for the definition) and mean 
interaction strength $\overline{\gamma}$ ($\overline{\gamma}=\sum_{ij}\gamma_{ij}$). 
Based on these variables, one can consider a generalized linear model for the probability $\omega$ that a
randomly selected species $i$ is feasible (i.e., $N_i^*>0$) defined by the equation~\cite{rohr2014structural}
\begin{equation}
\log{\Biggl(\frac{\omega}{1-\omega}\Biggr)}\sim \log{(\eta_A)}+\log{(\eta_P)}+\overline{\gamma}+\overline{\gamma}^2+\overline{\gamma}\,\mathcal{N}+\overline{\gamma}\,\mathcal{N}^2+\overline{\gamma}\,\delta+\overline{\gamma}\,\delta^2,
\label{regression}
\end{equation}
where the quadratic terms account for possible non-linear effects. 
The common factor $\overline{\gamma}$ in the contributions of nestedness and 
mutualistic trade-off reflects the fact that if the mutualistic interaction 
strength is zero, then nestedness and mutualistic trade-off cannot contribute to the fraction of surviving species.
The joint impact of nestedness, mutualistic-trade off, and mutualistic interaction strength
on the feasibility domain size is quantified through the partial fitted values
\begin{equation}
SS=\hat{\beta}_1 \,\overline{b}+\hat{\beta}_2\,\overline{b}^2+\hat{\beta}_3 \,\overline{b}\,\mathcal{N}+\hat{\beta}_4\,\overline{b}\,\mathcal{N}^2+\hat{\beta}_5 \,\overline{b}\,\delta+\hat{b}_6 \,\overline{\gamma}\,\delta^2.
\label{partial}
\end{equation}
where $\hat{b}_1,\dots\,\hat{b}_6$ are the coefficient of the respective terms that maximize 
the likelihood of the model described by Eq.~\eqref{regression}. The larger this contribution
is, the larger the size of the feasibility domain as a consequence of these factors.

The results by Rohr \textit{et al.}~\cite{rohr2014structural} (see Fig.~\ref{fig:structural_stability}) show that 
a larger degree of nestedness tends to be associated with a larger feasibility domain. The most structurally stable systems are characterized by maximal nestedness, small mutualistic trade-off, high mutualistic-interaction strength (compatible with the constraint that all feasible solutions are 
globally stable)~\cite{rohr2014structural}. Besides, for each empirical network, one can determine the largest theoretical feasibility domain, and compare it with the feasibility domain of the observed networks.
The results by Rohr \textit{et al.} (Fig. 6 in \cite{rohr2014structural}) suggest that the structure of observed 
empirical networks tends to maximize the size of the feasibility domain.

While these results point out the positive impact of nestedness on the size of the feasibility domain, 
a larger feasibility domain is not necessarily associated with a more globally stable system. It turns out
that while nestedness is indeed associated with larger feasibility domains~\cite{rohr2014structural,saavedra2016nested},
it also tends to decrease the global stability of the system (i.e., the upper value of mutualistic strength $\hat{\gamma}$
up to which global stability is guaranteed is negatively correlated with nestedness, see Fig.~2 in~\cite{saavedra2016nested}). 
These findings suggest that ``nestedness tends to promote feasibility over stability"~\cite{saavedra2016nested}.
Hence, it is natural to ask: 
does the dynamics of real communities tend to promote feasibility or stability in the long term? To 
address this question, Saavedra \textit{et al.}~\cite{saavedra2016nested} analyzed a time-stamped empirical dataset of an
Arctic pollinator community located in Greenland (1996-1997, \cite{olesen2008temporal}).
Their results indicate that newcomer species in the system tend to promote feasibility and nestedness over stability.
Saavedra \textit{et al.}~\cite{saavedra2016nested} stress that the assumption that dynamical stability is essential to the persistence of a community
might not be justified, and feasibility might be a more fundamental property for species co-existence. We expect additional empirical studies to further address this fundamental point for our understanding of the persistence of ecological communities.

\subsection{Local stability of mutualistic systems in the random-matrix theory approach}
\label{sec:rmt}

The results presented in Section~\ref{sec:population} concerned the impact of nestedness on the size of the parameter regions that are compatible with feasible and globally stable equilibrium points.
They were obtained by studying a population dynamics model.
The need for a detailed knowledge of the dynamics is bypassed by the random-matrix theory approach. In this alternative approach, one generates a random community matrix $\mat{M}$ with some given properties, and studies the local stability~\cite{allesina2012stability} and feasibility~\cite{grilli2017feasibility} of its equilibrium points. This means that instead of postulating a dynamics and deriving its community matrix $\mat{M}$ through Eq.~\eqref{jacobian}, one directly generates $\mat{M}$ and studies its properties. 

The advantage of this approach is that it allows us to estimate analytically the stability conditions for a variety of interaction topologies, including a random, a mutualistic, and a nested topology -- the corresponding results are presented in Sections \ref{sec:random_mat}, \ref{sec:random_mut}, and \ref{sec:random_nest}, respectively.
As this review focuses on nestedness, we will not cover the results for other interaction topologies (such as predator-prey interactions, competitive matrices, etc.); we refer the interested reader to~\cite{allesina2012stability} for these results.

While random-matrix theory allows for various analytic developments, it is important to be aware of its assumptions. Bascompte and Ferrera~\cite{bascompte2018structural} pointed out that the generality of the results obtained in~\cite{allesina2012stability} is limited by two factors: (1) The mutualistic bipartite matrices studied in~\cite{allesina2012stability} (see Section~\ref{sec:random_mut}) do not feature interspecific competition which is instead a critical component of mutualistic systems~\cite{bascompte2018structural}; (2) The local-stability criterion adopted in~\cite{allesina2012stability} is purely based on the community matrix's eigenvalue $\lambda^{M}$ with the largest real part, which provides us with a necessary condition for systemic instability. On the other hand, it is more complex to determine the system's \emph{distance-to-instability} -- i.e., the maximal perturbation
 that the system can absorb before losing local stability -- which also depends, in general, on the type of perturbation~\cite{bascompte2018structural}. Nevertheless, random-matrix theory plays a central role for many analytic results on the impact of complexity and network structure on stability and feasibility~\cite{allesina2012stability,allesina2015stability,grilli2017feasibility}, which is why it is important for the reader to understand its rationale and basic elements.

\subsubsection{Stability-complexity relation: May's result for random matrices}
\label{sec:random_mat}

In the random-matrix theory approach, the first step is the generation of a random $S\times S$ community matrix $\mat{M}$, where $S$ is the number of species in the system.
Each element $M_{ij}$ of the community matrix is, therefore, a random variable $X$ with mean $\braket{X}$ and standard deviation $\sigma$.  
We further denote by $C$ the matrix connectance, and by $-d$ the value of the diagonal elements.
The simplest topology is arguably the random one~\cite{may1972will}. To generate a random-topology matrix, we extract, for each off-diagonal matrix element $M_{ij}$ ($i\neq j$), a random number $p$ from the uniform distribution $\mathcal{U}[0,1]$. If $p\leq C$, we extract a random number $X$ from $\mathcal{N}(0,\sigma^2)$, and set $M_{ij}=X$. By contrast, if $p>C$, we set $M_{ij}=0$. The diagonal terms are fixed: $M_{ii}=-d$ for all species $i$.

As we have seen in Section~\ref{sec:feasibility}, the equilibrium points of the dynamics ruled by a community matrix $\mat{M}$ is locally stable if the real part of all $\mat{M}$'s eigenvalues is negative.
One can prove\footnote{This results is a consequence of Girko's circular law~\cite{girko1985circular,tao2010random}. Let us start by setting all the diagonal elements $d$ equal to zero. For any distribution of $X$ with zero mean and finite variance $\sigma^2<\infty$, in the limit $S\to\infty$, the eigenvalues of $\mat{M}$ are then uniformly distributed inside the circle of radius $\sigma\,\sqrt{S\,C}$ centered at $(0,0)$ in the complex plane. Non-zero diagonal elements $-d$ shift the circle's center from $(0,0)$ to $(-d,0)$. In order to have all the eigenvalues' real parts smaller than zero, we need the radius of the circle to be smaller than d, i.e., $\sigma\,\sqrt{S\,C}<d$.} that for the random matrices considered in this Section, this holds with high probability when~\cite{allesina2012stability}
\begin{equation}
\sqrt{S\,C}<\theta=\frac{d}{\sigma}.
\label{stability_may}
\end{equation}
This result poses a constraint to the number of species to a system that aims to be stable against small external perturbations. If, for instance, $C=0.1, X\sim\mathcal{N}(0,\sigma^2),\theta=2$, Eq.~\eqref{stability_may} poses $S<41$~\cite{allesina2012stability} as the necessary condition for local stability.
Besides, Eq.~\eqref{stability_may} suggests a stability-complexity dualism: complexity (to be interpreted as a large number $S$ of interacting species with a large fraction $C$ of interacting species) begets instability. 

Derived in 1972 by Robert May~\cite{may1972will}, this celebrated result has deeply influenced subsequent developments in theoretical ecology. 
Subsequent works have explored the stability-complexity relationship in more complex models of interaction~\cite{pimm1979complexity,allesina2012stability,stone2016google} and, more recently, the relation between feasibility and complexity~\cite{grilli2017feasibility}. For example, Allesina and Tang have generalized these relations to competitive, mutualistic, predator-prey, and mixed interactions; Stone~\cite{stone2016google} has generalized them to competition matrices where each element has an additional competition term that is uniform across the nodes, and to mutualistic systems.
Reviewing the literature on the stability-complexity relationship falls out of the scope of this article; in the following, we focus on the impact of interaction topology on the stability-complexity relationship for bipartite mutualistic interactions.

\subsubsection{Stability of mutualistic networks}
\label{sec:random_mut}

The random matrices by May are highly unrealistic: each pair of species interacts with the same probability. For sufficiently large matrices, this randomness results in fixed frequencies for each type of interaction: for example, predator-prey interactions ($M_{ij}>0$ and $M_{ji}<0$) are twice as frequent as mutualistic ($M_{ij}>0$ and $M_{ji}>0$) and competitive ($M_{ij}<0$ and $M_{ji}<0$) interactions -- see~\cite{allesina2012stability} for details.
Forty years after May's seminal work~\cite{may1972will}, Allesina and Tang~\cite{allesina2012stability} generalized his result to different interaction topologies.
We focus on their results for random (bipartite) mutualistic networks and nested networks, respectively. The comparison between the stability conditions for the two topologies will allow us to gauge the impact of nestedness on the system's local stability.

To generate mutualistic monopartite matrices, for each pair $(i,j)$ of species ($i\neq j$), Allesina
and Tang~\cite{allesina2012stability} extracted (independently) $M_{ij}$ and $M_{ji}$ from the half-normal distribution $|\mathcal{N}(0,\sigma^2)|$ with probability $C$. The diagonal elements $M_{ii}$ are always set to $-d$ ($d>0$) which represents intraspecific competition. Mutualistic bipartite matrices are constructed in a similar way, with the difference that the $S$ species are first split into two equally-sized groups. In bipartite networks, only pairs of species that belong to different groups are connected. The probability to connect a pair of species is given by $\rho=2\,C\,(S-1)/S$: this choice preserves the expected connectance $C$ of the network. It is essential noticing that both the unipartite and the bipartite mutualistic systems only consider intraspecific competition ($M_{ii}$) and mutualism ($M_{ij}>0$ when $i\neq j$), but not interspecific competition.

The stability condition for monopartite mutualistic matrices $\mat{M}$ is derived based 
on two observations: (1) the local stability of the matrix is achieved when the largest 
real part of $\mat{M}$'s eigenvalues is negative; (2) the eigenvalue $\lambda_1^M$ of $\mat{M}$ 
with the largest real part is approximately equal to the row-sum $\sum_j M_{ij}$ of the community matrix\footnote{Such result 
follows from the property that $E[M_{ij}]=C\,E[|X|]$ for $i\neq j$ and, therefore, $\sum_{j=1}^{S}M_{ij}\simeq -d+(S-1)\,C\,E[|X|]$. 
In the matricial notation, the previous equation can be rewritten as $\mat{M}\,\vek{e}=(-d+(S-1)\,C\,E[|X|])\,\vek{e}$, 
where $\vek{e}$ is the vector whose elements are all equal to one. This means that $\vek{e}$ is an eigenvector of $\mat{M}$ 
with eigenvalue $-d+(S-1)\,C\,E[|X|]$; one can further prove that this eigenvalue is the one with the largest real 
part, i.e., $\lambda_1^M\simeq-d+(S-1)\,C\,E[|X|]$. We refer to the Supplementary Material of~\cite{allesina2012stability} for more details.} 
\begin{equation}
\lambda_1^{M}\simeq \sum_{j}M_{ij}\simeq -d+(S-1)\,C\,E[|X|].
\end{equation}
The stability criterion is, therefore,~\cite{allesina2012stability}
\begin{equation}
\frac{(S-1)\,C\, E[|X|]}{\sigma}<\theta.
\end{equation}
In the numeric example above ($C=0.1, X\sim\mathcal{N}(0,\sigma^2),\theta=2$), we obtain that the 
upper bound to the community's size is given by $S< 16$: random mutualistic matrices are less likely to be locally stable than random unstructured matrices. 
The stability condition for mutualistic bipartite networks is the same as the one for mutualistic monopartite networks.

\subsubsection{Stability of nested mutualistic networks}
\label{sec:random_nest}

\begin{figure}[t]
\centering
\includegraphics[scale=0.8]{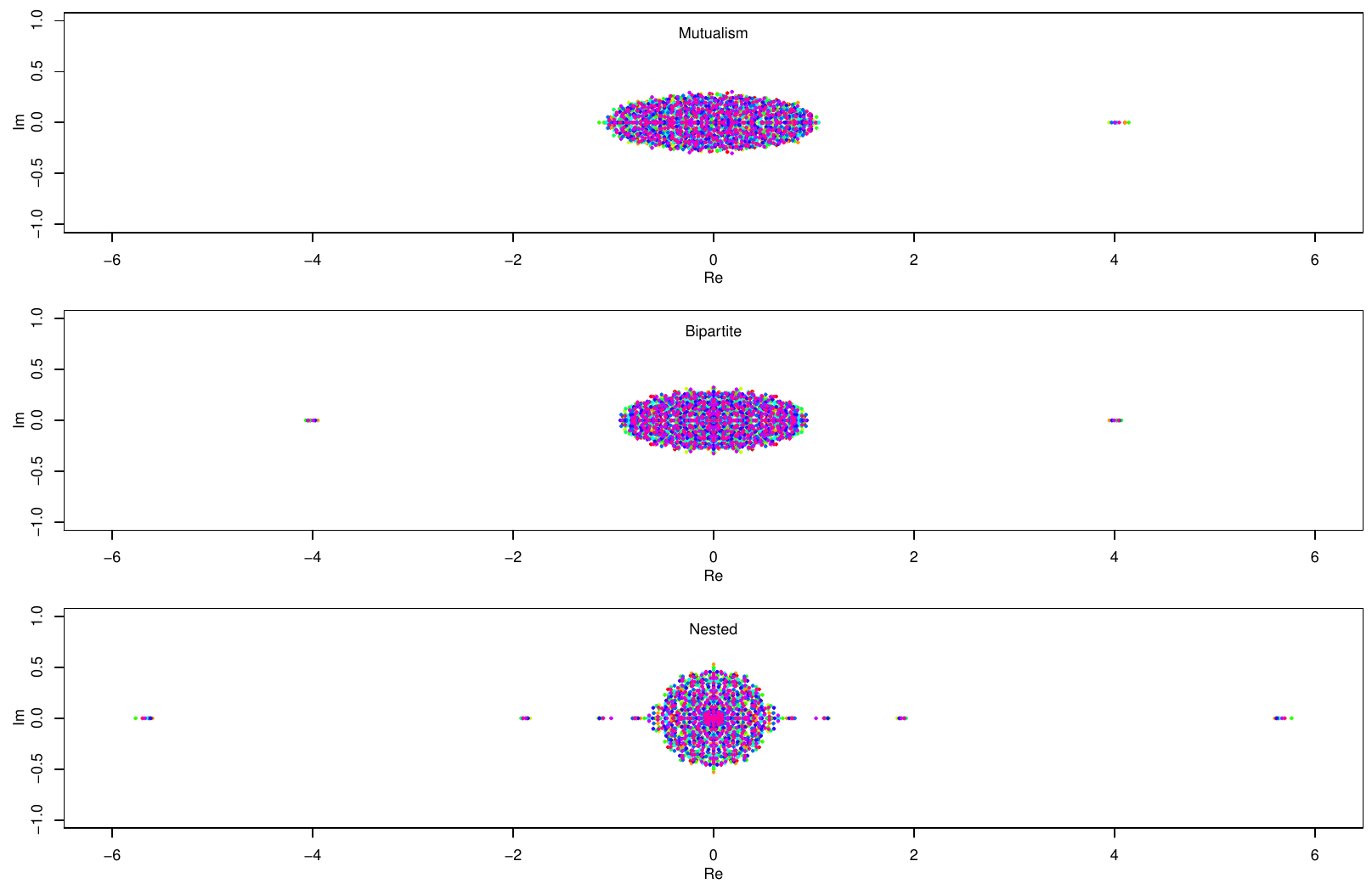}
\caption{Local stability in the random-matrix theory approach: real and imaginary parts of the eigenvalues of the community matrix $\mat{M}$ for an unstructured interacting population (a), a mutualistic bipartite network (b), and a nested network of interactions (c). Nested interactions have eigenvalues with larger real parts than unstructured and mutualistic interactions.
Reprinted from~\cite{allesina2012stability}.}
\label{fig:allesina2}
\end{figure}

Allesina and Tang~\cite{allesina2012stability} also investigated the impact of nestedness on the 
local stability of a mutualistic community. To generate nested mutualistic 
matrices $\mat{M}$, they fixed the number of links $E$ necessary to achieve the desired
connectance $\rho=C$. Then, they ran a filling algorithm that first fills the first row 
and column of the community matrix, and then arranges the remaining links in such
a way that the matrix is perfectly nested. The filled elements $M_{ij}$ of the
community matrix are then assigned a value that is randomly extracted from the 
half-normal distribution $|\mathcal{N}(0,\sigma^2)|$; the diagonal elements are again set to $-d$ ($d>0$).

Allesina and Tang~\cite{allesina2012stability} argued that in nested matrices, stability
is undermined by the existence of generalists species that yield larger-than-average 
row-sum $\sum_{j}M_{ij}$. Their numerical findings confirmed this intuition: nested 
bipartite matrices have larger eigenvalues than unstructured bipartite mutualistic 
matrices (see Fig.~\ref{fig:allesina2}). In the numeric example 
above ($C=0.1, X\sim\mathcal{N}(0,\sigma^2),\theta=2$), the upper bound 
to the community size is given by $S<10$: nested mutualistic matrices are 
less likely to be locally stable than unstructured mutualistic 
matrices~\cite{allesina2012stability}.

An analytic explanation of the local instability of nested structures under the random-matrix approach can be obtained by exploiting
the connection between perfectly nested structures and the maximal spectral radius of the adjacency matrix~\cite{staniczenko2013ghost}
(see Section~\ref{sec:eigenvalue}). The community matrices $\mat{M}$ introduced above can indeed be represented as
the sum of an interaction matrix $\mat{M}'$ with zero diagonal elements, and a diagonal matrix $\mat{D}$.
For large matrices, the spectral radius of $\mat{M}$ is given by $\rho(\mat{M})=\rho(\mat{M}')-d$, where $d$ is the 
average of the elements of $\mat{D}$~\cite{allesina2012stability};
the system is therefore locally stable only if $\rho(\mat{M}')<d$. As perfectly nested structures exhibit the largest possible spectral radius
(see Section~\ref{sec:eigenvalue}), they are also yield the equilibrium states that are the most unstable against small 
perturbations~\cite{staniczenko2013ghost}.

It is important to keep in mind that this result builds on the hypothesis that there are no interspecific competitive interactions. Analytic results on the  metapopulation model defined by Eq.~\eqref{mutualistic_model} also indicate that mutualistic interactions tend to destabilize the system as interspecific competitive interactions vanish~\cite{bastolla2009architecture,bascompte2018structural}; however, introducing interspecific competitive interactions leads to a substantially different scenario, where nestedness has a positive impact on biodiversity~\cite{bastolla2009architecture}. The careful reader needs to keep in mind that different assumptions on the nature of interactions lead to different conclusions on the impact of network topology on stability and biodiversity.

\subsection{Bottom-line: What is the impact of nestedness on robustness, feasibility, and stability?}
\label{sec:bottom2}

To summarize, scholars have investigated the implications of nestedness for various systemic properties related to stability and persistence. Importantly, conclusions obtained by the corresponding studies depend critically on whether interspecific competition is considered, whether feasibility (and not only stability) is ensured in analytical approaches, whether the role of model parameterization is taken into account, and the dimension of stability (linear stability, resilience, local stability) considered. 
 We summarize below some of the main conclusions:

\begin{itemize}
 \item \textbf{Topological robustness} [Section~\ref{sec:robustness}]. More nested networks are more (less) robust when less (more) connected nodes are more likely to fail and be removed from the system~\cite{burgos2007nestedness}. In bipartite networks, for simple co-extinction cascade processes, rankings of the nodes that produce more nested arrangements of the adjacency matrix better reproduce the structural importance (or vulnerability) of the nodes in both mutualistic networks~\cite{dominguez2015ranking} and country-product networks~\cite{mariani2015measuring}.
 \item \textbf{Feasibility in systems with competition and mutualism} [Section~\ref{sec:population}]. In systems with both competition and mutualism, nestedness enhances the increase of biodiversity due to mutualism~\cite{bastolla2009architecture}, and more nested networks are associated with larger feasibility domains, i.e., the network's nodes can co-exist for a larger range of environmental conditions~\cite{rohr2014structural,saavedra2016nested}.
 \item \textbf{Local stability in mutualistic systems without interspecific competition} [Section~\ref{sec:rmt}]. For purely mutualistic systems (i.e., systems without interspecific competition), under the assumptions of the random-matrix theory approach~\cite{allesina2012stability}, the maximal size for a nested network such that the system is locally stable is smaller compared to that for a bipartite, unstructured mutualistic network~\cite{allesina2012stability}. Even more, when one considers interaction weights, perfectly nested structures (i.e., topologies with the largest spectral radius) are maximally unstable~\cite{staniczenko2013ghost}.
\end{itemize}

Beyond the presented results, scholars have also found that more centralized topologies can attenuate the reach of propagating perturbations in the network~\cite{suweis2015effect}.
A rich scenario emerges where the impact of nestedness depends critically on the systemic property one is interested in.
This constellation of potentially contradictory effects of nestedness for different systemic properties (e.g., for systems with competition and mutualism, nestedness tends to favor feasibility but penalize global stability) makes it essential to assess the empirical relevance of different ways of looking at robustness for the actual persistence of real communities.
A recent study~\cite{saavedra2016nested} indicated that at short timescales, empirically observed network dynamics tends to promote feasibility and not stability, yet more research is needed in this direction.
We conclude by mentioning that while Saavedra \textit{et al.}~\cite{saavedra2014structurally} have already extended the structural stability framework to socio-economic networks, this avenue of research remains still largely unexplored.

\clearpage 

\section{Observing nestedness at the mesoscopic scale}
\label{sec:mesoscopic}

\begin{figure}[t]
\centering
\includegraphics[scale=0.6]{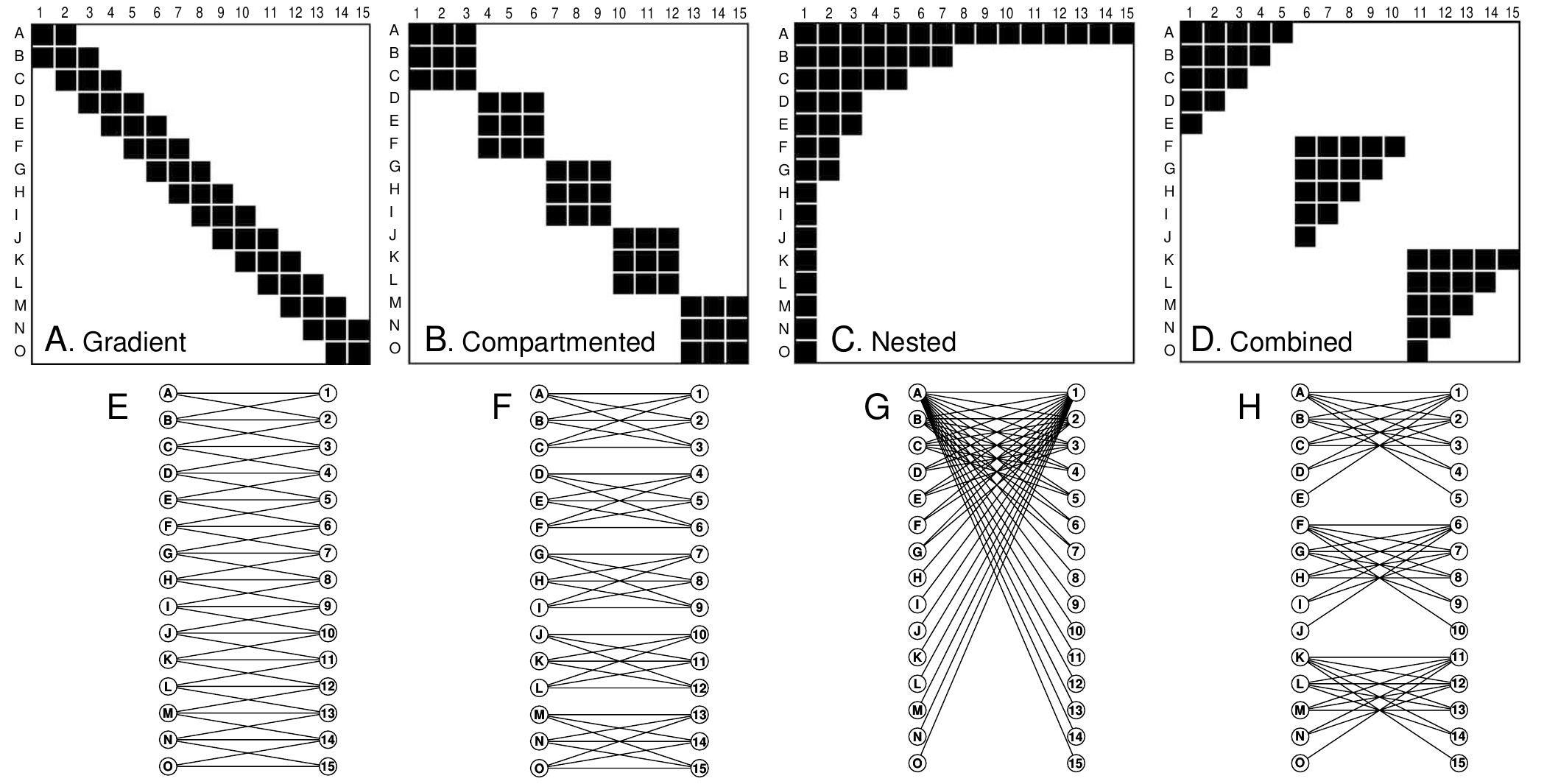}
\caption{Illustration of different topological patterns. \emph{Panels A-E} illustrate a gradient structure where each node only interacts with adjacent nodes in the matrix. While these patterns have been investigated in vegetation ecology~\cite{lewinsohn2006structure}, they are rarely found in plant-animal interaction networks and socio-economic networks. \emph{Panels B-F} illustrate a ``compartmented"~\cite{raffaelli1992compartments,dicks2002compartmentalization} (or ``modular"~\cite{fortunato2010community}) structure where the network can be partitioned into well-defined ``compartments", or ``modules", or ``blocks", or ``communities". There exist many strategies to detect this kind of structures~\cite{fortunato2010community,fortunato2016community}; extant detection techniques usually do not take into account the possible internal topology of the compartments. \emph{Panels C-G} illustrate a (perfectly) nested structure, which is the main focus of this review and is defined in the Introduction section. \emph{Panels D-H} illustrate a ``combined'' (or ``nested-modular"~\cite{beckett2013coevolutionary}, or ``in-block nested"~\cite{sole2018revealing}) structure where the network is compartmented, but each compartment (or ``block") exhibits an internally nested structure.
Reprinted from~\cite{lewinsohn2006structure}.}
\label{fig:all_the_structures}
\end{figure}

In the previous Sections, we have studied nestedness at a \emph{macroscopic} level by defining and measuring it as a property related to the connectivity patterns of \emph{all} nodes (see Figs.~\ref{fig:all_the_structures}C-G).
Besides, we have pointed out that ecological and socio-economic networks can exhibit other interesting structural patterns. Some of them (like heterogeneous degree distribution, disassortativity, core-periphery structure) are often observed together with nestedness (see Section~\ref{sec:network_properties}), whereas others -- like gradient structure (Figs.~\ref{fig:all_the_structures}A-E), modular or compartmental structure (Figs.~\ref{fig:all_the_structures}B-F) -- seem to be incompatible with nestedness.

On the other hand, real ecological processes that lead to the observed structural patterns are typically confined by inherent boundaries due to diverse reasons~\cite{lewinsohn2006structure}. For example, for ecological datasets that include observations made in distinct spatial regions, it is interesting to assess not only the degree of nestedness of the complete network, but also the degree of nestedness of subgraphs that correspond to the different geographical regions~\cite{fattorini2007non,strona2013protocol}. Indeed, variations in nestedness across different geographical locations might reflect different geographical histories of the subgroups of species that inhabit them~\cite{fattorini2007non}.

Boundaries and constraints play a significant role in socio-economic systems as well. For instance, the probability that two countries have a trade relationship decreases with their geographical distance~\cite{fagiolo2010international}, and geographical effects may affect the compartmentalized structure of trade~\cite{zhu2014rise} and communication networks~\cite{expert2011uncovering}. Both in ecological and socio-economic systems, if the mechanisms that drive the internal dynamics inside network compartments is ruled by one of the processes that lead to nested structures (see Section~\ref{sec:emergence}), one might observe as a result ``combined'' compartmentalized structures where each compartment exhibits, internally, a nested structure (see Figs.~\ref{fig:all_the_structures}D-H). 

Lewinshon \textit{et al.}~\cite{lewinsohn2006structure} already recognized the potential importance of the combined structure depicted in Figs.~\ref{fig:all_the_structures}D-H, emphasizing that processes that generate ecological networks typically ``operate
within a framework of boundaries set by morphological,
functional or phylogenetic constraints", which implies that ``most species, but not necessarily all, will preferentially establish links in a given compartment and, within that
compartment, their host range or fauna will  be  conditioned by more proximate factors"~\cite{lewinsohn2006structure}.
Lewinshon \textit{et al.}~\cite{lewinsohn2006structure} applied correspondence analysis techniques to discriminate between the different structural patterns showed in Fig.~\ref{fig:all_the_structures} -- we refer to~\cite{lewinsohn2006structure} for details.
Subsequent studies used standard nestedness and community-detection tools~\cite{fortunato2016community} to measure the level of nestedness for different geographical subregions~\cite{fattorini2007non}, to assess the overall correlation between nestedness and modularity~\cite{fortuna2010nestedness}, and to measure the degree of nestedness of the modules detected by modularity-maximization algorithms~\cite{flores2013multi,beckett2013coevolutionary}.

Here, we refer to \emph{nestedness at the mesoscopic scale} whenever nestedness is studied not as a property of the whole network (macroscopic scale), but of one or more than one of its subgraphs.
Methodologies specifically aimed at detecting nested patterns at the mesoscopic scale have been introduced only in the last  five years~\cite{strona2013protocol,grimm2016detecting,grimm2017analysing,sole2018revealing,kojaku2017finding}, in relation to three main problems: (1) quantifying the level of nestedness of a given subgraph~\cite{strona2013protocol} (see Section~\ref{sec:subgraph}): (2) detecting, for a given network, the largest subgraph that exhibits an internally nested topology~\cite{grimm2016detecting,grimm2017analysing} (see Section~\ref{sec:largest}); (2) partitioning the network into a set of compartments, such that each compartment exhibits, internally, a nested~\cite{sole2018revealing} or a core-periphery-like~\cite{kojaku2017finding,kojaku2018core} topology of interactions (see Section~\ref{sec:detecting_ibn}).

In this Section, we review the recent progress that has been made in relation to these three problems.
In spite of the massive amount of works on nestedness, our understanding of nestedness at the mesoscopic scale is still at its infancy.
Nevertheless, the obtained results in this direction suggest that future research might shift the focus from detecting system-wide, ``macroscopic'' nestedness to detecting confined, ``mesoscopic'' nested structures. Detecting these structures, understanding the mechanisms behind their emergence and their implications may lead to new insights on the organizing principles of ecological and socio-economic networks.

\subsection{Quantifying the degree of nestedness of a subgraph}
\label{sec:subgraph}

Given two subgraphs $\mathcal{S}_1 \subset \mathcal{G}$ and $\mathcal{S}_2 \subset \mathcal{G}$ of a network $\mathcal{G}$ of interest, one might be tempted to assess their relative level of statistical significance by simply comparing the $z$-scores of their degree of nestedness (as determined by a suitable nestedness metric, like NODF or matrix temperature). However, in principle, the resulting $z$-scores may depend on matrix size.
To overcome this obstacle, Strona \textit{et al.}~\cite{strona2013protocol} developed a procedure to compare the levels of nestedness of subgraphs of different size, and subsequently applied it to detect structural differences between Anatolian and Cyclades islands in the Aegean islands.

Suppose that we are interested in assessing the statistical significance of nestedness for a given subgraph $\mathcal{S}\subset\mathcal{G}$. To fix ideas, we consider NODF as our nestedness metric, yet in principle, the steps below apply to any nestedness metric.
The procedure by~\cite{strona2013protocol} requires five steps (see Fig. 1 in~\cite{strona2013protocol} for a diagrammatic representation of the procedure): (1) We calculate the overall level of nestedness of the network $\mathcal{G}$, i.e., the $z$-score of NODF (based on the expected value and the standard deviation of NODF in an ensemble of suitably randomized networks); (2) We randomly partition the network into matrices of random size, and we calculate the $z$-score of NODF for each of them (in the same way as in (1)); (3) We regress the obtained subgraphs' $z$-score against subgraph size; (4) The obtained regression coefficients determine, for each subgraph size, the expected NODF for subgraphs of that size; (5) The observed $z$-scores for any subnetwork can be compared with the random expectation for subgraphs of the same size. We refer to~\cite{strona2013protocol} for additional details on these five steps.

The procedure outlined above can be used to detect differences between different subgraphs of ecological interest.
Strona \textit{et al.}~\cite{strona2013protocol} applied it to species distribution patterns in the Aegean islands, a group of islands in the Aegean sea. Interestingly, based on ecological considerations~\cite{fattorini2007non}, the Aegean islands can be divided into two groups: Anatolian and Cyclades. The procedure by Strona at al.~\cite{strona2013protocol} revealed that the Cyclades and the Anatolian islands tend to be substantially more and less nested, respectively, than expected for a randomly-extracted subgraph of equal size. Importantly, as different processes can lead to different degrees of nestedness in the resulting species distribution patterns, variation in nestedness across subgraphs indicate that different parts of the systems have undergone different ecological processes~\cite{fattorini2007non,strona2013protocol}.

\subsection{Finding the largest nested subgraph}
\label{sec:largest}

Besides evaluating the degree of nestedness of specific subgraphs, one might be interested in finding the largest subgraph (referred to as \emph{nested component}~\cite{grimm2016detecting,grimm2017analysing}) that exhibits a significantly nested structure.
The relevance of this problem stems from the fact that, with increased data availability or longer observational periods, not necessarily all nodes in the dataset considered are to be part of a nested arrangement. Furthermore, traditional approaches to nestedness detection are top-down: It is up to the researchers to determine the realm of observation, in some cases in an arbitrary manner. Instead, it might be more appropriate to include all the available data into the analysis, and determine algorithmically the largest subset of the data that exhibit a nested structure. 
This problem has been recently tackled by Grimm and Tessone~\cite{grimm2016detecting,grimm2017analysing} who introduced an algorithm called NESTLON (NESTedness detection based on LOcal Neighborhood).
Below, we describe the algorithm for unipartite networks, yet the algorithm can be applied to bipartite networks as well.

Essentially, NESTLON sequentially ``constructs'' the nested component $\Gamma_{nest}\subseteq \mathcal{G}$. The nested component is initialized with an empty set of nodes. We start from the node $i_1$ with the largest degree\footnote{To simplify the discussion, let us assume that the node with degree equal to the largest degree $k_{max}$ in the network is unique. The extension to networks where more than one node have a degree equal to $k_{max}$ is straightforward~\cite{grimm2017analysing}.}; this node is a candidate to belong to the nested component. The main idea of NESTLON is to assess whether the neighborhood of this node includes the neighborhood of lower-degree nodes (up to a ``confirmation ratio'' $\theta_{nest}$ which allows for a tunable level of deviation from a perfectly nested structure). If this is true for a sufficiently large fraction of $i_1$'s neighbors (where such minimum fraction is a parameter of the algorithm), then the neighborhoods of $i_1$'s neighbors are nested inside $i_1$'s neighborhood (up to an acceptable level), and $i$ is added to the nested component, which now includes one node. $i_1$'s neighbors are now candidates for the same evaluation: one considers their neighbors, and assess whether, for a sufficient fraction of them, their neighborhoods are sufficiently nested inside $i_1$'s neighborhood.
If the neighborhoods of $i_1$'s neighbors are not sufficiently nested inside $i_1$'s neighborhood, then we do not add $i_1$ to the nested component, and we evaluate the neighborhood of the neighbors of the node $i_2$ with the second largest degree. The procedure continues iteratively until it is not possible to add more nodes to the nested component.

Clearly, in a perfectly nested network, $\Gamma_{nest}=\mathcal{G}$. For networks that are not perfectly nested, NESTLON provides us not only with the nested component, i.e., with the largest subgraph that exhibits a nested structure, but also a new metric for nestedness: the relative size $\mu_{nest}=|\Gamma_{nest}|/N$ of the nested component. Indeed, $\mu_{nest}=1$ corresponds to a perfectly nested network, whereas $\mu_{nest}=0$ indicates a maximally non-nested network where no pair of nodes respects the nestedness condition. Results on synthetic networks suggest that this metric may be better suited than NODF and BINMATNEST for nestedness quantification in sparse and high-density matrices. 
It is worth noticing that the algorithm operates based on local information, which makes the algorithm scalable with system size.

\subsection{Detecting in-block nestedness}
\label{sec:detecting_ibn}

Think to the communication patterns of the students in a given school. Likely, the students of a given class tend to communicate more often with the students of the same class than with the students of a different class.
Let us assume that the communication patterns within a class are nested -- i.e., the friends of a given student $i$ tend to form subsets of the friends of students who have more friends than $i$.
This pattern of interactions can be described as an \emph{in-block nested structure}: differently from nested and modular structures, in-block nested structures are characterized by compartments of nodes that \textit{internally} exhibit a nested pattern of interactions (Fig.~\ref{fig:all_the_structures}D-H). 

In relation to the previous example, we have no a priori guarantee that the communication patterns within the individual classes are nested. A natural question emerges: how to properly quantify the level of in-block nestedness in a given network?
To address this question, we need to simultaneously assess two properties: (1) How well the network can be partitioned into compartments; (2) How nested is the communities' internal topology.

\begin{figure}[t]
\centering
\includegraphics[scale=0.8]{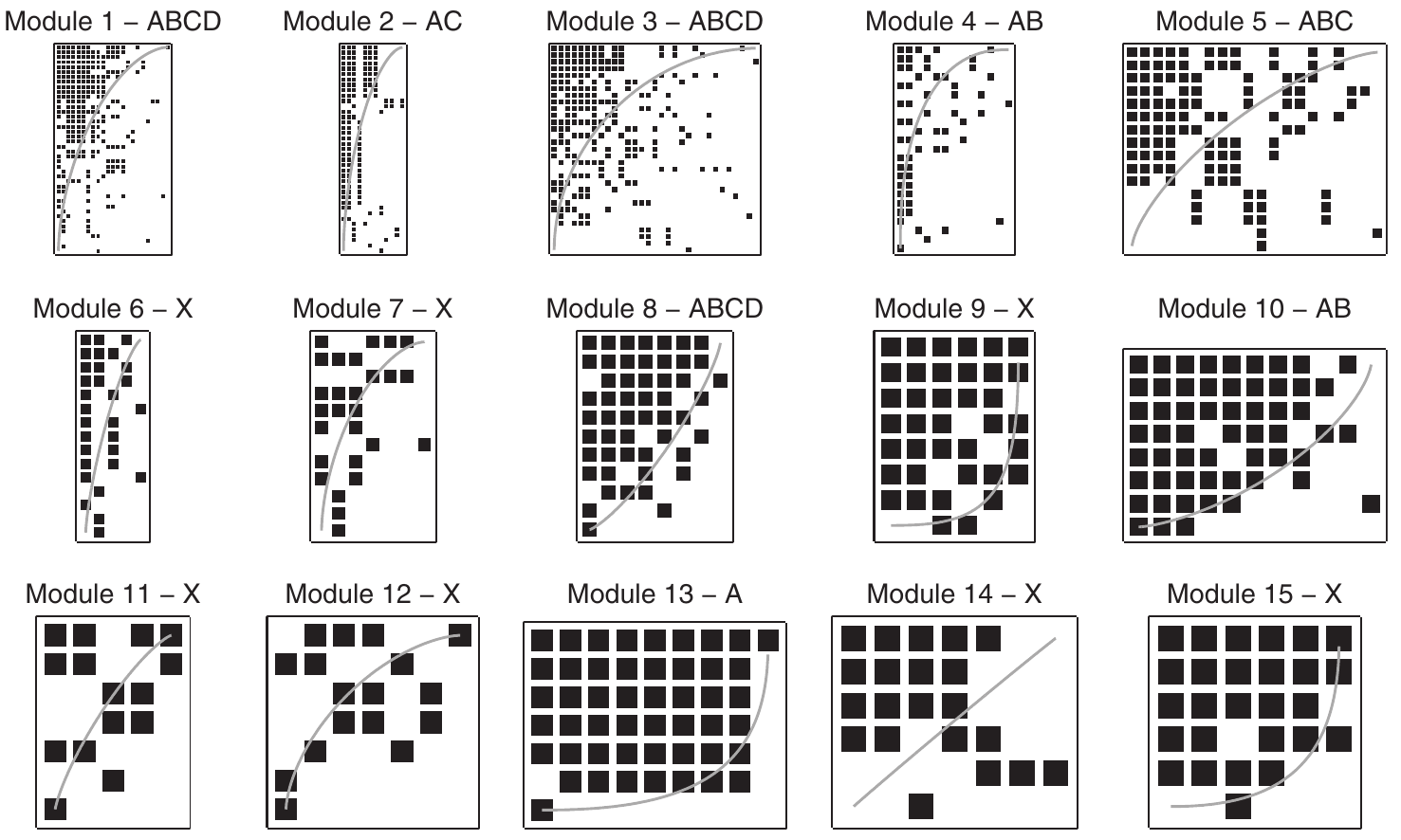}
\caption{Internal structure of the modules detected through modularity optimization in the phage-bacteria
infection network analyzed by Flores \textit{et al.}~\cite{flores2013multi}.
The gray line represents the isocline of perfect nestedness by the NTC algorithm (see Section~\ref{sec:distance}). Each module is marked with labels that denote the significance of its internal nested structure: A and B mean that the module is nested according to the NTC under the EE and PP null model, respectively (see Section~\ref{sec:nine}); C and D mean that the module is nested according to the NTC under the EE and PP null model, respectively; X means that no significance was found.
Reprinted from~\cite{flores2013multi}.}
\label{fig:nest_modules}
\end{figure}

A simple way to address both points would be to first identify the network's compartments, or communities, through a modularity-optimization algorithm, and then measure the level of nestedness of the detected communities (Section \ref{sec:nest_modules}). However, this approach can fail as modularity-optimization algorithms may be unable to detect blocks that exhibit a significantly nested structure (Section \ref{sec:nest_modules}). To overcome this limitation, we need methods that specifically aim to detect in-block nested structures; such methods are the main topics of Sections \ref{sec:ibn} and \ref{sec:multiple_cp}.

\subsubsection{Nestedness of network modules}
\label{sec:nest_modules}

A tempting procedure to evaluate the presence of communities with internally nested structure is to first identify the communities via some well-established community-detection algorithm~\cite{fortunato2010community,fortunato2016community}, and then measure the level of nestedness within each detected community.
Flores \textit{et al.}~\cite{flores2013multi} evaluated a large phage-bacteria interaction network composed of 215 phage types with 286 host types sampled from geographically separated sites in the Atlantic Ocean. They found that this interaction network is highly modular, and evaluated the level of (internal) nestedness of the individual modules.
They found that some of the modules are significantly nested, whereas others are not (Fig.~\ref{fig:nest_modules}).
They observed~\cite{flores2013multi} that even in situations where nestedness is unlikely at the global scale, one may still observe nestedness ``at the local scale'' (i.e., within modules), and they pointed out the need to develop metrics that can
disentangle nestedness and modularity in networks.
Beckett and Williams~\cite{beckett2013coevolutionary} referred to the presence of nestedness inside network modules as a ``nested-modular'' structure, and proposed a co-evolutionary model based on genetic matching to explain its emergence.

On the other hand, Solé-Ribalta \textit{et al.}~\cite{sole2018revealing} generate synthetic data with planted in-block nested structures (see Fig.~1 in~\cite{sole2018revealing}), and found that even in structures with no inter-community links, modularity maximization can fail to uncover the ground-truth communities. This happens because, in sparse nested structures, modularity maximization tends to cluster together the high-degree nodes due to the high density of links among them.
This result suggests that while sequentially applying two well-established community and nestedness detection techniques is appealing, such approach is inappropriate when we wish to detect communities that exhibit an internally nested structure, as modularity-optimization algorithms might incapable of detecting the ground-truth communities.
This finding calls for methods that take into account simultaneously the community structure of the network and the internal nested topology of the communities.

\begin{figure}[t]
\centering
\includegraphics[scale=0.35]{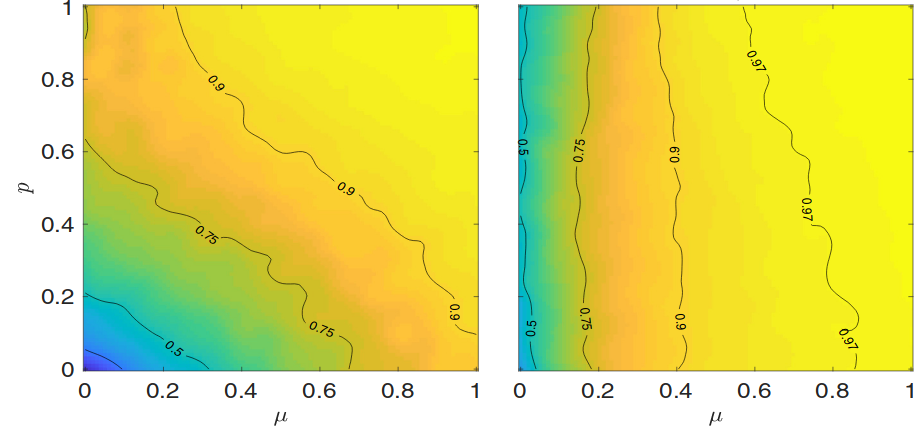}
\caption{An illustration of the ability of in-block nestedness optimization (left panel) and modularity optimization (right panel) to reconstruct planted compartments in networks with tunable fraction of inter-compartment links (parameter $\mu$) and level of nestedness within the compartments (parameter $p$ -- the smaller, the more internally nested the compartments). The reconstruction ability of the two methods is quantified by the Normalized Variation of Information (NVI)~\cite{fortunato2010community} between the partition detected by the method and the ground-truth partition of the nodes. In-block nestedness provides a significantly better reconstruction in the low-$p$ region. Adapted from~\cite{sole2018revealing}. }
\label{fig:ibn_vs_q}
\end{figure}

\begin{figure}[t]
\centering
\includegraphics[scale=0.9]{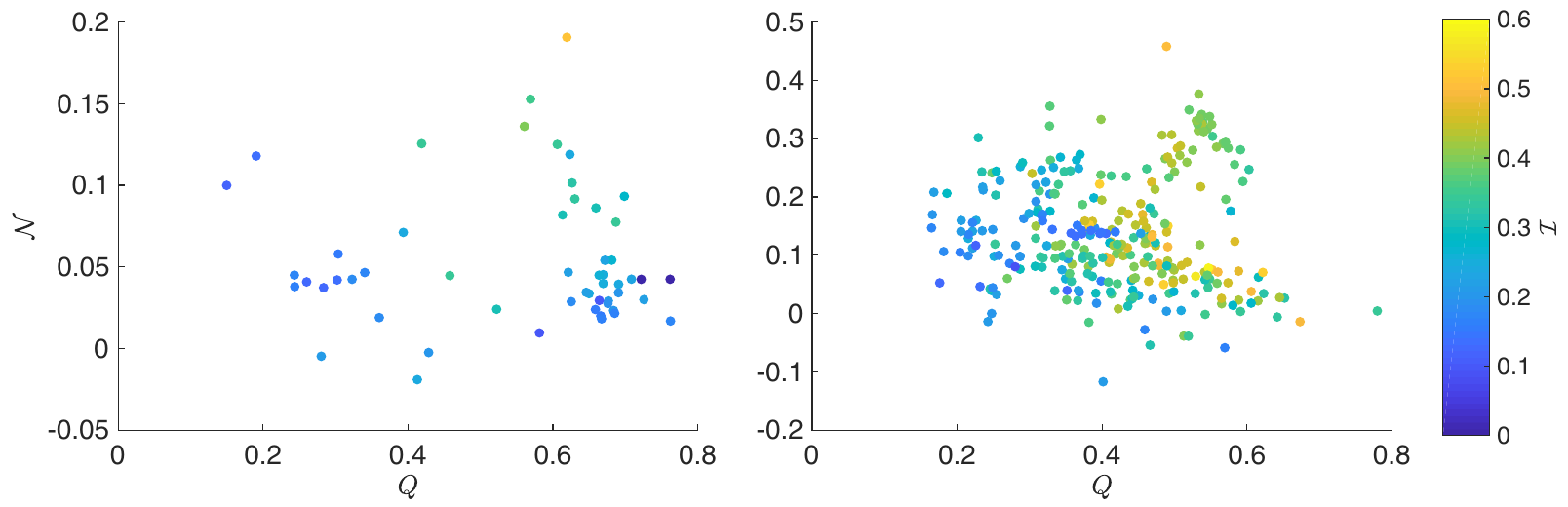}
\caption{The relation between nestedness $\mathcal{N}$, modularity $Q$, and in-block nestedness $\mathcal{I}$ in unipartite and bipartite empirical ecological, social or urban networks. Each dot represents a network, and its color represents the level of in-block nestedness $\mathcal{I}$. There is no evident pattern in the mutual dependence of these three properties. Reprinted from~\cite{sole2018revealing}. }
\label{fig:ibn_real}
\end{figure}

\subsubsection{Detecting in-block nested structures}
\label{sec:ibn}

To overcome the limitation of the procedure described in Section~\ref{sec:nest_modules}, Solé-Ribalta \textit{et al.}~\cite{sole2018revealing} introduced a quality function specifically aimed to detect in-block nested structures which takes into account the fact that the nodes can be partitioned into blocks, and each block exhibits an internally nested structure.
In bipartite networks, the resulting quality function, called \emph{in-block nestedness}~\cite{sole2018revealing}, is defined as
\begin{equation}
\mathcal{I}=\frac{2}{N+M}\Biggl\{\sum_{ij}\frac{O_{ij}-\braket{O_{ij}}}{k_j\,(|\Xi_i|-1)}\,\Theta(k_i-k_j)\,\delta(\Xi_i,\Xi_j) + \sum_{\alpha\beta}\frac{O_{\alpha\beta}-\braket{O_{\alpha\beta}}}{k_\beta\,(|\Xi_\alpha|-1)}\,\Theta(k_\alpha-k_\beta)\,\delta(\Xi_\alpha,\Xi_\beta) \Biggr\}.
\label{ibn1}
\end{equation}
In unipartite networks, the previous expression reduces to~\cite{palazzi2018antagonistic}
\begin{equation}
\mathcal{I}=\frac{2}{N}\sum_{ij}\frac{O_{ij}-\braket{O_{ij}}}{k_j\,(|\Xi_i|-1)}\,\Theta(k_i-k_j)\,\delta(\Xi_i,\Xi_j),
\label{ibn2}
\end{equation}
The $\mathcal{I}$ quality function can be interpreted as a generalization of the normalized NODF function defined by Eq.~\eqref{ibn0}. The contribution to $\mathcal{I}$ from a given pair $(i,j)$ of nodes has indeed the form $\Theta(k_i-k_j)\, O_{ij}/k_j$; the additional elements with respect to the NODF function are (1) the $\delta(\Xi_i,\Xi_j)$ term which restricts the contributing pairs of nodes to those whose nodes belong to the same block;
(2) the normalization factor $|\Xi_i|-1$ (where $|\Xi_i|$ denotes the number of nodes that belong to the block $\Xi_i$ to which node $i$ belongs) which ensures that the in-block nestedness $\mathcal{I}$ of a maximally nested structure tends to one in the thermodynamic limit~\cite{palazzi2018antagonistic}.
The maximization of $\mathcal{I}$ (performed through a biologically inspired optimization algorithm in~\cite{sole2018revealing}) provides us with both the optimal partition $\vek{\Xi}$ of the nodes into blocks, and the (optimal) degree of in-block nestedness $\mathcal{I}$ of the network. A large value of $\mathcal{I}$ indicates that it is possible to partition the network into compartments, each of them with an highly-nested internal structure.

Solé-Ribalta \textit{et al.}~\cite{sole2018revealing} validated the $\mathcal{I}$ function by applying it to synthetic networks with a planted in-block nested structure. The model which generates their benchmark graphs features two parameters: the level $p$ of nestedness within each block ($p=0$ and $p=1$ correspond to a perfectly nested and a random structure, respectively); the fraction $\mu$ of inter-block links (the correct reconstruction of the blocks is increasingly difficult as $\mu$ increases). One would expect a good method for the detection of IBN structures to correctly reconstruct the planted IBN blocks in the low-$p$ and low-$\mu$ parameter region.

They found (Fig. \ref{fig:ibn_vs_q}) that modularity-optimization fails to correctly reconstruct the planted IBN blocks in a large portion of the parameter space. For relatively sparse networks, modularity is confused even by a tiny amount of inter-block links.
By contrast, optimization of the $\mathcal{I}$ quality function leads to an accurate reconstruction of the planted IBN blocks for a broad portion of the low-$p$ and low-$\mu$ region of the parameter space. As $\mu$ and $p$ gradually increase, the performance of the $\mathcal{I}$-optimization algorithm gradually deteriorates, yet in-block nestedness maximization substantially outperforms modularity maximization for networks generated with low values of $p$.

In real data, they found (Fig. \ref{fig:ibn_real}) that nestedness and in-block nestedness are somewhat independent properties. The correlation between nestedness and modularity is indeed small for both unipartite and bipartite networks, and there is no specific range of nestedness values where in-block nestedness tends to be large. 
Besides, the modularity's resolution limit impairs its ability to identify modules that are smaller than a certain scale which depends on network properties~\cite{fortunato2007resolution}. A numerical investigation~\cite{sole2018revealing} suggests that differently from modularity, the in-block nestedness function might not exhibit a similar resolution limit.

\subsubsection{Detecting multiple core-periphery pairs}
\label{sec:multiple_cp}

In the previous Section, we have explored particular network structural patterns where the network can be partitioned into blocks of nodes, and the connectivity patterns within each block exhibit a nested organization. 
We have seen that the maximization of a suitable quality function, called in-block nestedness~\cite{sole2018revealing}, allows us to quantify the significance of this pattern.
Motivated by this observation, we can ask ourselves: can we detect other mesoscopic structural patterns with a similar approach?

In this Section, we consider the detection of multiple core-periphery pairs~\cite{kojaku2017finding}. The topic is relevant to the present review as a core-periphery structure can be considered as a special case of a nested structure (see Section~\ref{sec:core_periphery}).
As we have seen in Section~\ref{sec:core_periphery}, a core-periphery structure is composed of a ``core'' of nodes that are densely connected with all the other nodes, and a ``periphery'' of nodes that tend to be only connected with the nodes in the core. Kojaku and Masuda~\cite{kojaku2018structural} generalized the notion of core-periphery by building a quality function that considers the possibility of multiple core-periphery structures within the same network.

\paragraph*{Quality function for multiple core-periphery structures}
Kojaku and Masuda~\cite{kojaku2017finding} introduced a structural pattern where each node belongs to a given block,
and each block \textit{internally} exhibits a core-periphery structure.
In a similar way to Borgatti and Everett (see Section~\ref{sec:core_periphery}), they defined an ideal \emph{multiple} core-periphery structure $\mat{A}^{MCP}$ where the network is partitioned into disjoint blocks $\Xi\in\{1,\dots,B\}$, and each blocks exhibits an ideal core-periphery structure. Each node $i$ unambiguously belongs to one of the blocks, which is therefore denoted as $\Xi_i$. 
In a similar way to Section~\ref{sec:core_periphery}, for each node $i$, one introduces a binary variable $x_i$ such that $x_i=1$ and $x_i=0$ for core and peripheral nodes, respectively -- the core/peripheral role is now played by each node \textit{within} the block to which the node belongs.

In terms of the $\vek{\Xi}$ and $\vek{x}$ variables,
$A_{ij}^{MCP}=(x_i+x_j-x_i\,x_j)\,\delta_{\Xi_i,\Xi_j}$. The novel element with respect to $A_{ij}^{CP}$ (defined for a single, macroscopic core-periphery structure -- see Section~\ref{sec:core_periphery}) is the Kronecker delta $\delta_{\Xi_i,\Xi_j}$ that only allows nodes of the same block to be connected with each other.
The expected number of links that are present in both the original network $\mat{A}$ and the ideal multiple core-periphery network $\mat{A}^{MCP}$ is simply $\sum_{(i,j)}A_{ij}\,A_{ij}^{MCP}$.
By comparing this number with its expected value $p\,\sum_{(i,j)}A^{MCP}_{ij}$ for an Erdős–Rényi network, we obtain the $Q^{MCP--ER}$ quality function~\cite{kojaku2017finding} 
\begin{equation}
Q^{MCP;ER}(\vek{\Xi},\vek{x})= \sum_{(i,j)}(A_{ij}-p)\,A^{MCP}_{ij}(\vek{\Xi},\vek{x})=\sum_{(i,j)}(A_{ij}-p)\,(x_i+x_j-x_i\,x_j)\,\delta(\Xi_i,\Xi_j).
\end{equation}
By comparing $\sum_{(i,j)}A_{ij}\,A_{ij}^{MCP}$ its expected value $p\,\sum_{(i,j)}A^{MCP}_{ij}$ for a random network with the same degree sequence as the original one (i.e., with the expected value under the configuration model), we obtain the $Q^{MCP;CM}$ quality function~\cite{kojaku2018core} 
\begin{equation}
Q^{MCP;CM}(\vek{\Xi},\vek{x})=\frac{1}{2\,E}\sum_{(i,j)}\Biggl(A_{ij}-\frac{k_i\,k_j}{2\,E}\Biggr)\,(x_i+x_j-x_i\,x_j)\,\delta(\Xi_i,\Xi_j).
\end{equation}
The maximization\footnote{The corresponding code is available at: \url{https://github.com/skojaku/km_config/}.} of $Q^{MCP;ER}$ and $Q^{MCP;CM}$ (performed through a label switching heuristic~\cite{kojaku2017finding,kojaku2018core}) provides us with: (1) the set of optimal $\vek{\Xi}$ values -- i.e., the partition of the nodes into blocks; (2) the set of optimal $\vek{x}$ values -- i.e., the detected within-block cores. A large value of $Q^{MCP;ER}$ or $Q^{MCP;CM}$ indicates a structure that has large overlap with an ideal multiple core-periphery structure.

Interestingly, the $Q^{MCP;CM}$ reduces to the modularity function (defined by Eq.~\eqref{modularity}) when all nodes are core nodes. One can also define the quality of a detected block $\Xi$ as~\cite{kojaku2018core} 
\begin{equation}
q(\Xi)=\frac{1}{2\,E}\sum_{(i,j)}\Biggl(A_{ij}-\frac{k_i\,k_j}{2\,E}\Biggr)\,(x_i+x_j-x_i\,x_j)\,\delta(\Xi_i,\Xi)\,\delta(\Xi_j,\Xi).
\end{equation}
A block $\Xi$ is deemed as statistically significant if its quality $q(\Xi)$ substantially exceeds the expected value for blocks of the same size $|\Xi|$ in networks randomized with the configuration model.
In a similar way to the modularity function~\cite{fortunato2007resolution}, the $Q^{MCP-CM}$ function suffers from a resolution limit that impairs its ability to detect core-periphery pairs below a certain size~\cite{kojaku2018core}. Such a limitation can be overcome by introducing a multi-resolution quality function where the expected overlap between the observed and ideal multiple-core-periphery network is modulated by a resolution parameter $\gamma$ -- see~\cite{kojaku2018multiscale} for details.

\begin{figure}[t]
\centering
\includegraphics[scale=0.8]{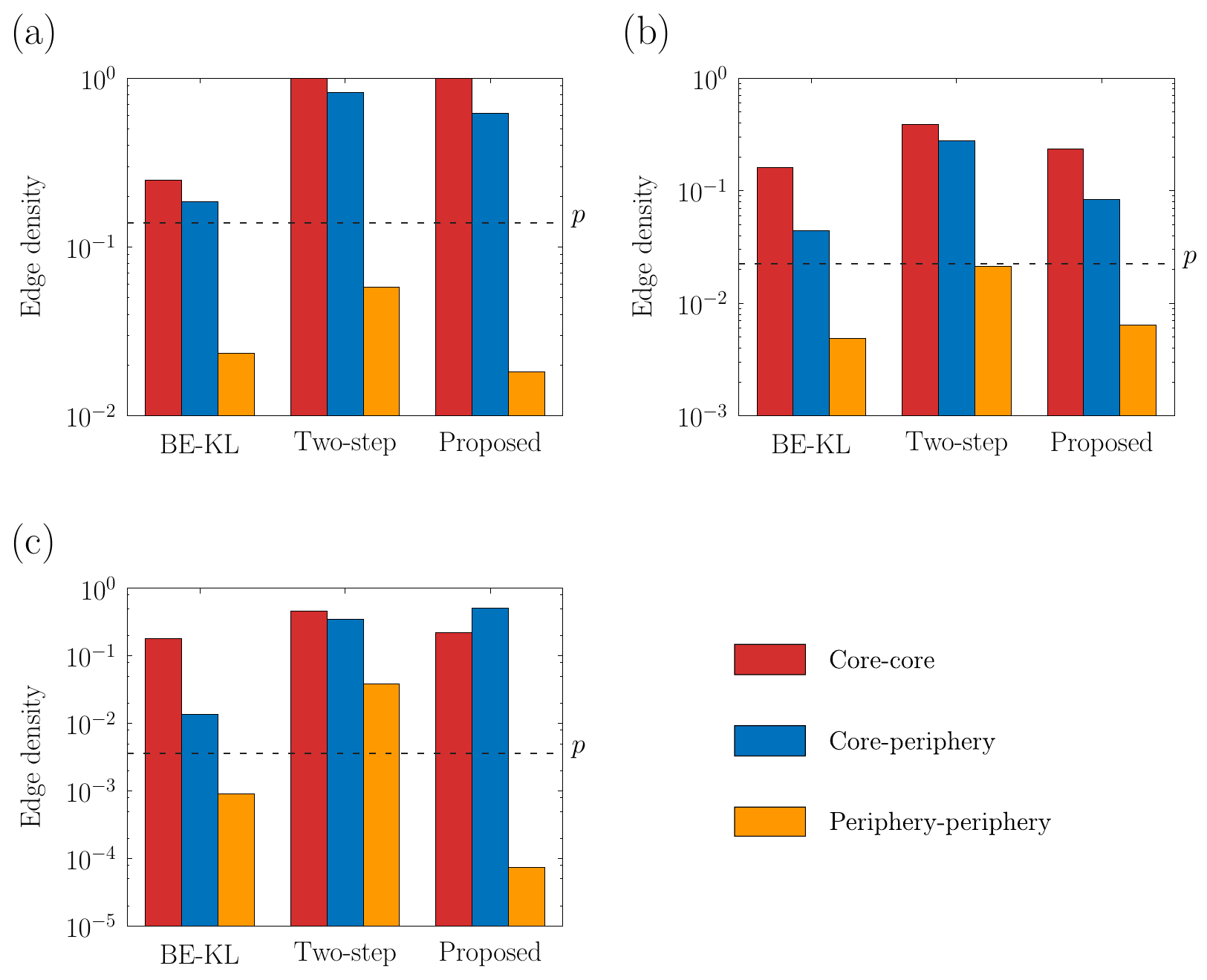}
\caption{A comparison of core-periphery detection methods in three empirical networks: (a) Zachary's karate club; (b) the network of political blogs on the 2004 US presidential elections; (c) the World air transportation network. The methods are compared with respect to the density of links between different classes (core/periphery) of nodes, where the two classes are determined by the three different methods described in the main text: BE-KL, Two-step, and $Q^{MCP}$-maximization method (referred to as ``proposed'' in the figure). Compared to BE-KL and Two-step, the $Q^{MCP}$-maximization produces core-periphery pairs that exhibit substantially lower densities of periphery-periphery edges, in agreement with the notion of core-periphery.
Reprinted from~\cite{kojaku2017finding}. }
\label{fig:mcp2}
\end{figure}

To validate the $Q^{MCP--ER}$ function, Kojaku and Masuda~\cite{kojaku2018core} compared its detected core-periphery structure with that from three alternative methods:
\begin{itemize}
\item The original Borgatti-Everett (BE) core detection algorithm which aims to detect one single core-periphery structure. As they maximized the corresponding quality function $Q^{CP}$ (defined by Eq.~\eqref{qcp}) through the Kernighan-Lin algorithm~\cite{kernighan1970efficient}, they referred to the method as the BE-KL method. 
\item A ``two-step algorithm'' where one first partition the nodes through modularity optimization (performed with the Louvain algorithm~\cite{blondel2008fast}) and, at the same time, applies the BE core detection algorithm to the whole network. The core nodes detected by the BE algorithm are then assumed to be the core nodes within each module.
\item A ``divisive algorithm'' where one first partitions the nodes through modularity optimization (performed with the Louvain algorithm~\cite{blondel2008fast}), and then applies the BE core detection algorithm to each of the detected modules, in a similar spirit to the procedures described in Section~\ref{sec:nest_modules}.
\end{itemize}
Results on three empirical networks (see Fig.~\ref{fig:mcp2}) show that with respect to these methods, the $Q^{MCP}$ maximization is the only one that produces core-periphery blocks such that -- consistently with the main idea behind core-periphery structures -- the density of intra-periphery edges is significantly smaller than the expected density for a random graph. Besides, the $Q^{MCP}$ maximization was found to be the only algorithm that reveals at the same time: (1) Groups of nodes that share the same property, as revealed by comparing the obtained node partitions with metadata; (2) Nodes' role within each group, as revealed by direct inspections of the core and the peripheral nodes within each detected block.
Recently, the $Q^{MCP;ER}$ maximization has been applied to detect structural changes in the interbank market~\cite{kojaku2018structural}.

Results on synthetic networks~\cite{kojaku2018core} indicate that the $Q^{MCP;CM}$-maximization algorithm outperforms both the divisive algorithm and the $Q^{MCP;ER}$-maximization algorithm in reconstructing planted core-periphery pairs. In particular, differently from the divisive and the $Q^{MCP;ER}$-maximization algorithm, the $Q^{MCP;CM}$-maximization algorithm is able to reconstruct ground-truth blocks even when a substantial fraction of core nodes are not hubs and a substantial fraction of peripheral nodes are hubs~\cite{kojaku2018core}.

\clearpage

\section{Outlook}
\label{sec:outlook}

This review has focused on a network structural property that has been introduced in ecological studies of species-site spatial networks~\cite{patterson1986nested} and has subsequently attracted great interest from diverse research fields: nestedness. 
It has examined existing studies on nestedness by following three steps: (1) introducing the main methodologies to detect nestedness (\textit{observation}), (2) presenting the main mechanisms proposed to explain the pervasiveness of nestedness in ecological and economic systems (\textit{emergence}), and (3) assessing the main implications of nestedness for network topological robustness, feasibility, and stability (\textit{implications}). In principle, this three-step framework provides us with a conceptual roadmap that can be equally adopted to survey existing studies on other structural patterns of real-world networks (e.g., modularity~\cite{fortunato2016community} and rich-club structure~\cite{van2011rich}).

Stimulated by our review, a natural question emerges: which are the next steps for research on nestedness and, more in general, on the architecture of ecological and socio-economic systems? As we have often recognized throughout this review, some questions related to the structure of ecological and socio-economic networks have not yet found a universal answer. Besides methodological questions (e.g., how to best measure the level of nestedness of a given network, and its statistical significance?), we argue that questions related to the emergence and implications of nestedness and other structural patterns of ecological and socio-economic systems need additional efforts in order to be answered.

As for the emergence of nestedness, we already explored (Section~\ref{sec:emergence}) various mechanisms that can, in theory, drive a network from a random to a nested topology of interaction, or growth mechanisms that can assemble nested networks. However, studies that adopt statistical techniques to validate such mechanisms on time-stamped data are non-existent.
As inferring the dynamical generative process from the structure of the final network is not possible, we point out a thorough temporal analysis of the dynamics of ecological and socio-economic networks as the only way to single out the mechanisms that are responsible for the empirically observed nestedness, and rule out those that do not take place. Toward this direction, experimental approaches~\cite{gilarranz2017effects} might play an increasingly central role.

It is essential to notice that our review has focused on the nestedness of unipartite and bipartite networks. The extension of nestedness to temporal and multilayer networks is still largely unexplored in the literature, in contrast with other structural properties that have been already generalized (e.g., centrality metrics~\cite{de2015ranking,liao2017ranking} and modularity~\cite{mucha2010community}).
While preliminary works in this direction can be already found~\cite{alves2019nested,sole2018disentangling}, we are still far from having a complete understanding of the causes and implications of nestedness in multilayer networks.
This is an important research gap because the fundamental role by different types of interactions is increasingly recognized in ecology~\cite{fontaine2011ecological,suweis2014disentangling,pilosof2017multilayer} and trade analysis~\cite{barigozzi2010multinetwork,alves2019nested}. We predict that several future studies will attempt to understand the role of nestedness (and related structural patterns) in multilayer and temporal networks.

Finally, while nestedness has traditionally been considered as a macroscopic network property that involves all the nodes of a given network (Section~\ref{sec:metrics}), a number of recent techniques have been developed to detect nestedness at a mesoscopic scale instead of the global one (Section~\ref{sec:mesoscopic}).
Nevertheless, our knowledge of the performance of these detection methods, and our understanding of the causes and implications of such mesoscopic structures are still at their infancy. This class of methods might play an increasingly central role in nestedness analysis as real systems are subject to various types of constraints~\cite{lewinsohn2006structure} which makes it natural to look for patterns of interaction organization (like nestedness) at a ``local'' scale and not at the global one.

We conclude by stressing that our review is far from being exhaustive. Besides introducing methodologies for the detection of nestedness, it mostly aimed to introduce the essential ``physical'' aspects of nestedness: stochastic mechanisms that lead to its emergence, and its impact on dynamical processes on the network.
We hope that this review will be a useful compendium not only for physicists, but also for ecologists and economists.
In particular, ecologists (economists) can gain inspiration from mechanisms and techniques introduced in economics (ecology) and physics, which might lead to unexpected applications of nestedness and network analysis which transcend traditional disciplinary boundaries.

\clearpage

\appendix
\renewcommand*{\thesection}{\Alph{section}}

\section{Software and relevant datasets}
\label{appendix:material}

We stress that the URLs provided below were active at the time when this article was published. There is no guarantee that these URLs will be still active at the time when this article will be read.

\subsection{Software for nestedness and network analysis}

Standard, general packages for network analysis are: \url{NetworkX}, \url{igraph}, \url{graph_tool}, among many others.
The R package \url{bipartite} was specifically designed for the analysis of bipartite networks, and it implements various nestedness metrics. Besides, the R package \url{vegan} was designed for the analysis of ecological communities, and it also incorporates various nestedness metrics.

\subsubsection{Metrics and null models}

Besides general packages for network analysis,
due to the massive popularity of nestedness analysis in ecology, scholars have developed specific packages to compute metrics for nestedness together with their statistical significance. Among them, we mention FALCON [\url{https://github.com/sjbeckett/FALCON}] and Nestedness for Dummies [\url{http://ecosoft.alwaysdata.net/download/}].
Both packages implement various nestedness metrics, null models, and significance tests. We refer the interested reader to~\cite{beckett2014falcon,strona2014nestedness} for more information.

\subsubsection{Nested network generation}

When we wish to generate perfectly nested or highly-nested networks, we have two options: (1) decide the shape of the separatrix in the adjacency/incidence matrix (e.g., Fig.~\ref{fig:nested-nets}), and add links above the chosen line; (2) adopt one of the generative mechanisms described in Section~\ref{sec:emergence}. As for (2), a function to generate threshold graphs (i.e., based on the discussion in Section~\ref{sec:nested_graph}, perfectly nested networks) based on the mechanisms described in Section~\ref{sec:threshold} can be found in the R package~\url{netrankr} [\url{threshold_graph} function].

\subsubsection{Community detection and nestedness at the mesoscopic scale}

The main network-analysis packages implement several algorithms for the detection of communities.
Among the modularity maximization algorithms, the Combo algorithm turned out to be more effective than other existing methods~\cite{sobolevsky2014general}; the respective code can be found at \url{http://senseable.mit.edu/community_detection/}.
For a recent and exhaustive list of available software for community detection, we refer the reader to Section 5 of the review article~\cite{fortunato2016community}.
Importantly to nestedness analysis, recent works have started investigating nestedness and core-periphery as a mesoscopic property of networks, as extensively discussed in Section~\ref{sec:mesoscopic}. The code that implements the maximization of the multiple core-periphery function by Kojaku and Masuda~\cite{kojaku2018core} (see Section~\ref{sec:multiple_cp}) can be found at \url{https://github.com/skojaku/km_config/}.

\subsection{Datasets of socio-economic and ecological networks}

\subsubsection{World Trade data}

World Trade datasets can be downloaded from several sources. 
The COMTRADE dataset [available at \url{https://comtrade.un.org/}] is curated by the United Nations
Statistics Division, and it includes bilateral trade flows declarations for several decades~\cite{tacchella2018dynamical}. World Trade datasets from various sources are also available at the Observatory of Economic Complexity collection [\url{https://atlas.media.mit.edu/en/resources/data/}] and the Economics Web Institute webpage [\url{http://www.economicswebinstitute.org/worldtrade.htm}]. It is essential to note that such datasets are usually noisy. For example, in the COMTRADE data, the declared yearly import volume for a given exchange often does not match the declared export volume, which makes it necessary to develop data sanitation procedure. Discussing the topic in detail goes beyond the scope of this review, but we refer the interested reader to~\cite{tacchella2018dynamical,angelini2018complexity} for relevant discussions.



\subsubsection{Ecological spatial networks}

The original Nestedness Temperature Calculator (NTC) software by Atmar and Patterson~\cite{atmar1993measure} can be downloaded at the link [\url{http://priede.bf.lu.lv/ftp/pub/TIS/datu_analiize/Nestedness/about.html}]. The program includes 294 presence-absence matrices taken from the ecology literature.
These matrices (together with the original references) can be readily obtained at the link [\url{http://wikieducator.org/Null_Model_Data}]. Spatial pattern datasets can be also found at the following online repositories: the  Geographic Information System repository [\url{https://freegisdata.rtwilson.com/}], the National Center for Ecological Analysis and Synthesis repository [\url{https://www.nceas.ucsb.edu/scicomp/data}].

\subsubsection{Ecological interaction networks}

\textbf{Mutualistic} network datasets can be found at the following online repositories: Web of Life [\url{http://www.web-of-life.es/}], Interaction Web DataBase [\url{https://www.nceas.ucsb.edu/interactionweb/resources.html}].
\textbf{Host-parasite} datasets can be found at the following online repositories: Interaction Web DataBase [\url{https://www.nceas.ucsb.edu/interactionweb/resources.html}].
\textbf{Food-web} network datasets can be found at the following online repositories: Web of Life [\url{http://www.web-of-life.es/}], Interaction Web DataBase [\url{https://www.nceas.ucsb.edu/interactionweb/resources.html}].

\section{Network-based metrics of node importance in bipartite networks}
\label{appendix:ranking}

In the following, we review some of the metrics that have been proposed in the literature to quantify node importance -- in particular, those that are relevant to Fig.~\ref{fig:dominguez2}.
We refer the interested reader to \cite{lu2016vital,liao2017ranking} for extensive reviews on centrality metrics and network-based ranking algorithms. Based on the methods and results in~\cite{allesina2009googling,dominguez2015ranking,mariani2015measuring}, in this review, centrality metrics for bipartite networks have been compared with respect to their ability to rank nodes by their structural importance (Section~\ref{sec:structural_nodes}).
We emphasize that centrality metrics are routinely used for different purposes (like the identification of influential nodes for diffusion processes~\cite{lu2016vital}, the identification of expert-selected significant nodes~\cite{liao2017ranking}, the prediction of the nodes' future popularity~\cite{ren2018structure}), and the results presented in Section~\ref{sec:structural_nodes} only refer to one particular application.

\paragraph{Degree}

The degree $k_i$ and $k_\alpha$ of row-node $i$ and column-node $\alpha$ can be considered as the simplest possible centrality metrics in bipartite networks: the two classes of nodes are ranked in order of decreasing degree, under the assumption that a node is central if it has many connections~\cite{newman2010networks}.

\paragraph{Closeness} The closeness centrality score $c_i$ of a row-node $i$ is defined as~\cite{borgatti2011analyzing}
\begin{equation}
    c_i=\frac{M+2\,(N-1)}{d_i},
\end{equation}
where $d_i=\sum_{j}D_{ij}+\sum_{\alpha}D_{i\alpha}$ is the average distance of $i$ from all the other nodes, and $D_{ij}$ denotes the shortest-path distance~\cite{newman2010networks} between $i$ and $j$.
Analogously, for a column-node $\alpha$, the closeness centrality $c_\alpha$ is defined as~\cite{borgatti2011analyzing}
\begin{equation}
    c_\alpha=\frac{M+2\,(M-1)}{d_\alpha},
\end{equation}
where $d_\alpha=\sum_{i}D_{i\alpha}+\sum_{\beta}D_{\beta\alpha}$.
We refer to~\cite{borgatti2011analyzing} for a detailed motivation of the normalization factor, and to the package \url{NetworkX} for the implementation [\url{https://networkx.github.io/documentation/networkx-1.10/reference/generated/networkx.algorithms.bipartite.centrality.closeness_centrality.html}]. 

\paragraph{Betweenness centrality} 
The betweenness centrality score of a row-node $i$ (column-node $\alpha$) is defined as the sum over all pairs of nodes of the share of shortest paths that pass through node $i$ (node $\alpha$). 
We refer to~\cite{borgatti2011analyzing} for a discussion about how to properly normalize the betweenness score in bipartite networks, and to the package \url{NetworkX} for the implementation [\url{https://networkx.github.io/documentation/stable/reference/algorithms/generated/networkx.algorithms.bipartite.centrality.betweenness_centrality.html}]. 

\paragraph{Eigenvector centrality} The vector of eigenvector centrality scores is given by the leading eigenvector of the network's adjacency matrix $\mathsf{A}$~\cite{bonacich1972factoring}. This definition implies that in a bipartite network, the score of a row-node (column-node) is proportional to the scores of its neighbors' column-nodes (row-nodes)~\cite{borgatti2011analyzing}.

\paragraph{Google's PageRank}
Google's PageRank~\cite{brin1998anatomy,allesina2009googling,liao2017ranking} is defined as the leading eigenvector of the $S\times S$ matrix
\begin{equation}
    \mathsf{G}=c\, \mathsf{P}+\frac{1-c}{S}
\end{equation}
where $P_{ij}=A_{ij}/k_i$. The parameter $\alpha$ is referred to as teleportation parameter or damping factor~\cite{liao2017ranking}; the results shown in Fig.~\ref{fig:dominguez2} were obtained by Dom{\'\i}nguez-Garc{\'\i}a and Mu{\~n}oz \cite{dominguez2015ranking} with $c=0.999$.

\paragraph{Contribution to nestedness}
The individual nodes' contribution to nestedness can be computed in various ways~\cite{saavedra2011strong,johnson2013factors}. The nestedness contributions used by Saavedra et al~\cite{saavedra2011strong} were described in Section~\ref{sec:survival}. Alternatively, one can take the perspective of overlap-based nestedness metrics, and use them to estimate the individual nodes' contribution. For instance, based on the JDM-NODF metric (Eq.~\eqref{eta}), one can assess the contribution of node $i$ to the overall nestedness as specified by Eq.~\eqref{individual}; this is the individual contribution to nestedness that was used by Dom{\'\i}nguez-Garc{\'\i}a and Mu{\~n}oz \cite{dominguez2015ranking} to obtain the results reported in Fig.~\ref{fig:dominguez2}.

\paragraph{Fitness-complexity algorithm and its reversed variant}

The algorithm and some of its variants are detailedly described in Section~\ref{sec:nonlinear}. 
Here, we only stress that the original algorithm ranks the row-nodes and column-nodes in order of decreasing fitness and complexity score, respectively.  One can also exchange the roles of row-nodes and column-nodes. In a country-product network, this corresponds to assign a fitness score to products and a complexity score to countries; countries and products are, therefore, ranked in order of increasing complexity and increasing fitness, respectively. We refer to the corresponding algorithm as reversed fitness-complexity.

Both algorithms (fitness-complexity algorithm and reversed fitness-complexity) are considered in Fig.~\ref{fig:dominguez2}. They are labeled as ``MUS'' and ``MUSrev'' in Fig.~\ref{fig:dominguez2}, as the fitness-complexity algorithm was relabeled as ``MUtualistic Species RANKing'' algorithm (MusRank) in~\cite{dominguez2015ranking}. As the fitness-complexity and MusRank algorithm are mathematically equivalent, to prevent ambiguity, we always referred to the algorithm as the fitness-complexity algorithm throughout this review.

\clearpage

\section*{Acknowledgements}

We wish to thank Javier Borge-Holthoefer, Mat{\'u}{\v{s}} Medo, Albert Solé-Ribalta for the many inspiring discussions on nestedness and the detection of structural patterns in networks.
We are grateful to Orazio Angelini, Tomaso Aste, Clàudia Payrató Borràs, Luciano Pietronero, Andrea Zaccaria for their feedback on the first draft of this article and their invaluable suggestions which helped us to improve several sections of the article.
This work has been supported by the Science Strength Promotion Program of UESTC and by the URPP Social Networks. MSM and CJT acknowledge financial support from the Swiss National Science Foundation (Grant No. 200021-182659). MSM acknowledges financial support from the UESTC professor research start-up (Grant No. ZYGX2018KYQD21). ZMR was partially supported by National Natural Science Foundation of China (Grant No. 61803137), Hangzhou Normal University Research Funding Project, Qiantang River Talents Plan (Grant No. QJD1803005),and Foundation of High level overseas returnees (team) in Hangzhou for Pioneering Innovation Program. JB acknowledges financial support from the Swiss National Science Foundation (Grant No. 31003A-169671).

\clearpage 

\bibliographystyle{elsarticle-num}

\end{document}